\documentclass[11pt,a4paper]{article}
\usepackage{jinstpub_mod}

\pdfoutput=1 

\usepackage[modulo]{lineno}
\usepackage{siunitx}
\DeclareSIUnit\clight{\text{\ensuremath{c}}}
\DeclareSIUnit[number-unit-product = ]\percent{\char`\%}

\usepackage{pdfpages}
\usepackage{subfigure}
\usepackage{float}
\usepackage{color}
\usepackage{ulem}
\usepackage{footnote}
\usepackage{textcomp}
\usepackage{soul}

\usepackage{xspace} 
\usepackage{booktabs}
\usepackage{multirow}
\usepackage{hyperref}
\usepackage{wasysym}

\newcommand{\murm}{%
  \ifmmode
    \mathchoice
        {\hbox{\normalsifze\textmu}}
        {\hbox{\normalsize\textmu}}
        {\hbox{\scriptsize\textmu}}
        {\hbox{\tiny\textmu}}%
  \else
    \textmu
  \fi
}

\newcommand{\pp}{\ensuremath{\mathrm {p\kern-0.05em p}}\xspace}
\newcommand{\PbPb}{\ensuremath{\mbox{Pb--Pb}}\xspace}
\newcommand{\pPb}{\ensuremath{\mbox{p--Pb}}\xspace}
\newcommand{\XeXe}{\ensuremath{\mbox{Xe--Xe}}\xspace}

\newcommand{\figref}[1]{Fig.~\ref{#1}}
\newcommand{\Figref}[1]{Figure~\ref{#1}}	
\newcommand{\secref}[1]{Section~\ref{#1}}

\newcommand{\tabref}[1]{Table~\ref{#1}}

\newcommand{\NeCOtwoNtwo}{Ne-CO$_2$-N$_2$ (90-10-5)\xspace}
\newcommand{\NeCOtwo}{Ne-CO$_2$ (90-10)\xspace}

\newcommand{\NeCFfourTwenty}{Ne-CF$_4$ (80-20)\xspace}
\newcommand{\ArCOtwo}{Ar-CO$_2$ (90-10)\xspace}
\newcommand{\ArCOtwoThirty}{Ar-CO$_2$ (70-30)\xspace}

\newcommand{\dEdx}{\ensuremath{\mathrm{d}E/\mathrm{d}x}\xspace}
\newcommand{\MeandEdx}{\ensuremath{\langle\dEdx\rangle}\xspace}\newcommand{\ibf}{ion backflow\xspace}
\newcommand{\Ibf}{Ion backflow\xspace}

\newcommand{\adc}{\ensuremath{\mathrm{ADC~count}}\xspace}
\newcommand{\adcs}{\ensuremath{\mathrm{ADC~counts}}\xspace}

\newcommand{\iroc}{IROC\xspace}
\newcommand{\oroc}{OROC\xspace}
\newcommand{\irocs}{IROCs\xspace}
\newcommand{\orocs}{OROCs\xspace}
\newcommand{\orocOne}{OROC\,1\xspace}
\newcommand{\orocTwo}{OROC\,2\xspace}
\newcommand{\orocThree}{OROC\,3\xspace}

\newcommand{\gemOne}{GEM\,1\xspace}
\newcommand{\gemTwo}{GEM\,2\xspace}
\newcommand{\gemThree}{GEM\,3\xspace}
\newcommand{\gemFour}{GEM\,4\xspace}

\newcommand{\xray}{x-ray\xspace}
\newcommand{\xrays}{x-rays\xspace}
\newcommand{\Xray}{X-ray\xspace}
\newcommand{\Xrays}{X-rays\xspace}
\newcommand{\theader}[3]{\multicolumn{#1}{#2}{#3}}

\newcommand{\kr}{\ensuremath{^{83}\mathrm{Kr}}\xspace}

\newcommand{\kalpha}{\ensuremath{\mathrm{K}_\alpha}\xspace}
\newcommand{\kbeta}{\ensuremath{\mathrm{K}_\beta}\xspace}

\newcommand{\vdrift}{\ensuremath{v_{\mathrm{d}}}\xspace}

\newcommand{\Eind}{\ensuremath{E_\mathrm{ind}}\xspace}

\title{The upgrade of the ALICE TPC with GEMs and continuous readout}
\abstract{The upgrade of the ALICE TPC will allow the experiment to cope with the high interaction rates foreseen for the forthcoming Run\,3 and Run\,4 at the CERN LHC. In this article, we describe the design of new readout chambers and front-end electronics, which are driven by the goals of the experiment. Gas Electron Multiplier (GEM) detectors arranged in stacks containing four GEMs each, and continuous readout electronics based on the SAMPA chip, an ALICE development, are replacing the previous elements. The construction of these new elements, together with their associated quality control procedures, is explained in detail. Finally, the readout chamber and front-end electronics cards replacement, together with the commissioning of the detector prior to installation in the experimental cavern, are presented. After a nine-year period of R\&D, construction, and assembly, the upgrade of the TPC was completed in 2020.
}

\collaboration{ALICE TPC Collaboration}
\affiliation[1]{Bose Institute, Department of Physics  and Centre for Astroparticle Physics and Space Science (CAPSS), Kolkata, India}
\affiliation[2]{Comenius University Bratislava, Faculty of Mathematics, Physics and Informatics, Bratislava, Slovakia}
\affiliation[3]{Department of Physics, University of Oslo, Oslo, Norway}
\affiliation[4]{Department of Physics and Technology, University of Bergen, Bergen, Norway}
\affiliation[5]{European Organization for Nuclear Research (CERN), Geneva, Switzerland}
\affiliation[6]{Faculty of Engineering and Science, Western Norway University of Applied Sciences, Bergen, Norway}
\affiliation[7]{Helmholtz-Institut f\"{u}r Strahlen- und Kernphysik, Rheinische Friedrich-Wilhelms-Universit\"{a}t Bonn, Bonn, Germany}
\affiliation[8]{Helsinki Institute of Physics (HIP), Helsinki, Finland}
\affiliation[9]{High Energy Physics Group, Universidad Aut\'{o}noma de Puebla, Puebla, Mexico}
\affiliation[10]{Horia Hulubei National Institute of Physics and Nuclear Engineering, Bucharest, Romania}
\affiliation[11]{Indian Institute of Technology Bombay (IIT), Mumbai, India}
\affiliation[12]{Institut f\"{u}r Kernphysik, Johann Wolfgang Goethe-Universit\"{a}t Frankfurt, Frankfurt, Germany}
\affiliation[13]{Instituto de Ciencias Nucleares, Universidad Nacional Aut\'{o}noma de M\'{e}xico, Mexico City, Mexico}
\affiliation[14]{Joint Institute for Nuclear Research (JINR), Dubna, Russia}
\affiliation[15]{Lund University Department of Physics, Division of Particle Physics, Lund, Sweden}
\affiliation[16]{Nagasaki Institute of Applied Science, Nagasaki, Japan}
\affiliation[17]{Niels Bohr Institute, University of Copenhagen, Copenhagen, Denmark}
\affiliation[18]{NRNU Moscow Engineering Physics Institute, Moscow, Russia}
\affiliation[19]{Oak Ridge National Laboratory, Oak Ridge, Tennessee, USA}
\affiliation[20]{Physics department, Faculty of science, University of Zagreb, Zagreb, Croatia}
\affiliation[21]{Physikalisches Institut, Ruprecht-Karls-Universit\"{a}t Heidelberg, Heidelberg, Germany}
\affiliation[22]{Physik Department, Technische Universit\"{a}t M\"{u}nchen, Munich, Germany}
\affiliation[23]{PINSTECH, Islamabad, Pakistan}
\affiliation[24]{Research Division and ExtreMe Matter Institute EMMI, GSI Helmholtzzentrum f\"ur Schwerionenforschung GmbH, Darmstadt, Germany}
\affiliation[25]{The Henryk Niewodniczanski Institute of Nuclear Physics, Polish Academy of Sciences, Cracow, Poland}
\affiliation[26]{The University of Texas at Austin, Austin, Texas, USA}
\affiliation[27]{Universidade de S\~{a}o Paulo (USP), S\~{a}o Paulo, Brazil}
\affiliation[28]{University of Houston, Houston, Texas, USA}
\affiliation[29]{University of Jyv\"{a}skyl\"{a}, Jyv\"{a}skyl\"{a}, Finland}
\affiliation[30]{University of Tennessee, Knoxville, Tennessee, USA}
\affiliation[31]{University of Tokyo, Tokyo, Japan}
\affiliation[32]{Variable Energy Cyclotron Centre, Kolkata, India}
\affiliation[33]{Wayne State University, Detroit, Michigan, USA}
\affiliation[34]{Wigner Research Centre for Physics, Budapest, Hungary}
\affiliation[35]{Yale University, New Haven, Connecticut, USA}

\author[15]{J.~Adolfsson,}
\author[12]{M.~Ahmed,}
\author[35]{S.~Aiola,}
\author[4]{J.~Alme,}
\author[12]{T.~Alt,}
\author[12]{W.~Amend,}   
\author[21]{F.~Anastasopoulos,}%
\author[10]{C.~Andrei,}
\author[15]{M.~Angelsmark,}
\author[21]{V.~Anguelov,}
\author[21]{A.~Anjam,}
\author[12,*]{H.~Appelsh\"{a}user,%
\note[*]{Corresponding authors}}\emailAdd{harald.appelshaeuser@ikf.uni-frankfurt.de}
\author[10]{V.~Aprodu,}
\author[22]{O.~Arnold,}
\author[21]{M.~Arslandok,}

\author[21,24]{D.~Baitinger,}
\author[7]{M.~Ball,}
\author[34]{G.G.~Barnaf\"{o}ldi,}
\author[12]{E.~Bartsch,}
\author[21,24]{P.~Becht,}
\author[28]{R.~Bellwied,}
\author[18,21]{A.~Berdnikova,}
\author[22]{M.~Berger,}
\author[12]{N.~Bialas,}
\author[12]{P.~Bialas,}
\author[1]{S.~Biswas,}
\author[21,24]{B.~Blidaru,}
\author[34]{L.~Boldizs\'{a}r,}
\author[12]{L.~Bratrud,}
\author[24]{P.~Braun-Munzinger,}
\author[27]{M.~Bregant,}
\author[19]{C.L.~Britton,}
\author[21]{S.~Brucker,}
\author[8]{E.J.~Br\"{u}cken,}
\author[12]{H.~B\"{u}sching,}

\author[9]{R.~Soto Camacho,}
\author[27]{A.L.~Campos,}
\author[10]{G.~Caragheorgheopol,}
\author[27]{D.D.~Carvalho,}
\author[30]{A.J.~Castro,}
\author[21]{P.~Chatzidaki,}
\author[15]{P.~Christiansen,}
\author[19]{L.G.~Clonts,}
\author[19]{T.M.~Cormier,}
\author[27]{A.L.D.~Couto,}
\author[27]{H.G.A.~Cubas,}

\author[21,24]{A.~Deisting,}
\author[11]{P.~Dhankher,}
\author[12]{S.~Dittrich,}
\author[10]{V.~Duta,}

\author[35]{R.~Ehlers,}
\author[7]{M.~Engel,}
\author[19]{M.N.~Ericson,}
\author[19]{N.B.~Ezell,}

\author[22]{L.~Fabbietti,}
\author[28]{F.~Flor,}
\author[21]{G.~F\"{o}hner,}
\author[24]{U.~Frankenfeld,}
\author[34]{E.~Futo,}

\author[17]{J.J.~Gaardh{\o}je,}
\author[27]{M.G.~Munhoz,}
\author[24,*]{C.~Garabatos,}\emailAdd{c.garabatos@gsi.de}
\author[22,24,*]{P.~Gasik,}\emailAdd{p.gasik@gsi.de}
\author[12]{T.~Geiger,}
\author[34]{\'{A}.~Gera,}
\author[21]{P.~Gl\"{a}ssel,}
\author[16]{D.J.Q.~Goh,}
\author[33]{O.~Grachov,}
\author[12]{A.~Grein,}
\author[23]{M.~Gul,}
\author[31]{T.~Gunji,}

\author[24]{M.~Habib,}
\author[16]{H.~Hamagaki,}
\author[34]{G.~Hamar,}
\author[17]{J.C.~Hansen,}
\author[21,24]{A.~Harlenderova,}
\author[35]{J.W.~Harris,}
\author[23]{S.~Hassan,}
\author[7]{P.~Hauer,}
\author[31]{S.~Hayashi,}
\author[22]{S.T.~Heckel,}
\author[24]{J.~Hehner,}
\author[8]{J.K.~Heino,}
\author[12]{E.~Hellb\"{a}r,}
\author[6]{H.~Helstrup,}
\author[10]{A.~Herghelegiu,}
\author[27]{L.~Hernandes da Costa Porto,}
\author[27]{R.A.~Hernandez,}
\author[27]{H.D.~Hernandez Herrera,}
\author[21]{T.~Herold,}
\author[8]{T.E.~Hilden,}
\author[22]{B.~Hohlweger,}
\author[21,24]{S.~Hornung,}
\author[30]{C.~Hughes,}
\author[21]{S.~Hummel,}

\author[24]{M.~Ivanov,}

\author[12]{J.~Jung,}
\author[12]{M.~Jung,}
\author[12]{D.~Just,}

\author[8]{E.~Kangasaho,}
\author[24]{L.~Karayan,}
\author[7]{B.~Ketzer,}
\author[12]{S.~Kirsch,}
\author[12]{M.~Kleiner,}
\author[22]{T.~Klemenz,}
\author[21]{S.~Klewin,}
\author[28]{A.G.~Knospe,}
\author[8]{E.~Koskinen,}
\author[25]{M.~Kowalski,}
\author[24]{L.~Kreis,}
\author[12]{M.~Kr\"{u}ger,}
\author[21]{N.~Kupfer,}

\author[22]{R.~Lang,}
\author[22]{L.~Lautner,}
\author[22]{M.~Lesch,}
\author[5]{Y.~Lesenechal,}
\author[12]{F.~Liebske,}
\author[24,*]{C.~Lippmann,}\emailAdd{c.lippmann@gsi.de}
\author[8]{V.~Litichevskyi,}
\author[15]{M.~Ljunggren,}
\author[33]{W.J.~Llope,}

\author[3]{S.~Mahmood,}
\author[7]{T.~Mahmoud,}
\author[35,\dagger]{R.D.~Majka, \note[$\dagger$]{Deceased}}
\author[26]{C.~Markert,}
\author[28]{J.~Martinez,}
\author[27]{T.A.~Martins,}
\author[24]{S.~Masciocchi,}
\author[22]{A.~Mathis,}
\author[15]{O.~Matonoha,}
\author[16]{Y.~Matsuyama,}
\author[25]{A.~Matyja,}
\author[2]{M.~Meres,}
\author[22]{D.L.~Mihaylov,}
\author[24]{D.~Mi\'{s}kowiec,}
\author[21]{T.~Mittelstaedt,}
\author[27]{L.S.~Montali,}
\author[27]{D.M.~Moraes,}
\author[22]{C.~Mordasini,}
\author[24]{T.~Morhardt,}
\author[21]{S.~Muley,}
\author[35]{J.~Mulligan,}
\author[12,*]{R.H.~Munzer,}\emailAdd{robert.muenzer@cern.ch}
\author[31]{H.~Murakami,}
\author[7]{K.~M\"{u}nning,}

\author[15]{A.~Nassirpour,}
\author[27]{H.~Natal da Luz,}
\author[30]{C.~Nattrass,}
\author[12]{R.A.~Negrao De Oliveira,}
\author[27]{H.G.~Neves,}
\author[17]{B.S.~Nielsen,}
\author[27]{W.A.V.~Noije,}

\author[16]{M.~Ogino,}
\author[30]{A.C.~Oliveira Da Silva,}
\author[15]{A.~Oskarsson,}
\author[7]{J.~Ottnad,}
\author[16]{K.~Oyama,}

\author[15]{A.~\"Onnerstad,}
\author[15]{L.~\"Osterman,}

\author[27]{A.A.~Pabon,}
\author[21]{Y.~Pachmayer,}
\author[13]{G.~Pai\'{c},}
\author[8]{J.~Parkkila,}
\author[28]{S.~Pathak,}
\author[32]{R.N.~Patra,}
\author[12]{V.~Peskov,}
\author[10]{M.~Petris,}
\author[10]{M.~Petrovici,}
\author[20]{M.~Planinic,}
\author[33]{F.~Pompei,}
\author[10]{L.~Prodan,}

\author[10]{A.~Radu,}
\author[10]{L.~Radulescu,}
\author[29]{J.~Rak,}
\author[19]{J.~Rasson,}
\author[7]{V.~Ratza,}
\author[19]{K.F.~Read,}
\author[4]{A.~Rehman,}
\author[12]{R.~Renfordt,}
\author[3]{K.~R{\o}ed,}
\author[4]{D.~R\"ohrich,}
\author[21]{E.~Rubio,}
\author[24]{T.~Rudzki,}
\author[19]{A.~Rusu,}

\author[33]{M.~Saleh,}
\author[27]{B.C.S.~Sanches,}
\author[26]{J.~Schambach,}
\author[12]{S.~Scheid,}
\author[24]{C.~Schmidt,}
\author[30]{A.~Schmier,}
\author[12]{H.~Schulte,}
\author[24]{K.~Schweda,}
\author[31]{D.~Sekihata,}
\author[31]{N.~Shimizu,}
\author[21]{S.~Siebig,}
\author[27]{R.W.D.~Silva,}
\author[15]{D.~Silvermyr,}
\author[19]{D.~Simpson,}
\author[2]{B.~Sitar,}
\author[35]{N.~Smirnov,}
\author[8]{T.~Snellman,}
\author[21]{H.K.~Soltveit,}
\author[30]{S.P.~Sorensen,}
\author[24]{F.~Sozzi,}
\author[21]{J.~Stachel,}
\author[2]{A.~Szabo,}

\author[22]{L.~\v{S}erk\v{s}nyt\.{e},}

\author[16]{Y.~Takeuchi,}%
\author[4]{G.~Tambave,}
\author[16]{Y.~Tanaka,}%
\author[31]{K.~Terasaki,}
\author[8]{R.J.~Turpeinen,}

\author[4]{K.~Ullaland,}
\author[22]{B.~Ulukutlu,}
\author[28]{E.~Umaka,}
\author[20]{A.~Utrobicic,}

\author[34]{D.~Varga,}
\author[9]{A.~Vargas,}
\author[29,34]{M.~Vargyas,}
\author[11]{R.~Varma,}
\author[15]{O.~Vazquez Rueda,}
\author[4]{A.~Velure,}
\author[14]{S.~Vereschagin,}
\author[9]{S.~Vergara Lim\'on,}
\author[15]{L.~Vergara Urrutia,}
\author[21]{O.~Vorbach,}
\author[24]{B.~Voss,}
\author[24]{D.~Vranic,}

\author[19]{R.J.~Warmack,}
\author[27]{T.O.~Weber,}
\author[12]{C.~Weidlich,}
\author[12,*]{J.~Wiechula,}\emailAdd{jens.wiechula@ikf.uni-frankfurt.de}
\author[21]{B.~Windelband,}
\author[22]{S.~Winkler}
\author[30]{and W.~Witt}

\dedicated{Dedicated to the memory of Richard Daniel Majka}

\begin{document}
\maketitle
\raggedbottom

\section{Introduction}
\label{sec:intro}

This paper describes the upgrade of the large Time Projection Chamber (TPC) of the ALICE detector at the CERN LHC. The TPC upgrade is an essential pillar of the ALICE upgrade strategy for operation during the proton and heavy-ion beam campaigns at the LHC in Run\,3 and Run\,4, from 2022 to 2030~\cite{ALICELOI}. 
The chief goal of the ALICE upgrade is to exploit the increased \PbPb luminosity delivered by the LHC, which corresponds to a minimum-bias interaction rate of up to \SI{50}{\kilo\Hz}. This requirement, driven by the focus of the future ALICE physics program on rare probes at low momentum, implies a paradigm shift for the operation of the TPC, where the concept of selective event triggers must be replaced by continuous readout of all detector data. To  this end, the TPC upgrade implies the complete replacement of the previous readout system, including the readout wire chambers for gas amplification of ionization electrons, and the front-end electronics. 

As a result of a comprehensive R\&D campaign, readout chambers (ROCs) using Gas Electron Multipliers (GEMs)~\cite{Sauli:97} in an optimized multilayer configuration were identified to stand up to the technological challenges imposed by continuous TPC operation at high rate, both in terms of detector resolution and operational stability.

Additionally, the new readout scheme requires a complete redesign of the TPC front-end and readout electronics. 
Besides having to accommodate for the negative signal polarity of the GEM detectors, the new electronics must support the continuous readout scheme which implies that the detector signals have to be sampled continuously while concurrently the acquired data is transferred off-detector at a rate of 3.28\,TB\,s$^{-1}$. This leads to an increase of the data throughput with respect to the original system by almost two orders of magnitude.

The basic concept of the TPC upgrade and its main components are detailed in~\cite{TDR:tpcUpgrade}. 
The use of Micro-Pattern Gas Detectors (MPGDs) for a large-scale application like the ALICE TPC demands well-defined assembly and QA protocols and standardized transport and testing procedures to ensure high quality and reliability of the detectors assembled in a distributed production process of several years.
In the present paper, we describe the final design and the mass production of the GEMs and the new readout chambers, the new high-voltage (HV) system, and the new front-end electronics. Special emphasis is put on results from production QA, installation and commissioning. A large part of the original TPC, such as the gas system and the field cage, are reused after the upgrade, but some important modifications of existing components became necessary. In particular, the optimization of the electrostatic matching between the field cage and the new readout chambers is also described in this paper.

\subsection{ALICE TPC}
\begin{figure}[t]
\centering
\includegraphics[width=0.7\linewidth]{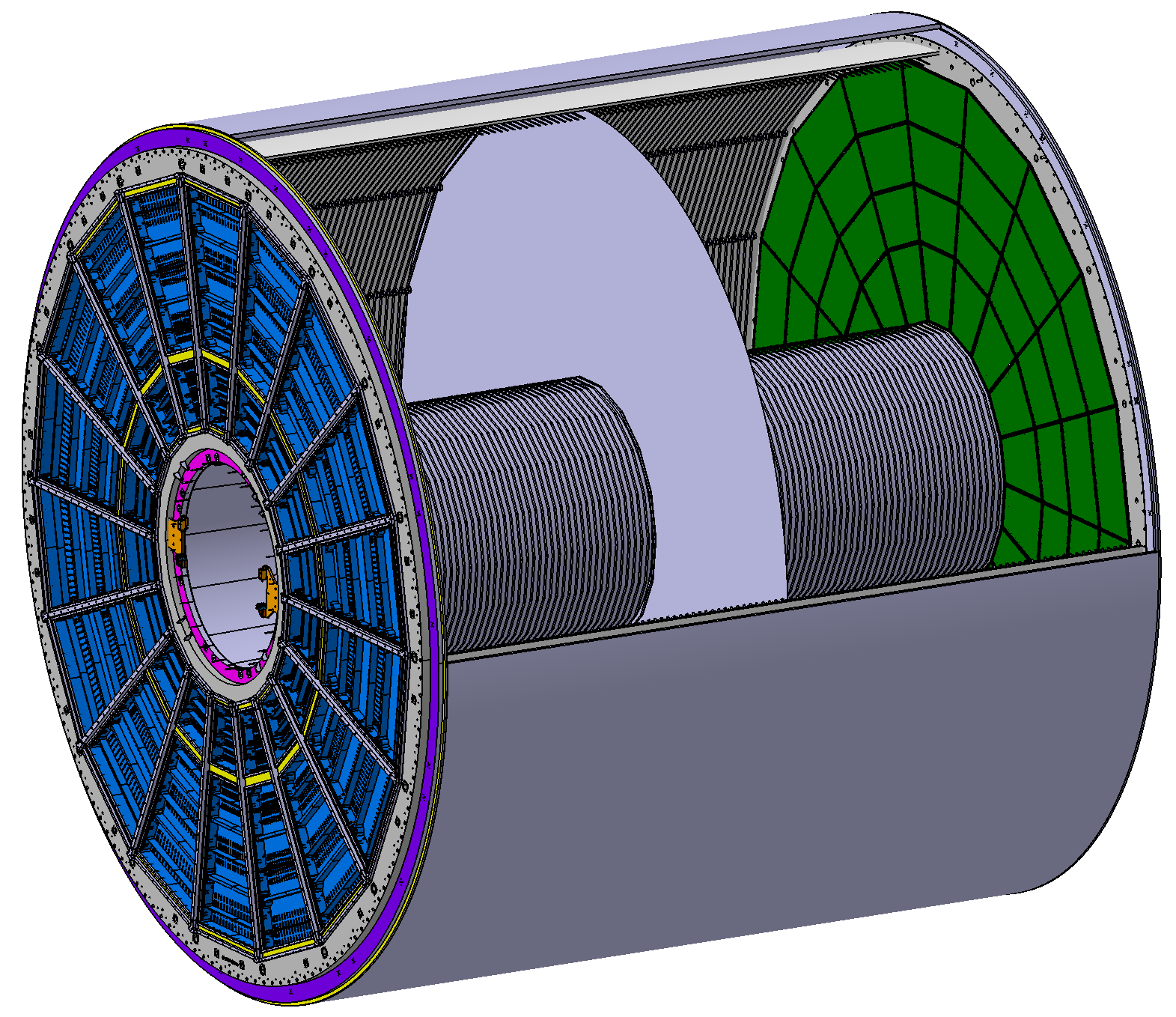}
\caption{Schematic view of the ALICE TPC.}
\label{fig:alitpc}
\end{figure}

The  TPC is the main tracking and particle identification device in the central barrel of ALICE. It covers a symmetric pseudorapidity interval around midrapidity ($|\eta|<0.9$) at full azimuth.
The active volume of the TPC has a cylindrical shape with length and outer diameter of about \SI{5}{\m}, resulting in a total active volume of \SI{88}{\cubic\m}.
It employs a cylindrical field cage with a high-voltage electrode in its center, dividing the active volume into two halves (see \figref{fig:alitpc}). Each of the two endplates houses 18 inner and outer readout chambers (\irocs and \orocs), respectively, which are arranged in pairs to form 18 equal azimuthal sectors. In the original TPC, the readout chambers were based on multiwire proportional chamber (MWPC)-technology. The TPC was successfully operated in \pp, \pPb, \PbPb and \XeXe collisions at a variety of collision energies during LHC Run\,1 and Run\,2. For further details see~\cite{TPCnim,aliceperf}.

\subsection{Upgrade concept}
The main objective of the TPC upgrade is to provide sensitivity to a minimum-bias interaction rate of \SI{50}{\kilo\Hz} in \PbPb collisions, as foreseen for LHC operation in Run\,3 and beyond. This goal requires elimination of the intrinsic trigger rate limitation  of the original MWPC-based TPC~\cite{TDR:tpc,TPCnim}. The  limitation to about \SI{3}{\kilo\Hz} was imposed by the operation of an active ion gating grid, which is used to  collect ions from the amplification region and  prevent them from drifting back into the drift volume, where they would lead to substantial space-charge distortions of the drift field. 
Further  limitations in MWPC-based readout chambers arise from space-charge effects at the amplification wires and ion tail accumulation, both resulting in substantial rate-dependent non-linearities of the signal response.
Despite the tightened operational demands, the ambitious ALICE physics program for Run\,3 and Run\,4 does not allow a degradation with respect to the excellent momentum and \dEdx resolution of the original TPC. The design considerations emerging from these challenging requirements and their technical solutions were worked out in an extensive R\&D program~\cite{TDR:tpcUpgrade,TDR:tpcUpgradeAddendum} and will be briefly outlined below. 

Operation of the TPC at a collision rate of \SI{50}{\kilo\Hz} implies that on average five collision events pile up within the  TPC  readout time window of about \SI{100}{\micro\s}, as given by the typical electron drift time over the maximal drift length of \SI{2.5}{\meter}. 
This excludes triggered operation and defines the need for continuous readout, demanding  novel gas amplification techniques which provide sufficient ion blocking without an active gate. The requirement to keep the ion-induced space-charge distortions at a tolerable level leads to an upper limit of \SI{2}{\percent} for the fractional \ibf, i.e.\ the ion escape probability per effective electron-ion pair produced in the gas amplification stage, at the operational gas gain of 
2000 in a \NeCOtwoNtwo gas mixture~\cite{TDR:tpcUpgrade}. At the same time, the readout system must ensure that the  \dEdx resolution of the TPC is preserved, which translates into a required local energy resolution better than \SI{14}{\percent} at the $^{55}$Fe-peak~\cite{TDR:tpcUpgradeAddendum}.

\begin{figure}[t]
  \centering
   \includegraphics[width=0.7\linewidth]{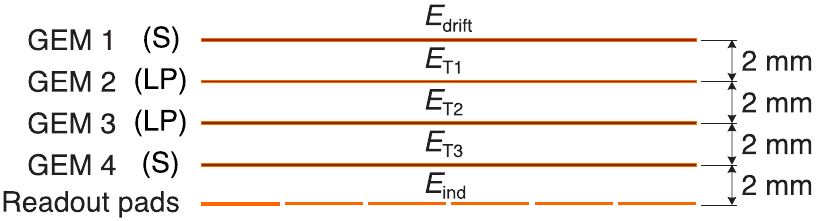}
  \caption[Short caption]{Schematic setup with four GEMs.}
  \label{fig:4GEM-schematic}
\end{figure}

GEMs~\cite{Sauli:97}  provide a viable solution to this challenge \cite{TDR:tpcUpgrade, BOHMER2013101}. They can be easily stacked in layers,  allowing separation of several amplification stages. After a careful optimization of the gain share among the GEMs and the electric transfer fields between them, the drift path of back-drifting ions that emerge from subsequent layers can be efficiently blocked \cite{SauliIBF}. R\&D studies have demonstrated that sufficient ion blocking can be achieved when stacks of four GEMs are used, and foils with standard (S, \SI{140}{\micro\m}) and large (LP, \SI{280}{\micro\m}) hole pitch are combined to an \mbox{S-LP-LP-S} configuration, as shown in \figref{fig:4GEM-schematic}. The main feature of this highly optimized setting is 
a very low transfer field $E_{\text{T3}}$ between \gemThree and \gemFour of only \SI{100}{\volt\per\centi\meter}, whereas the other transfer fields and the induction field \Eind are kept at typical values around \SI{3500}{\volt\per\centi\meter}.

\begin{figure}[htbp]
  \centering
   \includegraphics[width=0.7\linewidth]{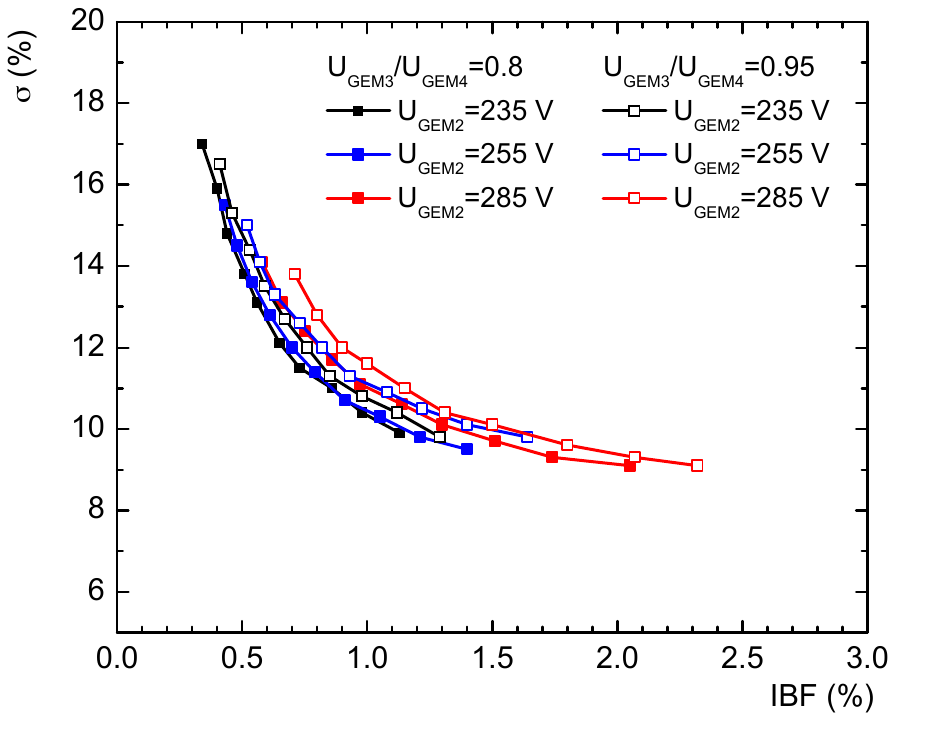}
  \caption[Short caption]{Energy resolution $\sigma(^{55}\rm{Fe})$ as a function of \ibf (IBF) in a 4-GEM stack (S-LP-LP-S)  in \NeCOtwoNtwo. The gas gain is kept at 2000 in all measurements by adjusting the voltages on \gemThree and \gemFour at a fixed ratio of 0.8 or 0.95 (from~\cite{TDR:tpcUpgrade}).}
  \label{fig:sigmavsIBF}
\end{figure}

The result for the optimized \mbox{S-LP-LP-S} configuration is shown in \figref{fig:sigmavsIBF} for different GEM voltage settings at the nominal gas gain of 2000 (from~\cite{TDR:tpcUpgrade}).
The data exhibit a characteristic anti-correlation between \ibf and the relative energy resolution $\sigma(^{55}\rm{Fe})$. Detailed simulations and studies with prototypes have demonstrated that the performance limit is largely determined by the operational conditions of the first GEM (\gemOne) in the stack. While the drift path of ions created at the following GEMs can be efficiently blocked by the upper GEM layers, provided a proper adjustment of the GEM voltages and electric transfer fields, those ions emerging from \gemOne have a large probability to escape into the drift volume. The number of ions produced at  \gemOne can be minimized by reducing the gas amplification, however, this leads to a loss of effective primary ionization and therefore to a degradation of the energy resolution. 

As shown in \figref{fig:sigmavsIBF}, an extended operational region with \ibf below \SI{2}{\percent} and\break $\sigma(^{55}\rm{Fe})<\SI{14}{\percent}$ can be found with an \mbox{S-LP-LP-S} setup, well in line with the required detector performance.
The final choice of the working point must also be optimized
with respect to  operational stability  under the radiation 
load expected in Run\,3. 
The minimization of the ion space-charge density requires also the operation of the readout chambers at the lowest possible gas gain, leading to a challenging front-end noise requirement of ENC of around 670\,$e$. A summary of the most important TPC parameters is given in Table~\ref{tab:tpc.sum}.

\begin{savenotes}
  \begin{table*}[htbp]\footnotesize
      \caption{Parameters of the upgraded TPC.}
      \begin{center}
      \begin{tabular}{ll}
        \toprule
        Detector gas	&	\NeCOtwoNtwo  \\
        Gas volume	&	\SI{88}{\meter\cubed} \\
        Drift voltage	&	\SI{100}{\kilo\volt} \\
        Drift field		&	\SI{400}{\volt\per\centi\meter} \\
        Maximal drift length & \SI{250}{\centi\meter} \\
        Electron drift velocity & \SI{2.58}{\cm\per\micro\s}\\
        Maximum electron drift time &\SI{97}{\micro\s} \\
        $\omega\tau$ ($B=\SI{0.5}{\tesla}$) & 0.32\\
        Electron diffusion coefficients & $D_{\rm
          T}=\SI{209}{\micro\meter\per\sqrt{\centi\meter}}$, $D_{\rm L}=\SI{221}{\micro\meter\per\sqrt{\centi\meter}}$\\
        Ion drift velocity & \SI{1.168}{\centi\meter\per\milli\s}\\
        Maximum ion drift time & \SI{214}{\milli\s}\\
        \midrule
        {\bf Readout chambers} & \\[0.25ex]
        Total number	&	$2\times2\times18 = 72$\\ 
        Readout technology	&	4-GEM stack, single mask, standard (S, \SI{140}{\micro\meter}) and \\
        ~			&	large (LP, \SI{280}{\micro\meter}) hole pitch GEMs in \mbox{S-LP-LP-S} configuration\\
        Effective gas gain	&	2000\\
%
        {\bf Inner  (\iroc)} & \\[0.25ex]
        Total number	&	$2\times18=36$\\ 
        Active range	&	$848.5<r<1321\,\si{\milli\meter}$ \\
        Pad rows 	&	63\\
        Total pads (\iroc)	&	5280\\
        $S$:$N$ (MIP)	&	20:1 \\[0.5ex]
        {\bf Outer  (\oroc)} & \\[0.25ex]
        Total number & $2\times18=36$\\ 
        Active range & $1347<r<2464\,\si{\milli\meter}$ \\
        Total pads (\oroc) & 9280\\
        $S$:$N$	(MIP)   &	30:1\\
        Pad rows &	89\\[0.5ex]
        {\bf \orocOne} & \\[0.25ex]
        Active range & $1347<r<1687\,\si{\milli\meter}$\\
        Pad rows & 34\\
        Number of pads & 2880\\[0.5ex]
        {\bf \orocTwo} & \\[0.25ex]
        Active range & $1708<r<2068\,\si{\milli\meter}$\\
        Pad rows & 30\\
        Number of pads & 3200\\[0.5ex]
        {\bf \orocThree} & \\[0.25ex]
        Active range & $2089<r<2464\,\si{\milli\meter}$\\
        Pad rows & 25\\
        Number of pads & 3200\\[0.5ex]
        \midrule
        {\bf Readout electronics} & \\[0.25ex]
        Number of channels & 524160\\
        Signal polarity & negative \\
        Average system noise (ENC) & 670\,$e$\\
        Conversion gain & \SI{20}{\milli\volt\per\femto\coulomb}\\
        Dynamic range & \SI{100}{\femto\coulomb} \\
        Peaking time & \SI{160}{\nano\s}\\
        ADC number of bits & 10\\
        ADC sampling rate & \SI{5}{\MHz}\\
        Power consumption (total) & 56\,mW per channel\\
        \bottomrule
      \end{tabular}
    \end{center}
    \label{tab:tpc.sum}
  \end{table*}
\end{savenotes}

\section{Gas mixture: choice and handling}
\label{sec:gas}

\begin{figure}[ht]
\centering
\includegraphics[width=.75\linewidth]{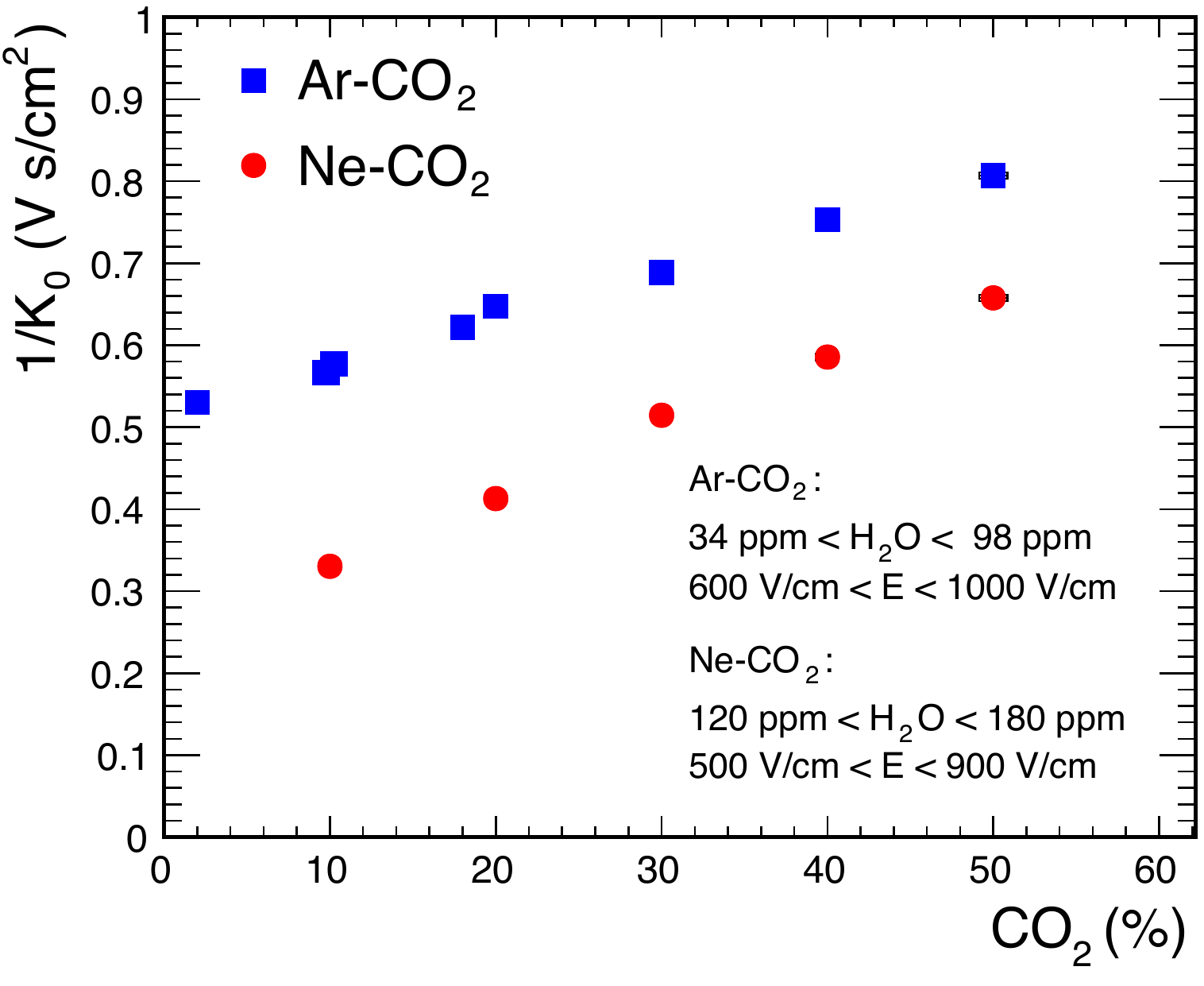}
\caption{The inverse of the ion mobility coefficient for \ArCOtwo and \NeCOtwo gas mixtures as a function of the CO$_2$ concentration. Data taken from~\cite{DEISTING20181}.}
\label{fig:gas:mobility}
\end{figure}

The gas mixture chosen for the operation of the upgraded TPC is \NeCOtwoNtwo (i.e.\ 90 parts of Ne, 10 parts of CO$_2$, and 5 parts of N$_2$). The main reason for this choice is the higher ion mobility of Ne mixtures compared to similar Ar-based mixtures (see \figref{fig:gas:mobility})~\cite{DEISTING2017215}. This directly reduces the size of the space-charge distortions by nearly a factor of 2.

The effective ion mobility of the \NeCOtwoNtwo mixture is \SI{2.92}{\centi\meter\squared\per\volt\per\s}~\cite{DEISTING20181}. At a drift field of \SI{400}{\volt\per\centi\meter} it takes ions \SI{214}{\ms} to reach the central electrode from the readout plane. Therefore, at \SI{50}{\kilo\Hz} up to 10$^4$ collisions partially contribute, with their particular multiplicity, angular, and energy-loss fluctuations, to the space-charge distribution at any time. 
Higher concentrations of CO$_2$ rapidly result in a decrease of the electron and also ion drift velocities. Other quenchers have been discarded: hydrocarbons because of aging and flammability considerations, CF$_4$ because of the extensive R\&D required to validate this gas for operation in an existing field cage.
In addition, Ne-based mixtures, in particular with the addition of N$_2$, have demonstrated high stability against primary discharges (see~\cite{GASIK2017116} and \secref{sec:discharge}). N$_2$ is a well suited quencher for Ne~\cite{Garabatos:2004iv}, and it has little impact on the electron and ion transport properties of \NeCOtwo, as shown in ~\tabref{tab:gas:gasproperties}.

Since the TPC has already been operated with this gas mixture, no changes to the existing gas hardware are necessary. The TPC gas system operates in a closed-loop mode. This allows for circulation of the gas through the \SI{88}{\meter\cubed} detector volume at high flows with little waste of expensive neon. In addition, in closed-loop systems the gas flow and the detector overpressure are decoupled. The TPC is operated at a very low overpressure relative to the atmospheric one, by means of a regulated bypass mechanism around the compressor. The compressor not only forces the gas to circulate, but also provides a high-pressure section in the system. This high pressure is used for gas storage and efficient gas cleaning and separation. 
Thus, a buffer is used to store gas in order to provide or to remove gas from the detector to keep its overpressure constant.
Oxygen is removed from the gas with cartridges filled with a copper catalyser agent. 

For filling the detector with the final gas mixture, it is first entirely flushed with CO$_2$ in an open loop, and then, in a closed loop, the necessary fraction of CO$_2$ is removed with cartridges filled with a molecular sieve, and replaced with neon. Therefore, only one system volume of neon is used to reach the correct concentration. The N$_2$ is injected at the end of this process. The system allows for the controlled addition of H$_2$O ($\sim$100\,ppm) in order to avoid highly non-conductive surfaces in regions of high electric fields. A low flow of gas mixture is continuously injected from a mixer into the loop. This extra gas allows for small streams of gas taken from various points to be vented through an analysis chain, for O$_2$ and H$_2$O measurements, and a gas chromatograph, for monitoring the gas composition. This information is also available to the data calibration software.
The ultimate protection for any over- or under-pressure in the TPC is provided by a high-capacity two-way safety bubbler, located at the detector, which ensures a relief pressure of \num{\pm 2}\,\si{\milli\bar}.

The two envelope volumes between the field cage vessels and the containment vessels are continuously flushed with CO$_2$ with an admixture of \SI{0.1}{\percent} of H$_2$O, in order to ensure proper HV insulation under high loads.

\begin{table*}[ht]\footnotesize
  \caption[Properties of gas mixtures]{Ionization and electron transport properties of a few gas mixtures used in modern TPCs, as calculated with the Magboltz~\cite{Biagi1018382} and Garfield~\cite{GARFIELD:1984} packages for an electric field of \SI{400}{\volt\per\centi\meter}.}
  \begin{center}
  \tabcolsep=0.11cm
   \begin{tabular}{lccccccc}
      \toprule
      Gas & Eff. ionization & \multicolumn{2}{c}{Number of electrons per MIP} & Drift velocity & \multicolumn{2}{c}{Diffusion coeff.} & \\
      & energy $W_{\rm i}$ & $N_{\rm p}$ (primary) & $N_{\rm t}$ (total) & \vdrift & $D_{\rm L}$ & $D_{\rm T}$ & $\omega \tau$ \\
      & (eV) & (e\,cm$^{-1}$) & (e\,cm$^{-1}$) & (cm\,\textmu s$^{-1}$) & (\si{\micro\meter\per\sqrt{\centi\meter}}) & (\si{\micro\meter\per\sqrt{\centi\meter}}) & \\
      \midrule
      \NeCOtwoNtwo    & 37.3 & 14.0 & 36.1 & 2.58 & 221 & 209 & 0.32 \\
      \NeCOtwo        & 38.1 & 13.3 & 36.8 & 2.73 & 231 & 208 & 0.34 \\
      \ArCOtwo        & 28.8 & 26.4 & 74.8 & 3.31 & 262 & 221 & 0.43 \\
      \NeCFfourTwenty & 37.3 & 20.5 & 54.1 & 8.41 & 131 & 111 & 1.84 \\
     \bottomrule
  \end{tabular}
  \end{center}
  \vspace{-2mm}
  \label{tab:gas:gasproperties}
\end{table*}

\section{Readout chambers}
\label{sec:roc}
To fulfil the challenging requirements of the ALICE upgrade (see \secref{sec:intro}), a novel configuration of GEM detectors was developed. By stacking four GEM foils operated at an optimized field configuration, excellent particle identification and efficient ion blocking are achieved (see \figref{fig:sigmavsIBF}).

A GEM~\cite{Sauli:97} is a gaseous detector made of a \SI{50}{\micro\meter} thin insulating polyimide foil with \SIrange[range-units=single]{2}{5}{\micro\meter} thick Cu layers on both sides. The foil is perforated by photolithographic processing, forming a dense, hexagonal pattern of double-conical holes. The holes have an inner (polyimide) diameter of \SI{50}{\micro\meter}, and an outer (copper) diameter of \SI{70}{\micro\meter}. By applying a suitable potential across the foil, charge amplification can be produced in the high electric field region inside the holes. 
The baseline solution of the amplification structure for the upgrade employs GEM foils with standard (S) and large-pitch (LP) geometry, with the hole pitches of \SI{140}{\micro\meter} and \SI{280}{\micro\meter}, respectively, and corresponding optical transparency of $\sim$\SI{24}{\percent} and $\sim$\SI{6}{\percent}. The foils are stacked in the following order: \mbox{S-LP-LP-S}. A detailed description of the baseline configuration and its performance is presented in \secref{sec:roc:gem}.

\subsection{Mechanical structure}

\begin{figure}[!htbp]
  \centering
   \includegraphics[width=0.9\linewidth]{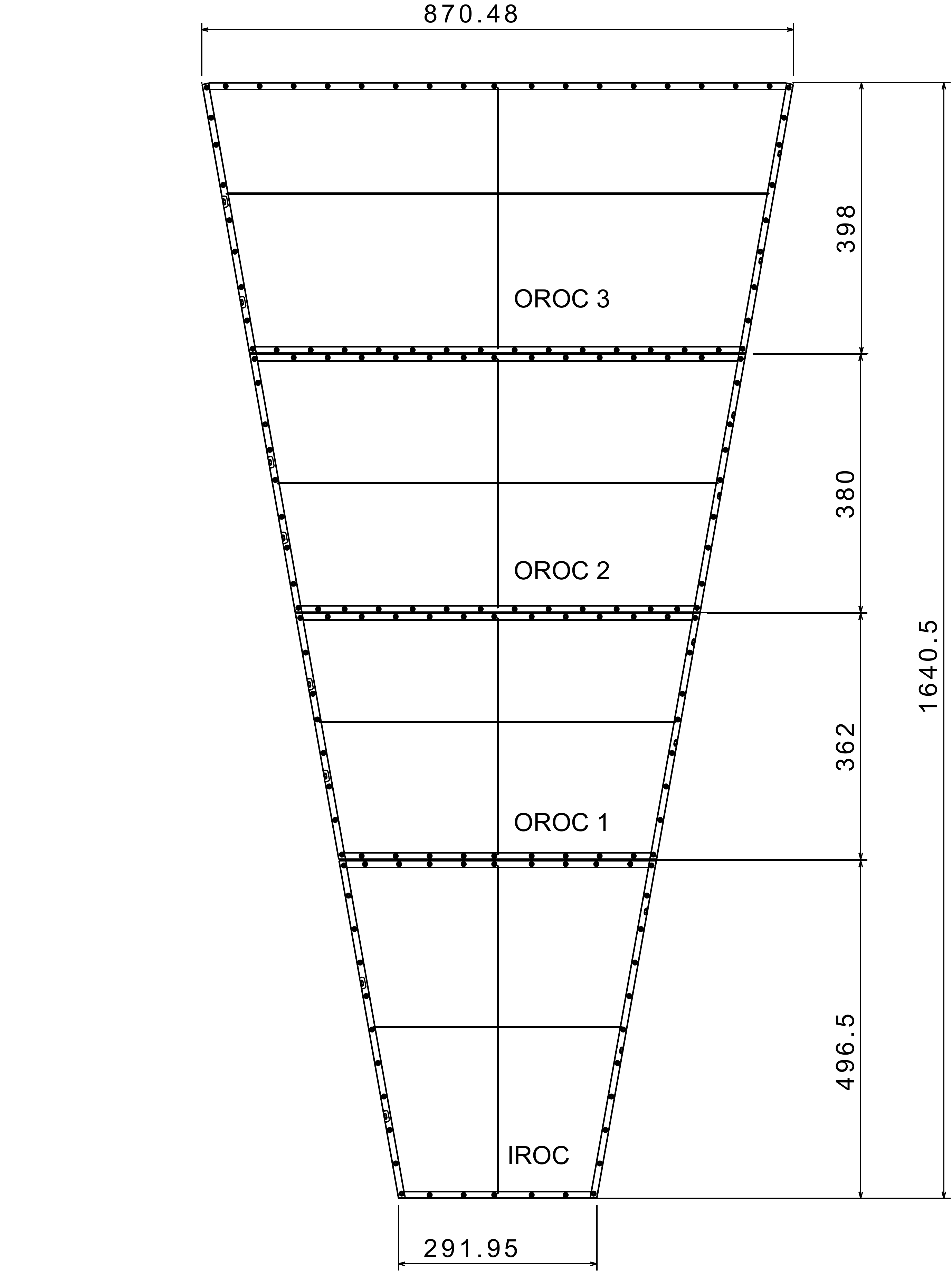}
  \caption[Readout chamber dimensions]{Dimensions (mm) of the ALICE TPC readout chambers.}
  \label{fig:dimensions} 
\end{figure}

\begin{figure}[htb]
  \centering
   \includegraphics[width=0.8\linewidth]{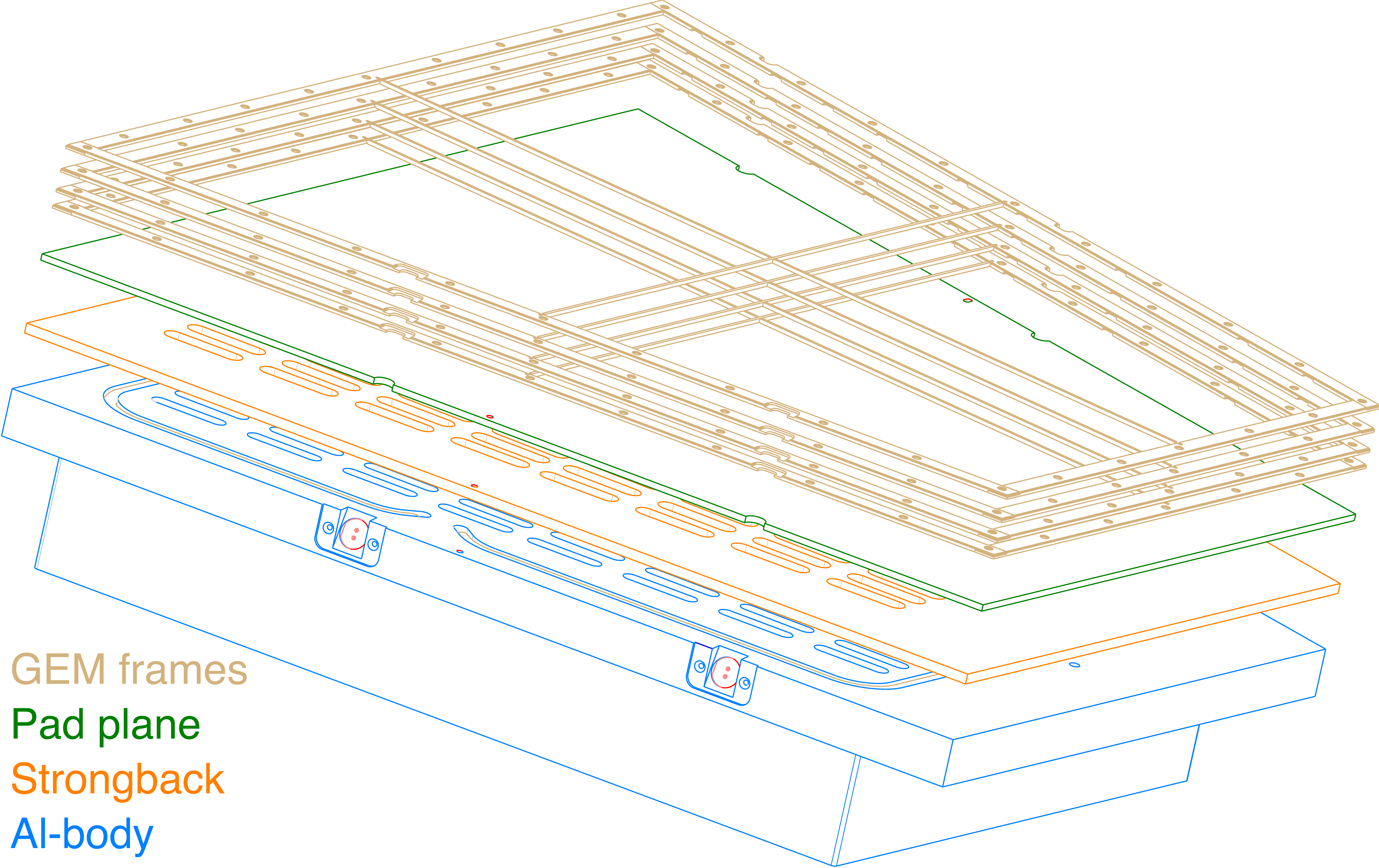}       
  \caption[Exploded view of a readout chamber]{Exploded view of an \iroc with chamber body components and GEM frames. See text for more details.}
  \label{fig:roc.exploded} 
\end{figure}

The TPC readout sectors, each covering 20$^{\circ}$ in azimuth, are segmented radially into two regions equipped with an inner and an outer readout chamber (\iroc and \oroc).
The mechanical structure of the readout chambers (ROCs) follows closely the design of the original MWPC-based detectors with the dimensions indicated in \figref{fig:dimensions}.
A chamber consists of: a trapezoidal aluminium frame (\textit{Al-body}), a \textit{pad plane} made of a multilayer Printed Circuit Board (PCB) and separated from the Al-body by a fiberglass plate (\textit{strongback}), and a stack of four GEM foils glued onto fiberglass epoxy frames, each containing a spacer cross (see \figref{fig:roc.exploded}).

An \iroc is assembled from one Al-body, the strongback, pad plane and GEM stack each. An \oroc, on the other hand, is assembled from one Al-body and strongback, and three pad planes plus three GEM stacks. The three stacks are labeled \orocOne, \orocTwo and \orocThree, where \orocOne is the stack closest to the \iroc (see \figref{fig:dimensions}). The total weight of an assembled chamber is \SI{\sim12}{\kilo\gram} and \SI{\sim46}{\kilo\gram} for \iroc and \oroc, respectively.

\subsubsection{GEM foils}
\label{sec:roc:gem}

The GEM foils for the ALICE TPC upgrade were produced by the CERN Micro-Pattern Technologies laboratory~\cite{CERN-MPT} using the so-called single mask technique~\cite{singlemask}. Their trapezoidal shape is defined by the size and shape of the readout chambers. Due to limitations on the size of the raw material, the \orocs are equipped with three independent GEM stacks, as summarized in \tabref{tab:GEMsize}. 

\begin{table}[t]\footnotesize
\caption{Size and geometrical characteristics of the GEM foils. The width is indicated by the long (short) side of the trapezoid.}
\begin{center}
\begin{tabular}{ c c c c c }
\toprule
GEM type & height & width &  \# HV  & avg. segment size  \\
 & (mm) & (mm) & segments & (cm$^{2}$)\\
 \midrule
 \iroc & 497 & 467\,(292) & 18 & 93 \\
 \orocOne & 362 & 595\,(468) & 20 & 87 \\
 \orocTwo & 380 & 730\,(596) & 22 & 105 \\
 \orocThree & 398 & 870\,(730) & 24 & 122\\
 \bottomrule
\end{tabular}
\end{center}
\label{tab:GEMsize}
\end{table}
The four GEMs in a stack are named, from top to bottom (i.e.\ from drift to pad plane): \gemOne, \gemTwo, \gemThree, and \gemFour. For each GEM a potential difference $\Delta V_{\text{GEM\,}i}$ (where $i=1,\dots,4$ is the number of the GEM in the stack) is applied. Transfer fields in the gaps between GEMs are called, from the top, $E_{\text{T1}}$, $E_{\text{T2}}$, $E_{\text{T3}}$. The last gap, between \gemFour and the pad plane, is the induction gap with the associated induction field \Eind. Typical HV settings applied to the stacks are discussed in \secref{hv:hv:overview}.

The baseline configuration of the detector was defined after an extensive R\&D phase. Several parameters were carefully optimized in order to minimize the \ibf of the final system and improve the uniformity across the area of individual readout chambers (see \secref{sec:roc:rnd} for more details). The resulting configuration for each GEM foil in a stack in terms of hole pitch and relative orientation of the GEM hole pattern is shown in \tabref{tab:foildet}.
\begin{table}[t]\footnotesize
    \caption{Specification of GEM foils in a quadruple-GEM stack.}
  \begin{center}
  \begin{tabular}{cccc}
    \toprule
       foil in a stack	& type & pitch (\textmu m)					& relative orientation \\\midrule
	\gemOne 	 & S   	& 140  & 0$^{\circ}$ \\
	\gemTwo & LP	& 280 & 90$^{\circ}$ \\
	\gemThree	& LP		& 280 & 0$^{\circ}$ \\
	\gemFour  & S    	& 140 & 90$^{\circ}$ \\\bottomrule
  \end{tabular}
  \end{center}
  \label{tab:foildet}
\end{table}

\begin{figure}[h]
  \centering
   \includegraphics[width=0.8\linewidth]{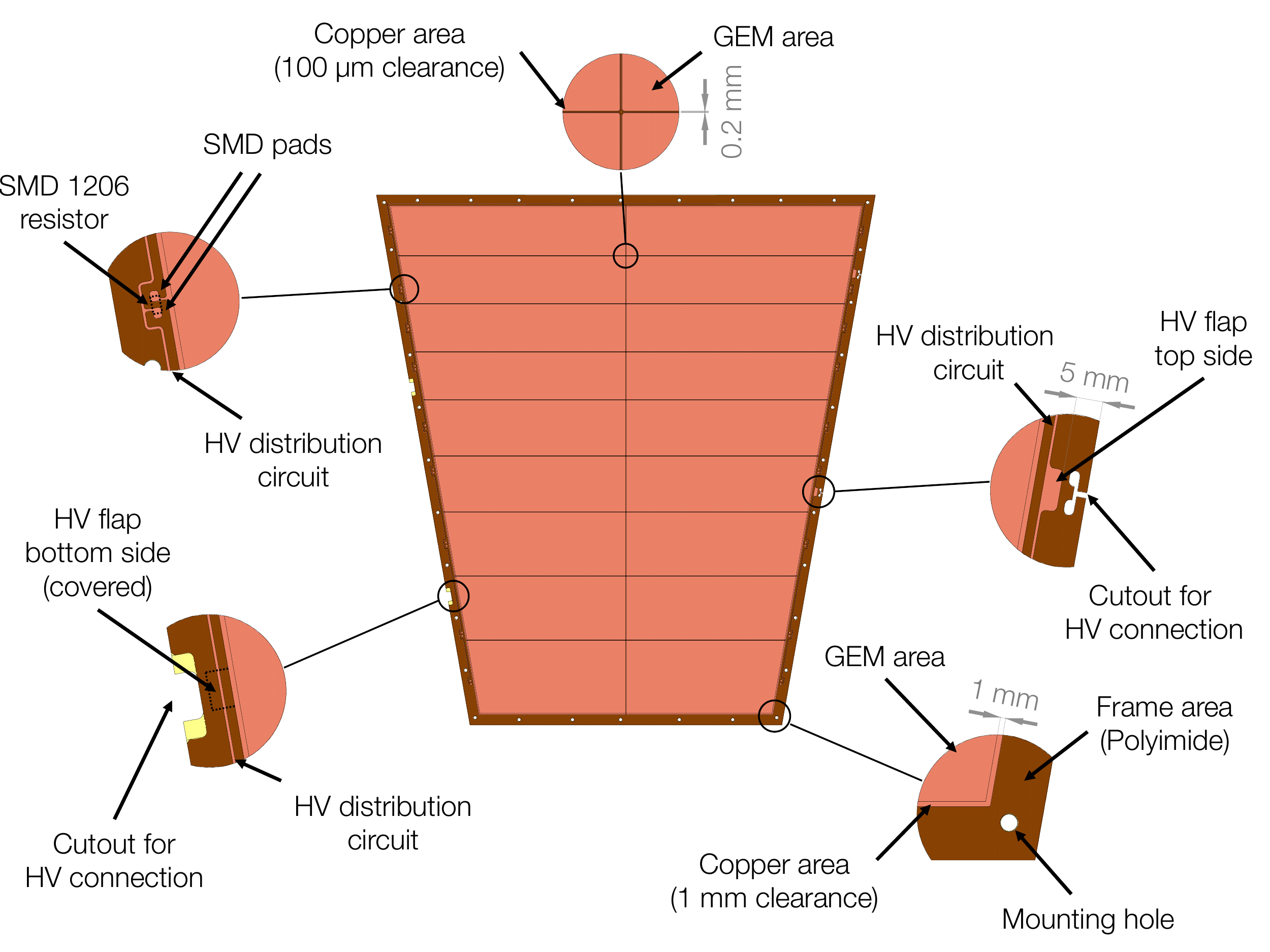}
  \caption[GEM design details]{GEM design details. Adopted from~\cite{Mathis:2018sjk}.}
  \label{fig:gem-design} 
\end{figure}

\Figref{fig:gem-design} shows the most important features of the GEM design. The top side of each foil is subdivided into individual HV segments with an area of $\sim$100\,cm$^{2}$, in order to limit the total charge in case of electrical discharges~\cite{Bachmann:radon} and to minimize the affected area of a stack in case a segment develops a short. The gap between the adjacent segments is \SI{200}{\micro\meter}, with an additional \SI{100}{\micro\meter} space between the segment boundaries and the GEM holes to accommodate a possible mask misalignment during foil production. Each segment is connected via a \SI{5}{\mega\ohm} SMD 1206,\footnote{Surface-Mounted Device, the number indicates the imperial code of the component size.} loading resistor to the common HV distribution, which is a \SI{1}{\milli\meter} wide copper trace running along three sides of the foil. The resistor footprints are located parallel to the segment boundary. Electric potentials are applied to a foil by wires soldered to HV flaps placed on the top and bottom sides of the foil (see \secref{sec:roc:roc:ass} for more details).

Special emphasis is put on the high-voltage stability and integrity of the readout chambers installed in the final position, including situations like a trip\footnote{A trip is defined as a controlled ramp down of a power supply channel upon detection of an over-current.} of individual GEM stacks or entire readout chambers. In such a case, a large potential difference of up to \SI{3.6}{\kilo\volt}, depending on the applied settings, may occur between the top electrodes of two neighboring stacks. %
Hence the distance between any conducting surface and the foil boundary is never less than \SI{5}{\milli\meter}, resulting in a \SI{13}{\milli\meter} gap between HV electrodes of neighboring stacks. This arrangement is robust against electrical discharges under the operating gas mixture. In order to reduce the electric field at all boundaries, all copper corners are rounded with a radius of at least \SI{0.5}{\milli\meter}.

\subsubsection{GEM frames}
\label{sec:frames}

Before mounting on a stack, the GEM foils are stretched and glued on \SI{2}{\milli\meter} thick frames made of Vetronite EGS 103~\cite{Vetronite}, an epoxy glass fabric laminate (G-11 type). Sets of four \SI{10}{\milli\meter} wide ledges form frames with the size of each GEM type (see \tabref{tab:GEMsize}). The ledges contain a number of precisely milled mounting holes, glue-excess grooves, and pockets to hold the SMD resistors soldered to the foil installed below. %
The frames contain also a \SI{1.5}{\milli\meter} wide, \SI{2}{\milli\meter} deep spacer cross (one longitudinal and one transverse bar) to prevent the foils from approaching each other due to electrostatic forces~\cite{GASIK2017222}. The bars are positioned on GEM segment and pad row boundaries in order to minimize their impact on the active area of the detector. The area of the largest quadrant is \SI{843}{\centi\meter\squared}.

Before machining, the frame material is fine-sanded to reach the final thickness, with a tolerance of \SI{\pm0.04}{\milli\meter}. No further treatment of the material, such as coating with a thin polyurethane layer as proposed e.g.\ in~\cite{Altunbas:02a, ABBANEO201967}, is performed before the GEMs are glued onto the frames. The gluing procedure is described in \secref{sec:roc:framing}.

\subsubsection{Pad planes and strongbacks}
\label{sec:roc:padplane}
\begin{figure}[ht]
\centering
\includegraphics[width=0.8  \linewidth]{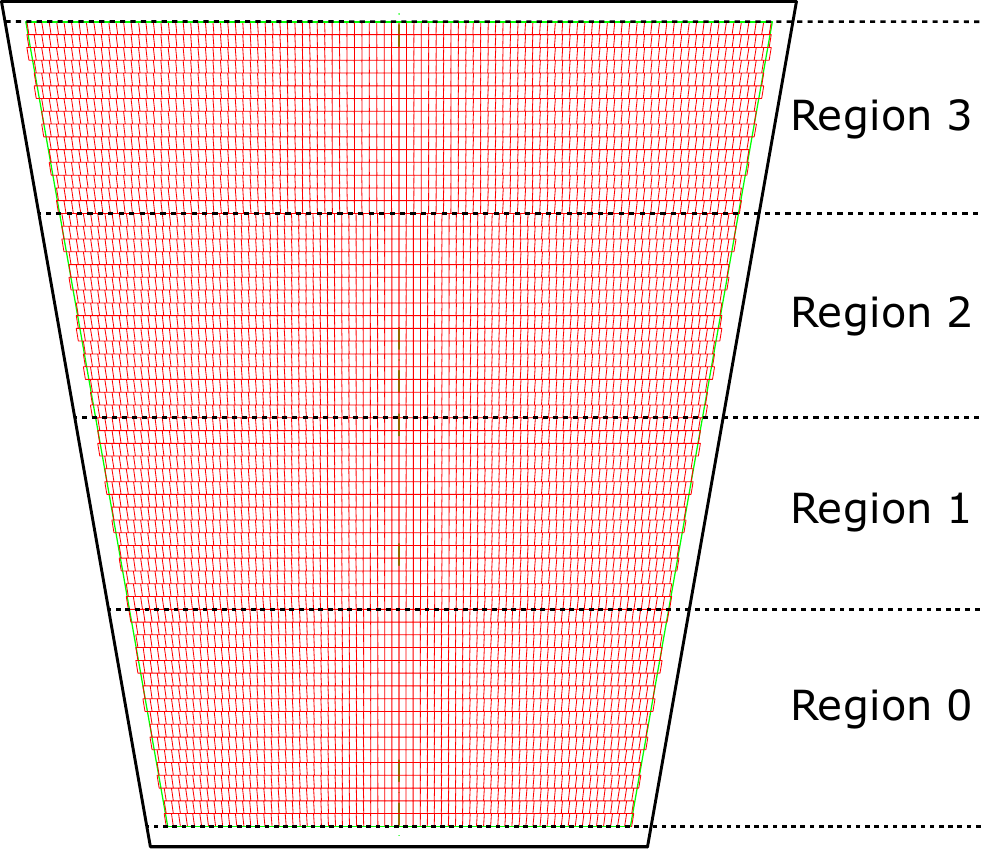}
\caption{\iroc pad plane layout. The active GEM area is indicated by the green lines. Also shown are groups of consecutive pad rows which are combined to readout regions.}
\label{fig:roc:padplane}
\end{figure}

The readout chambers are equipped with pad planes made of a 3.2\,mm thick FR4 multilayer board. The segmentation of the \oroc readout planes follows the segmentation of the GEM stacks. 
The top PCB layer consists of copper readout pads, arranged in pad rows (see \figref{fig:roc:padplane}).
The pad plane was modified with respect to the previous design with MWPC readout, roughly keeping the same pad sizes~\cite{TPCnim}. \tabref{tab:roc:padplane:pads} summarizes the dimensions and parameters of the readout planes and pads. The pad widths vary slightly around the mean value of $4.2\times 7.5$\,mm$^2$ for the \iroc and $6\times 10$\,mm$^2$, $6\times 12$\,mm$^2$ and $6\times 15$\,mm$^2$ for the regions corresponding to the three \oroc GEM stacks. This is done in order to achieve maximum geometrical coverage of the active area, while matching the total number of pads to the front-end electronics (160 channels per front-end card). 
To minimize angular effects, the pads are radially oriented according to the average local track angle (see \figref{fig:roc:padplane}).
In addition, the pad planes and the front-end cards are partitioned such that the data from all pads in the ten readout regions indicated in \tabref{tab:roc:padplane:pads} are sent to individual readout units (CRUs, see \secref{sec:fec.overview}). In particular, the data from full pad rows is sent to one CRU.
\begin{table}[t]\footnotesize
  \caption[Dimensions and parameters of readout planes and
  pads]{Dimensions and parameters of readout planes and pads. The active range refers to the distance of the respective regions from the beam axis, taken at the azimuthal center of the sector.}
  \begin{center}
  \begin{tabular}{lcccccc}
    \toprule
    region & \multicolumn{2}{c}{active range} & \multicolumn{2}{c}{pad size} & nr of & nr of \\
    & from & to & width & length & rows & pads \\
    & (mm) & (mm) & (mm) & (mm) & & \\[0.5ex]
    \midrule
    {\bf \iroc} \\[0.25ex]
    0 & 848.5 & 976 & 4.16 & 7.5 & 17 & 1200 \\
    1 & 976 & 1088.5 & 4.2 & 7.5 & 15 & 1200 \\
    2 & 1088.5 & 1208.5 & 4.2 & 7.5 & 16 & 1440 \\
    3 & 1208.5 & 1321 & 4.36 & 7.5 & 15 & 1440 \\[0.5ex]
    \midrule
    {\bf \orocOne} \\[0.25ex]
    4 & 1347 & 1527 & 6 & 10 & 18 & 1440 \\
    5 & 1527 & 1687 & 6 & 10 & 16 & 1440 \\[0.5ex]
    \midrule
    {\bf \orocTwo} \\[0.25ex]
    6 & 1708 & 1900 & 6.08 & 12 & 16 & 1600 \\
    7 & 1900 & 2068 & 5.88 & 12 & 14 & 1600 \\[0.5ex]
    \midrule
    {\bf \orocThree} \\[0.25ex]
    8 & 2089 & 2284 & 6.04 & 15 & 13 & 1600 \\
    9 & 2284 & 2464 & 6.07 & 15 & 12 & 1600 \\[0.5ex]
    \bottomrule
  \end{tabular}
  \end{center}
  \label{tab:roc:padplane:pads}
\end{table}

The expected (average) pad occupancies at \SI{50}{\kilo\Hz} \PbPb collisions are below \SI{30}{\percent}.
Despite the narrow pad response function, and leaving aside track inclination effects, a sufficiently wide distribution of the signals over 2 to 3 pads in a row is obtained from diffusion alone, for drift lengths larger than about \SI{30}{\centi\meter} for the small pads and 80\,cm for the larger pads.
For shorter drift distances, however, the appearance of single-pad clusters has a negative effect on the resolution.
It should be noted, however, that short drift distances correspond partially to regions outside the nominal acceptance of the ALICE central barrel (pseudorapidity interval $|\eta| < 0.9$).
Consequently, the good momentum resolution in the acceptance of the barrel detectors of ALICE will be maintained after the upgrade~\cite{TDR:tpcUpgrade}.

In order to minimize the capacitance to ground at the preamplifier input, the pad planes do not contain any ground electrode. The signals induced on the readout pads are routed to connectors on the backside of the pad plane. The pad vias are closed with a \SI{20}{\micro\meter} copper layer in order to avoid the collection of dust, which could pose a threat to the stability of the GEM foils. Beyond the pad layer, three further PCB layers are used to route the traces and place the connector footprints in an optimal way. The signal routing is optimized in order to minimize the trace length and thus capacitance. This is realized with an auto-router using the following boundary conditions: the trace width is 6\,mil\footnote,{1\,mil\,=\,0.001\,in\,=\,25.4\,\textmu m.} and the routing clearances are 10\,mil for the \iroc, \orocOne and \orocTwo, and 12\,mil  for the \orocThree pad planes. The results of the automatic procedure are further optimized with a manual pad-to-pin assignment in order to reduce the maximum trace length values. The final values, summarized in \tabref{tab:padplane:routing}, are on average shorter than in the old readout chambers~\cite{TPCnim}.
\begin{table}[b]\footnotesize
 \caption[Average and maximum trace lengths]{Average and maximum trace lengths for the TPC pad planes.}
  \begin{center}
  \begin{tabular}{lcc}
    \toprule
    & $\langle L\rangle$ (mm) & $L_{\mathrm{max}}$ (mm) \\[0.5ex]
    \midrule 
    \iroc  & 36.49 & 89.95 \\
    \addlinespace 
    \orocOne & 49.01 & 102.99 \\
    \addlinespace 
    \orocTwo  & 52.99 & 113.69 \\
    \addlinespace 
    \orocThree  & 57.04 & 115.97 \\[0.5ex]
    \bottomrule 
  \end{tabular}
  \end{center}
  \label{tab:padplane:routing}
\end{table}
A vertical 40-pin female connector\footnote{ERNI SMC 1.27\,mm. Product number 354065} is chosen to connect the front-end electronics. Four connectors in radial direction are grouped to connect 160 pads to one front-end card (see \secref{sec:fec}). %

The pad planes are glued onto the Al-bodies (see \secref{sec:alubody}) together with the 5\,mm thick strongback (see \figref{fig:roc.exploded}), forming an integral, compound \textit{chamber body}. The strongbacks are fabricated from Vetronite EGS 103 and provide electrical insulation between pad plane and Al-body, reducing pad capacitance to ground. The same design was used in the original version of the ALICE TPC readout chambers~\cite{TPCnim} and has shown to provide very good gas tightness and mechanical stability. 

\subsubsection{Al-bodies}
\label{sec:alubody}

The Al-bodies are the main readout chamber support structures holding the pad planes and the GEM stacks. The overall design of the Al-bodies is very similar to the one used with the MWPCs~\cite{TPCnim} and follows the dimensions shown in \figref{fig:dimensions}. They are produced from aluminium alloy\footnote{EN AW-5083 (AlMg4.5Mn0.7)}, providing sufficient mechanical stability to prevent deformations due to gravitational forces and the tension of the GEM foils. The chambers are milled from blocks of cast alloy so that the mechanical stress and deformations are minimized. According to finite-element method (FEM) calculations, the resulting maximum deformation of the \oroc Al-body at an operating overpressure of \SI{0.4}{\milli\bar} is less than \SI{2}{\micro\meter}. This assures a precise support and mounting base for the pad planes and the GEM foils. 

The detailed design of the Al-bodies includes cut-outs for the front-end card connectors, feedthroughs for high-voltage cables providing the electric potential to the GEM electrodes, precision holes for mounting the chambers in the TPC endplates, survey targets, and the infrastructure (tapered holes, O-ring grooves) for the final sealing of installed ROCs. As in the previous design (MWPC-based), a copper pipe is installed within the Al-body for water cooling and temperature control~\cite{TPCnim}.

\subsection{Tests with prototypes}
\label{sec:roc:rnd}
The ALICE TPC employs a novel mode of operation for a GEM detector system. The choice of quadruple-GEM stacks with foils characterized by different geometrical properties (such as pitch and relative orientation of the hole pattern) is dictated by the unprecedented challenges in terms of load and performance expected from operation at the LHC. In order to keep space-charge distortions to a tolerable level, the backflow of ions produced in the amplification regions is to be minimized. At the same time, good energy resolution has to be maintained, and stable operation in the harsh LHC environment has to be ensured. In addition, sufficient long-term gain stability has to be guaranteed. As a consequence, dedicated aging and charging-up studies are required.

The baseline solution for the new readout chambers was evaluated and optimized in a series of measurements with small and full-size prototypes during an extensive R\&D phase. In the following, we summarize the relevant outcome from this process, with references to the corresponding publications.

\subsubsection{\Ibf and energy resolution}
\label{sec:rdibf}
In the course of the R\&D campaign, a multiparameter optimization of various triple- and quadruple-GEM systems has been performed~\cite{TDR:tpcUpgrade, TDR:tpcUpgradeAddendum, Ball_2014}. In order to fulfil the requirements, the electric field configuration of the GEM stack and the sharing of the gain among the four amplification stages were optimized. In addition, the optical transparency of the GEMs in a stack has been varied by employing foils with various hole pitches: small (\SI{90}{\micro m}), standard (\SI{140}{\micro m}), medium (\SI{200}{\micro m}) and large pitch (\SI{280}{\micro m}).

Our measurements show that a suitable working region in terms of \ibf and local energy resolution can be achieved by utilizing a quadruple-GEM system in the configuration \mbox{S-LP-LP-S}. This arrangement enables efficient ion blocking by employing asymmetric transfer fields and foils with low optical transparency. Contrary to common practice~\cite{BOHMER2014214}, an increasing sequence of gas gains down the GEM stack is used~\cite{BERGER2017180}. In this way, the majority of ions is produced in regions where they can be effectively blocked. On the other hand, the efficiency for electron transmission, in particular in the first two layers, is also affected by this configuration. Therefore, a parallel optimization of \ibf and energy resolution is mandatory.

After optimization of the voltage settings, a working point was identified with an \ibf of about \SI{0.7}{\percent} at an energy resolution of \SI{12}{\percent} (see \figref{fig:sigmavsIBF}), which is well within the design specifications. The typical HV settings follow the scheme: $\Delta V_{\mathrm{\gemOne}}>\Delta V_{\mathrm{\gemTwo}}<\Delta V_{\mathrm{\gemThree}}\leq\Delta V_{\mathrm{\gemFour}}$. The first two transfer fields ($E_{\mathrm{T1}}$ and $E_{\mathrm{T2}}$) and the induction field (\Eind) are $\mathcal{O}$(\SI{4}{\kilo\volt\per\centi\meter}), whereas $E_{\mathrm{T3}}$ is kept low, $\mathcal{O}$(\SI{0.1}{\kilo\volt\per\centi\meter}), in order to maximize the blocking of the majority of the ions, which are produced in \gemFour. Having a relatively large safety margin for choosing the final HV configuration (see \figref{fig:sigmavsIBF}), all voltages can be further optimized for the stability against electrical discharges (see \secref{sec:discharge}).

Special emphasis was given to the uniformity over the relatively large surface of the detectors.
Simulations indicate that there can be a strong dependence of the \ibf on the relative orientation
of the hole patterns in consecutive GEM layers~\cite{TDR:tpcUpgradeAddendum}. In particular, a notable increase of \ibf may occur if
the GEM holes are aligned. This behaviour can be understood in terms of the low diffusion of ions which makes them follow a given electric field line. In case of displacement, ions emerging from a GEM hole end on the bottom electrode of the GEM above, while they escape through its holes in case of alignment. In the case of electrons, the much larger diffusion diminishes such correlations, and no effect on the gas gain is observed. Therefore, a misalignment of the GEM holes is preferred in terms of minimal \ibf, or at least precautions should be taken to avoid accidental alignment. 

While accidental alignment of two GEM layers over the whole surface is quite improbable, a slight relative rotation of two consecutive layers in the GEM plane can lead to macroscopic interference patterns (Moir\'{e} patterns), featuring regions of high and low optical transparency. Such patterns may lead to variations of the GEM characteristics over the detector surface. On the other hand, relative rotation of two consecutive layers by 90$^{\circ}$ prevents interference patterns, because the GEM hole pattern obeys a 60$^{\circ}$ rotational symmetry. This results in a randomization of relative hole positions over short distances and enhances the uniformity of the detector performance across the active area. For this reason, the masks of subsequent GEMs in the TPC ROCs are rotated by 90$^{\circ}$ with respect to each other (see \tabref{tab:foildet}).

\subsubsection{Discharge stability}
\label{sec:discharge}
Triple-GEM systems operated with so-called standard settings, i.e.\ voltage settings in which the spark discharge probability is minimized, were used successfully in high-rate experiments~\cite{Altunbas:02a, Bencivenni:2002jr, 1462231}. The gain of subsequent GEMs is, in such a case, decreasing towards the bottom of a stack, and thus the probability of exceeding maximum charge limits is significantly reduced~\cite{Bachmann:radon}.

The HV settings optimized for low \ibf values (see \secref{sec:rdibf}), however, violate these rules. Application of such settings in a triple-GEM stack yields a noticeable decrease in detector stability by more than three orders
of magnitude~\cite{Gasik:2018mfu}. However, the measurements reported in~\cite{Gasik:2018mfu} also show that the utilization of a fourth GEM in a stack operated in low \ibf mode restores the stability.

Upper limits for the discharge probability of an \mbox{S-LP-LP-S} system upon irradiation with alpha particles, operated at a gas gain of 2000 with voltage settings that are optimized for low \ibf, are found to be of the order of 10$^{-10}$ per particle~\cite{TDR:tpcUpgradeAddendum}. This number is compatible with the results for a triple-GEM detector operated in standard settings~\cite{Gasik:2018mfu}, which is used as a reference that has proven to work reliably under high-rate conditions. In a test-beam campaign at the CERN SPS fixed target facility, a full-size quadruple-GEM \iroc was exposed to hadron showers from a thick iron target irradiated with a high-intensity secondary pion beam. The measured discharge probability of $(6\pm4)\times10^{-12}$ translates into about 650 expected spark discharges in the whole TPC, or 5 discharges per GEM stack, during one month of \PbPb running at \SI{50}{\kilo\Hz}~\cite{TDR:tpcUpgradeAddendum}. This is expected to cause no damage to the detectors and is compatible with efficient and safe operation of the TPC in Run\,3 and beyond.

In the course of the discharge stability R\&D program, the new detectors have also been optimized with respect to the occurrence of secondary discharges. It was observed that a primary spark discharge in a GEM hole may trigger a secondary discharge in the gap below the GEM~\cite{Bachmann:radon, peskov2009research}. The latter occurs already at electric-field values lower than the amplification field in the given gas mixture with a delay of several microseconds~\cite{DEISTING2019168}.

In order to minimize the probability of a secondary discharge to occur, the recommendations presented in~\cite{Lautner:2019kas} have been taken into account while designing the final HV scheme for the new readout chambers. The most important ones are i)\,implementation of a decoupling resistor $\mathcal{O}(100)$\,k$\Omega$ to each power line providing HV to the GEM electrodes and ii)\,reduction of the transfer and induction field values from \SI{4}{\kilo\volt\per\centi\meter} to \SI{3.5}{\kilo\volt\per\centi\meter}.

The influence of the secondary discharge mitigation on the final HV design and settings and the implications for the TPC operation are discussed in \secref{sec:hv:hv}. The inputs of the preamplifiers of the front-end electronics are also protected against secondary discharges towards the pad plane with a circuitry described in \secref{sec:fec.fec.prot}. 

\begin{figure}[ht]
\centering
\includegraphics[width=\linewidth]{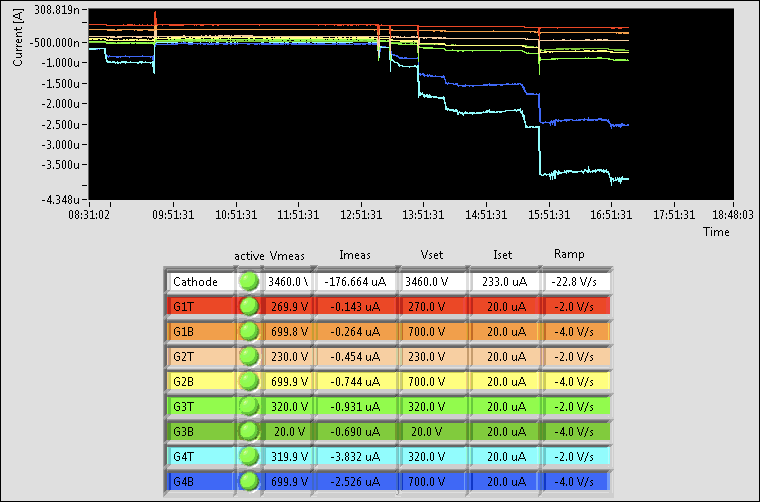}
\caption{Current measurement of \orocOne GEMs during a high-luminosity test at the LHC. The currents of the different HV power supply channels are shown in different colors. The current spikes correspond to the increase of the GEM potentials during ramp-up. The steps in the current readout are related to increases of the luminosity.
The table below shows instantaneous values of the measured voltages (Vmeas) and currents (Imeas), and set values of voltage (Vset), trip limit (Iset) and ramp speed (Ramp).
}
\label{fig:lhc_lumi}
\end{figure}

The overall stability test of the system was performed in the ALICE cavern at the CERN LHC, a few meters away from the interaction point. A preproduction \oroc was installed in a dedicated vessel equipped with a drift electrode and a \SI{11}{\milli\meter} drift gap (used later for the ROC quality assurance with the production chambers, see \secref{sec:roc:rocqc}). The baseline HV settings and the final hardware for the HV supply and distribution were used (see \secref{hv:hv:overview}). The pad plane of the chamber was grounded and all currents on the GEM electrodes were monitored by the built-in current meters of the HV power supplies. \Figref{fig:lhc_lumi} shows the trend of the currents in one stack during a luminosity scan performed with proton beams, where an interaction rate of 4.3\,MHz was reached in steps. The GEM voltages were also ramped up in steps, leading to the clearly visible current excursions. Otherwise, the currents on the GEM electrodes follow the luminosity, the largest ones being, as expected, those on the \gemFour top electrode. 
The particle loads impinging on the chamber at this large rapidity were substantially higher than those expected in LHC Run\,3~\cite{TDR:tpcUpgrade}, and thus GEM currents reached, with a short drift gap, \SI{4}{\micro\ampere}. These currents are indeed comparable with what the chambers would draw at midrapidity with the full TPC drift. The chamber behaved stable during the irradiation at the highest luminosity over several hours. This confirms that the chosen detector configuration and operational conditions are well suited for stable operation in the LHC environment.

\subsubsection{Aging}
Aging of gaseous detectors is well documented to happen if the gas mixture and/or the materials in contact with the gas are not chosen appropriately. In particular, outgassing from insulating materials may lead to aging effects in the form of gain loss, deterioration of the energy resolution, or Malter currents. Therefore,
long-term irradiation tests on $10\times10$\,\si{\centi\meter\squared} quadruple-GEM stacks were carried out with samples of various construction materials in an outgassing box placed upstream of the detector using an \ArCOtwoThirty gas mixture~\cite{jung:bsc:19}. The tests consisted of irradiating a spot of the detector for over a month in order to accumulate a charge of several tens of \si{\milli\coulomb\per\cm\squared}. No aging effects were observed with the tested samples. These include Araldite 2011 epoxy~\cite{ref:araldite}, SMD resistors soldered to a piece of GEM foil material, Vetronite pieces used for GEM frames, and polyamide studs and nuts for GEM fixing. Tests with the nominal mixture \NeCOtwoNtwo were also successfully carried out. However, it was noted that at substantially higher gains (order of 20000 and above) removal of the kapton in the GEM holes takes place as the irradiation proceeds, resulting in a deterioration of the detector performance. It is not yet fully understood which physical mechanism leads to this effect.

\subsubsection{Charging-up after ramping}
\label{sec:rnd:chargeup}
GEMs experience charging-up effects when powered up, leading to gain variations with a given time constant~\cite{BOUCLIER199750}. This is due to the rearrangement of charges within the polyimide upon power-up, and further charge-up processes due to the irradiation itself.
Therefore, measurements were carried out to study the gain behaviour of GEMs when ramped up from an intermediate to the full voltage under irradiation, which is the \textit{modus operandi} of the TPC at the LHC.
These measurements~\cite{yannik:bsc:15} show that in a $10\times10$\,\si{\centi\meter\squared} quadruple-GEM stack, after ramping up the GEM voltages from \SI{70}{\percent} to \SI{100}{\percent} of the nominal settings under an ALICE-equivalent irradiation, a gain increase of \SI{25}{\percent} occurs in about ten minutes. More detailed studies~\cite{Hauer:2019lms} demonstrated the charge rate dependence of this time scale, which in any realistic scenario is sufficiently slow for the foreseen online calibration procedures to correct for this effect.

\subsubsection{Particle identification performance}
\begin{figure}[ht]
\centering
\includegraphics[width=0.85\linewidth]{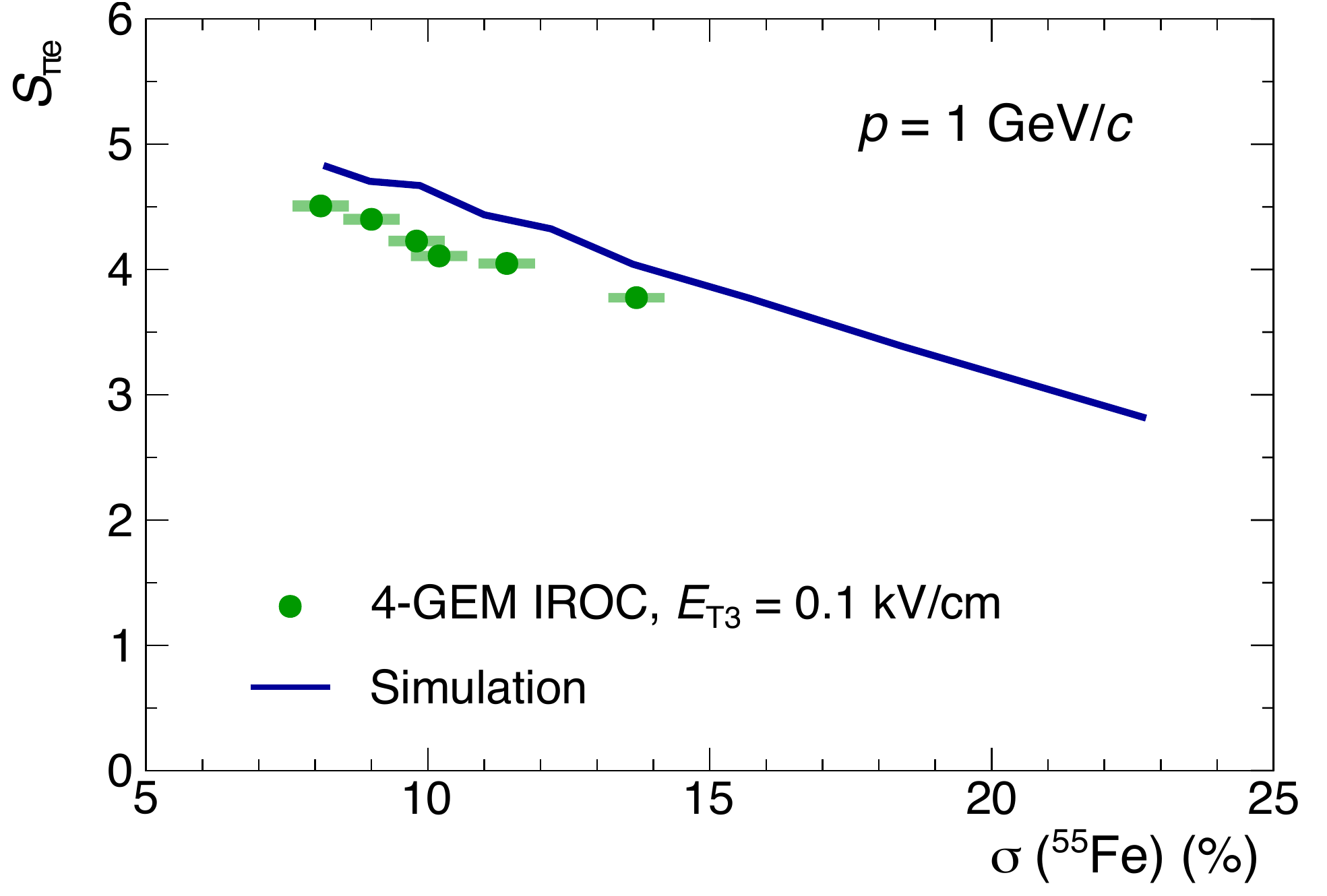}
\caption{Separation power for pions and electrons at $p=\SI{1}{\giga\electronvolt/\clight}$ measured with an \iroc prototype at a gain of 2000. The data points are recorded at different GEM HV settings that correspond to different values of $\sigma(^{55} \mathrm{Fe})$ and \ibf. The data points (from left to right) cover a range in \ibf between about \SI{2.5}{\percent} and \SI{0.5}{\percent} (see \figref{fig:sigmavsIBF}). Horizontal error bars correspond to the uncertainty of $\sigma(^{55} \mathrm{Fe})$ measurement with small prototypes. Data points and simulation are taken from~\cite{AGGARWAL2018215}.}
\label{fig:PIDprotoype}
\end{figure}

In order to validate the performance of a large-size detector equipped with a stack of four GEMs in terms of \dEdx resolution, a prototype \iroc was tested at the CERN Proton Synchrotron~\cite{TDR:tpcUpgradeAddendum, AGGARWAL2018215}. The main result of this campaign is shown in \figref{fig:PIDprotoype}.

The detector performance in terms of the separation power $S_{\mathrm{\pi e}}$ between electrons and pions was studied for different GEM HV settings. Each HV setting provided a total gain of 2000, but resulted in different values of \ibf and local energy resolution $\sigma(^{55} \mathrm{Fe})$, as confirmed in measurements with small prototypes (see \secref{sec:rdibf} and Ref.~\cite{TDR:tpcUpgradeAddendum}). The separation power is defined as
\begin{equation}
  S_{\mathrm{AB}} = \frac{2\left|\MeandEdx_{\mathrm{A}}-\MeandEdx_{\mathrm{B}}\right|}
    {\sigma\left(\dEdx\right)_{\mathrm{A}}+\sigma\left(\dEdx\right)_{\mathrm{B}}}
  \label{eq:sepa}
\end{equation}
for two particle species $\mathrm{A}$ and $\mathrm{B}$. Here, $\langle\dEdx\rangle$ is the mean of the specific energy loss distribution and $\sigma(\dEdx)$ the standard deviation, both determined from Gaussian fits to the truncated energy loss spectra.

As expected, the \dEdx resolution for pions and electrons, and thus the separation power, degrades with worsening local energy resolution (increasing $\sigma(^{55} \mathrm{Fe})$ value), however, the dependence is rather weak and no abrupt change is observed. The measurement is compared with a microscopic simulation, in which the effective primary electron efficiency of the readout system is varied. This leads to a degradation of the $\sigma(^{55}\mathrm{Fe})$ and hence the pion-electron separation power, which reproduces the data trend very well. A small constant offset of about $\Delta S_{\mathrm{\pi e}} = 0.5$ between data and Monte Carlo points to a residual systematic contribution in the prototype data that is not described by the simulation.
The results demonstrate that the resolution is compatible with that of the previous version of the TPC~\cite{TPCnim}. The design thus fulfils the requirements for the upgrade.

\subsection{GEM foil production and quality assurance}
\label{sec:roc:gem_production}

The production and quality assurance of the GEM foils were embedded in the overall material and production flow that is described in \secref{sec:roc:roc_assembly}.
The GEM foils were produced by the CERN Micro-Pattern Technologies laboratory~\cite{CERN-MPT}.
After a Basic Quality Assurance (QA, see \secref{sec:roc:basicqa}), the GEM foils were extensively tested at two Advanced QA centers (see \secref{sec:roc:advanced_qa}).

\subsubsection{GEM production}
\label{sec:roc:gemprod}
GEM foils were produced in pairs on $600\times1700\,\text{mm}^2$ sheets combining $\text{\iroc}+\text{\orocThree}$ and $\text{\orocOne}+\text{\orocTwo}$ foils. Due to technological reasons, only the foils with the same hole pitch could be combined together on a single sheet. The initial production rate of 20 sheets per month was increased to 30 sheets per month during the final production phase. The sheets were cut into individual GEMs, which were labeled with a unique bar code. 
Each GEM was equipped with \SI{5}{\mega\ohm} SMD protection resistors (see \secref{sec:roc:gem} for more details). All soldering was done using leaded, no-clean solder wire\footnote{Almit KR 19SH RMA \SI{2.2}{\percent}, Sn-36.0Pb-2.0Ag}~\cite{almit:wire}. After soldering, the foils were cleaned in order to remove corrosive remains of the solder flux, and tested for HV stability. The mechanical quality of soldering was checked with plastic tweezers by gently pulling the SMD elements. 

\subsubsection{Basic QA}
\label{sec:roc:basicqa}
All GEM foils were subjected to optical inspection and a measurement of the leakage current ($I_{\mathrm{leak}}$) prior to each production step, in order to reject malfunctioning GEM foils at the earliest possible stage. In the former, an operator checked visually the quality of a foil on a light table. Defects that were visible by eye (such as scratches, over-etched holes, large areas of missing holes, etc.) were noted in a database (see \secref{sec:database}). Severe defects, such as cuts or protruding pieces of copper, led to the rejection of a foil. 

The leakage-current measurement was conducted using a single-channel HV source and a multichannel pA-meter system~\cite{UTROBICIC201521}. The test was performed in a gas-tight Macrolon box, in dry nitrogen (\SI{<0.6}{\percent} of absolute humidity). The HV was applied to each segment separately using spring-loaded pins. During the very first Basic QA procedure, just after the foils were received from the producer, \SI{550}{\volt} was applied directly to the foil. This was done in order to increase the current flowing through residual dust particles confined in the GEM holes, which eventually melt and evaporate (\textit{HV cleaning}). Afterwards, the voltage was decreased to \SI{500}{\volt} and the leakage current was measured for fifteen minutes. For any further leakage current measurement, performed in the course of the readout chamber production, the first step at \SI{550}{\volt} was skipped.

\begin{figure}[t]
\centering
\includegraphics[width=0.9\linewidth]{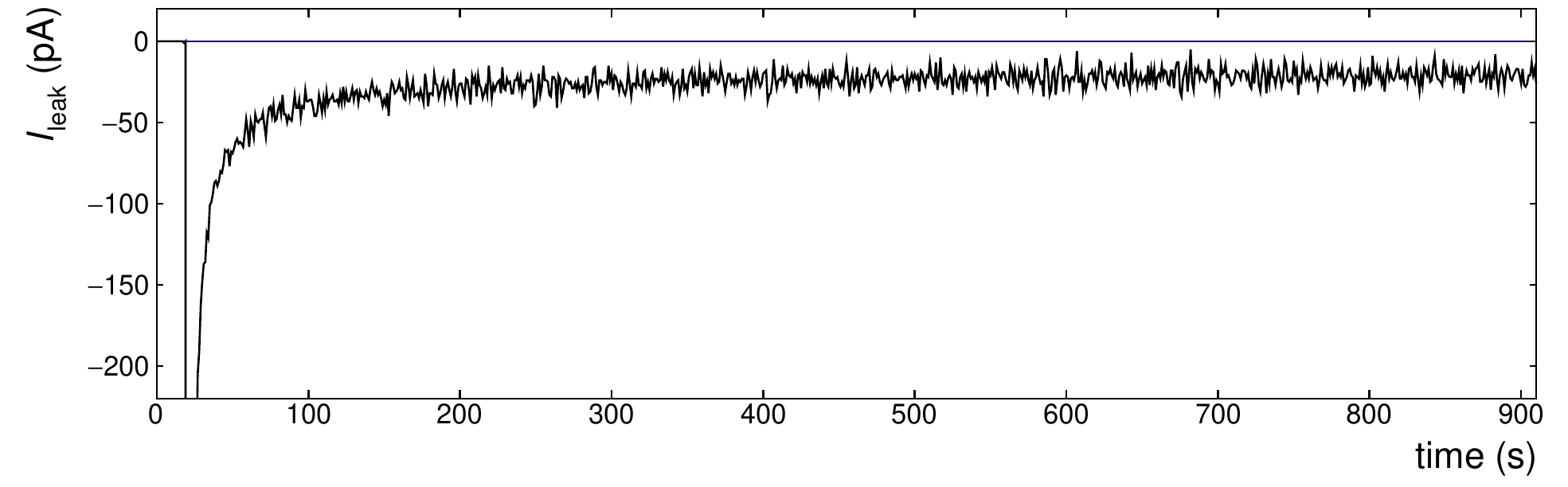}
\caption{Typical leakage current measurement of a GEM segment.}
\label{fig:ileak}
\end{figure}

\Figref{fig:ileak} shows an example of leakage current measurement for a single GEM segment. 
Each measurement was uploaded to a database for evaluation (see \secref{sec:database}). A foil was considered good if all of its segments showed an average leakage current in the interval [450\,s,~900\,s]  below \SI{0.5}{\nano\ampere}. It is worth mentioning that the typical values measured in the course of the production were well below \SI{0.1}{\nano\ampere} per segment.

If a foil did not pass the test, it was cleaned with a tacky roller to remove residual dust and the test was repeated. If the foil failed repeatedly, it was sent back to the producer for cleaning, which reduces the leakage current to a level well below the acceptance limit in most of the cases. Otherwise, the foil was discarded.

In addition to the leakage current, also the sparking rate was monitored during the test. The foils spark usually immediately after applying a potential difference, but the spark rate should drop to zero with time (although spurious discharges may occur at any moment). If a foil experienced continuous sparking, the test was interrupted and the foil was cleaned with a tacky roller. In case of further failures, the foil was sent back to the producer for cleaning (see above). Details of the Basic QA procedures, equipment and protocols are discussed in~\cite{Ball_2017}.

\subsubsection{Advanced QA}
\label{sec:roc:advanced_qa}

At the Advanced QA sites, a long-term HV test, high-definition (HD) optical scanning, and gain mapping of the GEM foils were performed. The main goal of these steps was to thoroughly monitor the GEM foil quality, i.e.\ to spot raw-material and production flaws at the earliest possible stage. As the Advanced QA required a significant amount of time to be performed, two QA centers were established to sustain the necessary production rate. A detailed discussion of the Advanced QA methods can be found in~\cite{Brucken:2018rej, Brucken:2018EPJ, Hilden:2018isz}.

A leakage-current test, equal to that executed in the Basic QA step (see \secref{sec:roc:basicqa}), but with increased duration of at least five hours, was performed as part of the Advanced QA. The same acceptance criteria were applied. The purpose of the long-term stability test was to maximize GEM stress under HV and reveal potential issues which can appear after longer operation (e.g.\ enhanced sparking rate). Foils which did not pass this QA step underwent the cleaning procedure (see \secref{sec:roc:basicqa}).

The HD optical scanning system for the GEM QA consisted of an $xyz$-robot mounted on a glass table, a high-definition camera with telecentric optics to avoid distortions and parallax errors, and lighting equipment~\cite{HILDEN2015113, Brucken:2018rej}. %
The GEM foils were scanned with a two-exposure mode, in which the background-light exposure was used for the reconstruction of the inner polyimide holes whereas the foreground-light exposure was used for the copper holes. For the largest foil, \orocThree, up to 3600 pictures were taken per side, which took approximately 2 hours~\cite{Brucken:2018rej}. Each foil was photographed from both sides. During the scanning process, the foils were stretched using a stencil frame system (see more in \secref{sec:roc:framing}) for efficient focusing and substantial reduction of the scanning time. 
\begin{figure}[t]
\centering
\includegraphics[width=\linewidth]{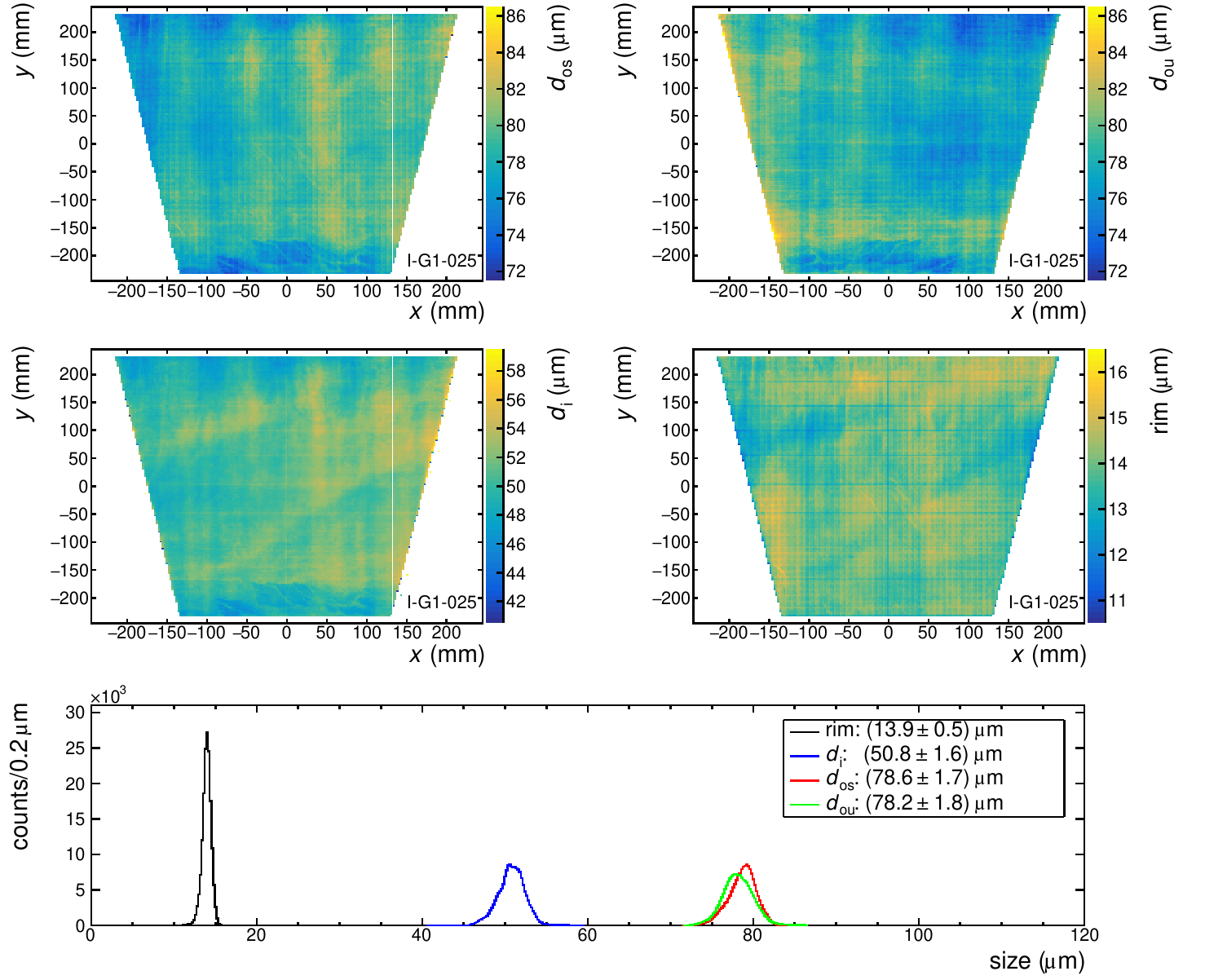}
\caption{High-definition scan of an \iroc \gemOne foil. The upper panels show the 2D distribution of the outer hole diameter for the segmented (left) and unsegmented (right) side of the foil. The middle panels show the inner hole diameter (left) and the rim width (right). The bottom panel shows the 1D projections of the measured values with mean and standard deviations. The regular structures visible in the 2D distributions are related to the production process. They have no influence on the quality of the foil.}
\label{fig:2Dholemap}
\end{figure}

The reconstruction of the images, the classification of the identified patterns into regular holes or defects, and the analysis of the data was done using a custom-made software package described in~\cite{HILDEN2015113, Brucken:2018rej}. From the resulting output, maps and distributions of the positions and dimensions (i.e.\ inner and outer hole diameter) of all GEM holes were generated. Example plots are presented in \figref{fig:2Dholemap} for an \iroc standard-pitch foil, where two-dimensional distributions of the outer ($d_{\text{os}}$ and $d_{\text{ou}}$, for the segmented and unsegmented side, respectively) and inner ($d_{\text{i}}$) hole diameters are shown together with their one-dimensional projections. In addition, the rim size, defined as the difference between the outer and inner hole diameter, is plotted for the segmented side. Average hole dimensions and standard deviations of their distributions are also evaluated. 

The scanning results were used to constantly monitor the quality of the foils. Large deviations from the mean values or structures visible in several foils of the same batch could potentially point to production flaws or problems with the raw material. Based on the scan results, fast feedback could be given to the producer.

\begin{figure}[t]
\centering
\includegraphics[width=0.8\linewidth]{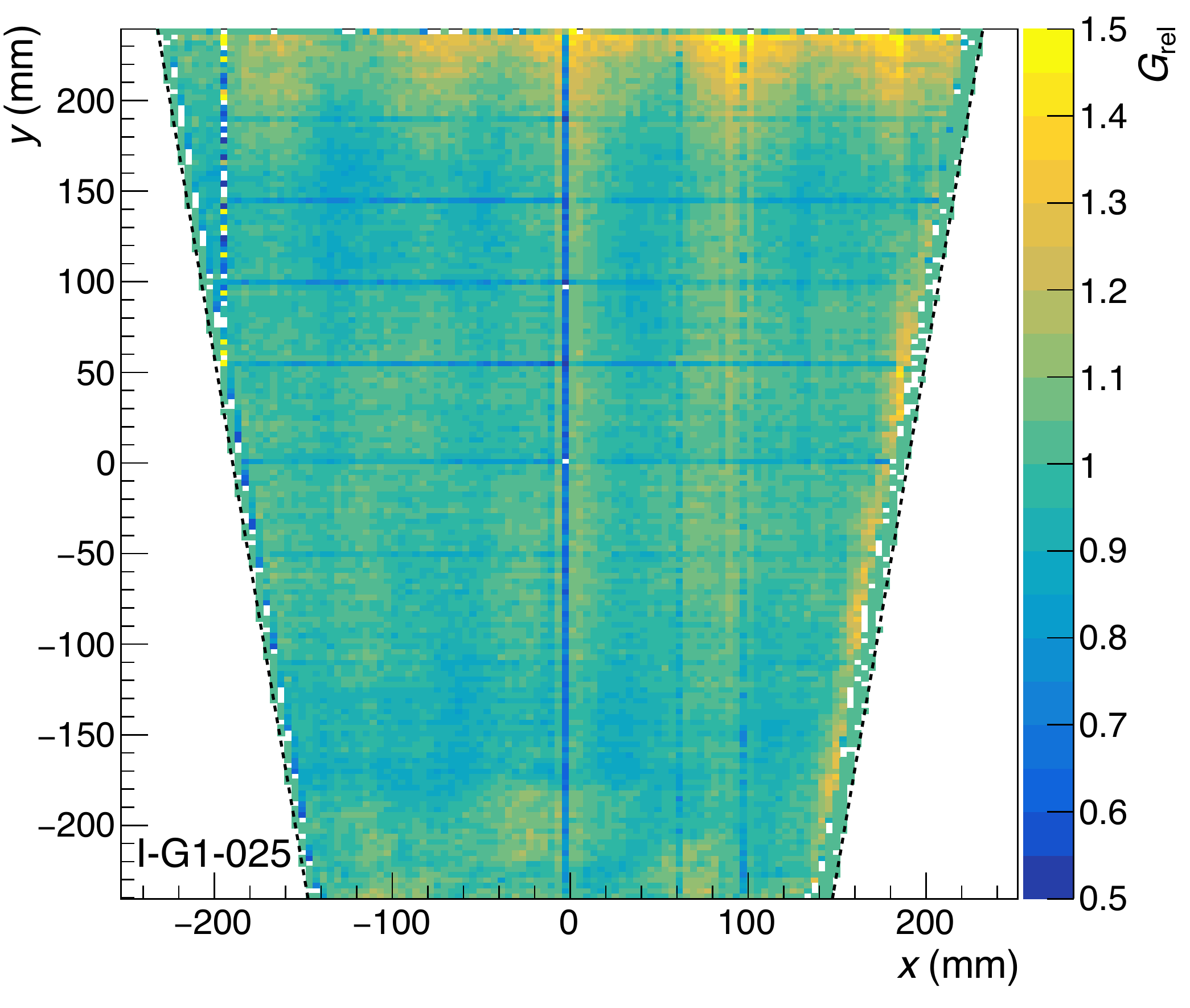}
\caption{Gain scan of an \iroc GEM foil, same as in \figref{fig:2Dholemap}.}
\label{fig:gain_scan}
\end{figure}

The two-dimensional hole-size distributions can also be used to predict the response of individual GEMs. It has been shown that a qualitative estimation of the behavior of the local variation in the gain across the GEM foil can be made based on the measured sizes of the outer and inner holes~\cite{HILDEN2015113}. For this reason, the gain uniformity of several foils from the same production batch was measured in order to further study this correlation. Gain uniformity scans were performed using a dedicated setup employing an MWPC with strip readout, following the procedure described in~\cite{Hilden:2018isz}. During a measurement, the entire GEM foil area was illuminated with a non-collimated $^{55}$Fe source. \Figref{fig:gain_scan} shows the relative gain ($G_{\mathrm{rel}}$) measured across the same \iroc GEM foil as in \figref{fig:2Dholemap}. 

\begin{figure}[t]
\centering
\includegraphics[width=0.74\linewidth]{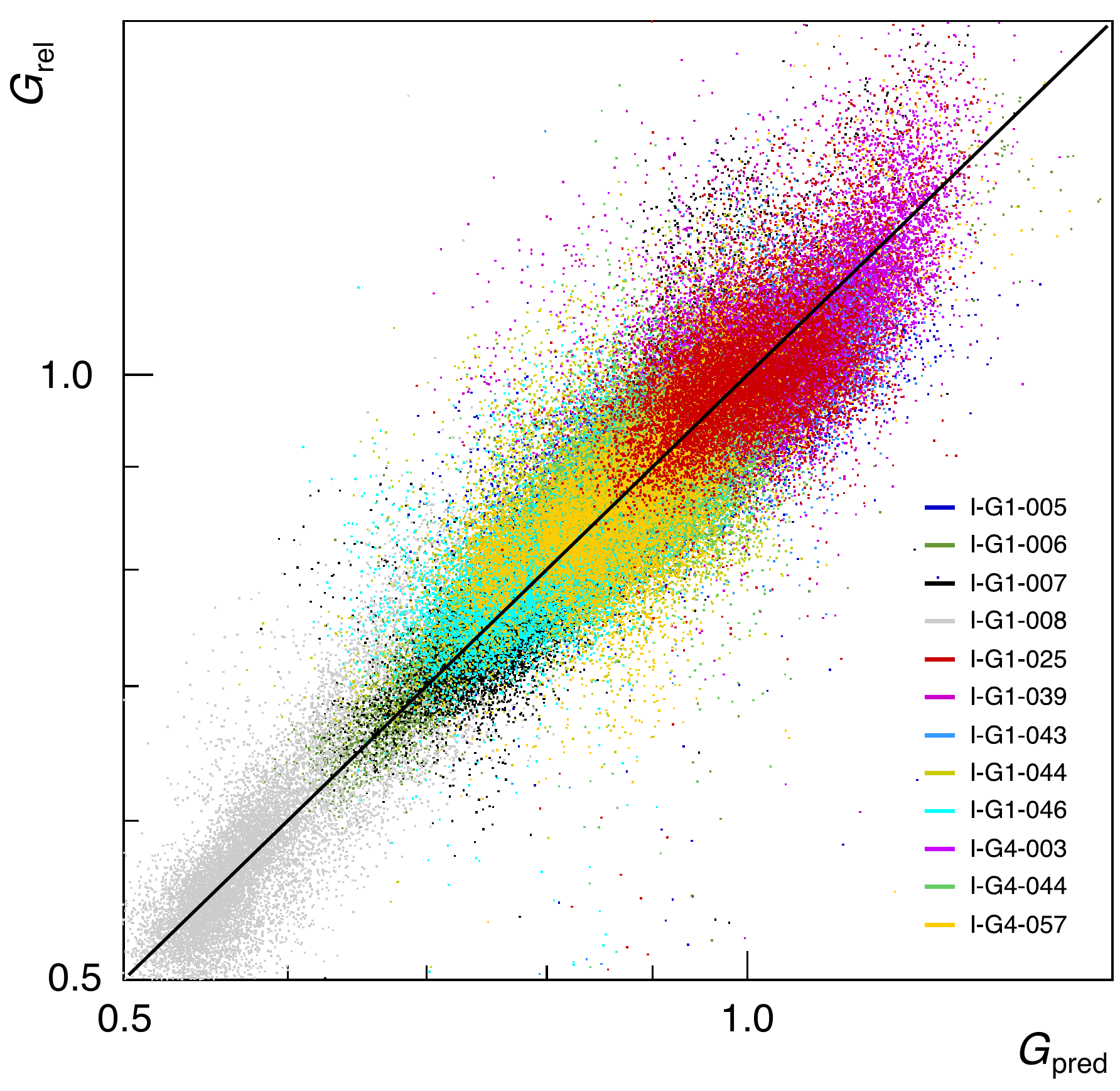}
\caption{Correlation between the measured relative gain and the gain prediction based on the hole-size measurements, given by Equation~\eqref{eq:gain:prediction} . Different colors represent data from different GEM foils. The solid line indicates $G_{\mathrm{rel}}=G_{\mathrm{pred}}$.
}
\label{fig:gain_correlation}
\end{figure}

Similar features can be observed in the distributions, which points to a correlation between gain and hole diameters. This is presented quantitatively in \figref{fig:gain_correlation}, where the measured $G_{\mathrm{rel}}$ values are plotted against the predicted gain ($G_{\mathrm{pred}}$) determined by a single slope parameter $G_0$ and a normalization constant $d_0$:
\begin{equation}
\text{ln}(G_{\text{pred}}) = -G_0(d_{\text{o-mean}} - d_0),
\label{eq:gain:prediction}
\end{equation}
where $d_{\text{o-mean}}=(d_{\text{os}}+d_{\text{ou}})/2$ is the average value of outer hole diameters measured from both sides of a foil and $G_0$ is found to be \SI{0.046}{\per\micro\m} for standard-pitch foils~\cite{VargyasPhd}. A clear correlation between the measured and predicted gains can be found for a number of different foils. This allows one to model the gain (and at the same time the gain uniformity) for any single foil. In principle, the gain uniformity of a quadruple-GEM stack can be modeled based on the data obtained from the optical scans of the single GEMs. However, further studies are required for a complete understanding of all effects.

The results of the optical scans were used to select the foils that form a stack. Namely, GEMs featuring similar non-uniformity patterns, especially in the same region, were not put together in order to avoid large gain variations in the assembled detector. Indeed, the gain uniformity of all produced chambers is well within the specifications (see \secref{sec:roc:gain_uniformity}).

\begin{figure}[t]
\centering
\vspace{0.5cm}
\includegraphics[width=0.8\linewidth]{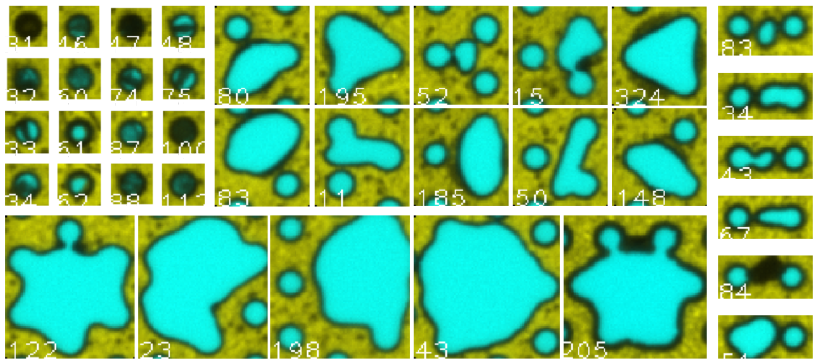}
\caption{Collage of typical defects identified during the Advanced QA, including blocked (16 images at top left) and over-etched holes.}
\label{fig:defectmap}
\end{figure}

In addition to the hole-size measurements, a convoluted neural network was used to recognize the defects. The latter were compiled into defect maps (examples are shown in \figref{fig:defectmap}) and uploaded to a database for each scanned GEM foil. However, these GEMs were not rejected from further processing because they did not show any limitation in terms of stability or performance. This confirms the efficiency of the visual selection of GEMs at identifying severe defects already at the Basic QA stage (see \secref{sec:roc:basicqa}).

\subsubsection{Foil yield}
In total, 829 GEM foils were produced in the course of the project. The final yield of good foils after all production and QA steps was $\sim$\SI{91}{\percent}. %
The average values of hole dimensions are summarised in \tabref{tab:holesize-rms}. The numbers are well within the specification given by the producer. It is, however, important to mention that no tight constraints were defined for an average hole diameter. The values were monitored throughout the production period to assure stability of the hole dimensions. The latter can be evaluated by looking at the foil-to-foil variation of the mean diameters and rim size. The variations in the mean values are mostly associated with different production batches. The variations within a given foil are significantly smaller. As the result of the single mask technology used for the GEM production, the variations of the standard deviations (for single foils) of the outer diameter and rim size values are slightly larger in the unsegmented (bottom) side than in the segmented (top).

\begin{table*}[tbp]\footnotesize
 \caption[Hole size and standard deviations]{Average hole size and standard deviations.}
  \begin{center}
  \begin{tabular}{ccccc}
    \toprule
    \multirow{2}{*}{GEM side} & \multirow{2}{*}{type} & \multirow{2}{*}{mean} & variation within foil & foil-to-foil variation\\
    & & & (std. dev.) & (std. dev.)\\[0.5 ex]
    \midrule 
    \multirow{3}{*}{segmented} & inner $\diameter$ (\textmu m) & 50.3 & 2.2 & 10.9 \\
    & outer $\diameter$ (\textmu m) & 74.4 & 2.1 & 15.4 \\
    & rim (\textmu m) & 12.1 & 0.6 & 3.1  \\[0.5ex]
    \midrule
    \multirow{3}{*}{unsegmented} & inner $\diameter$ (\textmu m) & 50.2 & 2.2 & 10.7 \\
    & outer $\diameter$ (\textmu m) & 75.4 & 2.8 & 15.4 \\
    & rim (\textmu m) & 12.6 & 0.8 & 3.3 \\[0.5 ex]
    \bottomrule 
  \end{tabular}
  \end{center}
  \label{tab:holesize-rms}
\end{table*}

\subsection{Assembly overview}
\label{sec:roc:roc_assembly}

\begin{figure}[b]
\centering
\includegraphics[width=0.9\linewidth]{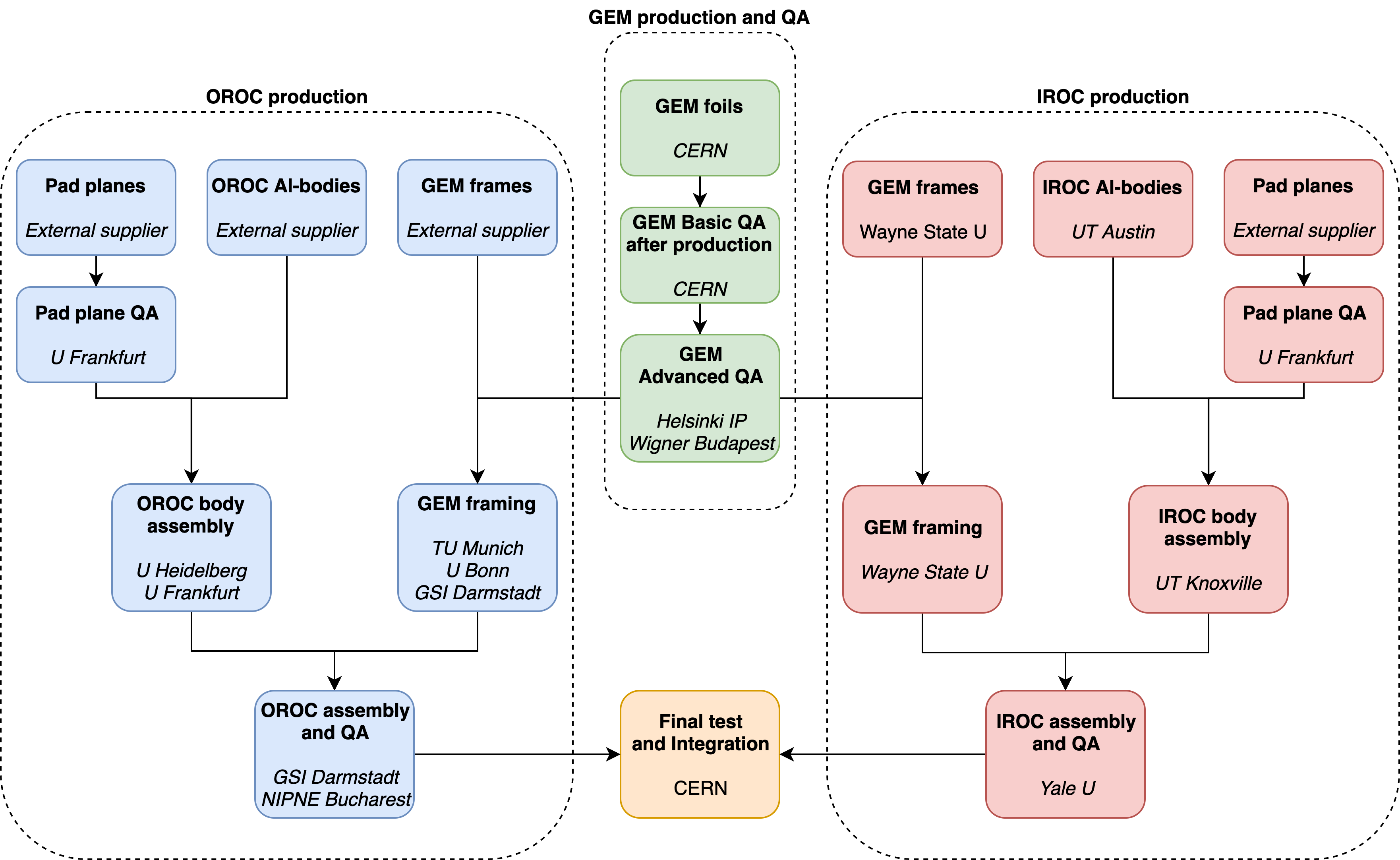}
\caption{Material flow of the readout chamber production.}
\label{fig:matflow}
\end{figure}

The mass production of the readout chambers followed strict protocols, which are outlined hereunder.
The assembly procedure was validated by producing several prototypes and, in particular, two \iroc and two \oroc preproduction chambers~\cite{Gasik_2014, GASIK2017222, Mathis:2018sjk}.
The overall material and production flow is presented in \figref{fig:matflow}.
After GEM production and QA (see \secref{sec:roc:gem_production}) the foils were framed (see \secref{sec:roc:framing}) and sent to the assembly sites, where they were mounted onto the preassembled readout chamber bodies (see \secref{sec:roc:roc:ass}).
The assembled detectors were then qualified (see \secref{sec:roc:rocqc}) and sent to CERN for acceptance tests, storage, and installation in the TPC (see \secref{sec:roc:at:CERN}).

\subsubsection{GEM framing}
\label{sec:roc:framing}
The GEM frames were assembled from four ledges and a spacer cross (see \secref{sec:frames}) in a precision jig. All parts were thoroughly cleaned in an ultrasonic bath and inspected under the microscope prior to the assembly. The pieces were glued together with an epoxy resin (Araldite 2011~\cite{ref:araldite}) and left for curing overnight.

\begin{figure}[h]
\centering
\includegraphics[width=0.8\linewidth]{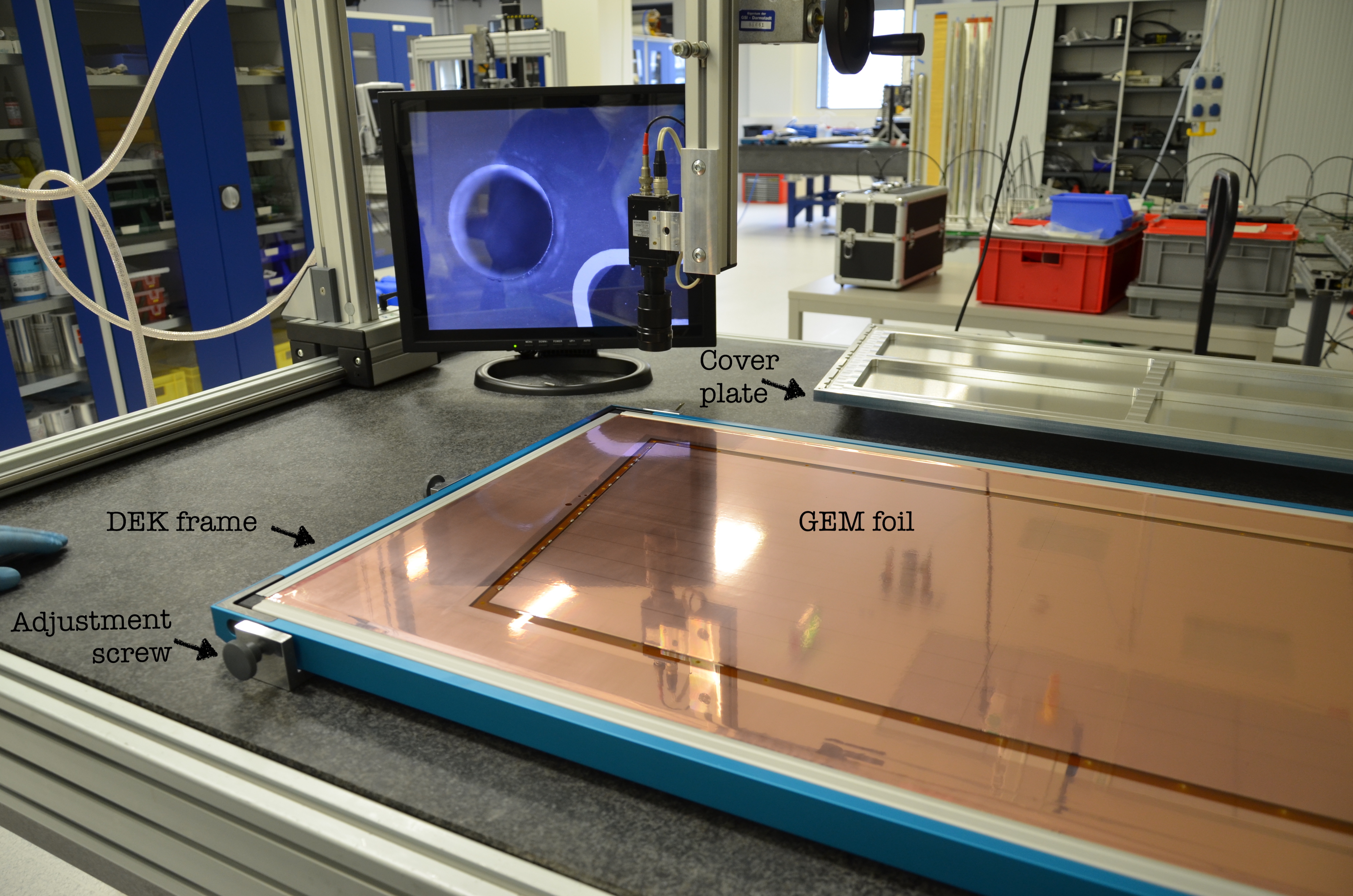}
\caption{Stretched GEM positioned on a framing jig.}
\label{fig:gem:framing}
\end{figure}

Before being glued onto the preassembled frame, the GEM foils were stretched with a tension of \SI{10}{\newton\per\centi\meter} using a modified DEK VectorGuard\textsuperscript{TM} stencil frame system~\cite{DEKframe}. The GEM frame was then covered with epoxy (same as for the frame assembly) using a rubber roller, and placed in the gluing jig. Afterwards, the foil was positioned \SI{0.5}{\milli\meter} above the frame (see \figref{fig:gem:framing}) and aligned using adjustment screws. 
The gluing jig was closed with a heavy cover plate milled such that the active area of the foil was not touched, and burdened with steel bricks to press the foil onto the frame. The epoxy was cured overnight in dry atmosphere (below \SI{30}{\percent} relative humidity). After that, the foil was removed from the DEK frame and tested (Basic QA procedure). Finally, a set of framed foils were packed and sent to the chamber assembly sites.

\subsubsection{Chamber body assembly}
\label{sec:roc:body:ass}
A chamber body consists of a pad plane and a strongback (see \secref{sec:roc:padplane}) glued together on an Al-body (see \secref{sec:alubody}). 
Before a chamber body was assembled, the pad planes--equipped with the front-end connectors (see \secref{sec:roc:padplane})--underwent a QA procedure. A connectivity test was performed by means of a  pad-to-ground capacitance measurement, using a custom-made test probe. In this way, bad contacts (low capacitance) or shorts between two pads (high capacitance) could be spotted and repaired. In order to increase the sensitivity of the measurement, the pad planes were put on a waterbed, making use of the high value of water relative permittivity, $\epsilon_r\approx80$. Only pad planes with all pads connected properly were accepted.

The gluing of the chamber body components was performed on a vacuum table with the pad plane facing the table surface. In this way, a very good flatness of the pad planes could be assured, in particular for the \orocs, which accommodate three pad planes on one Al-body. Precision holes and pins assured proper alignment of all components.

\begin{figure}[h]
\centering
\includegraphics[width=0.7\linewidth]{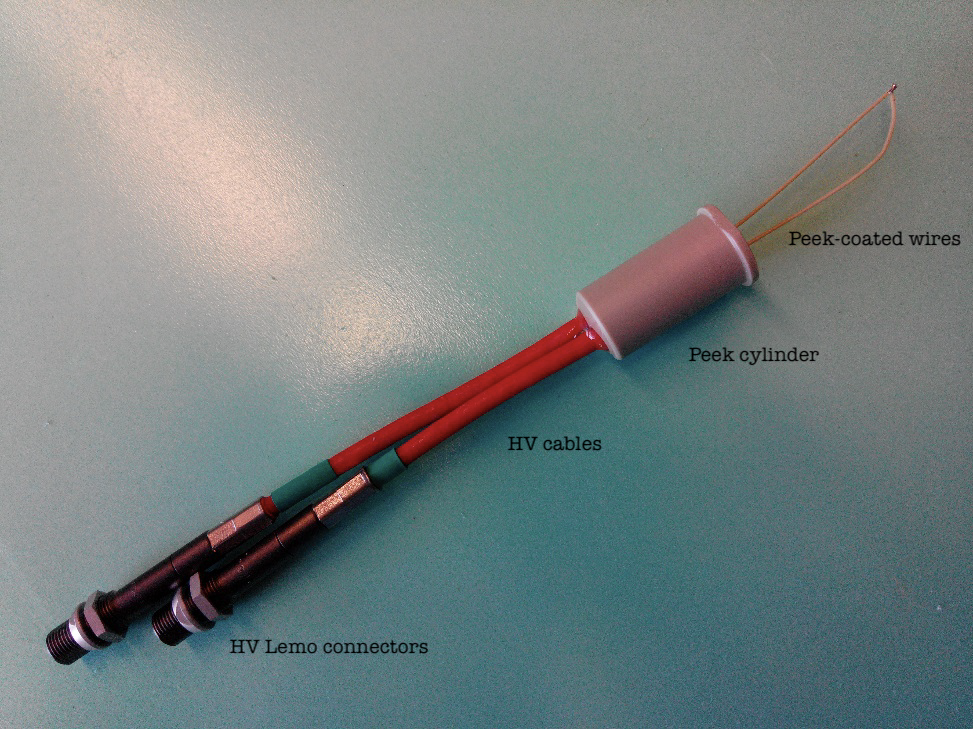}
\caption{HV plug prepared for the installation in a feedthrough.}
\label{fig:HVplug}
\end{figure}

The next step concerned the gluing of the HV plugs into feedthroughs machined into the Al-bodies. The plug is a peek cylinder with two HV cables glued inside in two separate channels (see \figref{fig:HVplug}). Each cable is made of two pieces soldered inside the peek tube to avoid gas leaks.
The HV cables provide the potentials to the GEM electrodes, thus four feedthroughs and plugs are necessary to provide potentials to all GEMs in a stack. High-voltage Lemo sockets installed on one end of the cables (see \secref{hv:hv:hvcomponents} for more details) were fixed to the Al-body after gluing. Peek-coated wires on the other side were soldered directly to the GEMs during the final ROC assembly step (see \secref{sec:roc:roc:ass} and \figref{fig:roc:gem:stack}).

After installation of the HV plugs into the feedthroughs, the chamber bodies were installed in gas-tight boxes and flushed with nitrogen for a leak test. The tightness was calculated from the O$_2$ content as measured at the gas outlet. Typical leak rates were below \SI{0.1}{\milli\litre\per\hour} at \SI{1}{\milli\bar} overpressure. This was the first of a series of four leak tests performed with each chamber (see \secref{sec:roc:gas:tightness} and \secref{sec:roc:qc:transport} for further details).

\begin{figure}[h]
\centering
\includegraphics[width=0.8\linewidth]{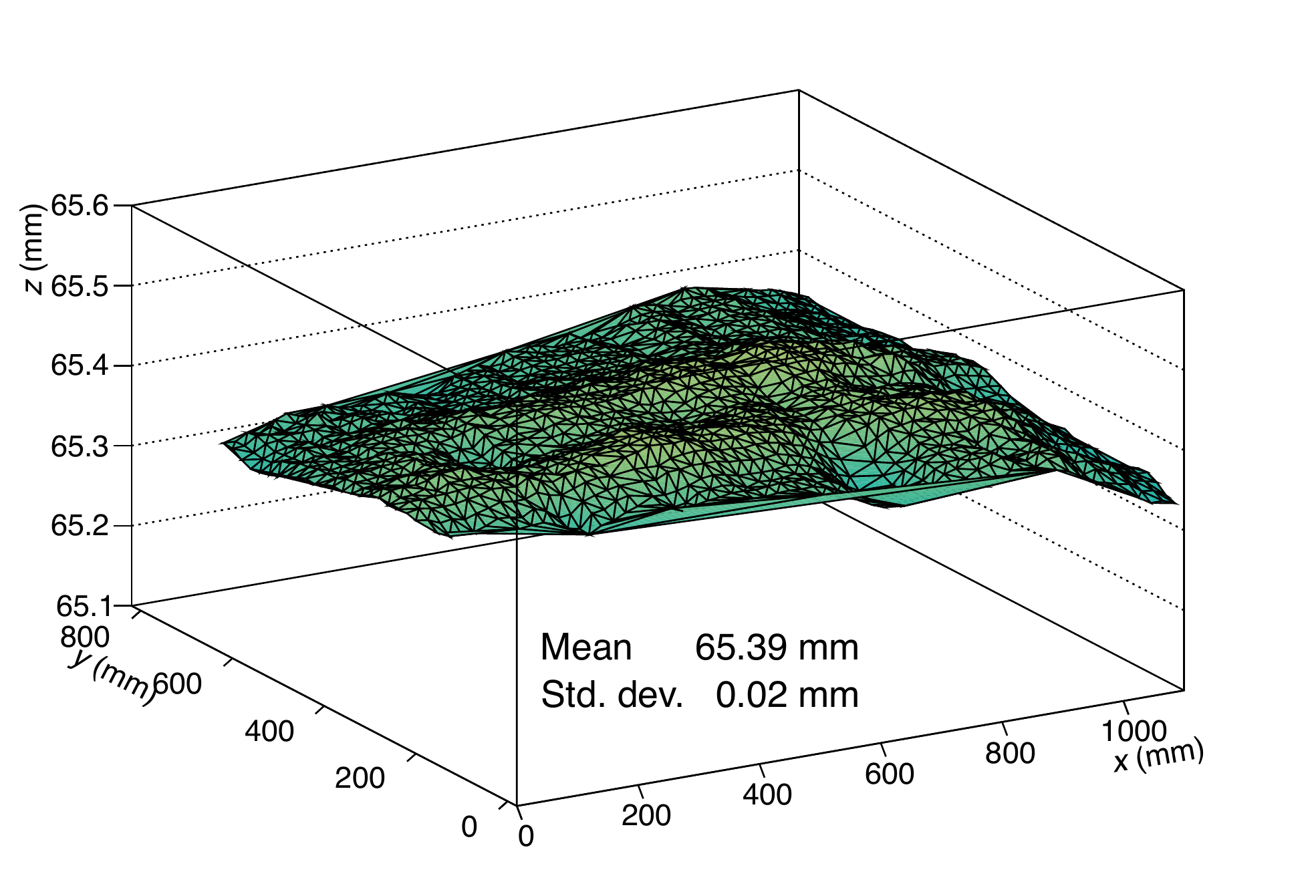}
\caption{\oroc pad plane planarity measurement in $z$-direction.}
\label{fig:roc:planarity}
\end{figure}

In the next step, the pad plane planarity was measured at predefined positions. For this purpose, reference points were precisely machined into the Al-bodies. These data were later used for the final survey of the TPC endplates and readout chambers (see \secref{sec:install:roc} for more details). In order to check the precision of the chamber body gluing procedure, the planarity of several \oroc pad planes was measured using a tactile coordinate measurement machine. The results for one chamber are presented in \figref{fig:roc:planarity}.
The width of the pad plane height distribution is \SI{20}{\micro\meter}, which is well within the specifications. 

In the last step of the chamber body assembly, 8\,mm deep holes with 3\,mm diameter were drilled into the pad planes along their circumference to house the threaded studs for the GEM stacks (see \secref{sec:roc:roc:ass}). These M3 studs, made of \SI{25}{\percent} glass-fiber reinforced polyamide, were glued into the holes. Finally, each chamber body, identified with a unique serial number, was mounted on a sealing flange, which in turn was closed with a gas-tight test/transport vessel (see \secref{sec:roc:trabox}) for shipment to one of the three ROC assembly sites (see \figref{fig:matflow}).

\subsubsection{ROC assembly}
\label{sec:roc:roc:ass}
The final assembly was preceded by a number of quality checks and cleaning procedures. Chamber bodies, from then on mounted on their flanges, were thoroughly cleaned and all pads were grounded for a pad connectivity test upon arrival. The HV integrity of the feedthroughs was tested at 5\,kV in air. After having passed the Basic QA procedure (see \secref{sec:roc:basicqa}), the framed GEMs were trimmed to their final size with the use of a scalpel. The surfaces of pad planes and GEMs were checked with a UV lamp, prior to the assembly. The residual dust was blown off with compressed air or N$_2$. 

\begin{figure}[ht]
\centering
\includegraphics[width=0.9\linewidth]{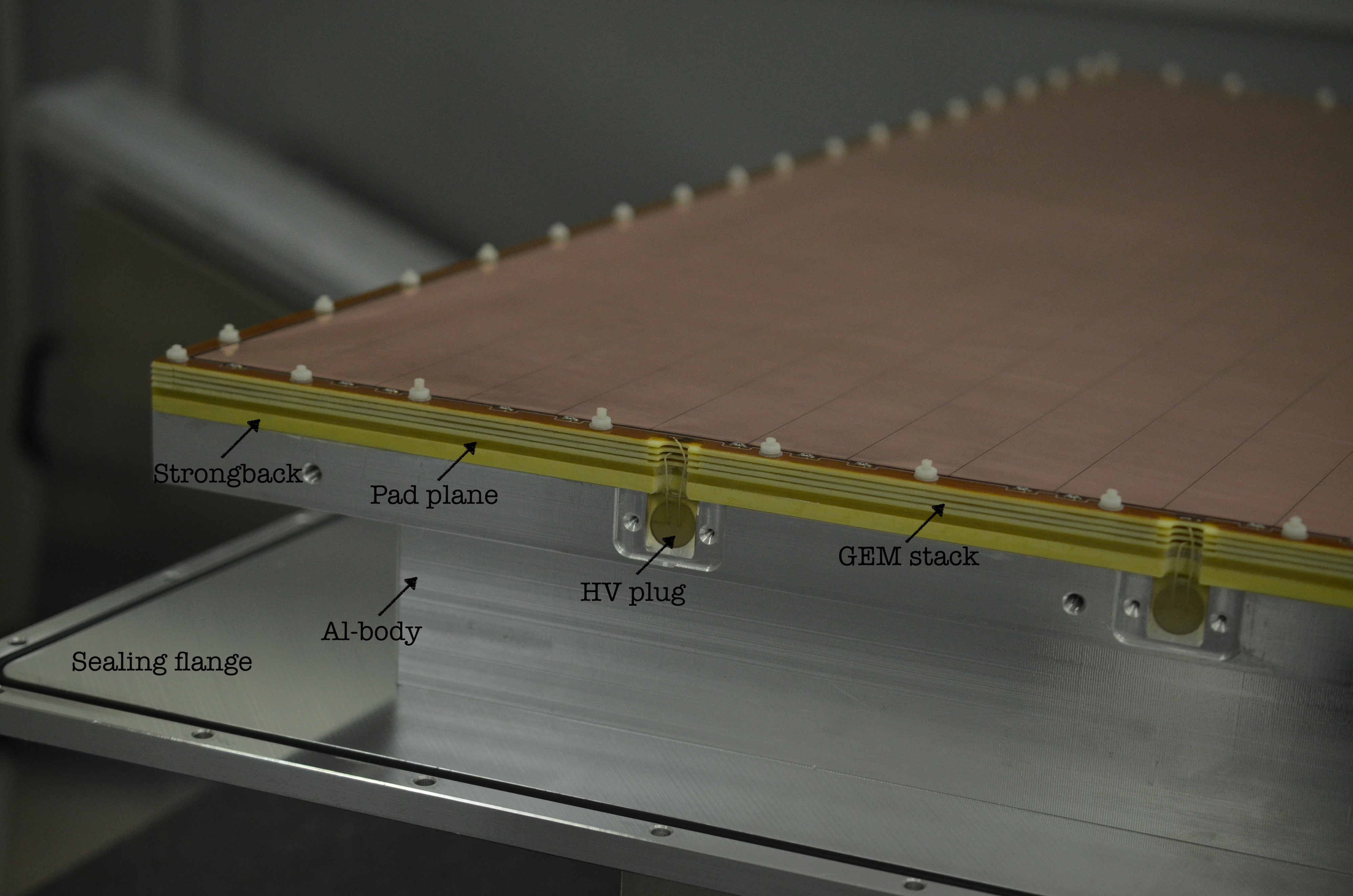}\\
\caption{Assembled \orocThree stack showing four GEMs installed on the chamber body and fixed on the studs, and HV wires soldered to the HV flaps of the foils.}
\label{fig:roc:gem:stack}
\end{figure}

The foils were subsequently mounted to the chamber bodies and the HV wires were soldered to the top and bottom HV flaps (see \figref{fig:gem-design} and \figref{fig:roc:gem:stack}). The wires were soldered using the same solder wire as for the soldering of the SMD loading resistors (see \secref{sec:roc:gemprod}).
After HV connection, the resistance and capacitance across each foil were measured in order to identify possible issues at the earliest possible stage.

The stack was fixed on the studs using reinforced polyamide nuts, tightened with a torque of \SI{6}{\newton\centi\meter}. A photograph of an assembled \orocThree stack is shown in \figref{fig:roc:gem:stack}. In case of the \orocs, the operation was repeated for all three stacks. After successful installation of all GEMs, the chamber was put in a dedicated gas vessel for the ROC QA measurements, described in detail in \secref{sec:roc:rocqc}.

\subsubsection{Clean room equipment}
All assembly activities involving GEM foils were performed in clean rooms with class ISO\,5 to 7, taking all precautions to avoid contamination of the GEMs. All tools and surfaces were regularly cleaned and the personnel was equipped with appropriate clean room gear (overalls, gloves, face masks, etc.).

The foils were stored in dry cabinets in which the relative humidity was kept at the level of \SIrange[range-units=single]{1}{2}{\percent}. For any activities in which the GEM surface was exposed to the environment, such as HV test, framing, or soldering, dedicated covers were used. When this was technically not possible (during optical scan, stretching, frame alignment, etc.), the activity took place directly under a filter fan unit.

\subsubsection{GEM and ROC transportation}
\label{sec:roc:trabox}
More than 300 individual shipments between the various production sites were necessary during the mass production. Thus, it was mandatory to assure safe transportation of the GEM foils and readout chambers. 

Immediately after production, each GEM was equipped with an aluminium frame which stayed attached to it until the trimming before ROC assembly (see \secref{sec:roc:roc:ass}). The aluminium frame was part of the pneumatic stretching system used for GEM framing (see \secref{sec:roc:framing}). It provided stability and made the foil handling more convenient. In addition, it allowed the GEMs to be hang vertically inside dry cabinets, and served as a basis for safe transportation employing a dedicated GEM Transportation System (GTS). 

\begin{figure}[t]
\centering
\includegraphics[width=0.7\linewidth]{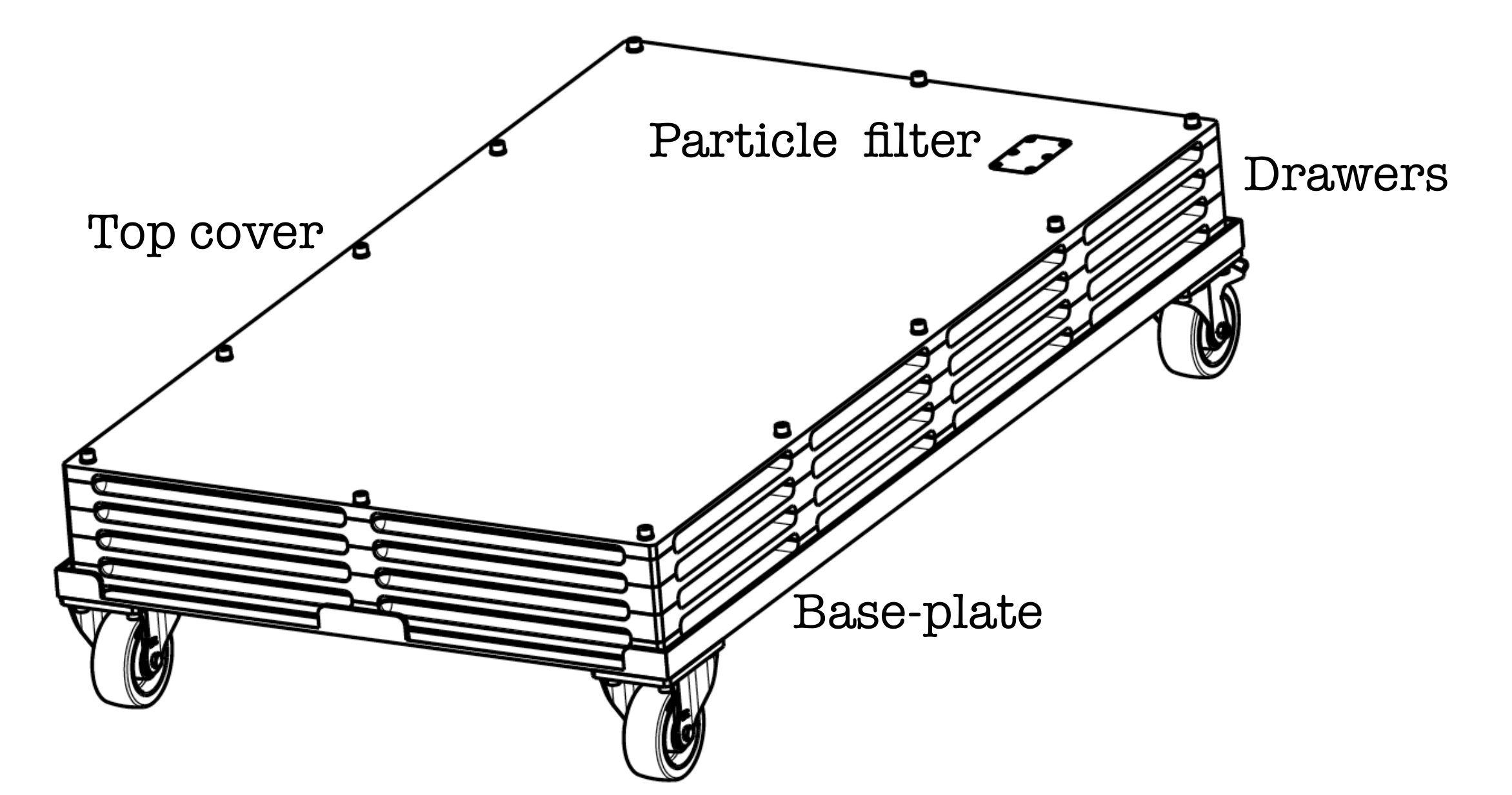}\\
\caption{A GEM Transportation System unit.}
\label{fig:roc:gts}
\end{figure}

The GTS was a custom-made modular system which consisted of a set of drawers, a wheeled base-plate, and a cover, all made of PVC material (see \figref{fig:roc:gts}). GEMs equipped with aluminium frames were packed into single-use clean room paper envelopes (non-sticky and non-abrasive) and secured inside the drawers. Up to three (stacked) GEMs of any \oroc size or up to six \iroc-size foils could be accommodated in a single drawer. O-rings between the GTS drawers assured gas tightness of the entire system. The top GTS cover was equipped with a particle filter to allow for pressure equalization during transport. Thanks to the aluminium frames, GEMs in the GTS could be easily fixed, thus preventing them from sliding or touching each other or any surface nearby. For additional safety during shipment by commercial carriers, all GTSs were packed into dedicated wooden boxes. It is worth noting that no GEM was damaged during any transport.

Chamber bodies and assembled ROCs were shipped in custom made test/transport boxes (TT-boxes) used for transport and storage until final installation in the TPC. The TT-box consisted of an aluminium plenum that was closed with a flange where the ROC was attached. O-rings provided gas tightness. The flange was equipped with gas fittings. For transportation, a particle filter was connected to the TT-box volume. In addition, the TT-boxes were placed in wooden boxes for extra safety. All \irocs were shipped from the US to CERN by plane, whereas \orocs were transported by road.

The TT-boxes were also used for acceptance tests before and after ROC transportation (see \secref{sec:roc:qc:transport}), and stability tests at CERN (see \secref{sec:roc:at:CERN}). They did not contain a drift electrode, however, the plenum (at ground potential) was placed at a safe distance from any ROC electrode to avoid electrical discharges.

\subsubsection{Database}
\label{sec:database}
The TPC upgrade campaign was managed via a custom-made database. It was used to record the QA results of all ROC parts from the QA tests of individual components like GEMs, pad planes, Al-bodies, to the ROC assembly and commissioning of completed ROCs (see \secref{sec:roc:rocqc}). QA data and measurements of all parts, as described above, were uploaded via a web interface at the responsible labs. In addition, the database was used to manage the flow of material between the participating labs. The control of stock, logistics and workflow for the ROC parts was a very important aspect of the database due to the fact that production and QA was distributed among several laboratories, requiring frequent shipping.

\subsection{ROC quality assurance}
\label{sec:roc:rocqc}
A fully assembled ROC was mounted in a vessel dedicated to QA measurements (QA-box) equipped with a drift electrode and a low material budget backplate (entrance window) to allow external irradiation with \xrays. In addition, a screen electrode was installed \SIrange[range-units=single]{2}{3}{\milli\meter} from the drift electrode and operated at the same potential as the latter to collect all positive ions created by radiation in the gas space between the entrance window and the drift electrode. The drift gaps in the utilized QA-boxes were \SIrange[range-units=single]{10}{25}{\milli\meter} long. The flange, to which the ROC was attached, fitted both the QA- and the TT-box. 
The chamber was flushed with the nominal detector gas mixture \NeCOtwoNtwo and a set of quality assurance tests were performed. They are described in the following.  

\subsubsection{Gas tightness}
\label{sec:roc:gas:tightness}
While flushing the chamber, the contamination of the exhausted gas with O$_2$ and H$_2$O was monitored. When the O$_2$ level reached its asymptotic value, the leak rate was determined. The requirement of a leak rate below \SI{0.5}{\milli\litre\per\hour} (e.g.\ \SI{<5}{ppm} O$_2$ at a flow of \SI{20}{\litre\per\hour}) at atmospheric pressure had to be fulfilled by all chambers. 

\subsubsection{Gain curve}
\label{sec:roc:gain:curve}
The HV configuration applied to the GEM stack in the course of this test, corresponding to a gain of $\sim$2000, is displayed in \tabref{tab:GEMHV}. The test settings were more demanding than the baseline HV configuration foreseen for operation, discussed in  \secref{hv:hv:overview}, since the former settings employ higher transfer and induction fields and, in particular, higher \gemFour voltage (see \secref{sec:discharge} for more details).
\begin{table}[b]\footnotesize
\caption{HV test settings used for ROC quality assurance.}
\begin{center}
    \begin{tabular}{ rcl }
    \toprule
    $\Delta V_{\text{\gemOne}}$ & = & \SI{270}{\volt} \\[0.5ex]
    $\Delta V_{\text{\gemTwo}}$ & = & \SI{230}{\volt} \\[0.5ex]
    $\Delta V_{\text{\gemThree}}$ & = & \SI{288}{\volt} \\[0.5ex]
    $\Delta V_{\text{\gemFour}}$ & = & \SI{359}{\volt} \\
    \midrule
    $E_{\text{drift}}$ & = & \SI{400}{\volt\per\centi\meter} \\[0.5ex]
    $E_{\text{T1}}$ & = & \SI{4000}{\volt\per\centi\meter} \\[0.5ex]
    $E_{\text{T2}}$ & = & \SI{4000}{\volt\per\centi\meter}\\[0.5ex]
    $E_{\text{T3}}$ & = & \SI{100}{\volt\per\centi\meter}\\[0.5ex]
    \Eind & = & \SI{4000}{\volt\per\centi\meter}\\
    \bottomrule
    \end{tabular}
\end{center}
\label{tab:GEMHV}
\end{table}
A potential $V_{\text{cathode}}$ was applied to the drift electrode and the GEM foils by means of a voltage divider. This simplified powering scheme, with predefined resistors was sufficient for the ROC QA measurements, allowing all chambers to be tested under the same conditions.
    
\begin{figure}[t]
\centering
\includegraphics[width=0.7\linewidth]{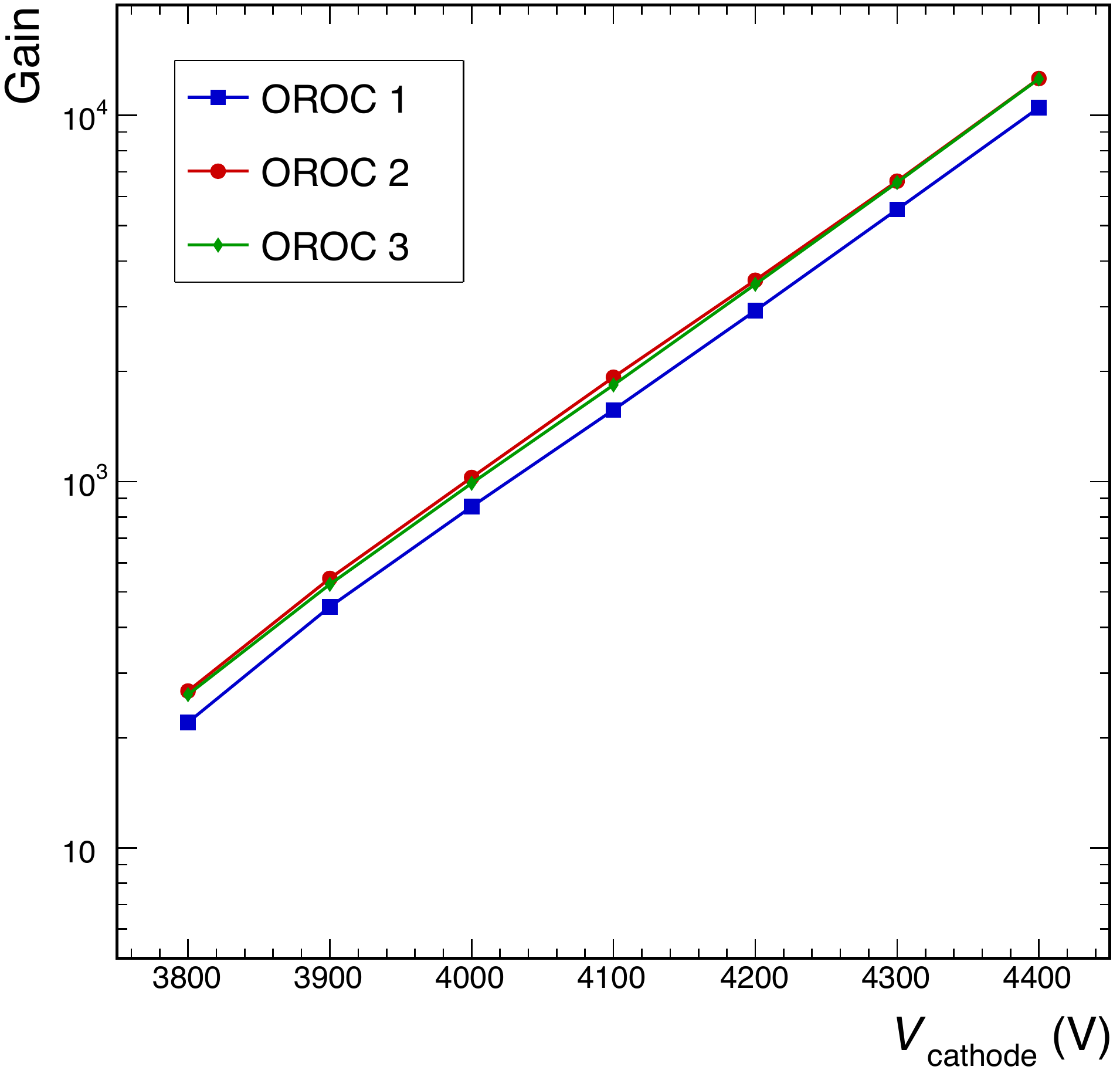}
\caption{Gain curves measured for the three \oroc GEM stacks of one chamber.}
\label{fig:OROCgain}
\end{figure}

The actual gain was determined by measuring the rate of absorbed \xrays from a collimated $^{55}$Fe source and the corresponding current induced on the readout pads. For a given chamber, the overall voltage applied to the voltage divider was varied such that the gain of the chamber was scanned between $\sim$1000 and $\sim$10000. A typical gain curve is shown in \figref{fig:OROCgain}.
Additionally, the $^{55}$Fe spectrum was recorded in one position of each stack at nominal gain in order to determine the corresponding energy resolution.

\subsubsection{Gain and \ibf uniformity}
\label{sec:roc:gain_uniformity}

In order to quantify the uniformity of gain and \ibf, the currents to the pad plane ($I_{\mathrm{pad}}$) and to the drift cathode ($I_{\mathrm{cathode}}$) were monitored while a collimated \xray source or a moderately powered \xray generator was moved over the surface of the readout chamber. For this purpose, all pads were connected together with the use of custom grounding cards. The detector was operated at an average gain of $\sim$2000.
Before the measurement, the full active area was irradiated until the GEMs were charged up, i.e.\ until the pad current reached its asymptotic value (see \secref{sec:rnd:chargeup}).
Then, the position of the radiation source was varied in steps of \SI{\sim 2.5}{\centi\meter} and the pad and cathode currents were measured, until the full active area was scanned.
The requirement for the gain uniformity across the ROC area is $\sigma (I_{\mathrm{pad}})/\langle I_{\mathrm{pad}} \rangle < 0.2$, where $I_{\mathrm{pad}}$ is proportional to the gain, and $\langle I_{\mathrm{pad}} \rangle$ and $\sigma (I_{\mathrm{pad}})$ are the mean and standard deviation of the single-point measurements. The upper limit for the \ibf of a ROC, given by the average over the single-point 
measurements of the $I_{\mathrm{cathode}}/I_{\mathrm{pad}}$ ratio, and its (absolute) standard deviation across the ROC area are\SI{1.0}{\percent} and \SI{0.2}{\percent}, respectively.

\begin{figure}[t]
\centering
\vspace{1cm}
\includegraphics[width=0.99\linewidth]{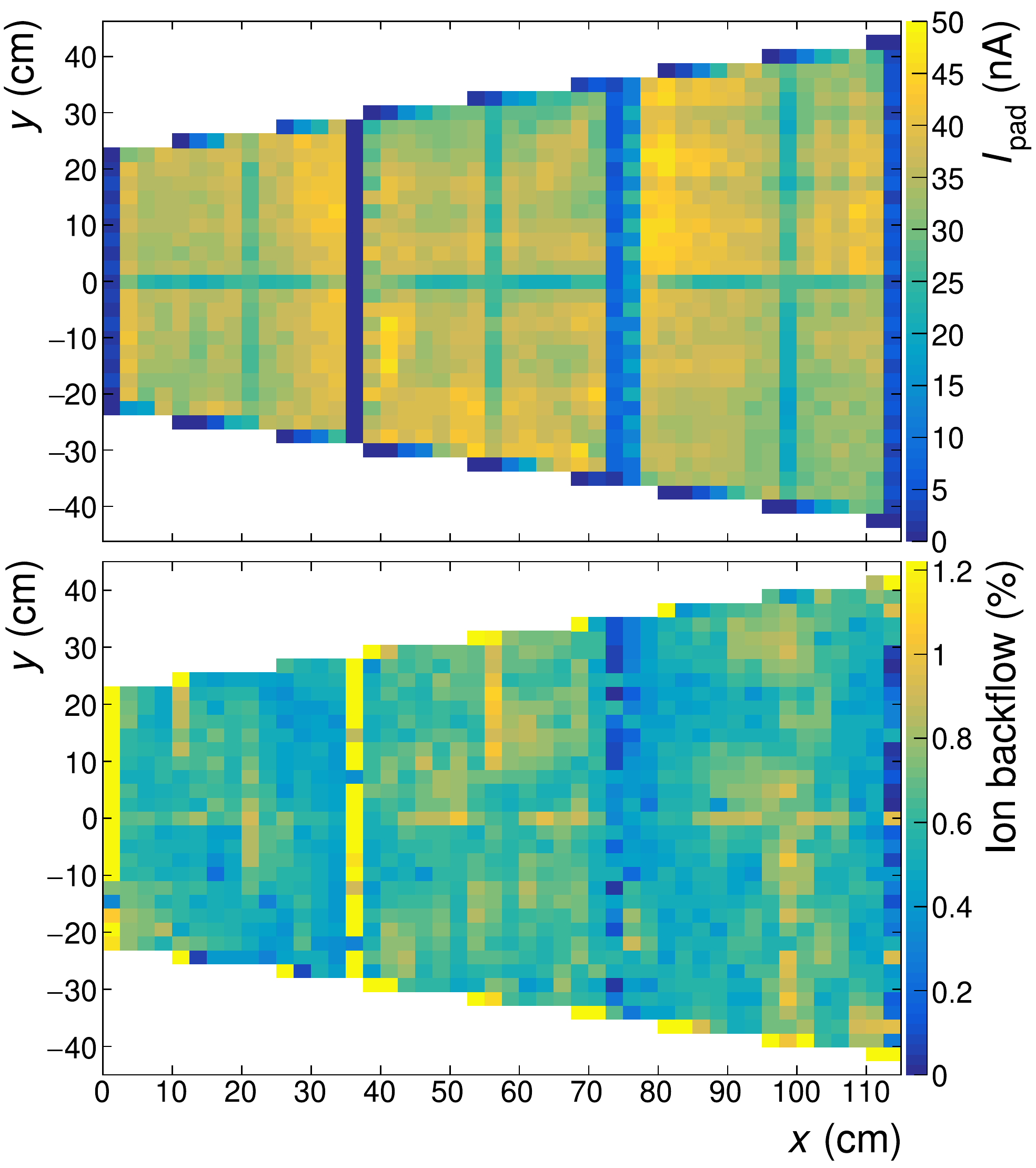}
\caption{Results of the pad current (upper panel) and \ibf (lower panel) uniformity measurement for an \oroc. The structures with low pad current and low or enhanced \ibf correspond to the spacer cross in the GEM frame and to the edge regions. The pad current uniformity, and thus the gain uniformity (in terms of relative standard deviation) for \orocOne, \orocTwo and \orocThree stacks, excluding cross and edge regions, is \SI{8.2}{\percent}, \SI{9.1}{\percent} and \SI{10.9}{\percent}. The average \ibf values are \SI{0.54 \pm 0.10}{\percent}, \SI{0.65\pm0.09}{\percent} and \SI{0.57\pm0.11}{\percent}, respectively.}
\label{fig:OROCresults}
\end{figure}

\Figref{fig:OROCresults} shows typical results obtained for one of the production \orocs. As described in \secref{sec:roc:advanced_qa}, the careful selection of single foils ensured that the required gain and \ibf uniformity was achieved in all production readout chambers.
The overall gain uniformity, in terms of relative standard deviation, measured for all assembled chambers, is \SI{10.3}{\percent}. The average \ibf and its absolute standard deviation are \SI{0.56}{\percent} and \SI{0.14}{\percent}, respectively.

\begin{figure}[h]
\centering
\includegraphics[width=0.73\linewidth]{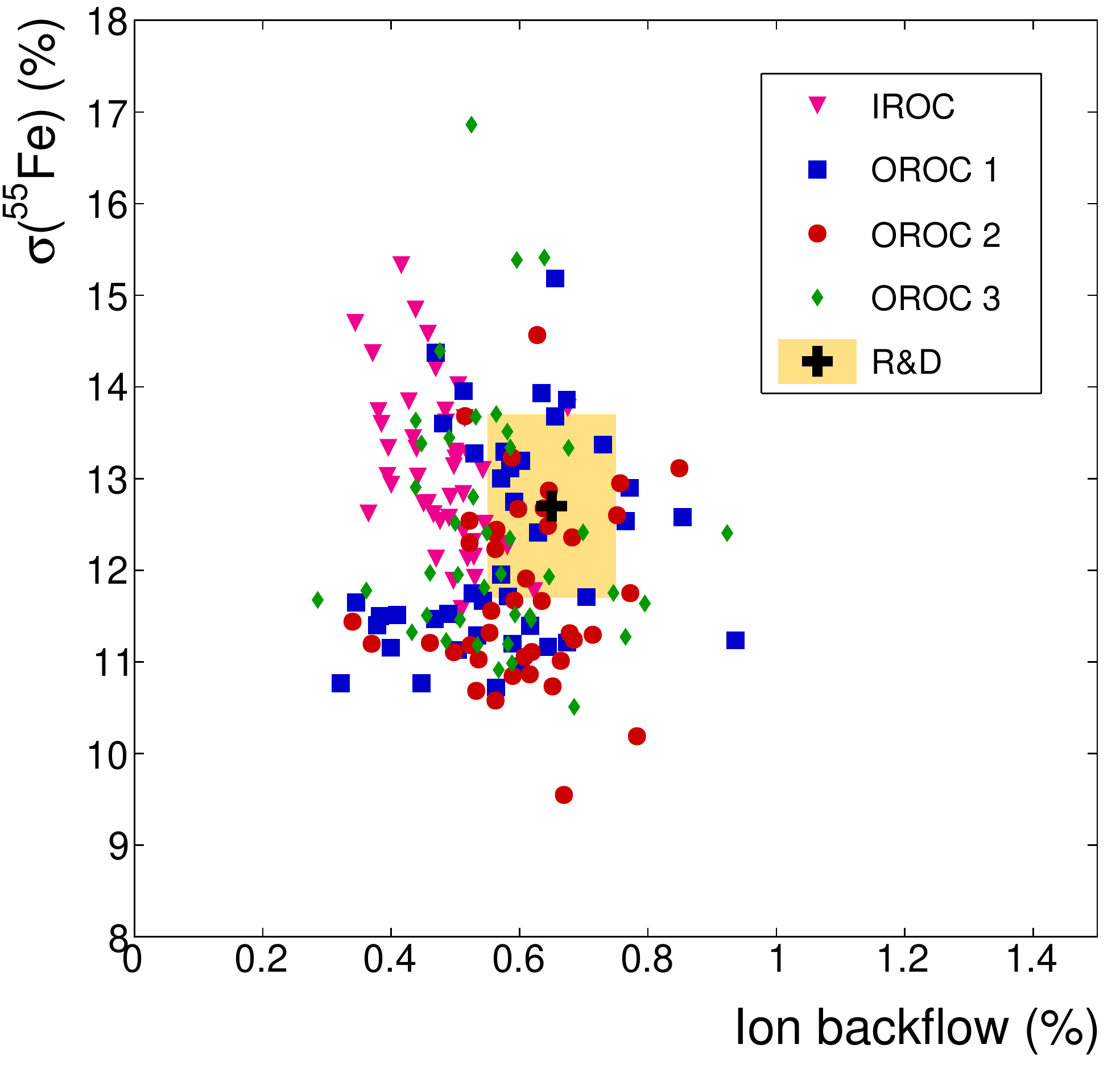}
\caption{Correlation of $\sigma(^{55}\mathrm{Fe})$ and \ibf for all produced \iroc and \oroc stacks. The black cross  shows the R\&D result obtained with a $10\times10$\,\si{\centi\meter\squared} prototype. The yellow area around the R\&D result indicates the typical systematic uncertainty assigned to each data point. }
\label{fig:all:ibf:gain:results}
\end{figure}

\Figref{fig:all:ibf:gain:results} shows the correlation of $\sigma(^{55}\mathrm{Fe})$ and \ibf for all produced \iroc and \oroc stacks. The black point corresponds to the value measured for the test HV settings (see \tabref{tab:GEMHV}) with a $10\times10$\,\si{\centi\meter\squared} prototype detector (see also \secref{sec:rdibf}). The scatter of the ROC measurements around the value expected from the small prototype is compatible with the systematic uncertainty of the single-point measurement, indicated by the error bars of the prototype data point, and the variation over the ROC surface. The values for $\sigma(^{55}\mathrm{Fe})$ and \ibf can be further adjusted by optimization of the voltage settings on a stack-by-stack basis.

\subsubsection{Full-area \xray irradiation}

The last qualification step was a stress test at the maximal current density of \SI{10}{\nano\ampere\per\centi\meter\squared} expected during nominal Run\,3 operation~\cite{TDR:tpcUpgrade}.
The detector was exposed for at least six hours to the corresponding \xray flux at the nominal gain, while monitoring the pad current for discharges. After the test, the leakage currents of all GEMs were measured at 250\,V and compared to the situation before the test. In both cases, the leakage current were required to be below be below 1\,nA (measured for the entire foil). 

In the course of the ROC production, an additional irradiation step was introduced at the assembly sites as a response to HV instabilities observed with a number of chambers during stability tests at CERN, described in details in \secref{sec:roc:at:CERN}. It included local irradiation of the soldering points, where the HV wires are soldered to the GEM foils (see \secref{sec:roc:roc:ass}). Problematic chambers were treated the same way as described in \secref{sec:roc:at:CERN}.

\subsubsection{Acceptance tests before/after transportation}
\label{sec:roc:qc:transport}
Once the chambers successfully passed the quality assurance at the assembly sites, they were installed back in their TT-boxes. Before shipping to CERN, acceptance tests were performed. These included: i)\,gas tightness, ii)\,leakage current at \SI{250}{\volt}, iii)\,GEM conductance, and iv)\,capacitance. The latter is an effective way to check whether any electrical contact (HV wire or SMD resistor on a foil) is broken. For these tests, the TT-boxes containing their ROCs  (see \secref{sec:roc:trabox}) were flushed with nitrogen. The purpose of these tests was to assure the integrity of the chambers before shipping.
Upon reception at CERN, the same tests were carried out again. In addition, each stack was operated at the nominal voltage settings for \SI{24}{\hour}, while the currents on all GEM electrodes were monitored.

\subsection{Stability tests at CERN}
\label{sec:roc:at:CERN}
All chambers delivered and accepted at CERN underwent a final stability test under irradiation in the ALICE cavern during LHC operation or at the CERN Gamma Irradiation Facility (GIF++)~\cite{Capeans-Garrido:1207380, Jaekel:2014yya}. The tests were performed with the chambers mounted in the TT-boxes, using the nominal gas mixture \NeCOtwoNtwo. The baseline HV settings and the final components of the HV system were used (see \secref{sec:hv:hv} for more details).
During the test, the pad plane was grounded via dedicated shorting cards. Since the TT-boxes had no drift electrode, the drift field had reversed polarity, i.e.\ the ionization electrons drifted towards the grounded plenum of the TT-box. However, the electrons liberated in the closest vicinity of \gemOne, or in the transfer gaps, could undergo amplification processes. Potential instabilities due to imperfections of the solder points or HV electrodes were effectively detected, and thus quality issues of the ROCs could be revealed. 
During the measurements, all currents on the GEM electrodes were monitored by the built-in current meters of the HV power supplies (see \secref{sec:hv:hv:powersupply}) and with an additional high-definition current meter installed in the power line of the top electrode of \gemFour (see \secref{sec:hv:hv:hdmeter}). 

All produced readout chambers (45 \irocs and 40 \orocs) successfully passed the final irradiation campaigns at the LHC and at GIF++, and were accepted for installation.

\subsubsection{Tests at the LHC}
\label{sec:roc:at:lhc}

\begin{figure}[t]
\centering
\includegraphics[width=0.75\linewidth]{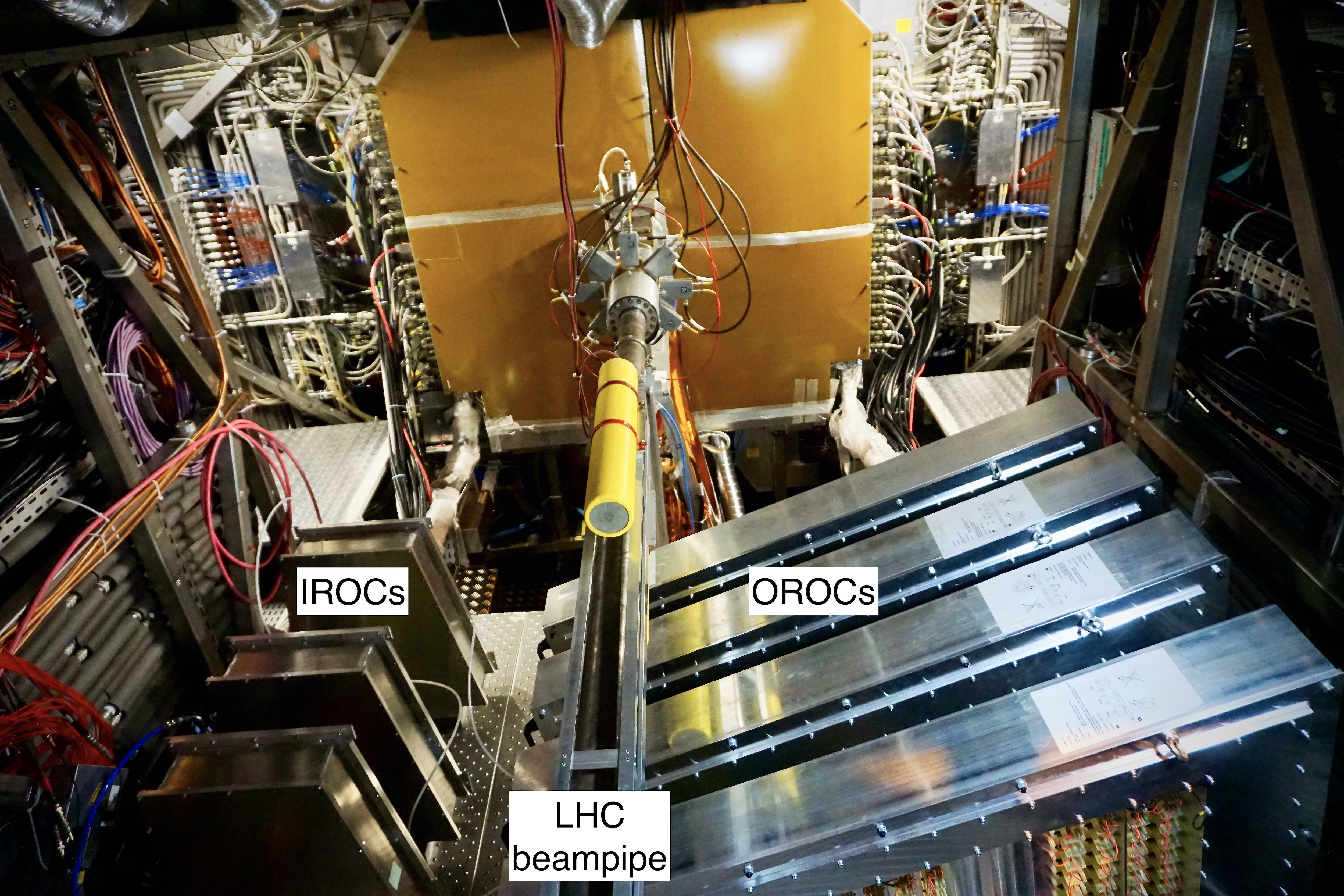}
\caption{\irocs and \orocs installed under the LHC beampipe, a few meters from the interaction point.}
\label{fig:roc:at:lhc}
\end{figure}

The tests were performed during the LHC Run\,2 during periods with \pp and \PbPb collisions.
Up to 4 \irocs and 4 \orocs were tested simultaneously in the ALICE cavern (see \figref{fig:roc:at:lhc}), a few meters from the interaction point in the forward pseudorapidity region, in which the conditions were comparable to those expected for the TPC in Run\,3.

All production chambers were tested in their TT-boxes. In total, 28 \irocs and 16 \orocs were irradiated at the LHC until the start of Long Shutdown 2 in December 2018. The voltages were ramped automatically according to the beam modes, and replaced during available time windows after several tens of hours of operation. Thus, these tests produced zero impact in the normal operation of the ALICE experiment and of the LHC. In \SI{10}{\percent} of the stacks, a systematic instability was identified in the form of an imperfect soldering of the HV wire to the \gemOne bottom electrode, which manifested itself as a discharge current $\mathcal{O}$(\SIrange[range-units=single]{0.1}{1.0}{\micro\meter}) developing from this electrode to ground or to the subsequent GEM electrodes. Passivation of the solder point by a small layer of epoxy completely solved this issue. All chambers treated this way were retested and showed no further signs of instabilities. After the end of Run\,2, the remaining ROCs were tested at GIF++.

\subsubsection{Tests at GIF++}
\label{sec:roc:at:gif}

The availability of the Gamma Irradiation Facility (GIF++) made it possible to complete the testing and certification of the entire chamber production.
The GIF++ facility is located at the CERN SPS North Area.
It features a 14\,TBq $^{137}$Cs source of 662\,keV photons with adjustable intensity~\cite{Capeans-Garrido:1207380, Jaekel:2014yya}. The gamma irradiation was performed at the maximum available rates at the upstream area of the GIF++ bunker. One \iroc and one \oroc could be tested at the same time. During irradiation, the chambers were positioned in front of the source at 2\,m distance inside of the illumination cone. The chambers were exposed for a one-hour period, enough to spot possible issues with respect to HV stability. The currents induced on the \gemFour electrodes were higher than those during the tests at the LHC by about a factor of 10.

It was confirmed, by testing several chambers both at the LHC and at GIF++, that the soldering imperfections of the \gemOne bottom electrode connection manifested themselves at GIF++ in a similar way as at the LHC. Thus, the test at GIF++ was considered equally sensitive as the measurements in the ALICE experimental cavern. In total, 18 \irocs and 27 \orocs were tested at GIF++. Again, about \SI{10}{\percent} of the stacks exhibited instabilities related to the solder point. All issues were fixed and chambers were retested successfully.
\section{High-voltage system}
\label{sec:hv}

In this chapter, the new TPC HV system is described. This includes the powering scheme of the new GEM readout chambers and the modifications of the TPC field cage circuit that were necessary in order to adjust the electrostatic matching between the field cage and the GEM ROCs.

\subsection{GEM HV system}
\label{sec:hv:hv}

The GEM HV system is a crucial aspect for the safe operation of the detector. In this section, we describe the powering scheme of the TPC ROCs and the employed components. They have been characterized during extensive R\&D and were successfully used during the readout chamber irradiation campaigns at the LHC and at GIF++.

\subsubsection{Overview and requirements}
\label{hv:hv:overview}

The HV settings for the operation of the GEM readout chambers are the result of an extensive optimization (see \secref{sec:rdibf}). The baseline settings for the voltages across the GEMs, for the transfer fields between the GEMs, and for the induction field between \gemFour and pad plane are listed in \tabref{tab:hv:baseline}. A first gain equalization across all TPC ROCs is described in \secref{sec:gain_calibration}.

\begin{table}[b]\footnotesize
\caption{Baseline settings for the voltages across the four GEMs, the transfer fields between the GEMs and the induction field between \gemFour and pad plane.}
    \begin{center}
    \begin{tabular}{rcl}
    \toprule
    $\Delta V_{\text{\gemOne}}$ & = & \SI{270}{\volt} \\[0.5ex]
    $\Delta V_{\text{\gemTwo}}$ & = & \SI{230}{\volt} \\[0.5ex]
    $\Delta V_{\text{\gemThree}}$ & = & \SI{320}{\volt} \\[0.5ex]
    $\Delta V_{\text{\gemFour}}$ & = & \SI{320}{\volt} \\
    \midrule
    $E_{\text{drift}}$ & = & \SI{400}{\volt\per\centi\meter}\\[0.5ex]
    $E_{\text{T1}}$ & = & \SI{3500}{\volt\per\centi\meter} \\[0.5ex]
    $E_{\text{T2}}$ & = & \SI{3500}{\volt\per\centi\meter}\\[0.5ex]
    $E_{\text{T3}}$ & = & \SI{100}{\volt\per\centi\meter}\\[0.5ex]
    \Eind & = & \SI{3500}{\volt\per\centi\meter}\\
    \bottomrule
    \end{tabular}
    \end{center}
\label{tab:hv:baseline}
\end{table}

For the operation of the GEM-based ROCs, a new high-voltage system was designed. It provides high flexibility for the adjustment of the potentials at the various electrodes, and a safe operation of the GEM stacks. Discharges in a GEM stack can lead to damages, in particular if the discharge leads to increased potential differences between adjacent electrodes. In order to maximize operational safety while providing the highest possible flexibility, a power supply system with cascaded channels is used in combination with properly selected decoupling resistors (\secref{sec:hv:hv:powersupply} and \ref{sec:hv:hv:prb}).

\begin{figure}[t]
    \centering
    \includegraphics[width=0.9\textwidth]{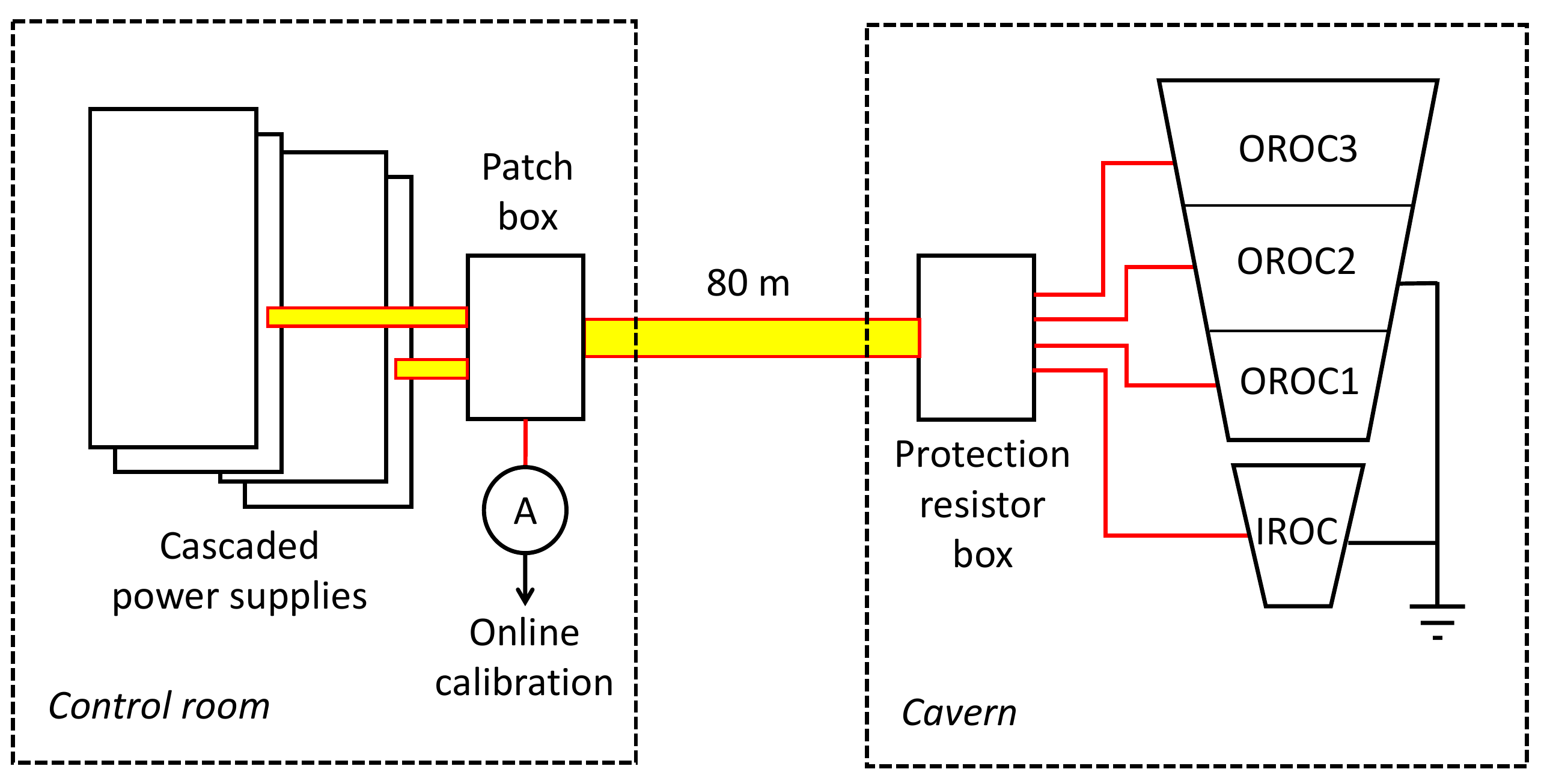}
    \caption{The modular HV scheme, shown for a single TPC sector. In the control room, the channels of two 2x8 channel power supplies are grouped together into a single cable in a patch box, which also houses the shunt resistors for the fast reading of the currents on the top electrode of \gemFour. At the detector in the cavern, i.e.\ at a distance of \SI{80}{\meter}, the multichannel HV line is split into individual coaxial cables inside the protection resistor box, in order to power the four stacks of the sector. The ground reference of the HV system is the aluminium structure of the TPC.}
    \label{fig:hv:hvscheme}
\end{figure}

 A schematic view of the HV system is shown in \figref{fig:hv:hvscheme}. The power supplies are located in the control room together with the patch boxes, in which the single HV lines are bundled to one thick \SI{80}{\meter}-long cable for each TPC sector. In each GEM stack, the current of one of the HV lines is measured with a high-definition current readout device (see \secref{sec:hv:hv:hdmeter}) and the information is digitised. In the cavern, close to the TPC, the bundled HV cable is split into single HV lines inside the protection resistor box (see \secref{sec:hv:hv:prb}). The protection resistor box contains also a \SI{100}{\kilo\ohm}  decoupling resistor ($R_{\text{dec}}$) for each single HV line (see \secref{sec:hv:hv:prb}). 
Additionally, each segment of the GEM top side is connected via a \SI{5}{\mega\ohm} loading resistor ($R_{\text{load}}$), which is soldered directly on the foil (see \secref{sec:roc:gem}). In \figref{fig:hv:caenpsscheme}, the complete powering scheme including all resistors is shown.

The resistor values are a compromise between minimal gain variations caused by voltage drops over the resistor under load and the possibility to supply excess currents in the case of shorts over one or several GEM segments. The chosen resistor value allows for a maximum of eleven shorted HV segments per GEM keeping the maximum leakage current below \SI{1}{\milli\ampere}. The change of the gain under load is expected to be at most \SI{14}{\percent} (\SI{10.8}{\percent} due to the loading resistor) in the innermost segments of the \iroc. These drops are foreseen to be corrected for in the data calibration procedures.

\subsubsection{Cascaded power supply}
\label{sec:hv:hv:powersupply}

\begin{figure*}[t]
    \centering
    \includegraphics[width=1.0\textwidth]{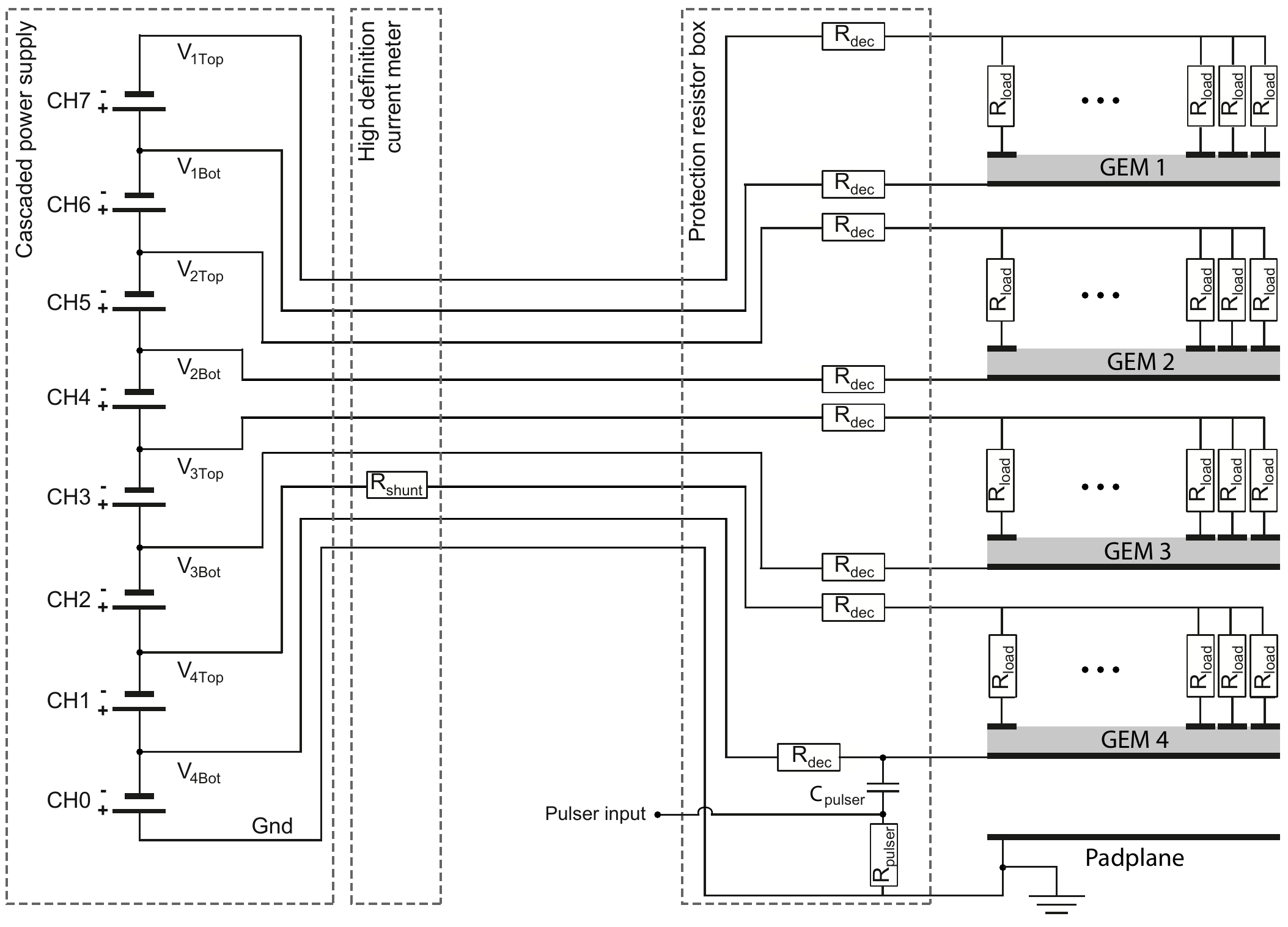}
   \caption{Detailed powering scheme of a GEM stack. Each subsequent high-voltage channel is stacked on top of the lower-lying channel. The ground reference is defined by a separate line connected to the ground of the detector. The line for \gemFour top is shunted with a resistor ($R_{\text{shunt}}$) inside the high-definition current meter. In the protection resistor box, each line is connected to the detector through a decoupling resistor ($R_{\text{dec}}$). The signal from the calibration pulser is coupled via a capacitor ($C_{\text{pulser}}$) to the line for \gemFour bottom. All segments on the top sides of the GEMs are equipped with individual loading resistors ($R_{\text{load}}$).}
    \label{fig:hv:caenpsscheme}
\end{figure*}

The main component of the HV system is a 
cascaded power supply unit\footnote{A1515QGA} from CAEN~\cite{caen:a1515}. It is designed for the operation of quadruple-GEM systems. The 16 channels are stacked in two groups of 8 channels. One unit can thus power two quadruple-GEM stacks. As a consequence, 72 units are used to supply all GEM stacks in the TPC. The current measurement resolution and accuracy were optimized for the TPC requirements. The unit has a floating ground reference that has to be defined by an external connection. The schematic arrangement of the eight HV channels used for a quadruple-GEM stack is shown in the left part of \figref{fig:hv:caenpsscheme}. 

The main advantages of cascaded power supplies over a simple voltage divider are the larger flexibility (each potential can be set individually) and the stability of the potentials on all GEM electrodes in case of significant changes in the leakage current on any electrode, e.g.\ due to a short. Moreover, they provide high sensitivity to any effect appearing in a single GEM foil, since the currents in all channels are measured individually and simultaneously.

Due to the stacked arrangement of the individual power lines, the risk of over-biasing single channels is minimised. This behaviour was studied carefully with small test setups and with the full HV system during the tests in the ALICE cavern and at the GIF++ (see \secref{sec:roc:at:CERN}). All tests confirmed safe operation using the cascaded power supply.
Each channel can deliver up to \SI{1}{\kilo\volt} and has a configurable current limit, which will cause the channel to trip in case the limit is exceeded for longer than a predefined time delay, which can be configured in steps of \SI{0.1}{\second}. The unit provides two distinct operational modes. The \textit{high-resolution mode} provides a maximum current of \SI{20}{\micro\ampere} per channel, and an accuracy for the measurement of the currents of \SI{400}{\pico\ampere}. This mode is used for the operation of the TPC. The unit provides also a \textit{high-power mode}, which allows currents up to \SI{1}{\milli\ampere} per channel with a current measurement accuracy of about \SI{20}{\nano\ampere}. This mode would be only needed in case the current consumption of a stack increased due to a short.
All units including spares of the final system, i.e.\ 1264 individual HV channels, were tested to fulfill the requirements. The tested criteria were the current offset ($\Delta I$), the current measurement resolution ($\sigma_{I,\text{res}}$), the current drift during 24-hour operation ($\sigma_{I,\text{drift}}$) in high-resolution mode without load, and the current measurement resolution in the high-power mode $\sigma_{I,\text{pow}}$.
Additionally, the correct trip delay time was measured with a dedicated trip emulation board that allows one to apply a current on the individual channels for a defined time. The deviation of the measured trip delay ($\sigma_{t,\text{trip}}$) with respect to the set value is required to be below a given limit. The parameters tested and the required limits are summarized in \tabref{tab:hv:hv:powersupply-masstest}.
\begin{table}[t]\footnotesize
    \caption{Limits of the performance parameters for single channels of the CAEN power supplies. The parameters correspond to the current resolution in high-resolution ($\sigma_{I,\text{res}}$) and high-power ($\sigma_{I,\text{pow}}$) mode, the offset in the current measurement ($\Delta_{I}$), the long-term drift of the current measurement ($\sigma_{I,\text{Drift}}$) and the maximum deviation from the set trip delay ($\sigma_{t,\text{trip}}$).}
    \begin{center}
    \begin{tabular}{ccc}
        \toprule
        Parameter & & Limit \\
        \midrule
         $\sigma_{I,\text{res}}$ & $<$ & \SI{1}{\nano\ampere}  \\[0.5ex]
         $\sigma_{I,\text{pow}}$ & $<$ & \SI{100}{\nano\ampere} \\[0.5ex]
         $\Delta I$ & $<$ & \SI{10}{\nano\ampere}   \\[0.5ex]
         $\sigma_{I,\text{drift}}$ & $<$ & \SI{10}{\nano\ampere}  \\[0.5ex]
         $\sigma_{t,\text{trip}}$ & $<$ & \SI{10}{\percent} \\
         \bottomrule
    \end{tabular}
    \end{center}
    \label{tab:hv:hv:powersupply-masstest}
\end{table}

\begin{figure}[ht]
    \centering
    \includegraphics[width=0.6\textwidth]{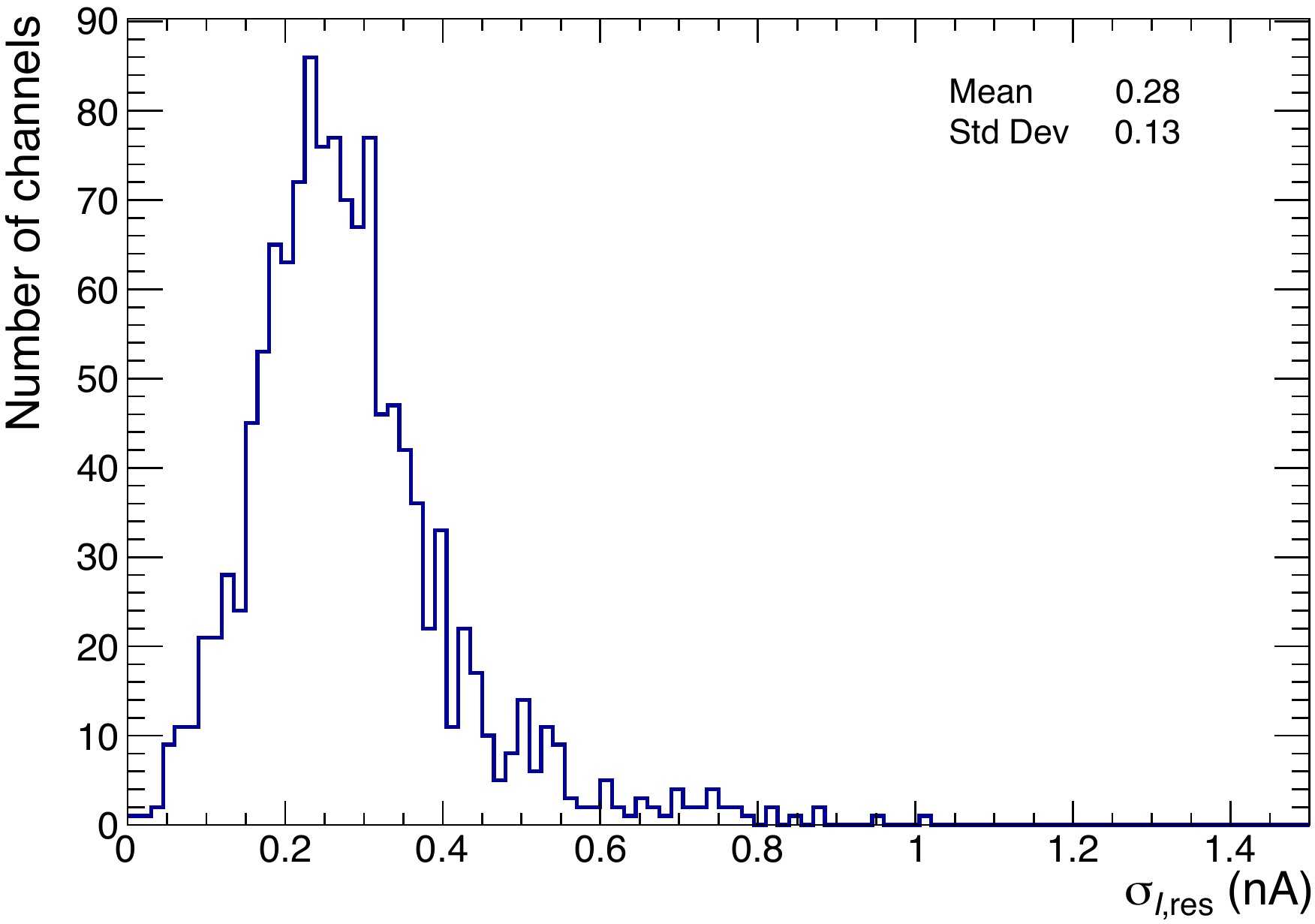}
    \caption{Resolution of the current measurement ($\sigma_{I,\text{res.}}$) in high-resolution mode for all HV-channels.}
    \label{fig:hv:caenpstestingcurrentresolution}
\end{figure}

As an example, the measured current resolution distribution ($\sigma_{I,\text{res}}$) for all tested channels is shown in \figref{fig:hv:caenpstestingcurrentresolution}. The mean value of \SI{282}{\pico\ampere} is in good agreement with the value stated by the producer (\SI{300}{\pico\ampere}).
Finally, a long-term test of each module at the final position inside the crate was performed.
The test results were collected in the database (see \secref{sec:database}).

\subsubsection{Protection resistor box}
\label{sec:hv:hv:prb}

Primary discharges in a GEM foil can be followed by violent secondary discharges in the transfer gap.
In order to reduce the probability for secondary discharges~\cite{GASIK2017116}, serial decoupling resistors 
($R_{\text{dec}}$) are installed close to the detector in all HV channels (see \figref{fig:hv:caenpsscheme}). 
The interplay of the cable length between the resistor and the GEM, as well as the value of the decoupling resistor, were carefully studied\cite{Lautner:2019kas}. As a result, decoupling resistors  are used for each individual HV line (see $R_{\text{dec}}$ in \Figref{fig:hv:caenpsscheme}). The chosen value is the lowest possible providing sufficient safety for the GEMs, while keeping the gain change under load, additional to the contribution of the loading resistor, at a level of about $\sim$\SI{3}{\percent}.

All decoupling resistors of a TPC sector are located inside the protection resistor box, mounted close to the sector on the TPC. The PRB is also used to split the thick multiline HV cable that arrives at the TPC into individual coaxial lines. This results in cable lengths of 1--\SI{3}{\meter} between PRB and GEM. The boxes are designed to allow for an easy modification of the resistor values. In order to prevent floating channels, but still have the flexibility to modify the resistance, each channel is equipped with three resistors of values \SI{50}{\kilo\ohm}, \SI{100}{\kilo\ohm} and \SI{350}{\kilo\ohm}, connected in series, and each of them can be shorted using parallel jumpers.

In addition, the PRBs contain also the input connection for an external pulser signal, which is coupled via a \SI{30}{\pico\farad} capacitor ($C_{\text{pulser}}$) to the HV line of each \gemFour bottom electrode (see \Figref{fig:hv:caenpsscheme}). The input is terminated by \SI{50}{\ohm}. For the generation of the pulser signal, the same device as for the MWPCs is used~\cite{TPCnim}, but with inverted polarity. To ensure equal timing of the pulse position relative to the trigger in all stacks, the length of the cables to all \gemFour bottom electrodes is equalized.

\subsubsection{High-definition current readout}
\label{sec:hv:hv:hdmeter}

In order to measure the average drift-field distortions due to space charge from back-drifting ions, but also the fluctuations of these distortions with time, the information about the local track density is used during the online calibration. Since track density variations are reflected in current variations in the GEM stacks, a fast current monitoring system was developed to read the current on the top electrode of \gemFour, which is the electrode with the highest current in the stack. The current reading is realized via the measurement of the voltage drop on a \SI{10}{\kilo\ohm} shunt resistor.

\begin{figure}[h]
    \centering
    \includegraphics[width=0.7\textwidth]{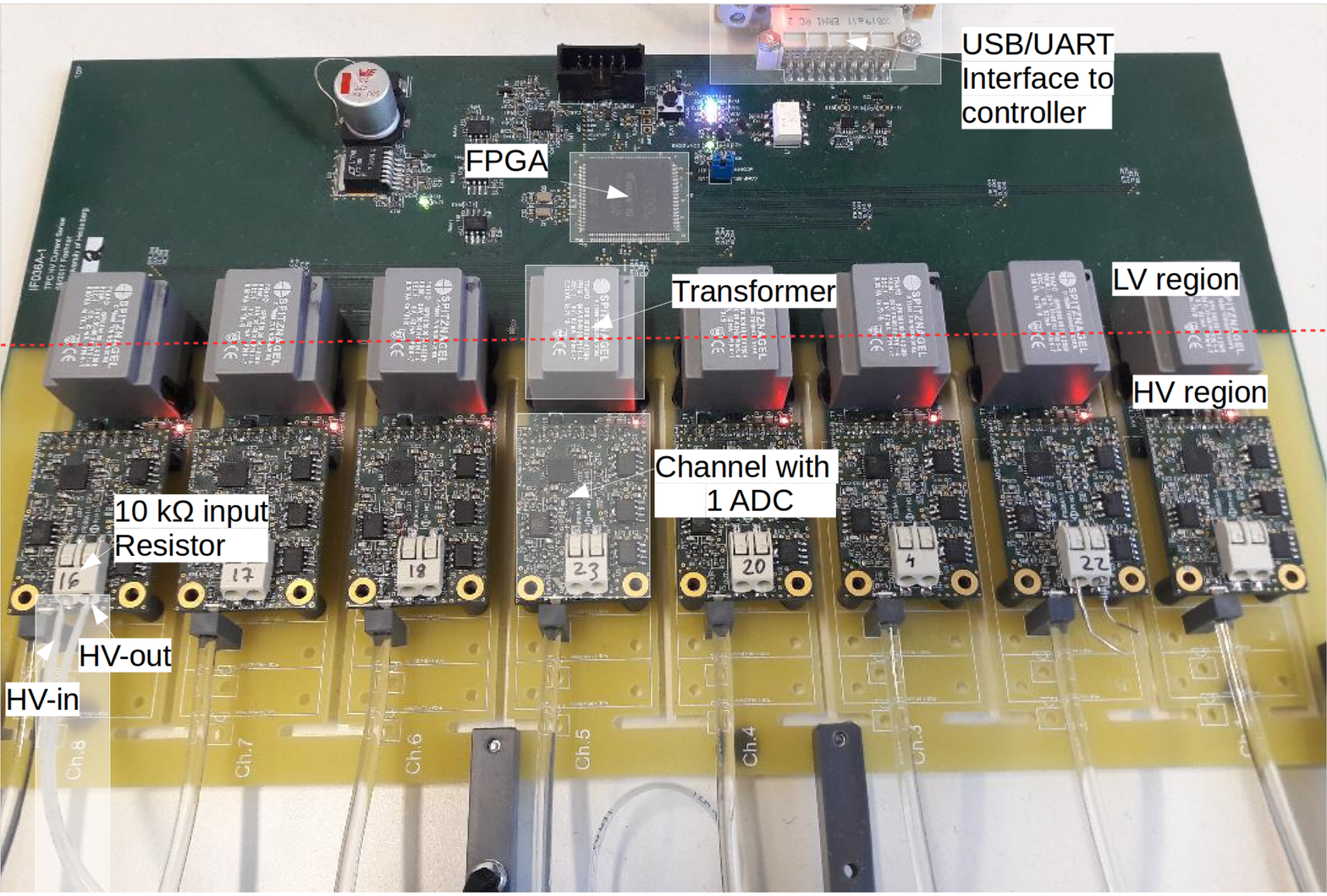}
    \caption{8 channel module for high-definition GEM current readout~\cite{Anastasopoulos:bsc:18}.}
    \label{fig:hv:hdcurrentmeterscheme}
\end{figure}

\Figref{fig:hv:hdcurrentmeterscheme} shows an 8 channel current readout module and its main components. The currents are digitized at a rate of \SI{1}{\kilo\hertz} by a 24-bit ADC\footnote{24-bit Sigma-Delta Analog Devices AD712404} with a programm\-able gain amplifier and a reference voltage. This is sufficient for a resolution of better than \SI{1}{\nano\ampere} over a dynamic range from 0 to \SI{125}{\micro\ampere}. The data is bit-encoded and sent through an E-Link to an on-board GBTx that transmits the data via an optical link to a receiving FPGA-based readout card (CRU) at the data acquisition side (these components are described in \secref{sec:fec.gbt}). The decoded data is then available for calibration purposes during data processing.

\subsubsection{High-voltage components}
\label{hv:hv:hvcomponents}
All high-voltage connectors are from the Lemo S- (51 contact connectors), K- (22 contact connectors) and 05 (single contact) series. The connectors are rated for voltages up to \SI{12}{\kilo\volt}~\cite{lemo} and, due to their small size, are very well suited for a large number of HV channels. For the production of the single-pin contacts, an advanced assembly procedure was established using an extra layer of polyurethane lacquer with an increased drying time (12 hours). This procedure reduced the leakage current in a single pin below \SI{200}{\pico\ampere} at a voltage of \SI{5}{\kilo\volt}, which corresponds also to the requirement for the validation of the HV stability.
After the assembly in the final connectors, the parts were tested again at \SI{150}{\percent} of the operational voltages in the TPC, and only accepted if they fulfil the requirement. The test results were collected in the database (see \secref{sec:database}).

\subsubsection{Control software}
\label{sec:hv:hv:dcs}
The detector control system (DCS) of the ALICE experiment is based on the commercial Supervisory Control and Data Acquisition (SCADA) system WinCC-OA~\cite{winccoa}. For the operation of the GEM HV system, the existing software was adapted to handle the new system. Since the used power supply modules are tailored for GEM operation, a channel grouping is realized in the firmware. This feature allows for the propagation of powering commands to all channels of the stack. Consequently, a trip of a single channel will also switch off all other channels of a given GEM stack. In the operational software, this grouping is also extended to the set-voltage command, which allows one to scale all set voltages relative to the nominal operational voltage. This way of operation was chosen, since the relative scaling of all voltages in the GEM stack is directly related to the gain of the stack.

The most important part in the software is the realization of several stages of protection mechanisms, which will eventually allow easy operation by a shift crew. This includes a proper setting of current and voltage limits and trip delays. Furthermore, a protection mechanism against faulty HV connections, such as swapped cables or disconnected electrodes, is implemented. Although these problems appear very rarely, they could create severe damage to a stack, thus they need to be detected during ramp-up already at low voltages. The proper observables to check for such issues are the ramping currents of all electrodes. During the ramping process, the GEM foils are charged up, resulting in a specific pattern of the charging currents in the two electrodes of every GEM and the electrodes of two subsequent GEMs. This pattern depends on the capacitance of the foils and scales with the ramp speed. Connectivity issues lead to significant deviations from these specific patterns and can be detected easily by software, ensuring to abort the ramping process before high voltages are reached.
With the use of a full GEM simulator board~\cite{Gemulator:2016}, the pattern of these charging currents, and the deviations from it in the event of any anomaly, were verified.

\subsection{Field cage modification}
\label{sec:hv:fc}
\subsubsection{Overview}
\label{sec:hv:fc:overview}

The field cage of the TPC is described in detail in~\cite{TPCnim}. The central electrode is biased to a potential of --\SI{100}{\kilo\volt}. The uniform electric field in the drift volume is generated by potential strips suspended close to the walls of the inner and outer field cage vessels and powered through a resistor chain (i.e.\ the resistor rod). In addition, separate resistor chains are mounted on the walls of the field cage vessels. They connect to aluminium strips glued to the walls (guard rings) to prevent local charging-up of their surface. Each field cage vessel is surrounded by a grounded containment vessel, and the gaps between them are flushed with CO$_{\text{2}}$. In total, the field cage contains four resistor rods and four guard ring chains. 

Due to the small distance of about \SI{4.5}{\milli\meter} between resistor rods and guard rings, the voltage difference between them must be kept sufficiently low. In general, the potential of the last strip of each resistor chain has to be set to a value similar to that of the ROC surface facing the drift volume. In the case of the previous, MWPC-based ROCs, the reference potential was that of the gating grid wires (\SI{-70}{\volt}), so the chains were conveniently terminated with a small resistor to ground mounted inside the vessels. 
In the case of the GEM detectors, however, the reference potential is defined by the potential of the \gemOne top electrode, which is \SI{-3.26}{\kilo\volt} for nominal configuration. This new scenario requires the setting of the potential at the last strips (both resistor rods and guard rings) with additional power supplies.

For an adjustment of the potential of the last strip, an active powering scheme of the last strip and of the last guard ring was implemented. The resistors that had defined the potential in the past were removed, and existing HV connections in blind flanges in the aluminium endplate were adapted to the new HV ratings. Furthermore, the connections to the last field cage strips and the last guard ring strips were adapted to the new HV ratings.

\begin{figure}[t]
    \centering
    \includegraphics[width=0.45\textwidth]{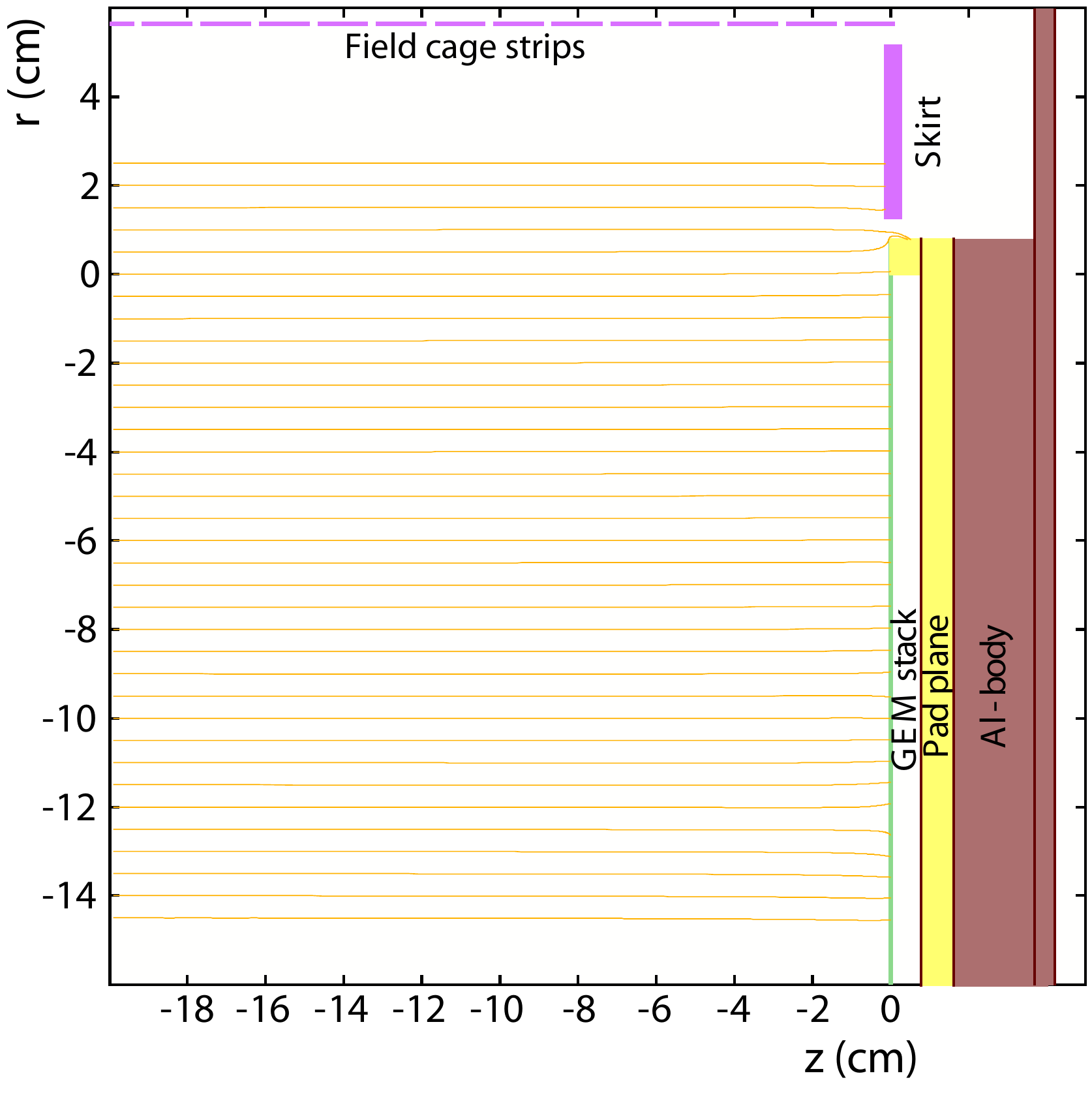}\hspace{5mm}
    \includegraphics[width=0.45\textwidth]{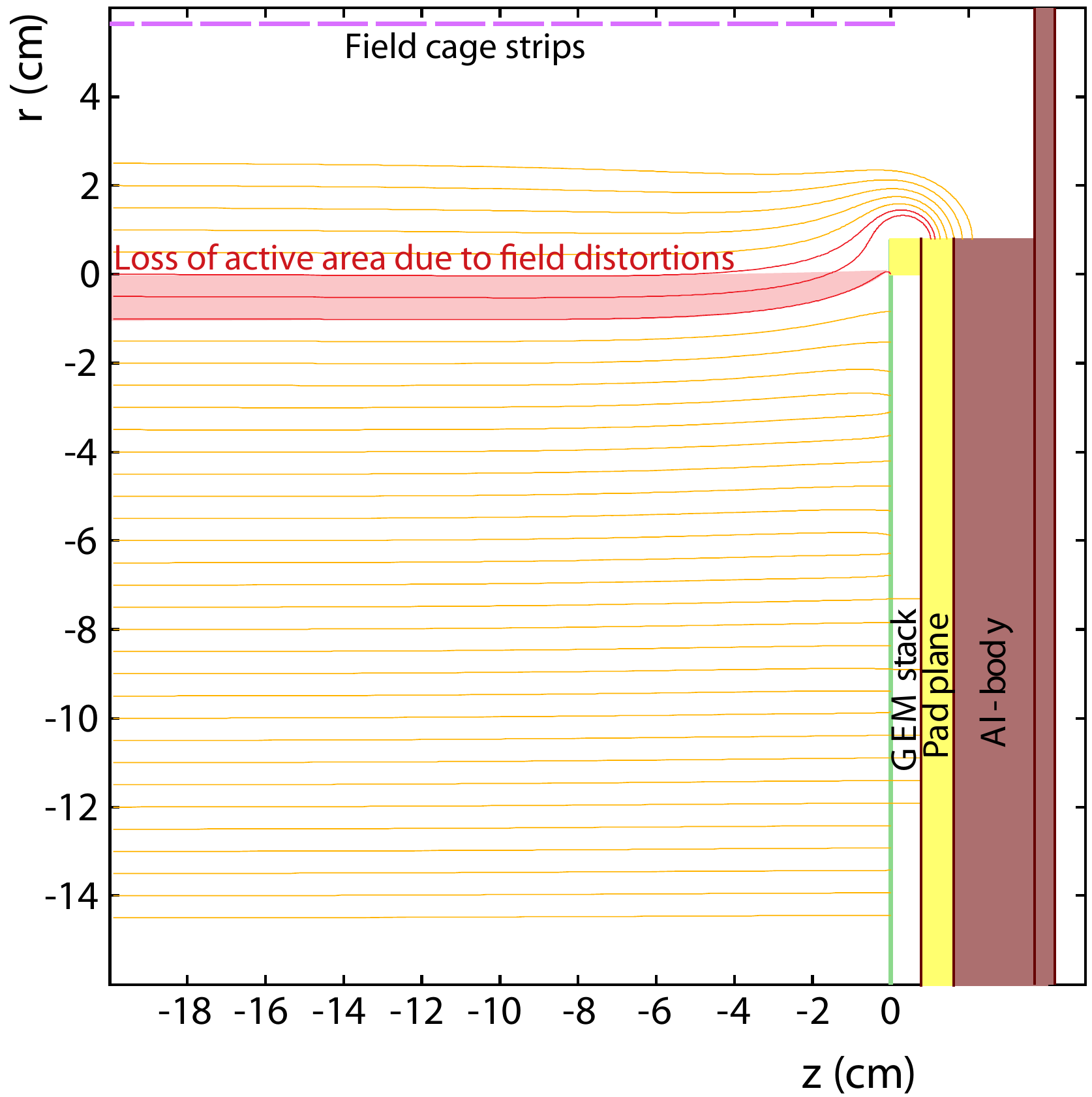}
    \caption{Simulations of the electron drift lines in the region of the outer field cage vessel close to the \oroc edge. In both cases, the active powering of the last strips of the field cage is implemented. The left panel corresponds to the presence of a properly powered skirt electrode, whereas in the right panel the skirt is removed. The area in which electrons are lost due to the distortions introduced by the missing skirt is highlighted.}
    \label{fig:hv:fc:skirtsimulation}
\end{figure}

The so-called skirt electrodes~\cite{TPCnim}, originally installed along the outer perimeter of the field cage in order to terminate the drift field lines in the gap between the \orocs and the field cage wall, were operated at a similarly low potential. Since the HV interconnections between neighboring skirt electrode plates are not graded for being powered at the \gemOne top potential, the skirt electrodes were removed. The effect on the electron drift lines was studied with Garfield simulations~\cite{GARFIELD:1984}. In \figref{fig:hv:fc:skirtsimulation}, the electron drift lines with (left) and without (right) skirt electrode in a region close to the outer vessel are shown. The active powering of the last strip is taken into account in both cases. The simulations show that drifting electrons from a small region of about \SI{1}{\centi\meter} are lost due to the removal of the skirt, as indicated in the right plot. This loss is negligible for the performance of the TPC. 

The temperature sensors installed in the skirt electrodes, necessary to monitor the temperature inside the TPC, were replaced by new sensors installed in various existing flanges around the endplate.

\subsubsection{Active powering scheme}
\label{sec:hv:fc:act}

\begin{figure}[ht]
    \centering
    \includegraphics[width=0.9\textwidth]{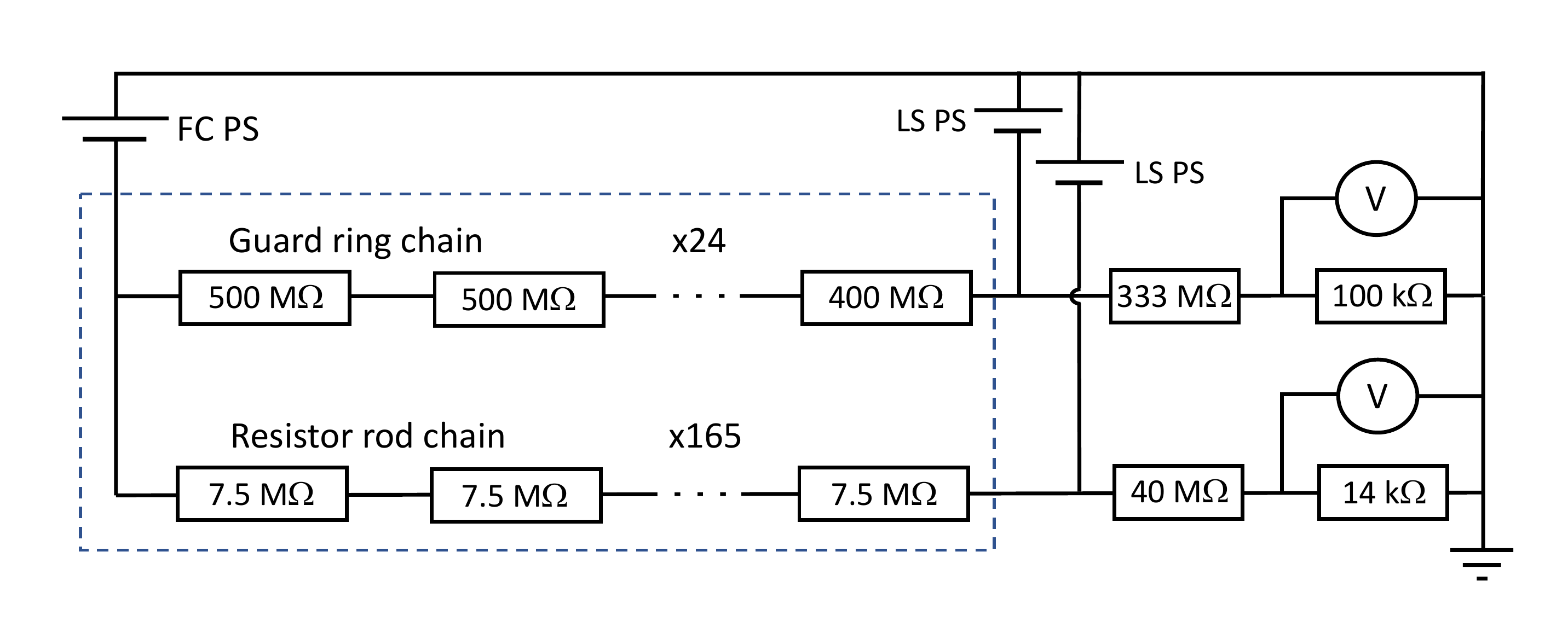}
    \caption{The modified powering scheme of the field cage: The resistor rod and guard ring chains are powered by the FC high-voltage power supply (FC PS). The last strips and guard rings are connected to ground through resistors with suitable values. In addition, they are powered by additional HV channels (LS PS) in order to define the required potentials. A very small resistor is inserted in each chain for the current measurements. These last resistors are installed outside the TPC volume (indicated by the dashed enclosure).}
    \label{fig:hv:activepowering}
\end{figure}

The active powering scheme that was implemented to achieve the proper potential settings of the resistor rods and the guard rings close to the ROCs is shown in \figref{fig:hv:activepowering}. A total of eight HV channels are used to be able to independently adjust the potential to terminate each chain, whereas suitable last resistors to ground are chosen to allow for the sinking of a small current to ground. The measurement of these currents provides information about the integrity of the system, while the last resistors ensure that the resistor chains are always properly terminated.

\subsubsection{Control software}
\label{sec:hv:fc:dcs}
While the control software of the field cage power supply (FC PS) remains the same as for the old system, new control software is added to drive the power supplies for the active powering of the last strips (LS PS). The voltages on the LS PS are always set according to the voltage on the FC PS, in particular while ramping.

An important safety feature is the early detection of any disconnected electrode which could result in high potentials in any last strip of the resistor rod or the guard ring chain. To avoid such cases, the currents  are continuously monitored  during ramp-up, which happens typically with a ramp speed of around \SI{130}{\volt\per\second}. In case of a significant deviation from the given limits, the ramp-up is automatically aborted.

\section{Front-end electronics and readout}
\label{sec:fec}

Charged particles traversing the TPC volume ionize the gas along their path.
The electrons liberated in this process drift towards the endplates where the readout chambers are mounted.
The signal amplification is provided through the avalanche effect in the GEM stacks.
The electrons created in the avalanche drift through the induction gap, inducing a negative current signal on the pad plane, composed of a few thousand readout pads.
The signal is characterized by very fast rise and fall times and carries a charge that, in the ALICE TPC, can be as low as 2.4\,fC to 3.2\,fC for a minimum ionizing particle (MIP), depending on the pad size (see \tabref{tab:roc:padplane:pads}).
It is picked up by one or a few pads on the pad plane, each connected to a front-end channel.
The front-end electronics and readout system for the upgraded TPC are described in the following.

\subsection{System overview}
\label{sec:fec.overview}

\begin{figure}[t]
\begin{center}
\includegraphics[width=0.9\textwidth]{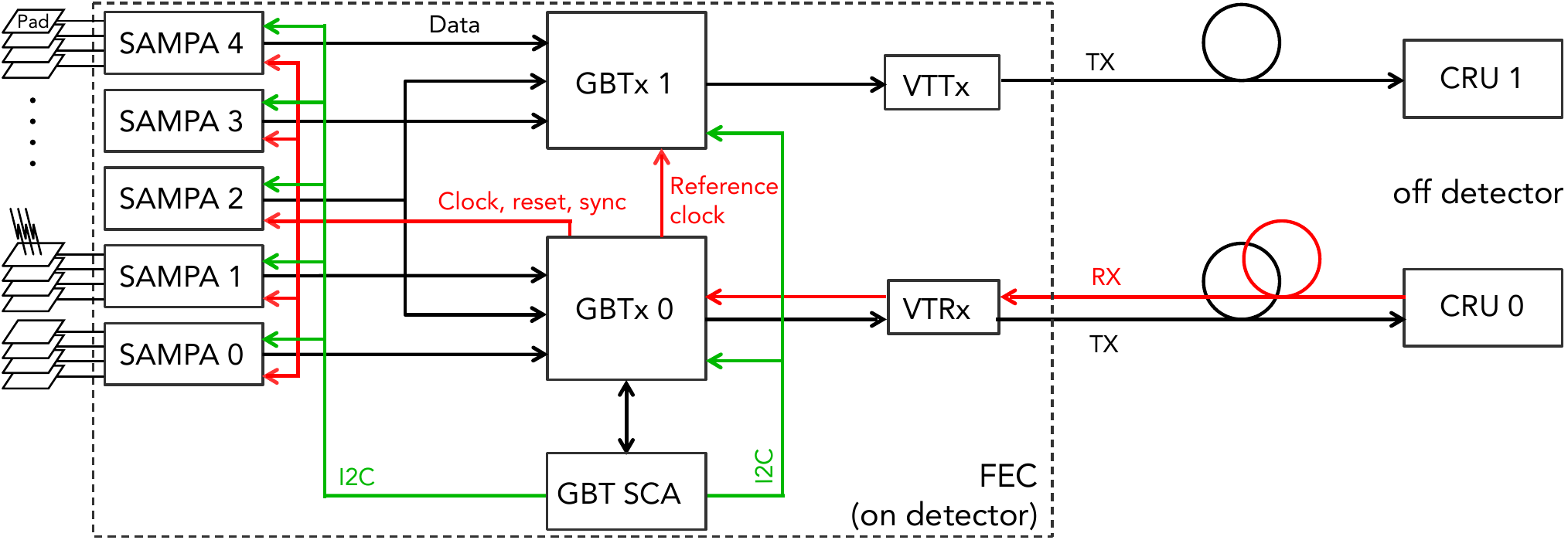}
\end{center}
\vspace{-0.5cm}
\caption{Schematic view of the TPC readout system. The different components are described in the text.}
\label{fig:fec:overview:blockdiagram}
\end{figure}

A schematic view of the front-end electronics and readout system is shown in \figref{fig:fec:overview:blockdiagram}.
160 front-end channels are combined in a single Front-End Card (FEC).
On each FEC, the signals are transformed into differential, semi-Gaussian voltage signals, and digitized in five custom-made ASICs (SAMPA, see \secref{sec:fec.sampa}).
The FEC further incorporates two optical readout links for the digitized data, and one control link (see \secref{sec:fec.gbt}).
The FEC is described in more detail in \secref{sec:fec.fec}.

The detector topology leads to a front-end electronics system consisting of 3276 FECs.
One TPC sector is read out using 91 FECs.
The FECs on both sides of the TPC are supported by a Service Support Wheel (SSW, see Sections \ref{sec:comm.ssw} and \ref{sec:install:fec}).
Forces on the endplates of the field cage, and on the readout chambers, are kept to a minimum by the implementation of flexible signal cables from the FECs to the readout chambers.
These signal cables are implemented as part of the FEC PCB (Rigid Flex, see \secref{sec:fec.fec.impl}), and allow for a compensation of a relative misalignment between SSW and end-plates of a few mm in all directions.

In total 6552 optical TX (transmit) links are needed to send the digitized data off-detector.
This leads to an unprecedented data throughput of 3.28\,TB\,s$^{-1}$.
360 FPGA\footnote{Field Programmable Gate Array}-based readout cards (Common Readout Unit, CRU) receive the data and perform online processing (common-mode correction, see \secref{sec:fec.cm}, and data reduction).
The FECs can be controlled through 3276 optical RX (receive) links, which also provide the reference clock to the digital circuitry on the FEC.
One sector is controlled and read out by ten CRUs.
The partitioning of the pad regions is realized in a way that each CRU receives data from complete rows of pads.
The pad dimensions and other parameters of the ten pad regions are described in \tabref{tab:roc:padplane:pads}.

\subsubsection{General requirements}
\label{sec:fec.overview.requirements}

The requirements for the front-end electronics system are listed in \tabref{tab:fec.specifications:fecs}.
With respect to the previous, MWPC-based readout system, the new FEC has to meet two new requirements:
continuous readout and negative input signal polarity.

\begin{table}[ht]\footnotesize
  \caption{Front-end electronics requirements.}
  \begin{center}
  \begin{tabular}{lc}
    \toprule
    Number of channels & 524160 \\
    Signal polarity & negative \\
    CSA saturation limit & $>$30\,nA\\
    Dynamic range &  30\,$\times$\,MIP\\
    Average system noise (ENC) & 670\,e (r.m.s.) \\
    Conversion gain & 20~(30)\,\si{\milli\volt\per\femto\coulomb} \\
    Cross talk & $<$2\,\% \\
    Peaking time & 160\,ns (FWHM: 190\,ns) \\
    ADC sampling rate & 5\,MHz \\
    ADC resolution & 10\,bit \\ 
    Readout mode & continuous \\ 
    Power consumption (total) & $<$100\,mW per channel \\
    \bottomrule 
  \end{tabular}
  \end{center}
  \label{tab:fec.specifications:fecs}
\end{table}

The requirement on system noise is based on the performance achieved with the previous system.
The amplifier peaking time is unchanged.
For the Charge Sensitive Amplifier (CSA), a requirement for the saturation limit of \SI{30}{\nano\ampere} is derived from the expected average rate of primary ionization clusters (up to about \SI{3}{\nano\ampere} per pad).
The factor of 10 accommodates for fluctuations due to local track multiplicity.
A dynamic range of \SI{100}{\femto\coulomb} allows for the measurement of the ionization signals of low-momentum particles that may produce signals 30 times larger than those of a MIP.
With respect to the previous system, the linear range was reduced for the benefit of a better resolution around the threshold level.
The conversion gain was increased from 12 to \SI{20}{\milli\volt\per\femto\coulomb}.
In order to approximately match the signal amplitudes in \irocs and \orocs, a second conversion gain setting of \SI{30}{\milli\volt\per\femto\coulomb} may be used to compensate for the smaller pad size in the \irocs.
In this case, the linear range is decreased from \SI{100}{\femto\coulomb} to \SI{67}{\femto\coulomb}.
To minimize the quantization error, the conversion to digital values takes place with a precision of 10 bits.

\subsubsection{Sampling frequency}
\label{sec:fec.overview.sampling}

In order to have maximum flexibility for online data processing (see \secref{sec:fec.cm}), all ADC values are transmitted off-detector.
The number of optical readout links (and thus the cost of the system) and the power consumption of the front-end electronics were reduced significantly by reducing the ADC sampling frequency from 10 to 5\,MHz.
6552 optical links (2 per FEC) are needed in this case.
In order to validate possible consequences on the physics performance, full microscopic simulations
were carried out~\cite{Appelshaeuser:2231785}.
Central Pb--Pb collisions were embedded into a background of minimum bias collisions that correspond to an interaction rate of 50\,kHz.
The results shown below correspond to various Signal-to-Noise ratios\footnote{The SNR is the ratio between the MIP signal in the largest pad-time bin ($Q_{\text{max}}$) in a cluster and the average noise.} (SNR) scenarios, including the nominal SNR.
At the operational point of the upgraded TPC, i.e.\ at an effective gas gain of 2000 and a noise level of 670\,e, a SNR of 20:1 will be achieved in the \irocs for minimum-ionizing particles.

\begin{figure}[h]
\begin{center}
\includegraphics[width=0.8\textwidth]{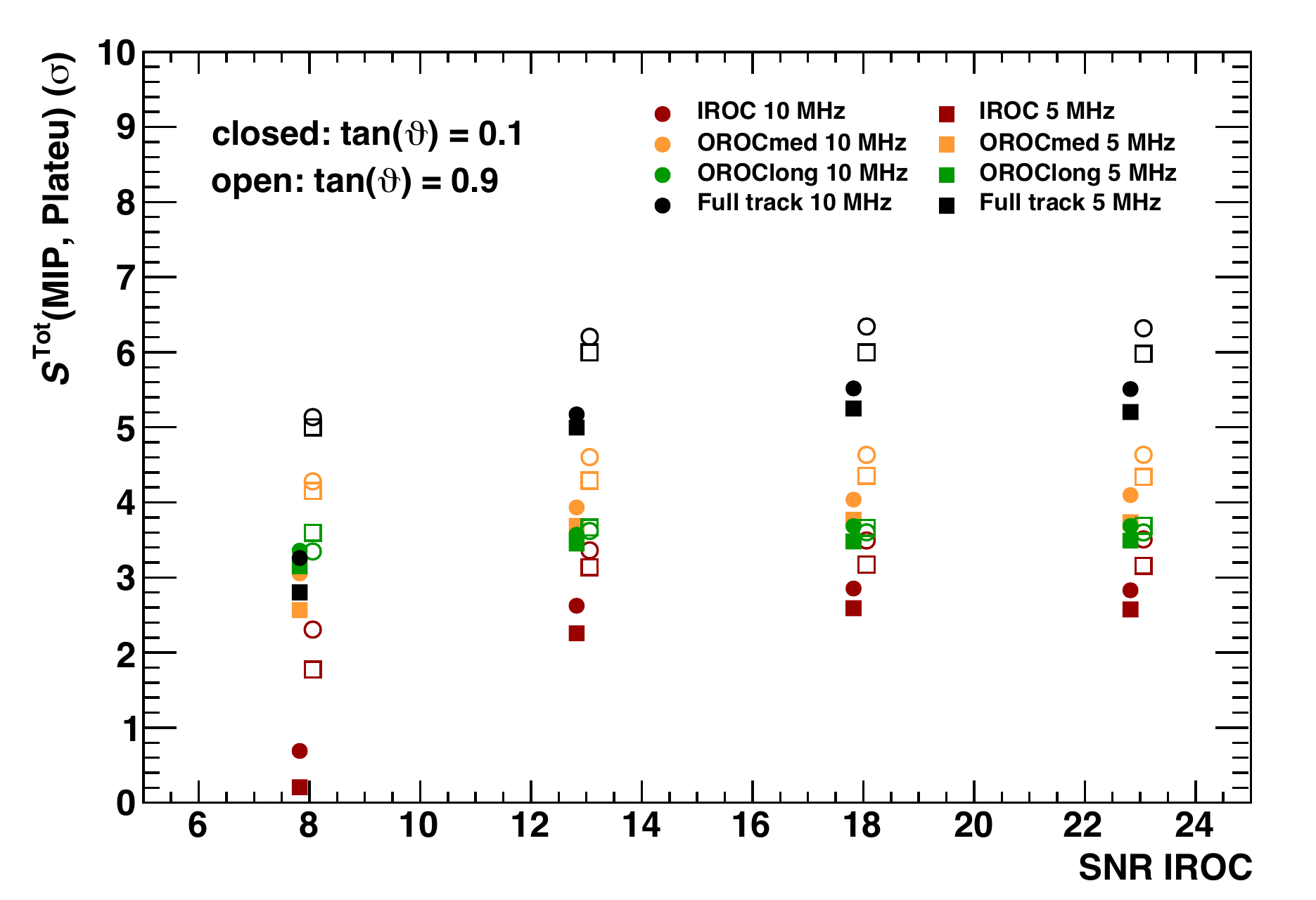}
\end{center}
\vspace{-0.3cm}
\caption{Separation power of minimum-ionizing particles with respect to particles at the plateau of the energy loss as a function of the SNR.
Closed symbols correspond to small pseudorapidities ($\eta$), open symbols to large pseudorapidities.
Pseudorapidities are expressed as the tangent of the dip angle, the angle between the particle momentum and the electron drift direction.
Circles correspond to the \SI{10}{\MHz} sampling case, squares to the \SI{5}{\MHz} sampling case.
The colors correspond to the different pad regions.
Black is for the full track length.}
\label{fig:fec:overview:MHz_dedx}
\end{figure}

In the \orocs, the SNR will be larger by a factor 1.3 to 2 at the same gas gain, because the larger pad sizes collect a larger number of primary electrons.

\Figref{fig:fec:overview:MHz_dedx} summarizes the \dEdx performance in terms of the separation power, as defined in Eq.~\ref{eq:sepa}, for minimum-ionizing pions and electrons in the plateau region of the Bethe-Bloch curve, using the total cluster charge as estimator.
Closed and open symbols correspond to small and large pseudorapidities, respectively (here shown in terms of the tangent of the dip angle).
Circles depict the \SI{10}{\MHz} sampling case, squares the \SI{5}{\MHz} sampling case.
The colors correspond to \dEdx calculated in different pad-length regions.
The studies were carried out with the pad plane design from the previous system, which has two different pad sizes for the \oroc: medium and long pads.
The black symbols indicate the result for the full track length.
The overall difference of the separation between 5 and 10\,MHz sampling frequency for reasonable SNRs is of the order of 3 to \SI{5}{\percent}.
It is expected that minor differences will be further reduced when a more detailed optimization of the reconstruction algorithms has been carried out.

\begin{figure}[htb]
\begin{center}
\includegraphics[width=\textwidth]{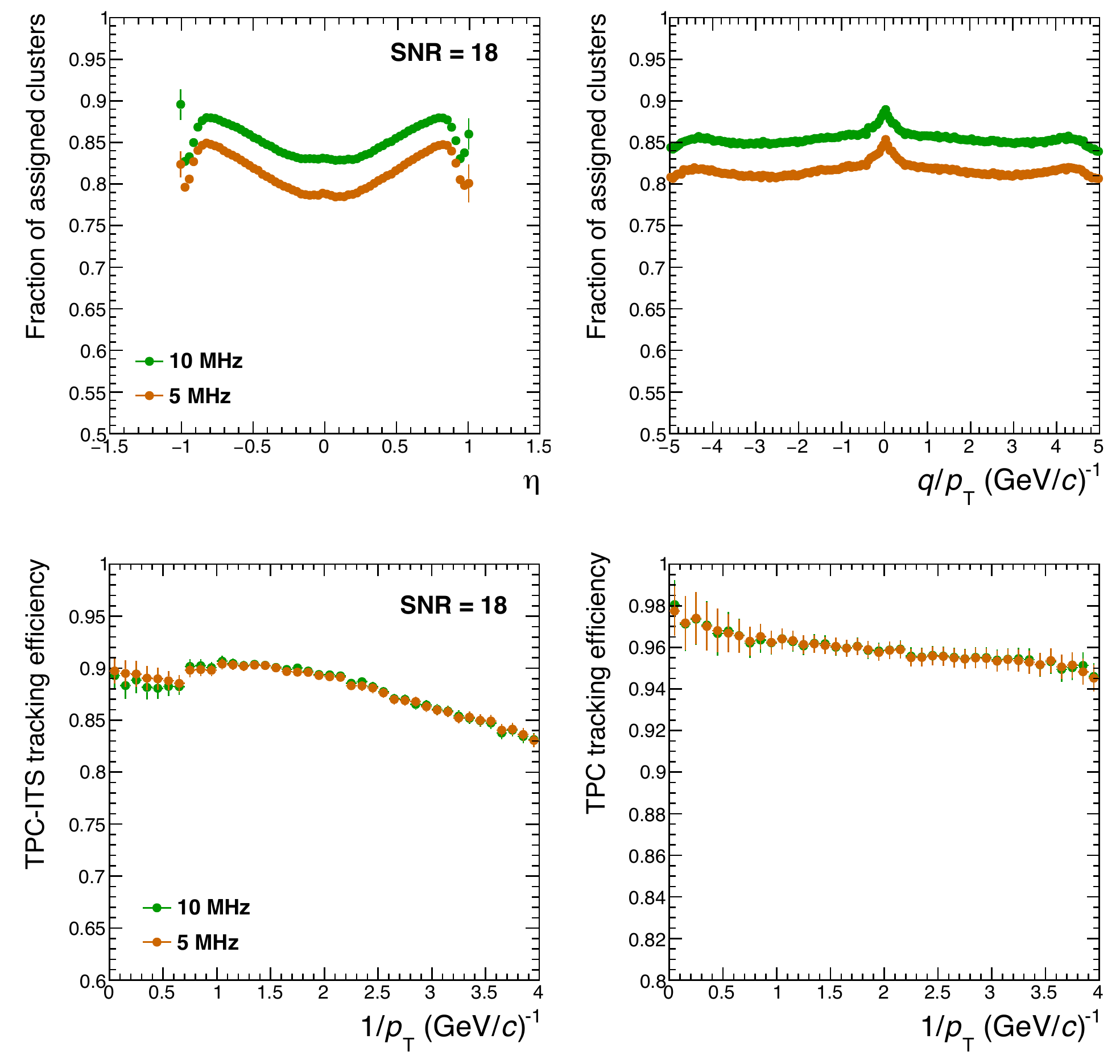}
\end{center}
\vspace{-0.3cm}
\caption{Cluster-to-track association probability as a function of pseudorapidity (top left) and charge-over-$p_{\mathrm{T}}$ (top right) for \SI{5}{\MHz} (orange circles) and \SI{10}{\MHz} (green circles).
The bottom left plot shows the tracking efficiency as a function of $1/p_{\mathrm{T}}$ for tracks where data from the ALICE Inner Tracking System (ITS) is included in the tracking procedure. 
The bottom right plot shows the tracking efficiency for TPC-standalone tracks.}
\label{fig:fec:overview:MHz_tracking}
\end{figure}

In \figref{fig:fec:overview:MHz_tracking} the track reconstruction performance of the 5 and \SI{10}{\MHz} scenario is displayed for TPC-standalone tracks and for TPC tracks combined with clusters from the ALICE Inner Tracking System\footnote{The previous design of the ITS (before the respective upgrade) was used for these studies.} (ITS).
The upper panels show the cluster-to-track association probability as a function of pseudorapidity $\eta$ and charge-over-$p_{\mathrm{T}}$, where $p_{\mathrm{T}}$ is the transverse momentum of the particles.
The lower panels show the tracking efficiency as a function of $1/p_{\mathrm{T}}$.
About \SI{5}{\percent} less clusters are associated to a track in case of sampling with the lower frequency.
The loss is slightly larger for longer drift lengths (small $\eta$) due to diffusion.
Nevertheless, the tracking efficiency is very comparable in both cases.

\begin{figure}[htp]
\begin{center}
\includegraphics[width=1.0\textwidth]{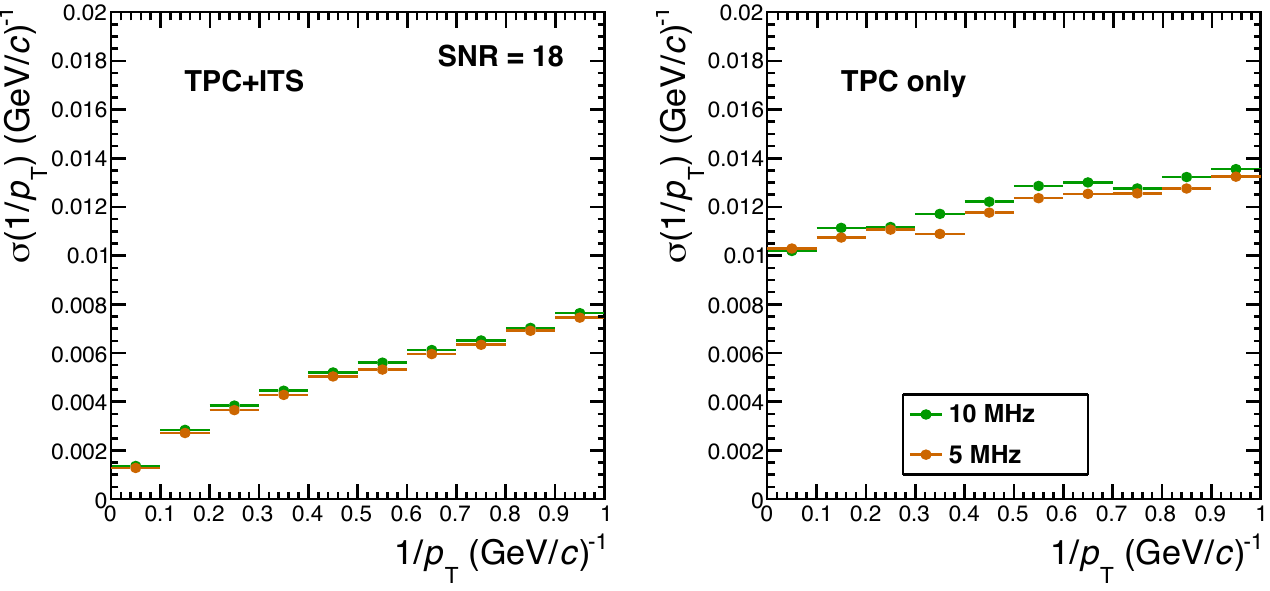}
\end{center}
\vspace{-0.3cm}
\caption{Transverse momentum resolution expressed as $\sigma(1/p_{\mathrm{T}})$ as a function of $1/p_{\mathrm{T}}$ for TPC-ITS-combined tracking (left) and TPC-standalone tracking (right) in central Pb--Pb collisions with pileup at 50\,kHz collision rate.
Green and orange circles correspond to the \num{10} and \SI{5}{\MHz} sampling case.
}
\label{fig:fec:overview:MHz_pT}
\end{figure}

\Figref{fig:fec:overview:MHz_pT} shows the transverse momentum resolution, expressed as the width of gaussian fits $\sigma(1/p_{\mathrm{T}})$, as a function of $1/p_{\mathrm{T}}$, for  TPC-ITS-combined tracks (left) and TPC-standalone tracks (right).
No notable dependence on the sampling frequency is observed.
This observation holds for all SNRs under study. 

These simulations demonstrate that the \dEdx performance shows only very little sensitivity  on the choice of the ADC sampling frequency between \num{5} and \SI{10}{\MHz}.
The same observation holds for the tracking efficiency and momentum resolution.
The choice of \SI{5}{\MHz} also does not reduce the safety margin with respect to the nominal setting at SNR\,=\,20:1.

\subsubsection{Common-mode effect and correction}
\label{sec:fec.cm}

\begin{figure}[htp]
\begin{center}
\includegraphics[width=0.7\textwidth]{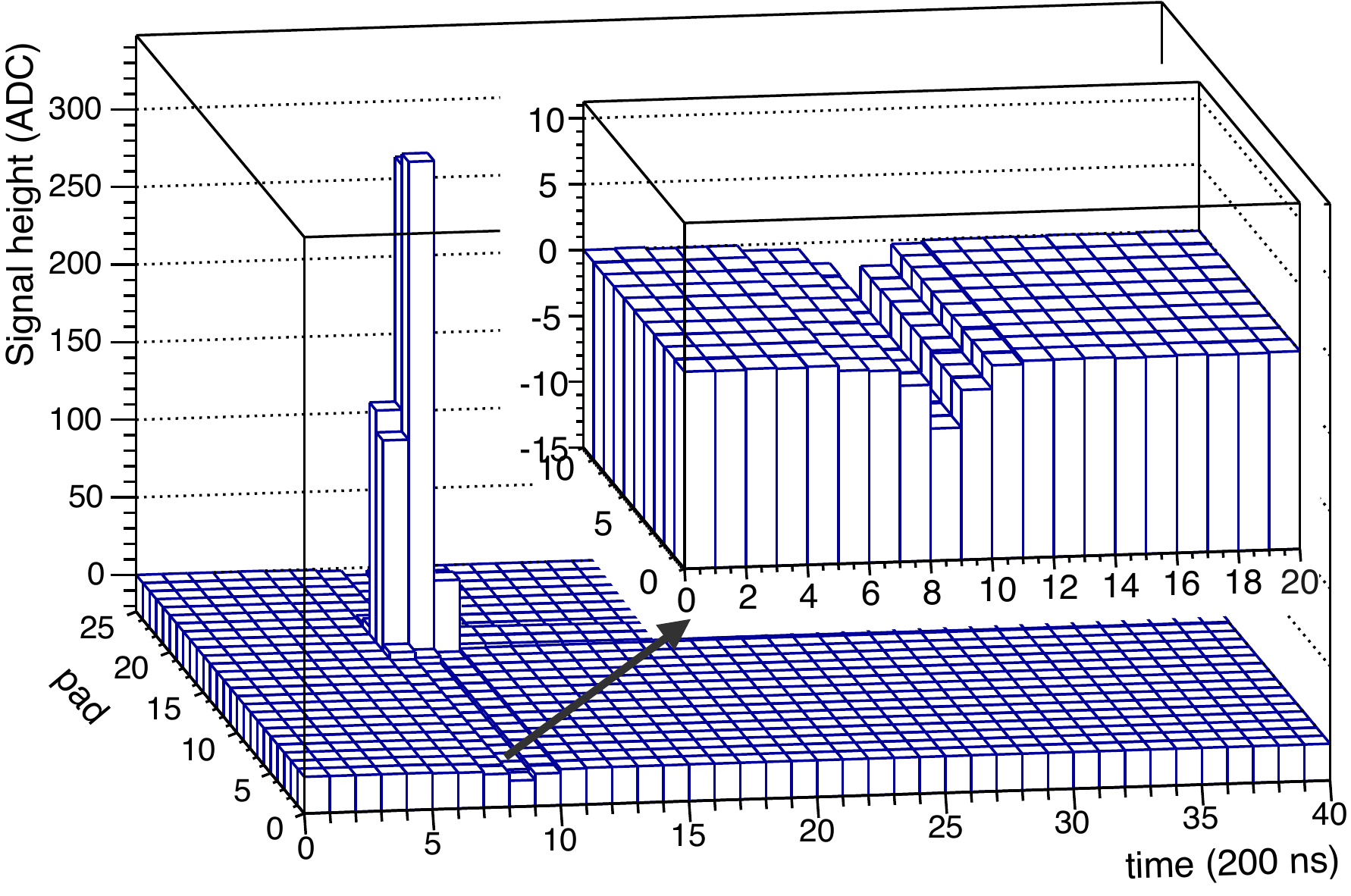}
\end{center}
\vspace{-0.3cm}
\caption[Common-mode effect principle]{Visualization of the common-mode effect in a data sample triggered by the TPC laser system (see \secref{sec:precom:laser_data_taking}). Several events are superimposed in order to remove noise fluctuations. Data on 25 selected pads under the same GEM stack are shown. A small signal of opposite polarity is visible on all pads at the same time position as the positive signal from a laser track. Note: Since laser signals are induced simultaneously on many pads, a calculation of the strength of the common-mode effect is not possible based on the data displayed here.}
\label{fig:fec:overview:cm}
\end{figure}

The common-mode effect occurs due to capacitive coupling of the amplification structure (i.e.\ the \gemFour bottom electrode) to the readout pads.
The currents due to charges drifting through the induction gap of a GEM system introduces a voltage drop on the GEM electrode, which leads to the induction of a correlated \textit{common-mode} signal with opposite polarity on all anode pads facing the \gemFour bottom electrode (see \figref{fig:fec:overview:cm}).
The amplitude of the common-mode signal on a given pad is suppressed, with respect to the original signal, by a factor related to the surface of the GEM electrode.
Common-mode signals pile up in a high-multiplicity environment, leading to an average baseline drop and an effective noise contribution.
The magnitude of the common-mode signals can be reduced significantly by adding capacitance between ground and the \gemFour bottom electrode.
However, this counteracts the effort to minimize the stored energy in the GEM system to improve operational stability and safety.
Alternatively, the common-mode effect can be corrected based on the available raw ADC data.

\begin{figure}[htp]
\begin{center}
\includegraphics[width=0.6\textwidth]{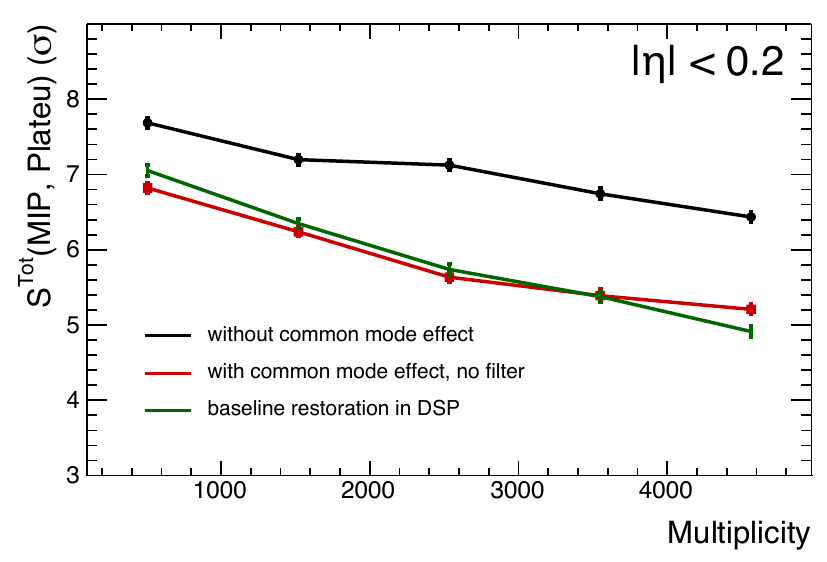}
\end{center}
\vspace{-0.3cm}
\caption[Separation power with baseline correction in SAMPA]{Separation power of minimum-ionizing particles to the plateau as a function of charged-particle multiplicity at midrapidity ($|\eta|<0.2$).
A multiplicity of 4500 corresponds to a central \PbPb collision at 50~kHz, i.e.\ superimposed on 4 minimum bias collisions.
The black curve shows the performance without common-mode noise.
The red curve shows the performance with common-mode noise, but without correction in the SAMPA DSP.
The green, blue, and magenta curves show the performance with common-mode noise and with different parameter settings for the filter algorithm in the SAMPA DSP.}
\label{fig:fec:overview.bc3perf}
\end{figure}

Different filtering algorithms were studied to treat the effect inside the SAMPA Digital Signal Processor (DSP, see \secref{sec:fec.sampa.desc}).
In these approaches, the baseline is restored on the single-pad level.
True particle signals are excluded from the common-mode calculation using a slope-based extrapolation approach and a peak-detection algorithm.
Quantitative performance studies showed that the average baseline can be well restored.
However, a significant correlated noise contribution is added to the data and leads to a sizable degradation of the detector performance (see \figref{fig:fec:overview.bc3perf}).

\begin{figure}[htp]
\begin{center}
\includegraphics[width=0.7\textwidth]{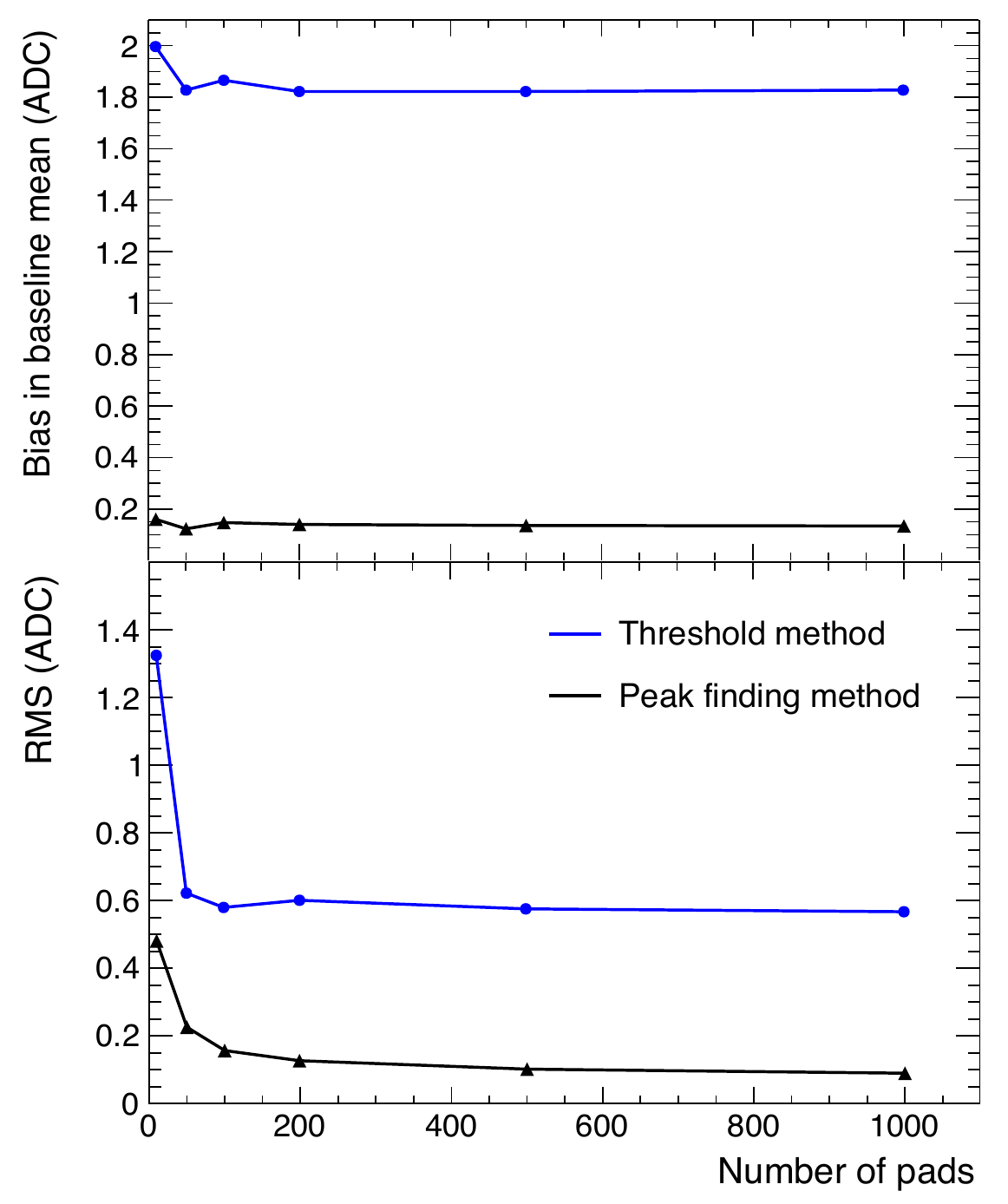}
\end{center}
\vspace{-0.3cm}
\caption{Residual average baseline shift (top) and effective common-mode noise contribution (bottom) as a function of the number of pads used for the average calculation.
Two different approaches are used to reject particle signals from the baseline calculation.
A constant threshold of 2.5 ADC counts (blue) is compared with a peak-finding algorithm, where the peak region is omitted from the baseline calculation (black).}
\label{fig:fec:overview.cmperf}
\end{figure}

A much better correction can be achieved by extracting the baseline shift due to the common-mode effect from a large number of pads.
This takes into account that the common-mode signal is, to first order, the same on all pads facing the same \gemFour bottom electrode.
If the true particle signals are detected by a simple peak detection algorithm, the baseline can be restored with negligible residual bias and noise contribution by averaging over at least 100 pads (see \figref{fig:fec:overview.cmperf}).
Such averaging over a large number of pads is not possible in the SAMPA DSP, where only data from a single pad is available at a time.
In order to achieve averaging over a large number of channels, the TPC readout system was designed such that all ADC raw data are sent off-detector without compression.
The data from 1200 to 1600 channels are merged in a large FPGA in a CRU, where efficient and flexible algorithms can be implemented in order to perform common-mode correction.

\subsection{SAMPA}
\label{sec:fec.sampa}

The SAMPA~\cite{sampa2020} is a custom integrated circuit containing 32 channels with selectable input polarity and five possible combinations of shaping time and sensitivity. The SAMPA was developed over a series of prototypes, and the v4 design was selected for mass production and instrumenting the ALICE TPC (about 20000 chips needed).

\subsubsection{Circuit description}
\label{sec:fec.sampa.desc}

\begin{figure}[htp]
\begin{center}
\includegraphics[width=0.7\textwidth]{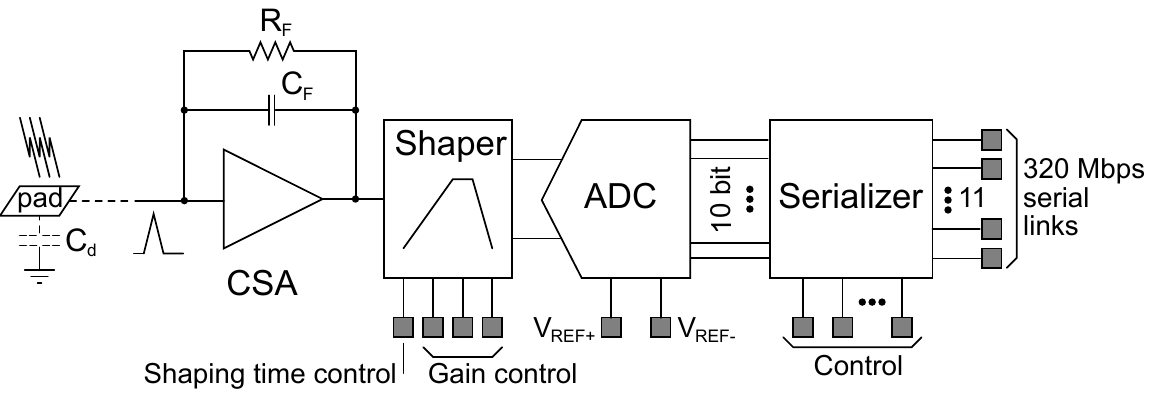}
\end{center}
\caption{SAMPA block diagram. The Digital Signal Processor block, which is bypassed in the operational mode selected for the TPC, is not shown.}
\label{fig:fec:sampa:blockdiagram}
\end{figure}

A simplified block diagram of the signal processing chain is shown in \figref{fig:fec:sampa:blockdiagram}.
In the following, short descriptions of the main building blocks are given, following the signal processing path.

\begin{figure*}[t]
\begin{center}
\includegraphics[width=0.8\textwidth]{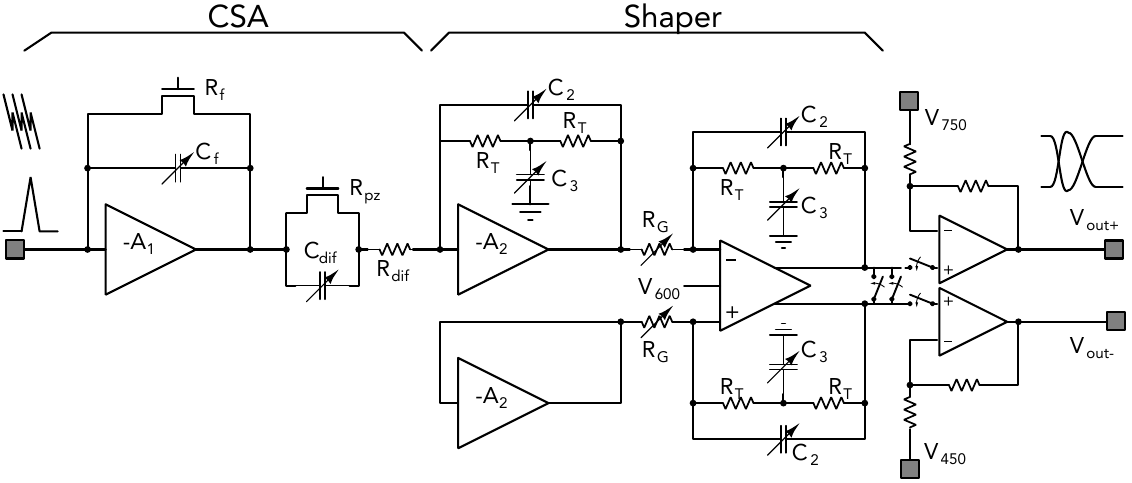}
\end{center}
\caption{A simplified block diagram of the SAMPA analog signal processing chain.}
\label{fig:fec:sampa:analog}
\end{figure*}

The analog part of the SAMPA integrates 32 identical Charge Sensitive Amplifiers (CSAs), followed by a pole-zero cancellation network and a pulse-shaping amplifier.
A simplified block diagram of the analog signal processing chain is shown in \figref{fig:fec:sampa:analog}.
The positive/negative polarity CSA, with capacitive feedback $C_{\text{f}}$ and resistive feedback $R_{\text{f}}$ connected in parallel\footnote{With respect to previous versions, the SAMPA v4 features an improved pileup robustness in order to satisfy the TPC specifications.
A faster discharge of $C_{\text{f}}$ is obtained by a reduction of the drain-to-source resistance $R_{\text{f}}$.
The pole-zero cancellation resistance $R_{\text{pz}}$ is reduced proportionally to maintain a proper cancellation.}, is followed by a pole-zero cancellation network, a high-pass filter, two bridged-T second-order low-pass filters and a non-inverting stage.

The first shaper is a scaled-down version of the CSA and generates the first two poles and one zero.
A copy of the first shaper, connected in unity gain configuration, is implemented in order to provide a differential-mode input to the next stage.
The second stage of the shaper is a fully differential second-order bridged-T filter and it includes a common-mode feedback network.
The non-inverting stage adapts the DC voltage level of the pulse-shaping amplifier output to utilize the full input dynamic range of the ADC.
It consists of a parallel connection of two similar Miller-compensated amplifiers.

\begin{figure*}[t!]
\begin{center}
\includegraphics[width=1.0\textwidth]{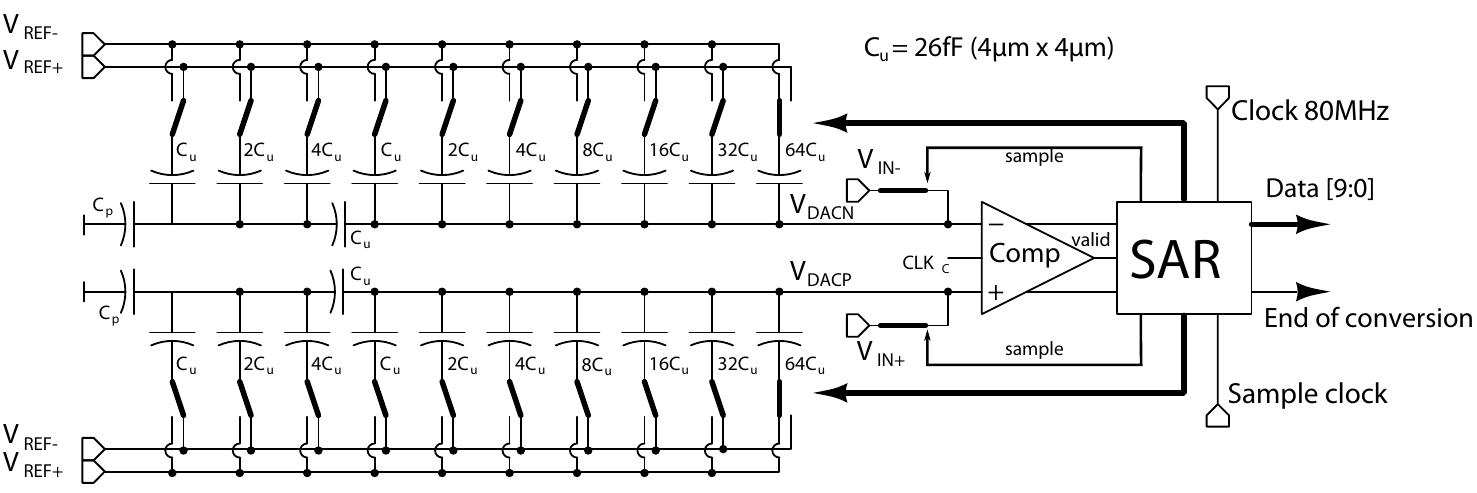}
\end{center}
\caption{SAMPA ADC block diagram.}
\label{fig:fec:sampa:adc}
\end{figure*}

The ADC is based on a differential SAR\footnote{Successive Approximation Register} architecture and is implemented with a low-power switching technique.
It has a resolution of 10 bit and a sample rate up to \SI{20}{\MHz}.
The block diagram of the ADC is shown in \figref{fig:fec:sampa:adc}.
The main parts of the circuit are: a capacitive array, switches, a comparator and the SAR control logic.
The capacitor array is used to perform sample\&hold and the digital-to-analog converter functions.

The Digital Signal Processor (DSP, not shown in \figref{fig:fec:sampa:blockdiagram}) block is composed of digital filters, a data formatting unit, a ring buffer, a trigger manager, a configuration register bank, and a control state machine.
Eleven serial electrical output links (E-Links) are available for data transmission at 80, 160 or 320\,Mbit\,s$^{-1}$.
The E-Links use the SLVS\footnote{Scalable Low-Voltage Signaling} standard.
Triggered and continuous operation with data processing in the DSP can be configured.
In continuous mode, also direct transmission of all ADC values to the output E-Links can be enabled.
In this case, the DSP is bypassed and the DSP processing blocks are shut down through clock gating.
The direct readout mode operates with a serialization speed of 32 times the ADC sampling speed, such that 10 E-Links suffice to send all ADC data.
The 11th E-Link transmits the sampling clock, which may be used for aligning the data and to monitor the sampling phase.
For synchronization to the data stream in direct readout mode, a dedicated 32-cycle pattern is generated upon reception of a dedicated trigger signal.

In direct readout mode, the 10 output bits of each ADC are sent over the 10 E-Links, respectively\footnote{In the first clock cycle (e.g.\ after sending the synchronisation pattern), the ten bits of the data word from channel zero are sent on the ten E-Links. On the consecutive cycle, the ten bits for channel one are sent, and so on.}. In addition, a split mode is available.
In split mode, the data for channels zero to 15 and for channels 16 to 31 can be sent to two different upstream receivers\footnote{When using the split mode, the ADC values from the lower 16 channels are transmitted through the lower five serial link numbers, and the ADC values from the higher 16 channels are transmitted through the higher five serial link numbers.
In the first clock cycle, the five lower bits of the data word from channel zero are sent on serial links zero to four, and the five lower bits for channel 16 are sent on links four to nine.
On the consecutive cycle the five upper bits for channel zero are sent on links zero to four and the five upper bits for channel 16 are sent on links four to nine. The readout then continues with channel one in the same manner.}. The split mode is used on the TPC FECs due to the particular configuration with five SAMPA chips and two data links per FEC.

\subsubsection{Physical implementation}
\label{sec:fec.sampa.impl}

\begin{figure}[ht]
\begin{center}
\includegraphics[width=1.0\textwidth]{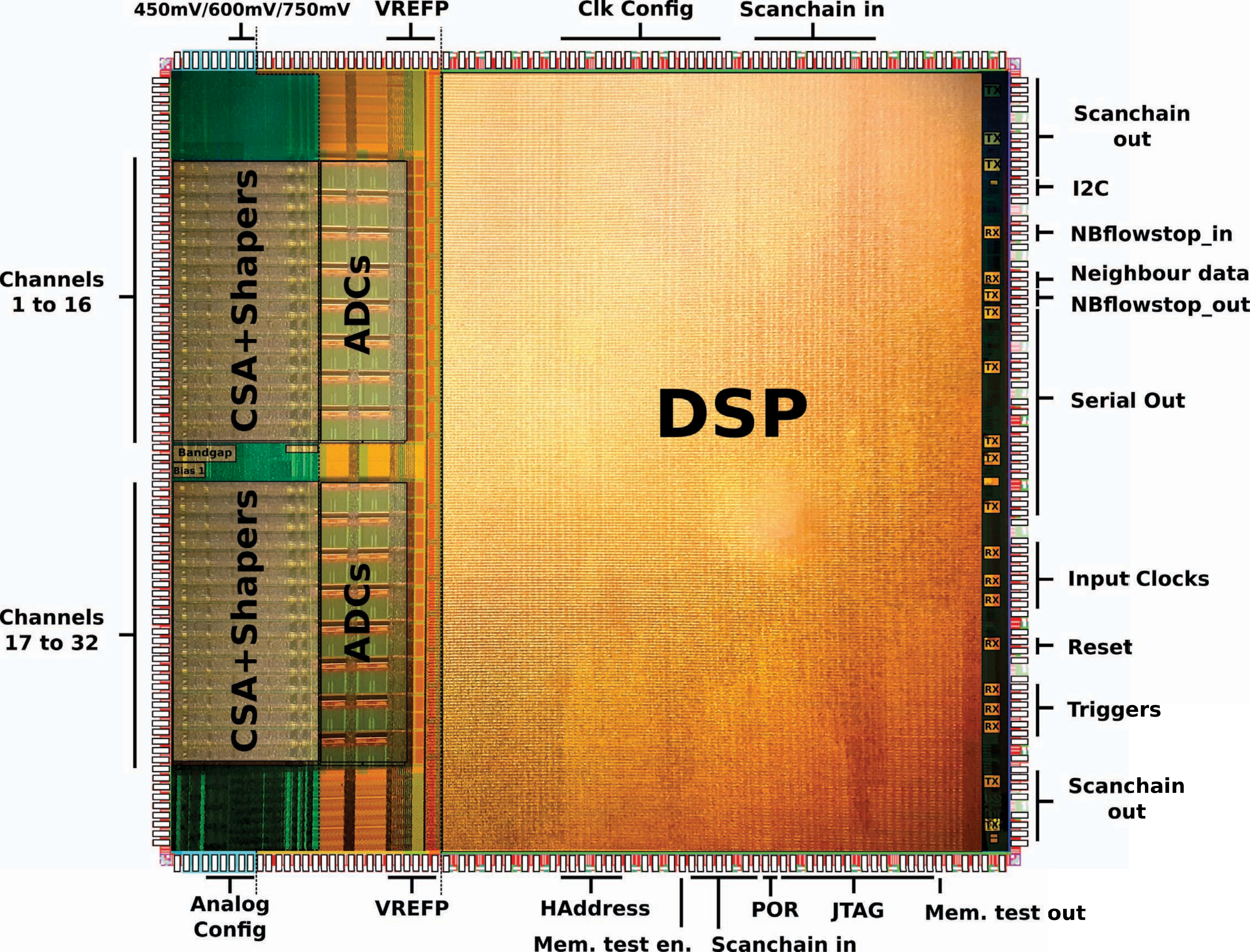}
\end{center}
\caption{SAMPA chip layout. The placement of the individual functional blocks on the die is indicated, as well as the positions of some of the various input and output pins along the edge of the die.}
\label{fig:fec:sampa:layout}
\end{figure}

The main physical characteristics of the circuit are summarized in \tabref{tab:fec.characteristics.sampa}.
The chip was manufactured in the 130\,nm TSMC CMOS technology.
It has a width of 8.9\,mm and a length of 9.5\,mm, with a total area of 85\,mm$^2$.
It works with a 1.25\,V nominal voltage supply.
The input and output pads are located on opposite sides of the chip, as shown in \figref{fig:fec:sampa:layout}.
The 32 analog CSA and shaper blocks are placed directly behind the input pads.
The 32 ADCs are arranged in two rows behind the analog channels.
The analog and ADC channels are divided in two groups of 16 channels each, separated by the bandgap and bias network in the middle.
The DSP uses \SI{60}{\percent} of the surface of the chip.
The E-Link SLVS RX and TX drivers are placed along the output side of the SAMPA chip.

The SAMPA features device-level ESD\footnote{ElectroStatic Discharge} protection for handling.
It consists of one 7\,$\Omega$ resistor, in series with the input pad, and one back-to-back diode pair per input.

\begin{table}[ht]\footnotesize
  \caption[SAMPA physical characteristics]{SAMPA physical characteristics.}
  \begin{center}
  \begin{tabular}{lc}
    \toprule
    Process & \SI{130}{\nano\meter} TSMC CMOS \\
    Die area & \SI{85}{\milli\meter\squared} \\
    Die size & 8944\,$\times$\,\SI{9534}{\micro\meter\squared} \\
    Package & TFBGA \\
    Package area & \SI{225}{\milli\meter\squared} \\
    Package dimensions & 15\,$\times$\,\SI{15}{\milli\meter\squared} \\
    Number of substrate layers & 4 \\
    Number of balls & 372 \\
    Ball pitch & \SI{0.65}{\milli\meter} \\ 
    Number of transistors & 36 million \\ 
    Supply voltage & \SI{1.25}{\volt} \\ 
    \bottomrule 
  \end{tabular}
  \end{center}
  \label{tab:fec.characteristics.sampa}
\end{table}

As detailed in \secref{sec:fec.testing}, about 90883 SAMPA chips were tested in the mass production step.
These comprise the chips needed for the production of the TPC front-end electronics, and those for the ALICE Muon system, that uses the same ASIC.
\SI{79.6}{\percent} of the chips were fully functional.
The general performance of the chip is listed in \tabref{tab:fec.sampa.perf}.
About 19500 good SAMPA chips were selected for the production of the TPC FECs, including spares.

\begin{table}[h!]\footnotesize
  \caption[SAMPA key performance figures]{SAMPA key performance figures for TPC settings (at conversion gain \SI{20}{\milli\volt\per\femto\coulomb}, peaking time 160\,ns). The acronyms stand for:
  Effective Number Of Bits (ENOB), Differential Non-Linearity (DNL) Integral Non-Linearity (INL), and Spurious-Free Dynamic Range (SFDR). %
  $^1$Measured on FEC at capacitance 12\,pF.
  The spread of the measured distributions (compare to \figref{fig:fec:fec:resultzz}) is given as r.m.s.
  $^2$Measured up to 95\,fC ($>$\SI{85}{\percent} of the full range).
  $^3$ADC simulation at 10\,MHz sampling frequency, input frequency of 2451171.875\,Hz, \SI{90}{\percent} input range, typical corner at 40 deg. Celsius.}
  \begin{center}
  \begin{tabular}{lcc}
    \toprule
    Parameter & Specification & Performance \\
    \midrule
    Noise$^1$ (ENC) & $<$\SI{570}{e} & (529\,$\pm$\,7)\,e \\
    Conversion gain$^1$ & \SI{20}{\milli\volt\per\femto\coulomb} & (19.8\,$\pm$\,0.3)\,mV\,fC$^{-1}$\\
    Baseline values$^1$ & $<$\SI{200}{\milli\volt} & (167\,$\pm$\,19)\,mV \\
    Crosstalk$^1$ & $<$\SI{2}{\percent} & (1.34\,$\pm$\,0.12)\,\% \\
    Peaking time & \SI{160}{\nano\second} & \SI{165.3}{\nano\second} \\ 
    CSA saturation limit & $>$\SI{30}{\nano\ampere} & \SI{33}{\nano\ampere} \\
    Max. non-linearity$^2$ & - & \SI{10}{\milli\volt} \\
    ADC ENOB$^3$ & $>$9.2 & 9.25 \\
    ADC DNL$^3$ & - & $<$\SI{0.2}{LSB} (r.m.s.)\\
    ADC INL$^3$ & - & $<$\SI{1}{LSB} (abs.) \\
    ADC SFDR$^3$ & - & \SI{-59.0}{dB} \\
    Power per channel (SAMPA only) & $<$\SI{35}{\milli\watt} & \SI{9}{\milli\watt} (direct readout mode) \\ 
    \bottomrule 
  \end{tabular}
  \end{center}
  \label{tab:fec.sampa.perf}
\end{table}

\subsection{Giga-Bit Transceiver optical link}
\label{sec:fec.gbt}

The data produced by the SAMPA ASICs are multiplexed into optical readout links.
For this  purpose, the radiation-hard Giga-Bit Transceiver optical link (GBT)~\cite{moreira2009gbt} system is used.
The GBT system was developed at CERN with the purpose of providing within a single optical link the simultaneous transfer of all types of information required by High-Energy Physics experiments: readout data, timing and trigger information, and detector control and monitoring information.
These different logical paths are merged into a single optical bi-directional link operating at a rate of 4.8\,Gbit\,s$^{-1}$.
The default user bandwidth of 3.2\,Gb\,s$^{-1}$ (plus 80\,Mb\,s$^{-1}$ for control and monitoring data) guarantees protection from bit errors, since it implements a Forward Error Correction mechanism. 
The bandwidth can be increased to 4.48\,Gb\,s$^{-1}$ in the TX link from the front-end electronics to the CRU by activation of the \textit{wide frame mode}~\cite{gbtmanual_016}.
In this case, the Forward Error Correction is not available.
This mode is enabled on the TPC front-end card.
In the following, the individual components of the GBT system are briefly described:

\begin{itemize}
\item 
The \textbf{GBTx} transceiver ASIC is a radiation tolerant link interface chip with a large number of programmable options~\cite{gbtmanual_016}.
It acts as an interface for up to 40 front-end modules, depending on the communication bandwidth selected, that can be connected to the device by means of electrical links (E-Links).
An E-Link consists of three differential signals: data input and output (operating at 80, 160 or 320\,Mbit\,s$^{-1}$), and a clock output at 40, 80, 160 or 320\,MHz.
In transceiver mode the GBTx recovers the LHC clock, which allows the GBTx to act as a clock fan-out using the E-Links and eight phase/frequency programmable output clocks.
The E-Ports are equipped with a phase-aligner for the data inputs in order to provide optimal sampling of the incoming data
stream.
The GBTx serializes and multiplexes these data streams and sends them to the optical data transmission components.
The data consists of frames with size 120\,bit sent at 40\,MHz, where the data from each E-Link is associated with a specific set of bits in the frames.

\item 
The Versatile Transceiver (\textbf{VTRx}) is a bi-directional custom-developed optical transceiver module loosely following the SFP+,\footnote{Small Form-factor Pluggable} form-factor~\cite{Soos_2013}.
The uni-directional Versatile Twin Transmitter (\textbf{VTTx}) module consists of two transmit channels in the same package as the VTRx. 

\item 
The \textbf{GBT-SCA} ASIC~\cite{Caratelli_2015, scamanual_2019} distributes control signals to the different components on the FEC, and performs monitoring operations (read voltages, currents and temperatures).
It provides various user-configurable interfaces capable of performing simultaneous operations.
In particular, the 12-bit ADC channel with 31 multiplexed, analog inputs, the I2C,\footnote{Inter-Integrated Circuit} serial bus channels, and the General Purpose digital IO lines (GPIO) are used on the TPC FEC (see \secref{sec:fec.fec.moni}).
\end{itemize}

\subsection{Front-End Card}
\label{sec:fec.fec}

The FEC comprises the following functionality: amplification, shaping and digitization of the TPC signals, multiplexing and sending of the digitized data, and monitoring and control of the individual components.

\begin{figure}[ht]
\begin{center}
\includegraphics[width=0.98\textwidth]{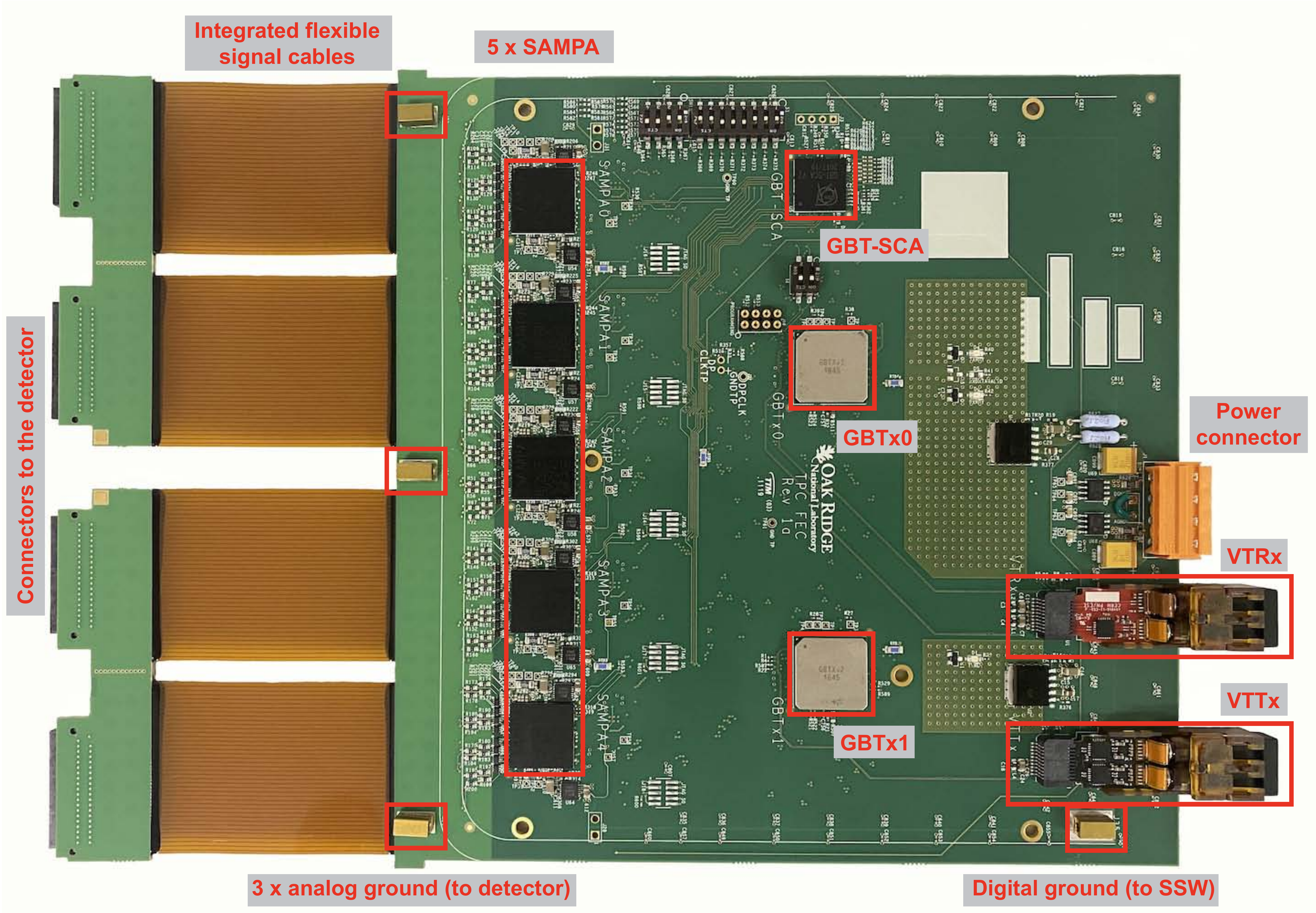}
\end{center}
\caption[Layout of the final TPC Front-End Card Rev.\,1a.]{Layout of the final TPC FEC PCB Rev.\,1a. The components are mounted on both sides of the board. The figure shows the top side with five SAMPAs, two GBTx, one GBT-SCA, one VTRx, one VTTx, and some other components. On the bottom side a few additional small components and the connectors to the detector are placed.}
\label{fig:fecrev1a}
\end{figure}

\subsubsection{Circuit description}
\label{sec:fec.fec.desc}

The final (Rev.\,1a) FEC is shown in \figref{fig:fecrev1a}.
The FEC receives 160 analog signals through 4 flexible cables and the corresponding connectors.
Each FEC includes five SAMPA chips, which amplify, shape and digitize the input signals in a continuous manner. 
The digital signal flow is shown schematically in \figref{fig:fec:overview:blockdiagram}.
The FEC includes two GBTx and one GBT-SCA chips, one VTRx and one VTTx.
The bi-directional VTRx is connected to GBTx0 and further to the GBT-SCA.
The clock is received from the CRU through this link, and distributed to the five SAMPA chips and to GBTx1.
The E-Links are operated at 160\,Mbit\,s$^{-1}$.
The ADC sampling clock of 5\,MHz is derived by division from the E-Link clock speed inside the SAMPA chips.

The SAMPA chips are operated in direct (split) readout mode (see \secref{sec:fec.sampa.desc}).
Each SAMPA produces a data rate of $32\times 10$\,bit\,$\times\,5$\,MHz = 1.6\,Gbit\,s$^{-1}$, which is output on ten of the SAMPA output E-Links.
The eleventh E-Link transmits the sampling clock, which is used for monitoring the sampling phase across all 16380 SAMPAs on the TPC.

In wide frame mode and at 160\,Mbit\,s$^{-1}$ E-Link speed, a total of 28 input E-Links are available per GBTx.
The E-Links from SAMPA\,2 are connected such that ADC values from channel numbers 0 to 15 are transmitted to GBTx0, while ADC values from channel numbers 16 to 31 are transmitted to GBTx1 (see \figref{fig:fec:overview:blockdiagram} and \secref{sec:fec.sampa.desc}).
The eleventh E-Link of SAMPA\,2 is split such that the sampling clock is transmitted to GBTx0 and GBTx1.
In this way, all 56 input E-Links across the two GBTx ASICs are utilized.

\subsubsection{Physical implementation}
\label{sec:fec.fec.impl}

The new FEC was designed implementing essentially the same form factor as the old one.
The 8-layer FEC PCB utilizes rigid-flex technology, a hybrid construction consisting of rigid and flexible substrates which are laminated together into a single structure.
The flexible signal cables towards the detector plane are realized in this way without the need for additional connectors.
Only signal traces are routed to the pad plane, ground traces were omitted in order to minimize capacitance.
The four flex extensions each terminate in 40-pin right-angle, male connectors\footnote{ERNI SMC 1.27\,mm. Product number 354097}, distributed on two rigid PCBs.

The physical layout of the board follows from the signal flow.
The rigid-flex inputs are all on one side and lead directly to the SAMPA chips.
To minimize the channel-to-channel crosstalk, the five SAMPA circuits were placed as close as possible to the input connectors.
The SAMPA digital outputs are routed into the GBTx ASICS.
The GBTx outputs are routed to the optical transceivers in a more or less linear fashion.
The power input and major digital voltage regulation is located on the optical-output side of the board, while the regulation for the SAMPA analog and digital supplies is located very close to the SAMPAs in order to ensure the supply voltages to be as clean as possible.

The board is divided into an analog section and a digital section.
The analog section is supplied with +2.25\,V and connects to 1.25\,V regulators for the analog functionality of the SAMPAs and to 1.1\,V regulators for the ADC reference voltage.
The digital section is supplied at +3.25\,V and connects to a 2.5\,V regulator for the GBTx ASICs and for the GBT-SCA, to a 1.5\,V regulator for the VTRx and VTTx, and to 1.25\,V regulators for the digital functionality of the SAMPAs.
The 2.5\,V regulator is heat-sunk through multiple internal layers and has a power-sharing dissipation resistor on the input, which reduces the amount of heat dissipated by the regulator itself.

The ground planes (analog and digital ground) are distributed over two PCB layers that are tied together with three pads, which are located at the power input to the board upstream of the voltage regulators, and close to the SAMPA ASICs.
After installation on the detector, the analog ground plane on the FEC is connected to the ROC Al-body via short cables.
Three ground lugs are available for this purpose close to the SAMPA chips.
The digital ground plane is connected to the SSW via similar cables.
One ground lug is available for this purpose in one corner of the board.

The planes are designed to take advantage of the SAMPA pinout locations regarding analog and digital sections.
The analog planes are large strips which are placed underneath the SAMPA front-end input sections and their associated components.
The digital planes are placed under the SAMPA backend and GBTx sections.
Even though the SAMPA substrate grounds are tied together inside the ASIC, an attempt to control return currents and to keep as much digital noise as possible away from the sensitive front-end inputs was made.
The closing of the grounds at the output of the board is required to maintain continuity of the power supply to all ASICs.
This approach, combined with the use of SLVS signals, has resulted in a very low-noise board.

\subsubsection{Power and cooling}
\label{sec:fec.fec.power}

In direct readout mode, where the DSP is bypassed, the SAMPA consumes 9\,mW per channel, which adds up to 1.5\,W for a 160 channel FEC.
A larger fraction of the total power consumption is due to the GBT components.
Additional power is also dissipated in the regulators.
The combined power consumption for a FEC is about 9\,W (56\,mW per channel).
The Low Voltage (LV) system from the previous system~\cite{TPCnim} is reused in order to supply the required power.
In particular, two low voltage channels are used per TPC sector (91 FECs) to supply the analog (2.25\,V, 85\,A) and digital (3.25\,V, 185\,A) power.

Following the implementation of the previous setup, each FEC is embedded into water-cooled copper envelopes.
These cooling envelopes are reused from the previous system.
Efficient heat transfer from the hottest components to the copper plates is realized by the addition of flexible heat-transfer pads.
They are mounted on the two GBTx ASICs, on the five SAMPA ASICs and on the voltage regulators.

\subsubsection{Monitoring and control}
\label{sec:fec.fec.moni}

FEC monitoring is done with the GBT-SCA 12-bit ADC and includes 14 measurements per FEC.
Five resistance temperature detectors (RTDs) are used to monitor temperatures at different positions on the FEC.
The resistance of the RTDs is extracted using the SCA-internal current source.
To improve the precision of the measurement, the current source itself has to be calibrated.
Such a circuit is realized by adding a low tolerance reference resistor on the FEC.
The SCA-internal temperature sensor is read out in addition.
The primary 1.5\,V and 2.5\,V supplies of the digital part of the board are voltage monitored.
Current monitoring at the input power connector is implemented for the 2.25\,V and 3.25\,V supplies via shunt resistors (two voltage measurements each).
Finally, the received signal-strength on the VTRx interface is monitored.

FEC control is achieved via the GBT-SCA as well, using the I2C protocol and GPIO connections.
The five SAMPA ASICs can be powered on/off by setting the appropriate bits in the GPIO output mask.
The GBTx0 ASIC is fuse programmed, such that the associated TX and RX links are always active after power to the FEC is established.
Tuning of the programmed parameters can be achieved via I2C.
The GBTx1 ASIC always needs to be configured after power-up. This is achieved via I2C.

\subsubsection{Input protection}
\label{sec:fec.fec.prot}

The inputs of the SAMPA may be exposed to discharges from the GEM stack.
In GEM detectors, discharges typically begin with a sudden, radiation-induced breakdown of the gas rigidity in one GEM.
The capacitance of one GEM segment is around 5\,nF, and the charge stored is around \SI{1.5}{\micro\coulomb}.
Secondary discharges may propagate to the pad plane (see \secref{sec:discharge}), such that a fraction or all of the stored charge is delivered to an electronics channel. 
The capacitance of the induction gap (\gemFour bottom to pad plane) is between 1 and 2\,nF, and the charge stored is around \SI{5}{\micro\coulomb}.

The SAMPA features device-level ESD protection for handling.
In addition, system-level protection is implemented on the FEC:
Each of the 160 channels is protected by a circuit containing a resistor directly at the input, followed by a clamping diode\footnote{Littlefuse SP3004 series, SOT563 package}.

The protection circuit was optimized in two steps.
In a first test, the SAMPA inputs were subjected to real discharges from a GEM detector. The GEM stack was operated at high gain, in order to increase the secondary discharge rate under radiation.
Standard axial resistors were used in the protection circuit.
A resistor value of 100\,$\Omega$ was found to protect the SAMPA inputs efficiently.
In a second test, different SMD\footnote{The input protection resistors have size 0603 (imperial).} resistors with resistance 100\,$\Omega$ were tested on the final FEC PCB and with a discharge generator.
Worst-case charge injections (0.32\,mJ into a single pad) were generated using a 1\,nF capacitor at 1\,kV, and a relay.
Many of the tested SMD resistors failed under the load from the discharges, most likely due to heat generated in the device.
A few SMD resistor types, mainly labeled "pulse withstanding" or "discharge resistant", survived a large number (about 200) of discharges without any observable damage.
A thick-film resistor for the automotive market\footnote{ROHM KTR03EZPF1000} was chosen for mass production of the FEC.

\subsection{Radiation tolerance}
\label{sec:fec.rad}

Exposure to energetic particles can produce instantaneous failures and, over time, degradation of electronic components.
The High-Energy Hadron (HEH) flux is the primary source of instantaneous failures: Single Event Effects (SEE) may occur when a highly energetic particle strikes the sensitive regions in the digital circuitry of a CMOS\footnote{Complementary Metal Oxide Semi-conductor} device, disrupting its correct operation.
The SEE are mainly characterized as soft errors (Single Event Upset,
SEU, where a memory cell can change its logical state) and hard errors (Single Event Latchup, SEL, where a conducting path through the substrate of the semiconductor creates a short between the supply rail and ground).
A slow gradual degradation of the device performance is on the other hand due to Total Ionizing Dose (TID) effects related to damages caused in the semiconductor lattice. 

The flux of HEH ($>$20\,MeV) at the TPC inner radius of the SSW is expected to reach\break $\Phi_{\text{HEH}}^{\text{inner}}=\SI{3.4}{\kilo\hertz\per\centi\meter\squared}$~\cite{TDR:CRU}.
At the outer radius the flux is $\Phi_{\text{HEH}}^{\text{outer}}=\SI{0.7}{\kilo\hertz\per\centi\meter\squared}$.
Both numbers include a safety factor of 2.
The TPC electronics located at the inner radius has to stand a dose of 2.1\,krad.
These requirements were tested via multiple radiation campaigns.
All components were shown to tolerate a TID of at least 10\,kRad.

The SAMPA v4 was tested to perform close to specifications to beyond a TID of 12\,kRad.
Based on the null observation of SEL, an upper limit for the SEL cross section of  $10^{-7}$\,cm${^2}$ was determined for ions with a linear energy transfer of \SI{16}{\mega\eV\centi\meter\squared\per\milli\g}.
Most of the circuitry in the SAMPA is TMR\footnote{Triple Modular Redundancy} protected.
However, the clock division circuit to derive the 5\,MHz ADC sampling clock from the 160\,Mbit\,s$^{-1}$ E-link clock is not TMR protected in order to avoid the risk of glitches on the voter output.
The cross section for flip flops was measured with SAMPA prototypes as $10^{-13}$\,cm$^2$ per bit.
With the number of flip flops used in this circuit (nine), the worst-case HEH flux $\Phi_{\text{HEH}}^{\text{inner}}$, and the total number of SAMPAs (16380), one can calculate an expected error rate of $5.01\times 10^{-5}$\,SEU\,s$^{-1}$, which corresponds to a mean time between failures (MTBF) of 5.5 hours.
Such an error will most likely lead to a phase shift of the ADC sampling clock with respect to the other SAMPAs on the TPC.
Since all sampling clocks are transmitted via the 11th SAMPA E-Link, and are thus being monitored, the error will be detected on the CRU.
The erroneous state can be resolved by resetting and re-configuration within the matter of a few seconds.

The voltage regulators and input protection diodes were tested to over 10\,kRad TID using 184\,MeV protons.
The performance of these devices was subsequently studied further by exposing entire FECs to \xrays with occasional operation and readout of the FEC during the exposure.

The components of the GBT link system were designed to operate in radiation environments much harsher than those seen in ALICE.
Consequently, advanced techniques for radiation hardening were applied.
The GBTx was radiation qualified by CERN to 400\,MRad~\cite{Lesma_2017}.
SEUs in the configuration registers are automatically corrected by the internal logic.
However, other soft errors have to be expected at a low rate~\cite{Leitao_2015}:
TX and RX \textit{lock errors} cause bursts of errors while the GBTx relocks;
TX \textit{frame errors} may cause a few bits to be wrong in one or two individual (consecutive) data frames.
Lock errors persist for some time, up to a few \si{\us}, until the PLL\footnote{Phase-Locked Loop} re-acquires lock. As a consequence, several consecutive frames will be affected.

For the full TPC (3276 RX links, 6552 TX links), and using the worst-case HEH flux $\Phi_{\text{HEH}}^{\text{inner}}$, the MTBF for lock errors is expected to be one every 4.9 hours for the RX path, and one every 14.1 hours for the TX path.
Since in transceiver mode the clock recovered by the receiver will be used as a reference to the transmitter, any receiver loss of lock will directly impact on the transmitter error rate.
As a consequence, the MTBF for all TX lock errors reduces to one every 3.6 hours.
The MTBF for TX frame errors is one every 1.8 hours.
RX frame errors are effectively corrected by the Forward Error Correction mechanism that is always active in the RX path.

Lock errors will be recovered automatically and will lead to an insignificant loss of data.
The consequence of a temporary link loss can be that the synchronization of the data transmission is not guaranteed anymore.
In this case, a sync command to all FECs can be transmitted automatically in order to re-synchronize data taking across all TPC links.

Loss of phase alignment of the ADC sampling clocks will be visible by a phase shift on the 11th SAMPA E-link.
In this case, a reset and re-configuration of all SAMPAs is required.

\subsection{SAMPA and FEC quality assurance}
\label{sec:fec.testing}

The testing and qualification procedure for the electronics was implemented in two consecutive stages.
First, all SAMPA chips underwent functional testing using a semiautomatic setup based on a robot.
In a second step, the assembled FECs were tested.
All test results were filled into a database (see \secref{sec:database}). 

The digital and analog parts of the SAMPA were subject to different acceptance tests.
The digital parts were required to be \SI{100}{\percent} functional. The analog parts were allowed a small variation in the most critical parameters (conversion gain, peaking time and DC offset (pedestal) and noise).
The test results classified by source of failure are displayed in \tabref{tab:fec.testresults.sampa}.
The final SAMPA production yield is \SI{79.6}{\percent}.

\begin{table}[htp]\footnotesize
  \caption[Failures during SAMPA testing]{Failures during SAMPA testing. Multiple counting occurs.}
  \begin{center}
  \begin{tabular}{lcc}
    \toprule
    Failure & \multicolumn{2}{c}{Frequency} \\ 
    \midrule
    None & 72\,333 & \SI{79.6}{\percent} \\ 
    \midrule
    Power & 3790 & \SI{4.2}{\percent} \\ 
    Pedestal & 8243 & \SI{9.1}{\percent} \\ 
    Noise & 8327 & \SI{9.2}{\percent} \\ 
    Gain & 10\,427 & \SI{11.5}{\percent} \\ 
    Peaking time & 8609 & \SI{9.5}{\percent} \\ 
    Cross talk & 469 & \SI{0.5}{\percent} \\ 
    Clock generation & 990 & \SI{1.1}{\percent} \\ 
    Data synchronization & 1173 & \SI{1.3}{\percent} \\ 
    I2C interface & 833 & \SI{0.9}{\percent} \\ 
    JTAG interface & 1346 & \SI{1.5}{\percent} \\ 
    Memories & 5707 & \SI{6.3}{\percent} \\ 
    Digital tests & 5229 & \SI{5.8}{\percent} \\ 
    \midrule
    Tested chips & 90\,883 & \SI{100}{\percent} \\
    \bottomrule 
  \end{tabular}
  \end{center}
  \label{tab:fec.testresults.sampa}
\end{table}

The FECs were required to satisfy the specifications listed in \tabref{tab:fec.specifications:fecs}.
Without the additional capacitance from the detector, an average noise value over all tested channels of 0.79\,LSB (530\,e ENC) was achieved.
A test station was developed to provide automated testing over the course of mass production.
The test station provides automated control of voltages and input waveform generator signals to allow for testing of basic functionality, noise, pedestals, gain, and crosstalk for all 160 channels.
Two custom test stations were operated simultaneously at the assembly vendor with a total testing time of 6 minutes per FEC per station.
All FECs were required to pass a temperature cycling test.
Here, the temperature of a thermal test chamber was cycled from 20$^\circ$\,C to 85$^\circ$\,C multiple times at 2$^\circ$\,C per minute.

\begin{figure}[h]
\begin{center}
\includegraphics[width=0.49\textwidth]{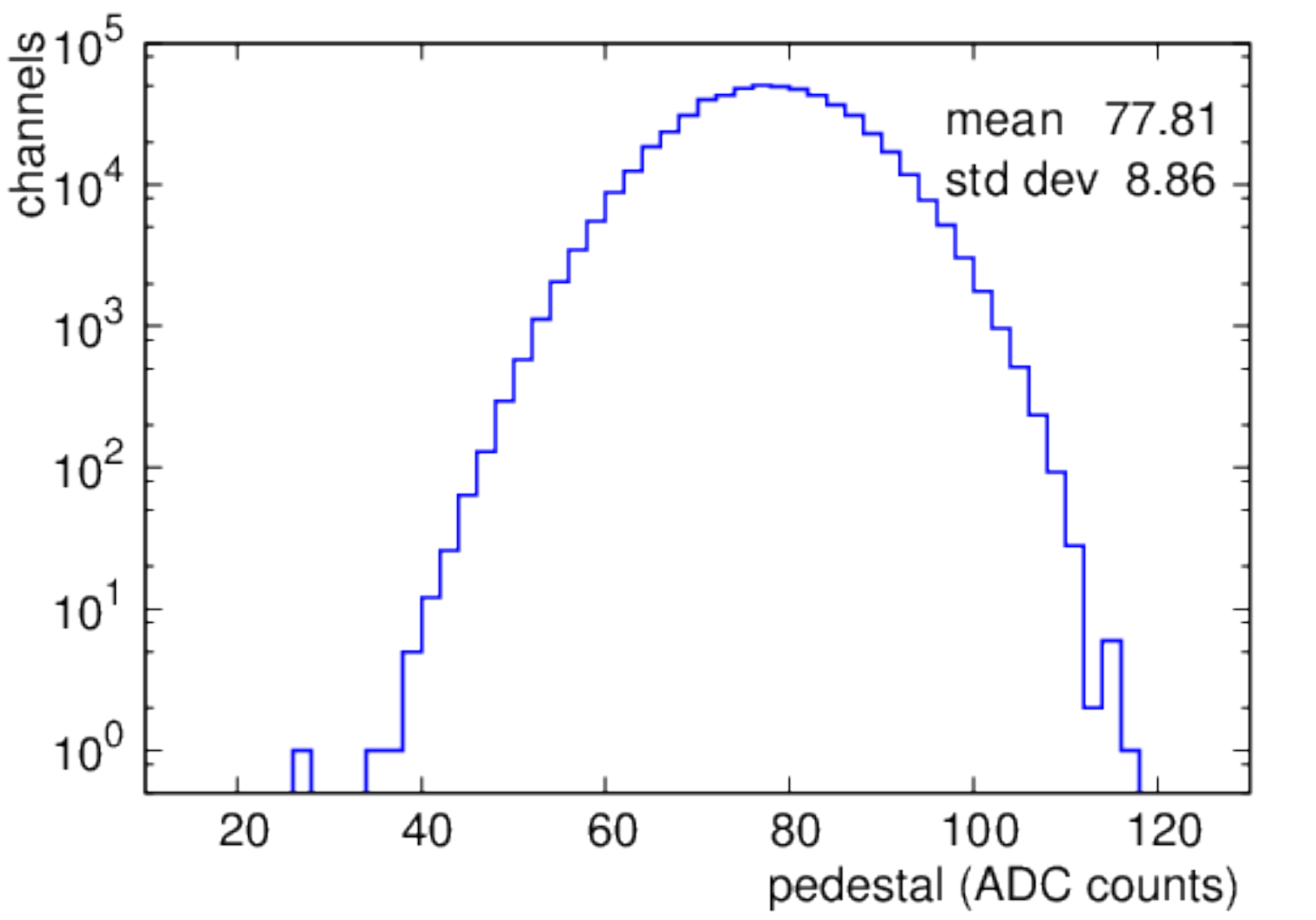}
\includegraphics[width=0.49\textwidth]{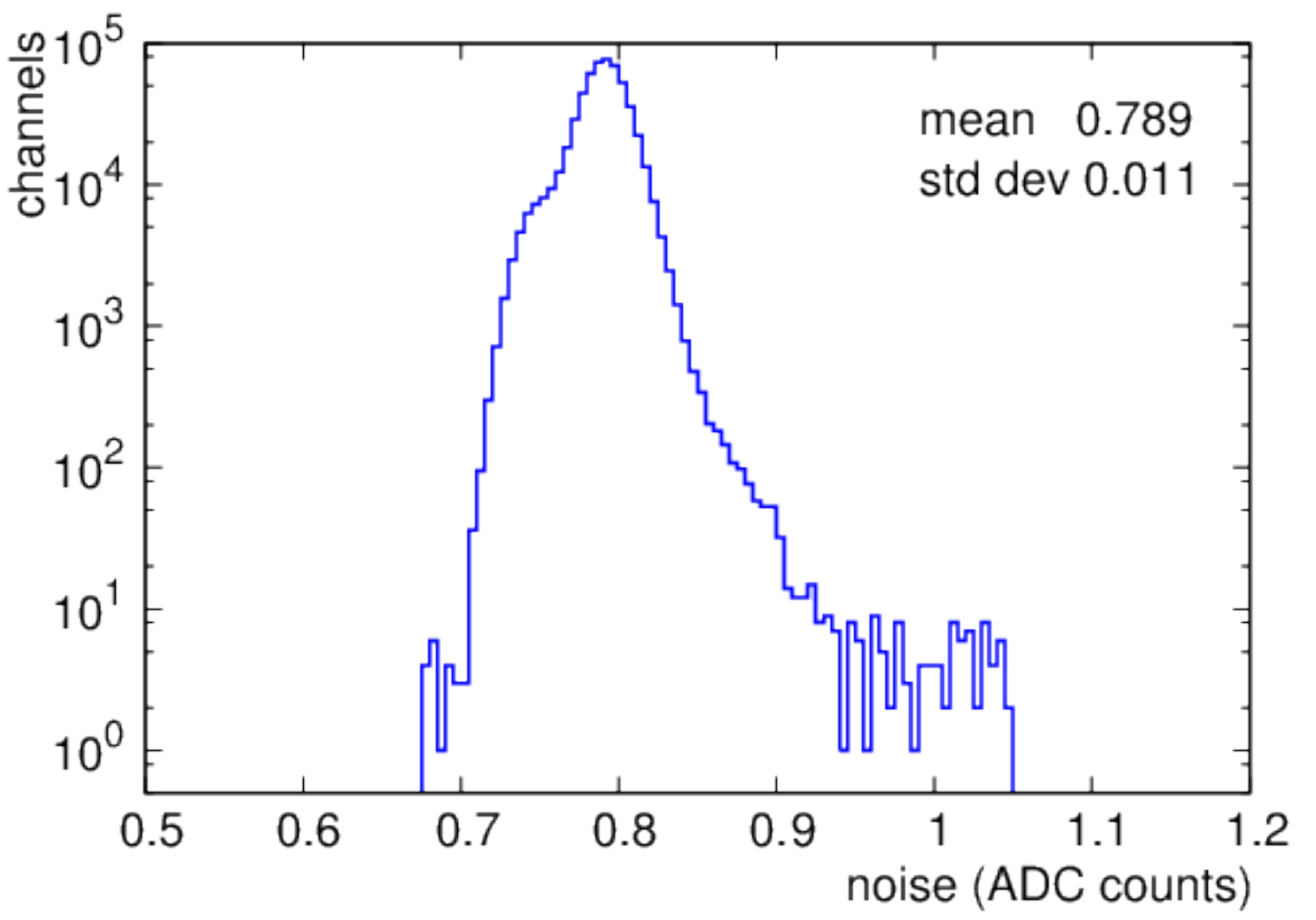}
\includegraphics[width=0.49\textwidth]{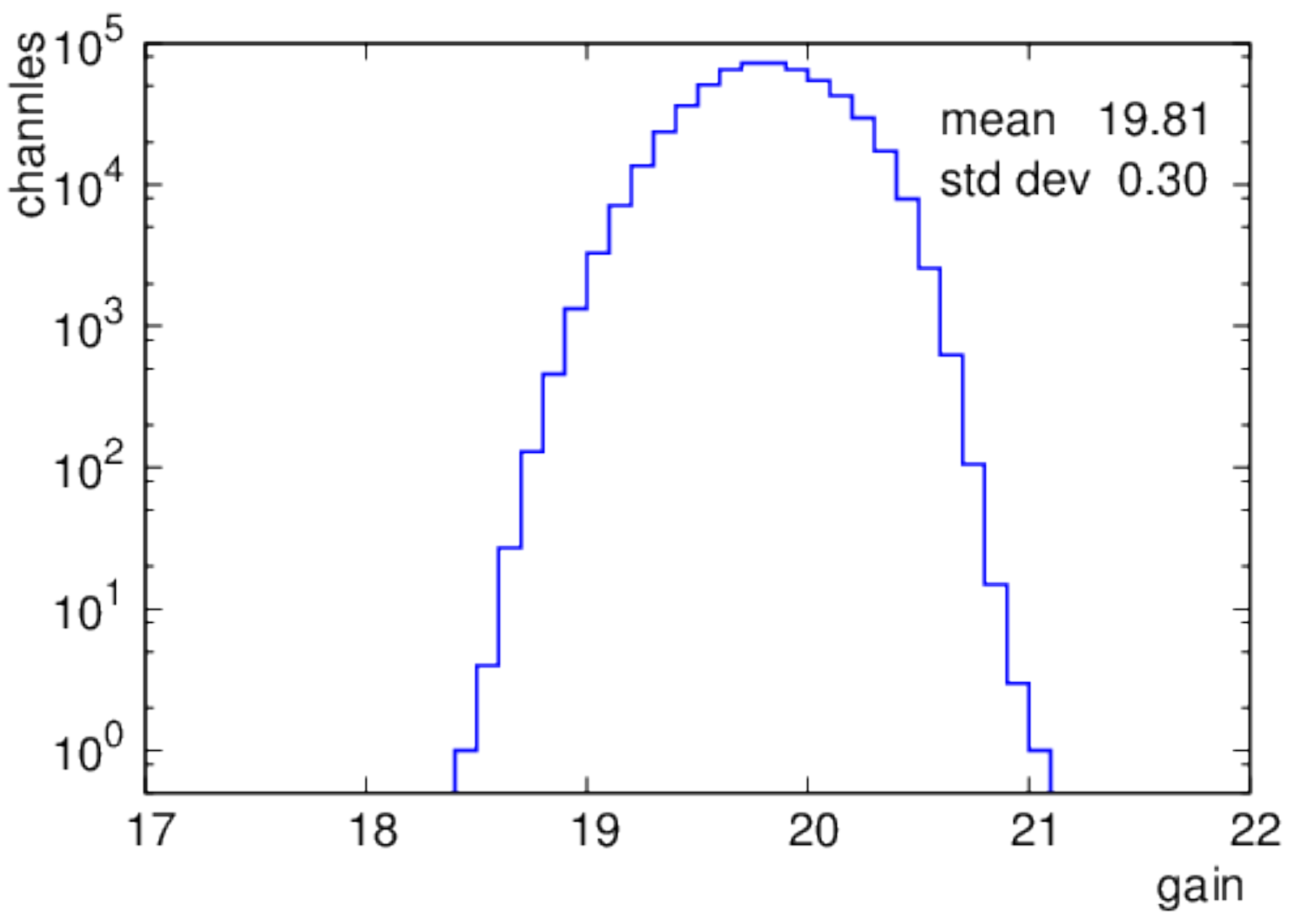}
\includegraphics[width=0.49\textwidth]{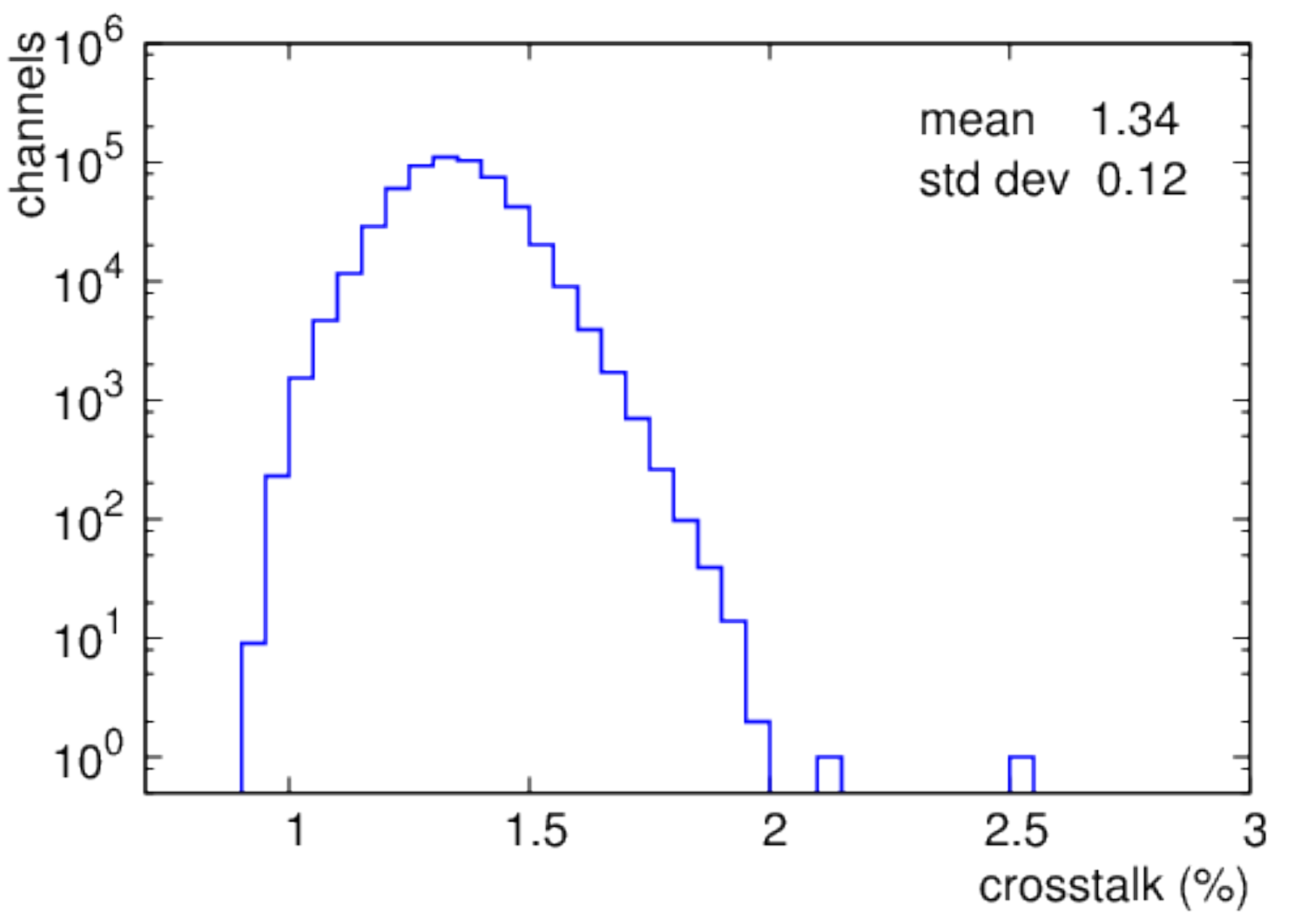}
\includegraphics[width=0.49\textwidth]{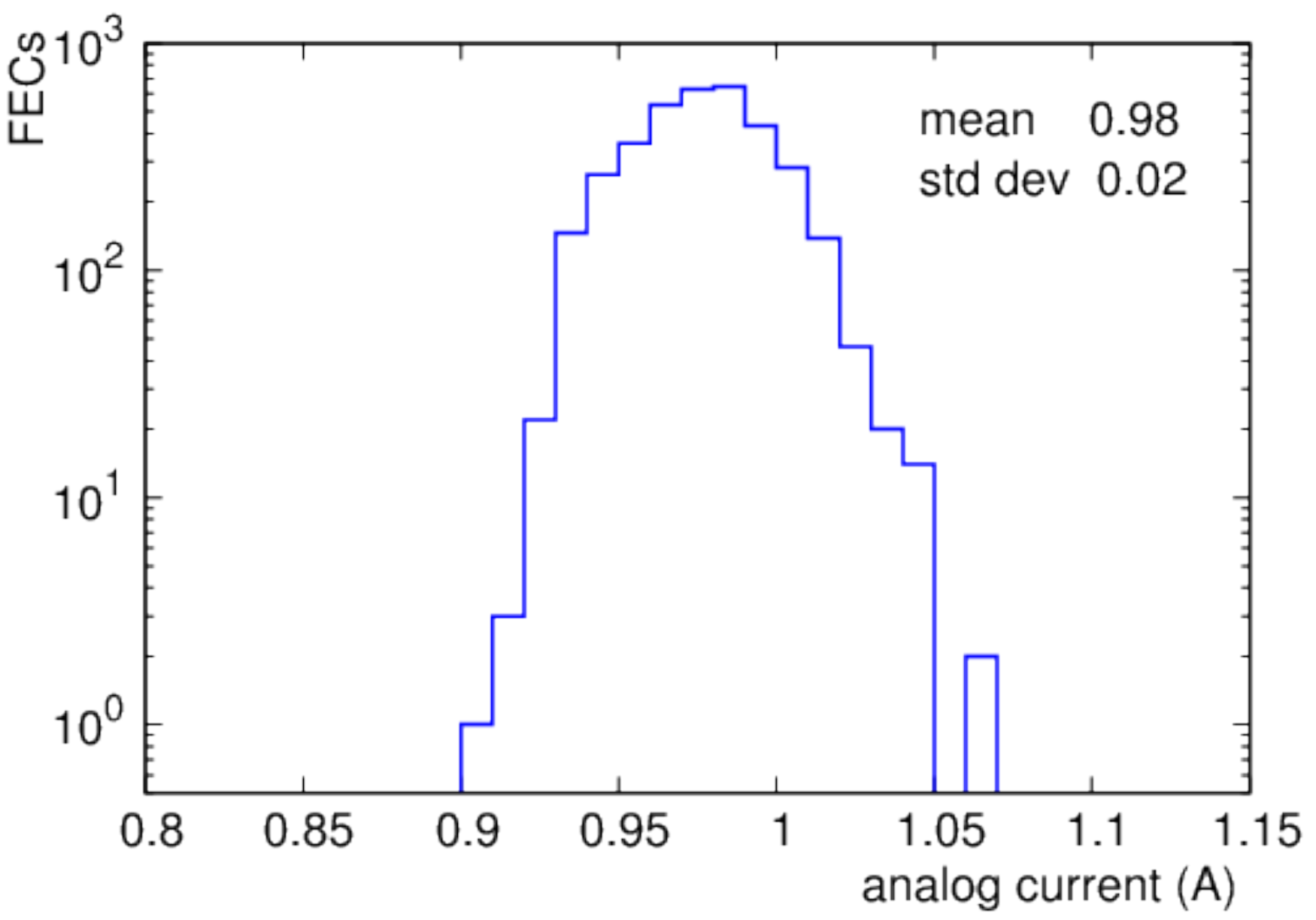}
\includegraphics[width=0.49\textwidth]{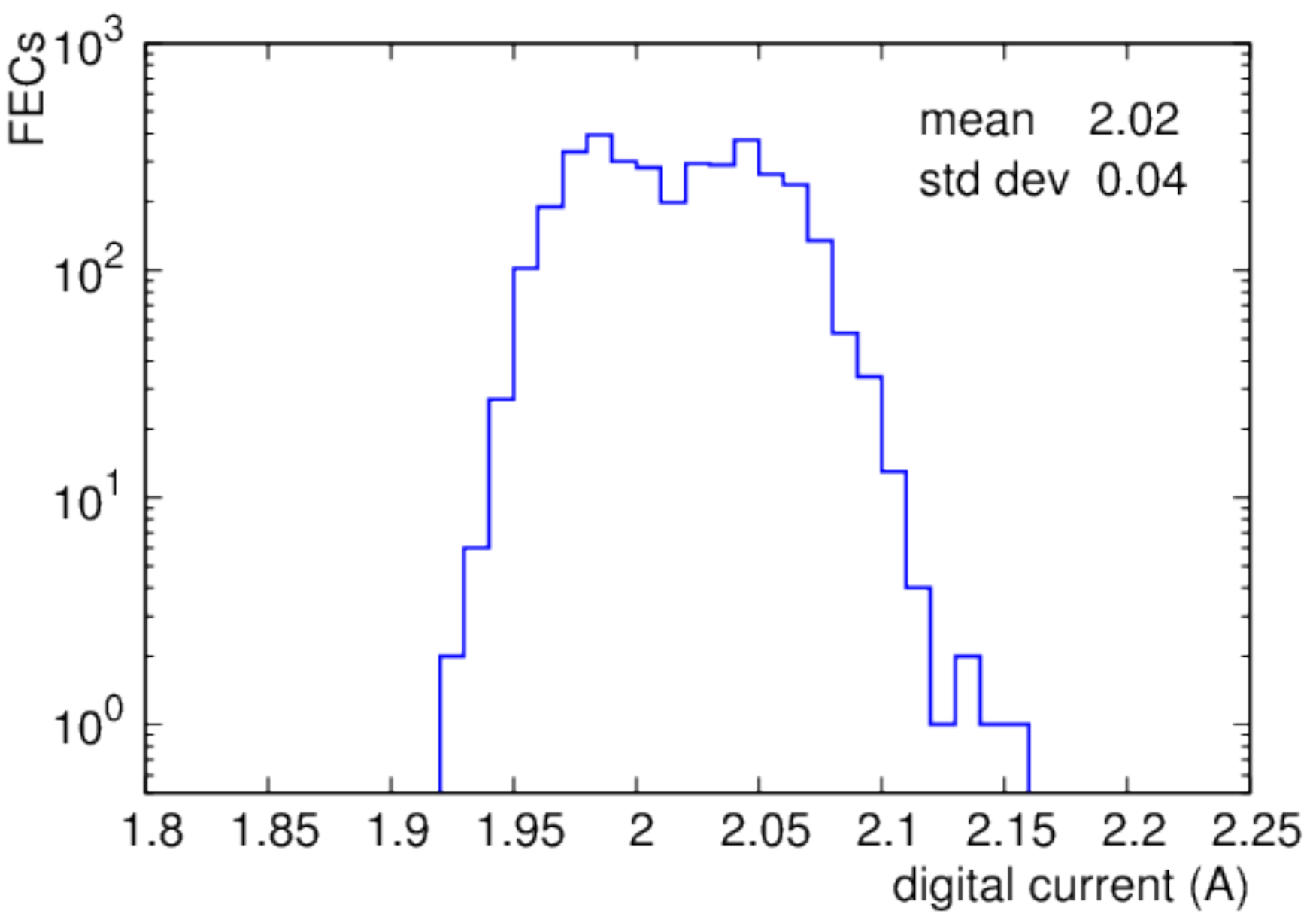}
\end{center}
\caption{Distribution of the DC offset (pedestal, all channels, top left), noise (all channels, top right), conversion gain (all channels, middle left), crosstalk (all channels, middle right) and analog (bottom left) and digital (bottom right) power consumption (per FEC) over all production FECs. The digital power consumption depends on the temperature of the FEC. A double-peak structure appears in the plot as some FECs were tested immediately after power up, while others were tested after some time in powered state.}
\label{fig:fec:fec:resultzz}
\end{figure}

The initial yield of FECs satisfying all requirements was \SI{94}{\percent}.
After completing repairs, the overall final FEC production yield for 3646 FECs (including \SI{11}{\percent} spares) was \SI{99.75}{\percent}.
The distributions of the key parameters are presented in \figref{fig:fec:fec:resultzz}.
The sources of failures are summarized in \tabref{tab:fec.testresults.fec}.

\begin{table}[htp]\footnotesize
  \caption[Failures during FEC testing]{Failures during FEC testing. Multiple counting may occur.}
  \begin{center}
  \begin{tabular}{lcc}
    \toprule
    Failure & \multicolumn{2}{c}{Frequency} \\ 
    \midrule
    None & 3637 & \SI{99.75}{\percent} \\ 
    \midrule
    Power & 0 & \SI{0}{\percent} \\ 
    Noise & 2 & \SI{0.06}{\percent} \\ 
    Crosstalk & 7 & \SI{0.19}{\percent} \\ 
    Gain & 1 & \SI{0.03}{\percent} \\ 
    \midrule
    Tested FECs & 3646 & \SI{100}{\percent} \\
    \bottomrule 
  \end{tabular}
  \end{center}
  \label{tab:fec.testresults.fec}
\end{table}

\section{Installation and commissioning}
\label{sec:commissioning}

In this section, the actual process of the TPC upgrade is described. This 15-month-long operation, executed in a dedicated clean room on the surface of the ALICE site, included the replacement of the old readout chambers by the new GEM detectors, the installation of the new front-end electronics and a series of basic functionality and performance tests during an extended commissioning campaign.
\subsection{ROC installation}
\label{sec:install:roc}
After being purged with clean, breathable, and oil-free air, the TPC was moved from the experimental cavern to the surface, cleaned, and installed in the same clean room where it had been assembled fifteen years before. As then, the detector was held inside a frame, and access to the endplates was provided by a large elevating platform on one side, and an aluminium scaffolding on the other side.
In order to meet the requirements for handling GEM detectors, the $\sim$1300\,m$^3$ clean room had been upgraded with new air handling units, equipped with coarse (G4), fine (F7), and high-efficiency particulate air (HEPA, H14) filters, capable of supplying air at a rate of \SI{15000}{\meter\cubed\per\hour} via induction diffusion using two textile half ducts.
Thus, an ISO\,7 air quality was reached in the full room.
Furthermore, three extra fan filter units with HEPA filters were mounted vertically on the elevating platform, in order to further improve the air quality to ISO\,6 in the area where the GEMs of the new chambers and the TPC drift volume were exposed.

Before installation, all new readout chambers were equipped with resistance temperature detectors (pt1000).
They allow temperature measurements at one (for the IROC) or two (for the OROC) positions on the chamber Al-body.
This information, together with the one provided by the internal sensors (see \secref{sec:hv:fc}), allows for the necessary homogenization of the  temperature across the detector.

\begin{figure}[p]
\centering
\includegraphics[width=0.7\textwidth]{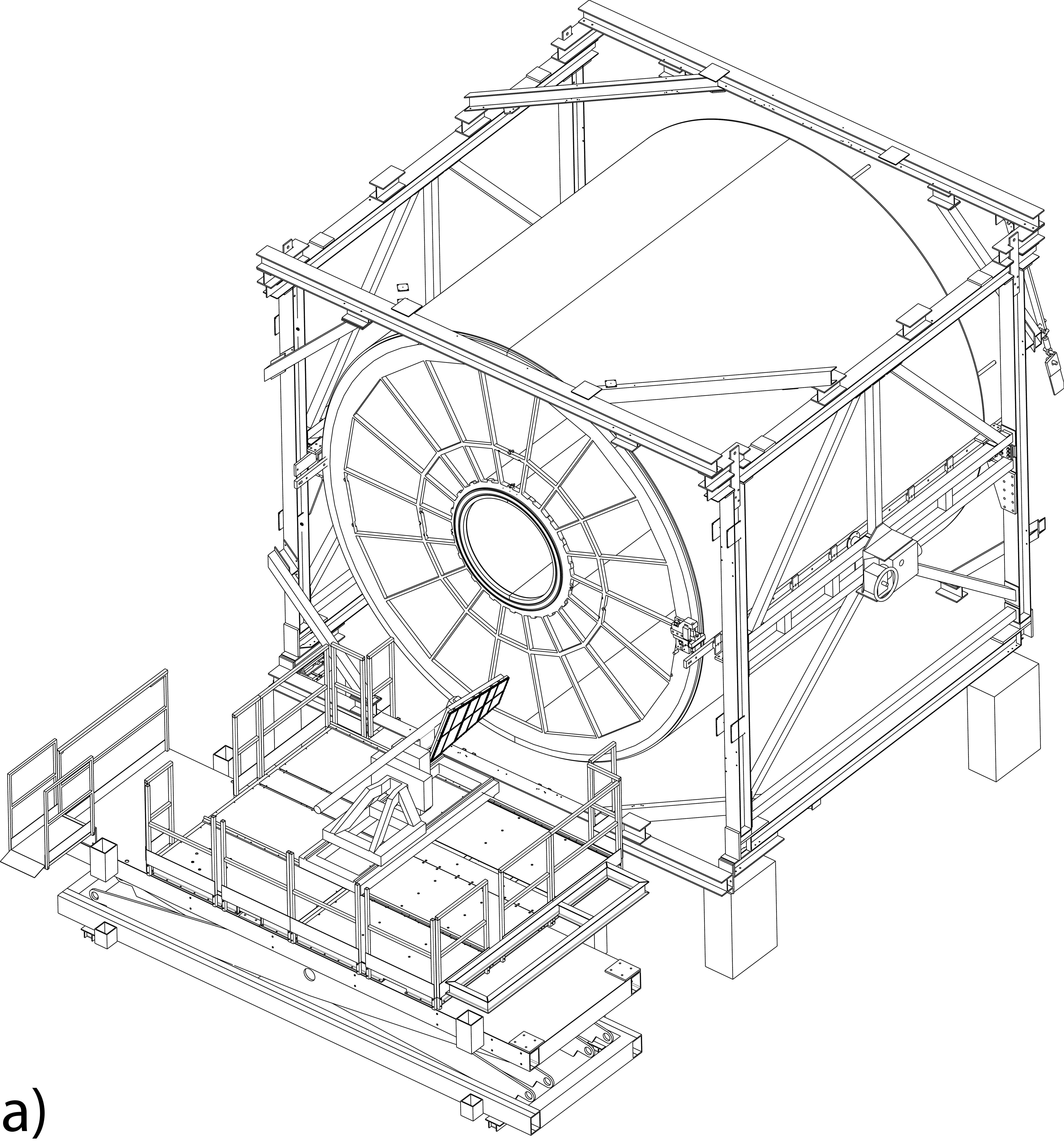}\\\vspace{3mm}
\includegraphics[width=0.3\textwidth]{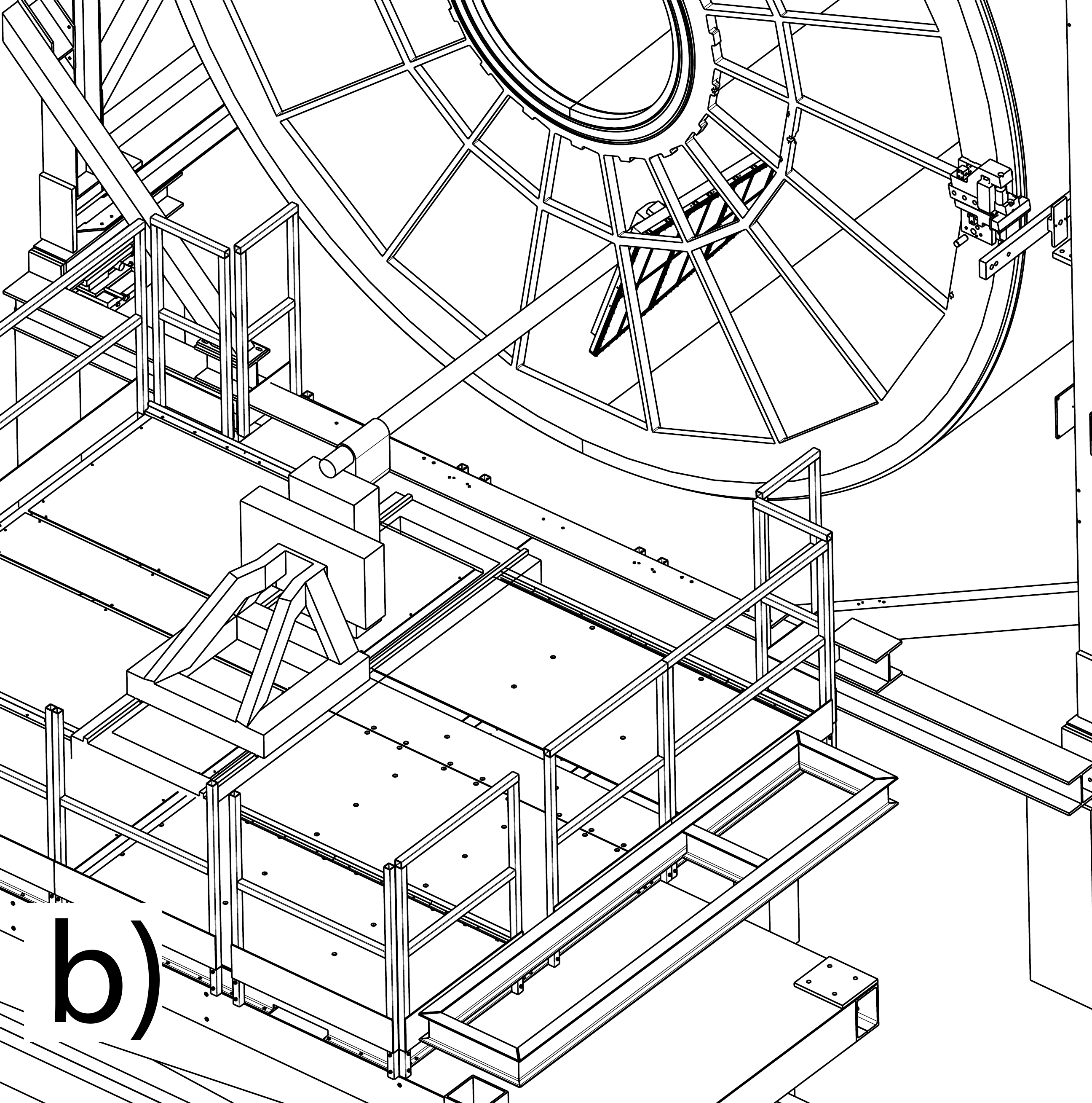}\hspace{3mm}
\includegraphics[width=0.3\textwidth]{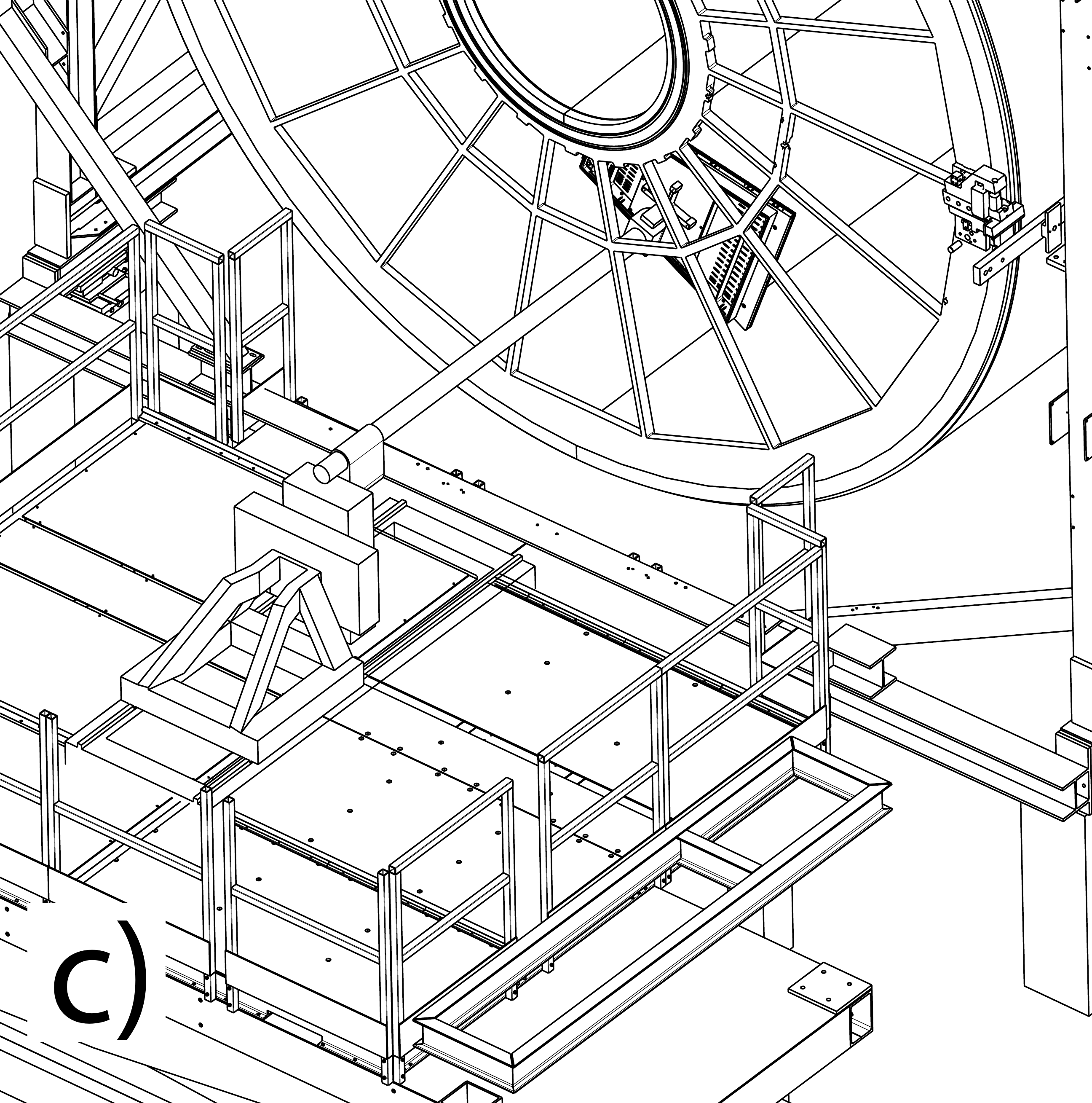}\\\vspace{3mm}
\includegraphics[width=0.3\textwidth]{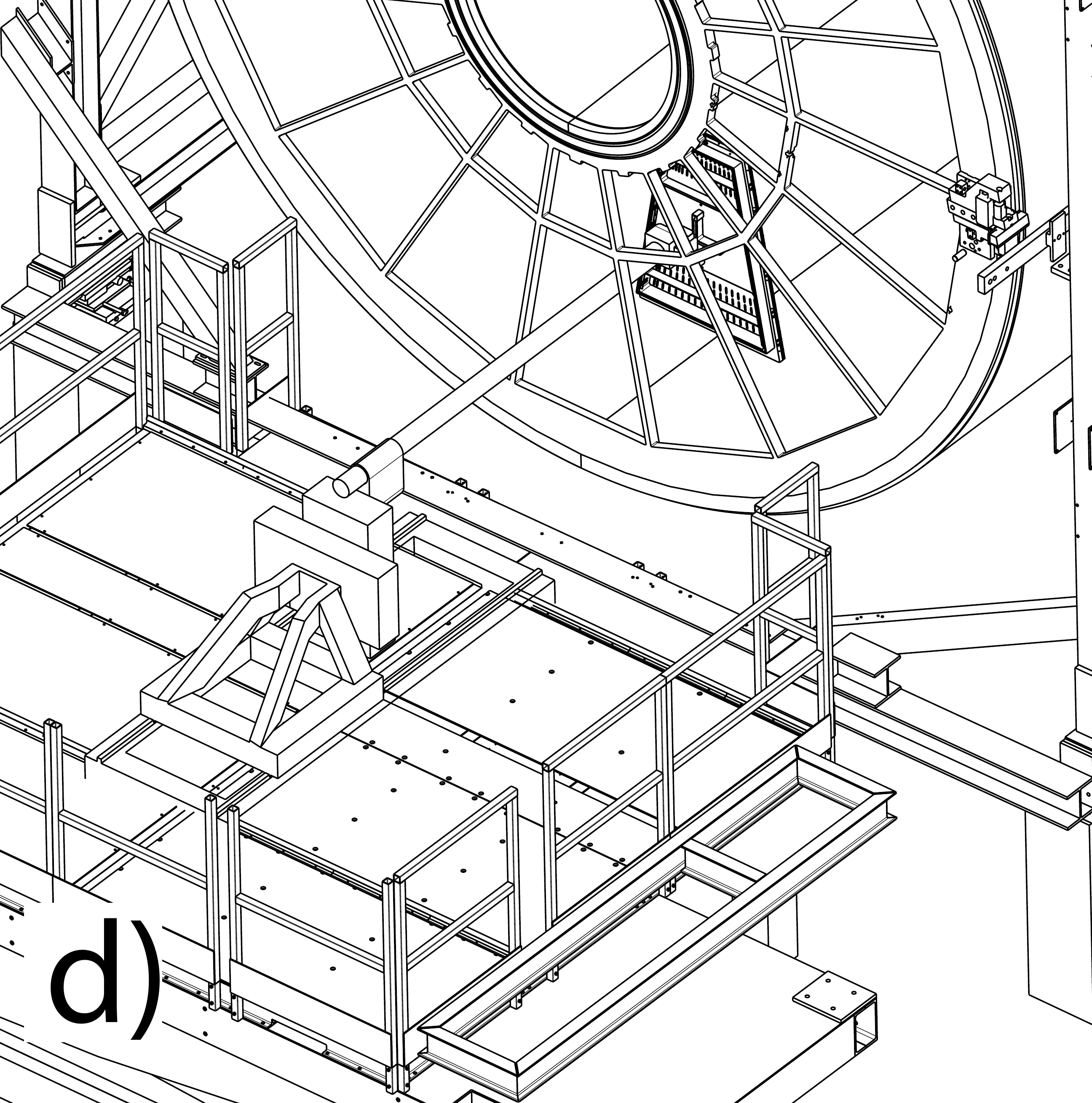}\hspace{3mm}
\includegraphics[width=0.3\textwidth]{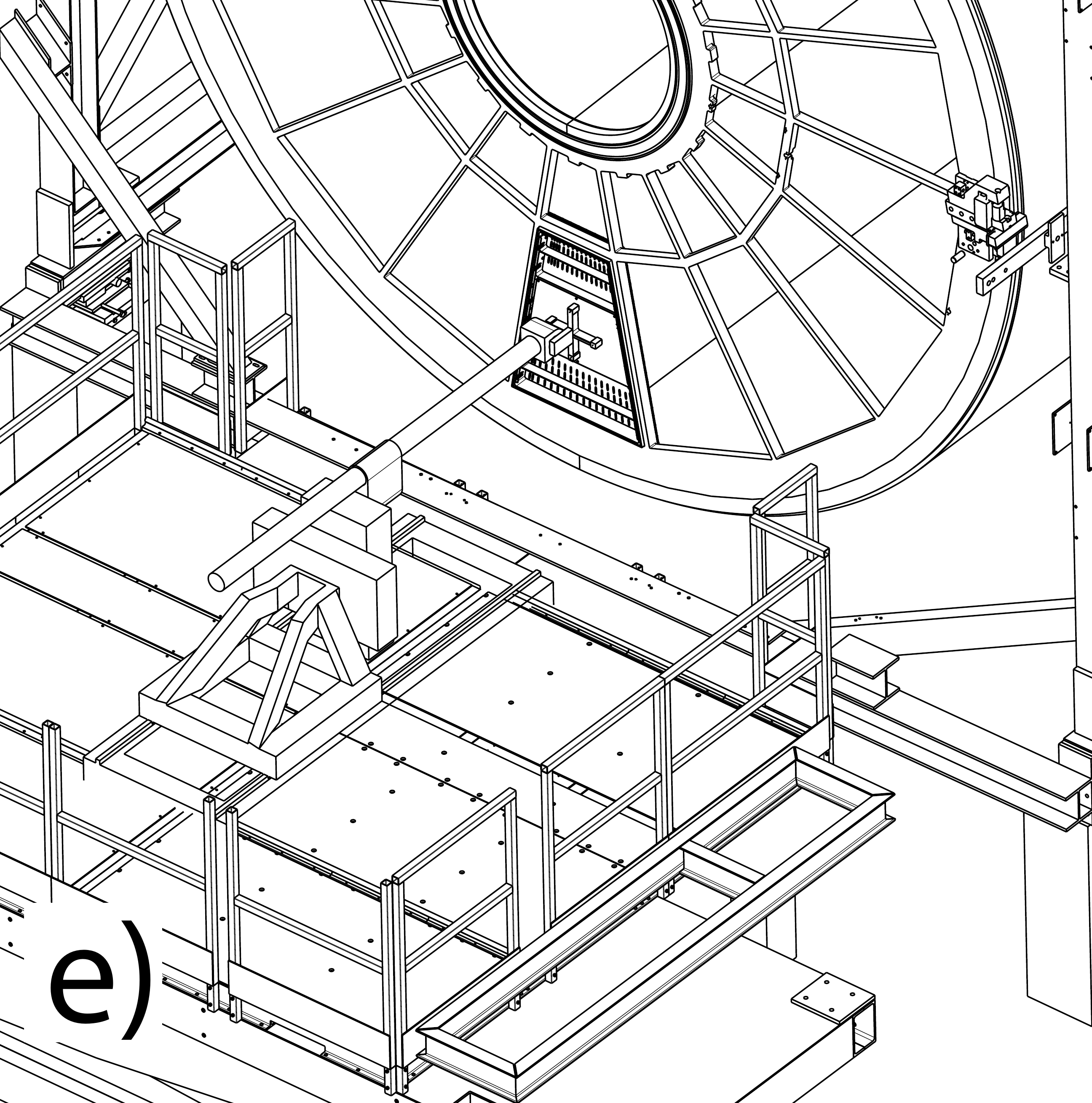}
\caption{Installation sequence of, in this case, an \oroc. The chamber is mounted on the mounting tool, oriented radially in front of its corresponding slot in the endplate (a). The arm of the mounting tool is extended in order to fully introduce the chamber into the field cage (b). It is then rotated to stand parallel to the endplate (c) and roughly aligned in front of its slot (d). Finally, it is brought back into the slot, along with fine-adjustments (e), bolted on the three fixation points, and temporarily sealed.}
\label{fig:install:manipulator_sketch}
\end{figure}

In order to work inside the clean room, personal equipment such as overalls, hair nets, masks and gloves were worn.
All material and tools entering the room were cleaned with compressed N$_2$ and alcohol (ethanol or isopropanol).
The entire clean room, including walls, was cleaned by a professional contractor right after the clean room was opened for operations with an overhead crane. In addition, a weekly cleaning session was scheduled during the ROC installation activities.

\begin{figure}[ht]
\centering
\includegraphics[width=0.9\textwidth]{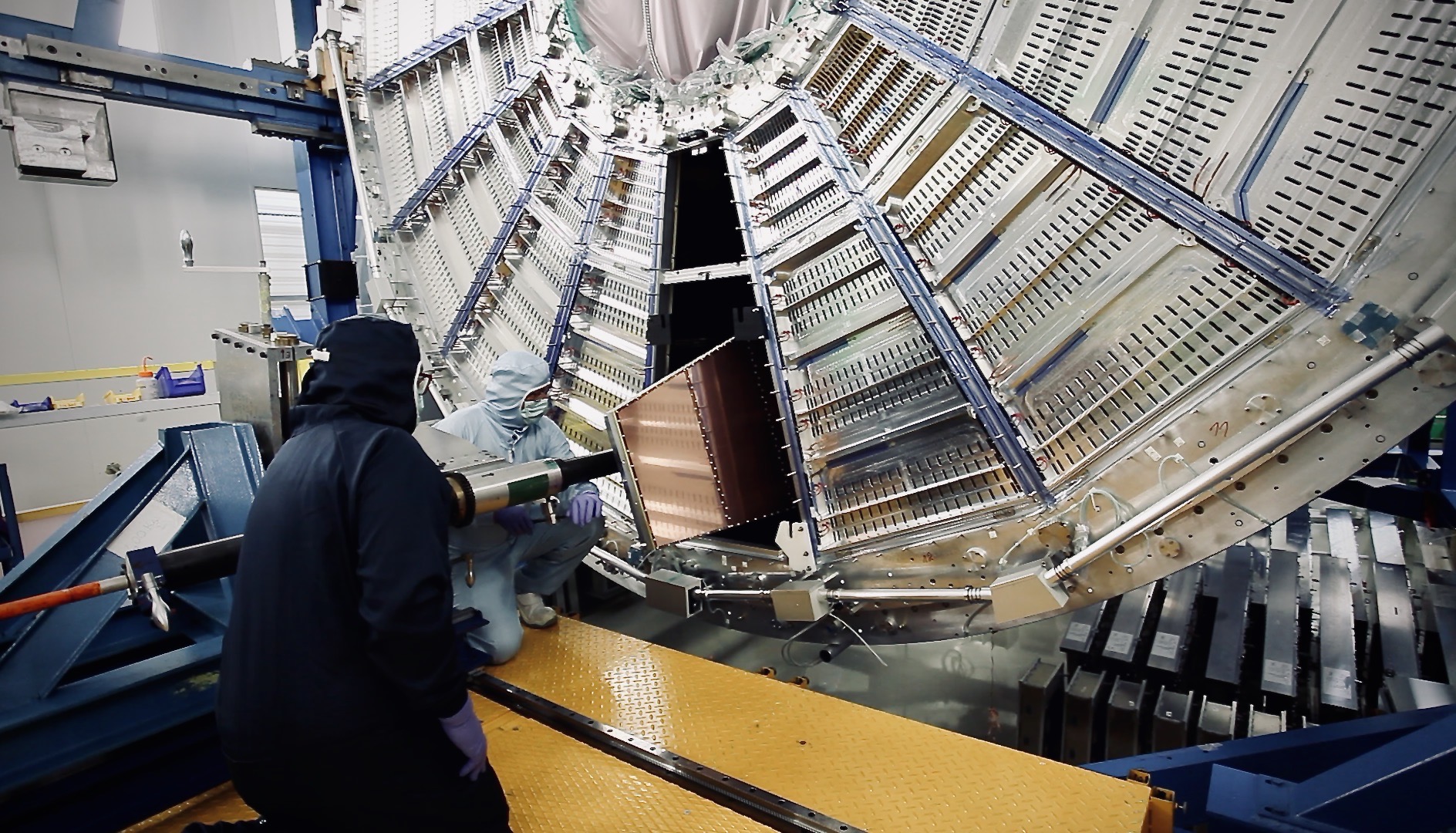}
\caption{Installation of an \oroc mounted on the mounting tool, which in turn stands on the elevating platform. The uninstalled MWPCs are stored underneath the TPC in TT-boxes, visible in the lower right corner.}
\label{fig:install:oroc}
\end{figure}

The active area of the readout chambers is larger than the corresponding cut-outs in the endplates, so as to maximize the active area between chambers.
Therefore, for (un)installing a ROC, it has to be moved into the drift volume and rotated about different axes at subsequent instances.
For this purpose, a mounting tool with 6 translational and rotational degrees of freedom was used. \Figref{fig:install:manipulator_sketch} sketches an \oroc as it is manipulated with the mounting tool at several stages of installation. \Figref{fig:install:oroc} shows an \oroc being inserted into the field cage. The mounting tool was installed on the elevating platform on horizontal rails. In this way it could be moved to the exact position for each chamber. A jib crane (not shown in the figures) mounted on the platform was used to move each chamber to or from the mounting tool. 

To install e.g.\ one \oroc, the chamber was fully inserted into the field cage using the main arm of the mounting tool. It was then rotated, brought back, aligned into its cut-out, and finally secured on its three mounting points. At every stage care was taken not to collide with the infrastructure of the field cage (strips, rod-holding brackets, etc). During the passage of the chambers through the cut-outs, the minimum clearance was less than 10\,mm for the \irocs. After installation, the perimeter of the chambers was temporarily sealed with mild tape.

\begin{figure}[ht]
\centering
\includegraphics[width=0.9\textwidth]{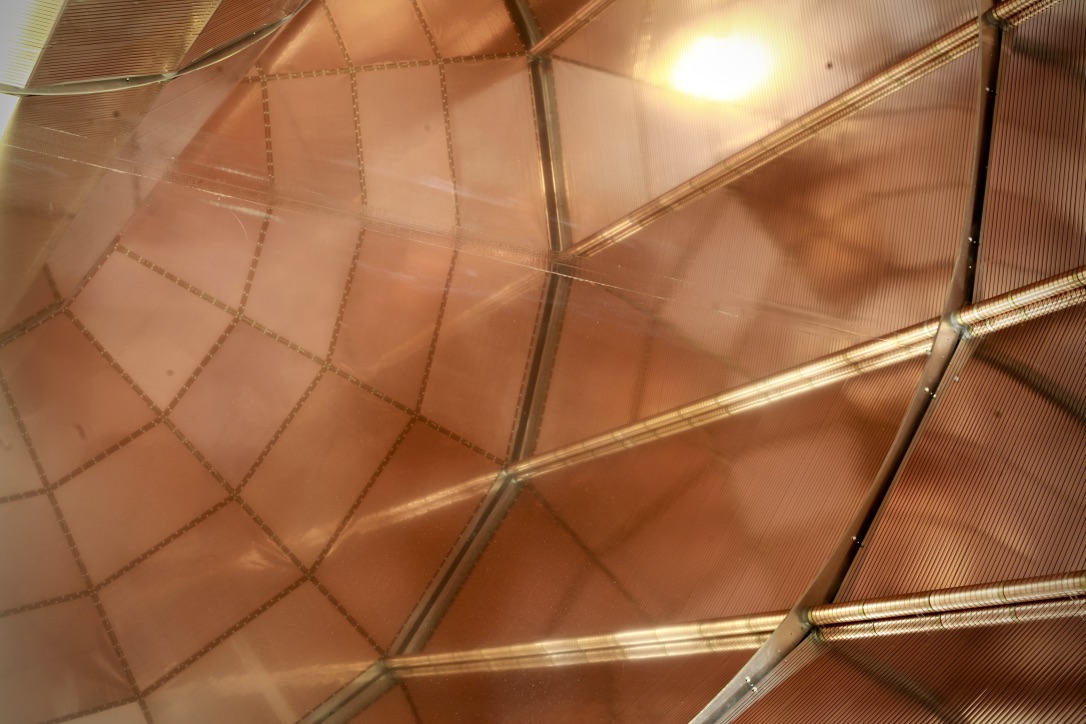}
\caption{An image inside the TPC. The central electrode reflects a view of the field cage and the readout chambers. The subdivision of each sector into four GEM stacks can be seen.}
\label{fig:install:reflect}
\end{figure}

The pace of (un)installation was 2 sectors (2 \irocs and 2 \orocs) per day. First, all MWPCs were extracted on one side, the endplate cut-outs progressively being covered with blind plates. 
Then the field cage modifications, described in \secref{sec:hv:fc}, were completed, and the GEM chambers were installed. Conductance and capacitance of each GEM foil was measured after the installation in order to spot any possible issues, such as disconnected HV elements or shorts in GEM segments, at the stage when the chambers could easily be replaced.
After completion of one side, the TPC was lifted and rotated by 180 degrees for chamber installation on the other side. \Figref{fig:install:reflect} shows the last view inside the drift volume before installation of the last readout chamber.

\begin{figure}[h]
\centering
\includegraphics[width=0.7\textwidth]{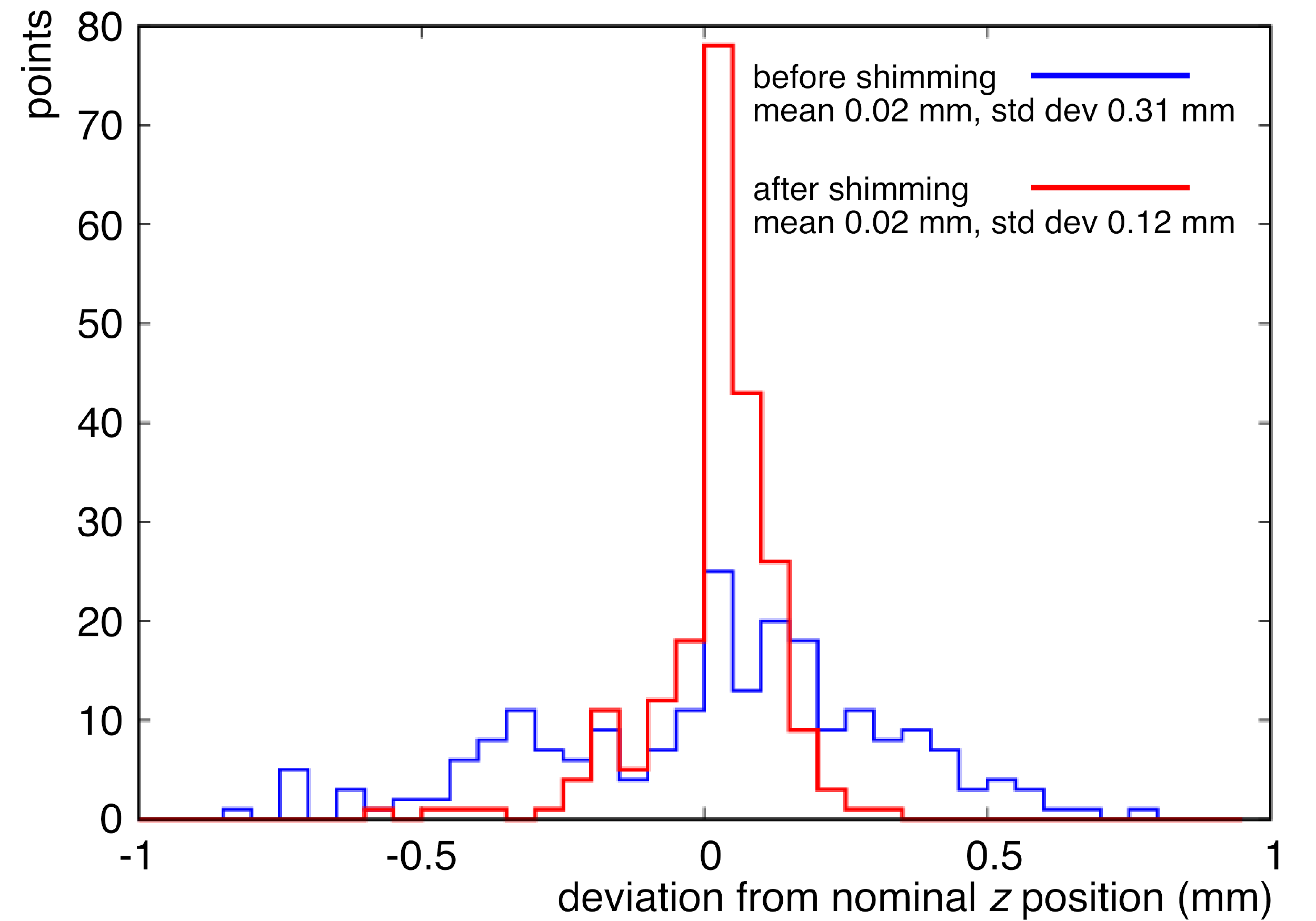}
\caption{Results of the photogrammetry before (blue) and after (red) the shimming procedure. Mean values and standard deviation of the distributions are indicated in the legend.}
\label{fig:install:survey}
\end{figure}

Each chamber is fixed on three points on the endplate. On these points, shims ensure that all chambers lie parallel to the central electrode. Since the tolerance in the thickness of the chamber bodies is $\sim$200\,\textmu m, once one full endplate was equipped with detectors, a photogrammetric survey was carried out to measure the orientation of each chamber relative to the endplate, and thus to the central electrode. These measurements were processed to calculate the correct length of each shim, which were dismounted, machined, and mounted back. Once a new survey gave satisfactory results, the chambers were sealed. \Figref{fig:install:survey} shows the results of the photogrammetry before and after shimming, for both sides of the TPC.

In order to decouple the chamber positioning from the gas tightness, the sealing employs a double O-ring, one around each chamber body, one around each sector in the endplate. An aluminized polypropylene foil is placed between endplate and chambers and pressed against the O-rings with square aluminium sealing bars.

\subsection{Service Support Wheels}
\label{sec:comm.ssw}

After ROC installation, the refurbished Service Support Wheels (SSW)~\cite{TPCnim} were installed on the two sides of the TPC.
Each SSW supports the front-end electronics and related services (LV and fibers), the HV infrastructure (protection resistor boxes), the manifolds for the various cooling circuits, and the drift gas manifolds.
The SSWs were reused with necessary modifications. In particular, the frames holding the FECs were redesigned, in order to take into account the new partitioning and card spacing. In addition, all fibers and HV cables were premounted and routed along the SSW circumference and spokes before mounting on the TPC support rails.

The gas infrastructure was then mounted on the SSWs so that the TPC could be connected to the gas system via temporary pipes laid through an existing network of trenches and shafts.
The procedure to fill the TPC with the nominal gas, after verification of its tightness, takes about three weeks.

\subsection{FEC installation}
\label{sec:install:fec}

\begin{figure}[h]
\centering
\includegraphics[width=0.9\textwidth]{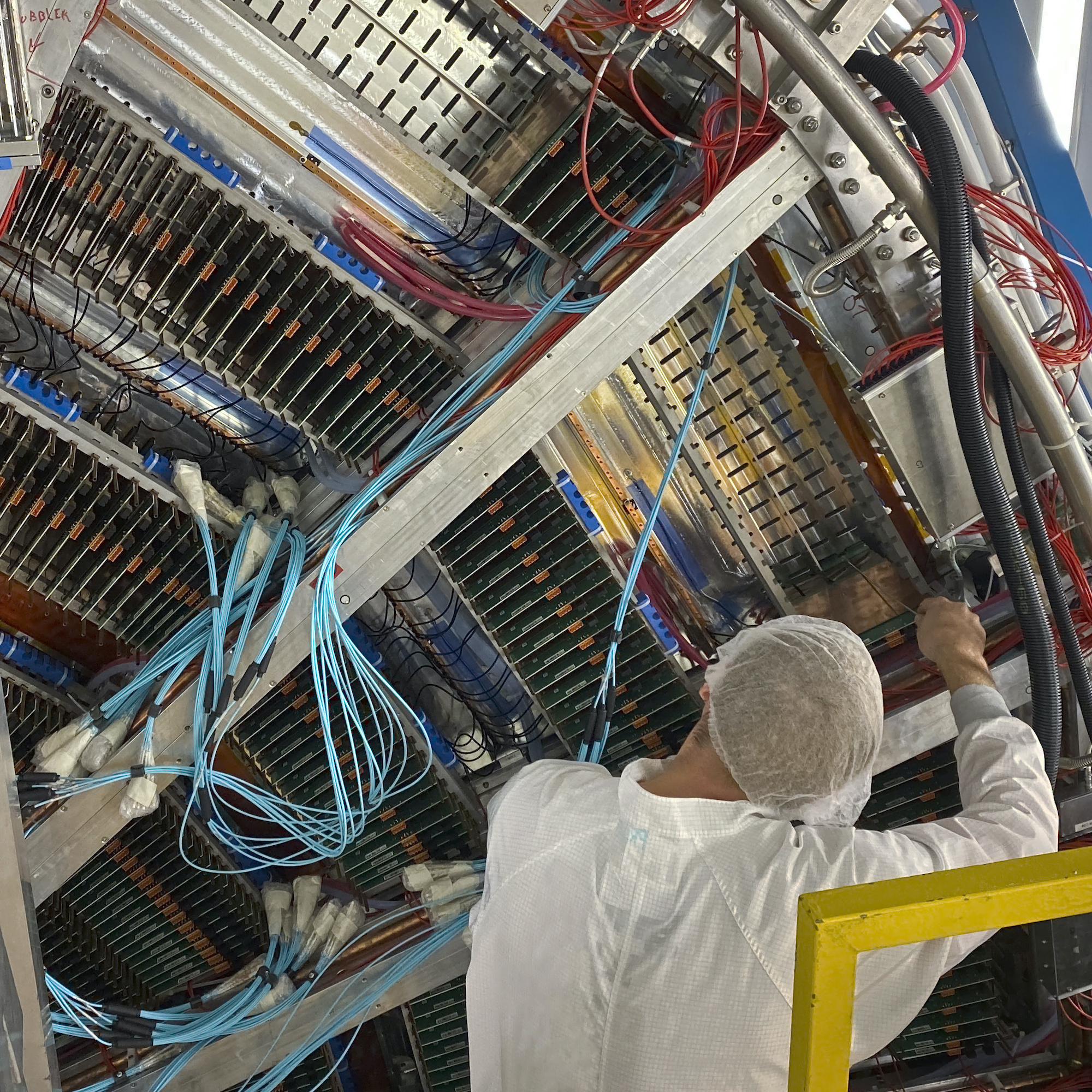}
\caption{Installing FECs in the outermost partition of one sector.}
\label{fig:install:fec}
\end{figure}

Before installation of the FECs, the FEC frames were aligned with respect to the front-end connectors on the corresponding ROC to a precision of better than 1\,mm.
A small misalignment of up to a few mm can be compensated by the flexible signal cables integrated on the FEC PCB (see \secref{sec:fec.fec.impl}).
The installation of an individual FEC was performed by inserting the FEC in the corresponding slot in the SSW, and subsequently inserting the connectors to their counterparts on the ROC (see \figref{fig:install:fec}) using a suitable tool.

The FECs are connected to a copper grounding bar screwed onto the ROC Al-body by one or two (for the outermost readout partition) copper cables with mini-banana connectors.
All FECs in one readout partition are held in place by an additional copper grounding bar screwed onto the SSW.
A second ground cable connects each FEC to this grounding bar.

After installation, each FEC was connected to a temporary readout system, powered up, and tested for connectivity to the pad plane.
For this purpose, a pulser signal was injected into the GEM stack through the input provided in the HV protection resistor box (see \secref{sec:hv:hv:prb}).
The pulse was synchronized (with a fixed delay of around \SI{50}{\micro s}) with the readout trigger.
About \SI{2}{\percent} of the FECs were replaced due to different faults found during these tests.
These faults included bent pins, damaged connectors and non-functional optical components.

After FEC installation, LV cable bundles were mounted on the copper bus bars running along the SSW spokes, and connected to the FEC.
Water cooling loops were defined by connecting the copper envelopes of groups of FECs to the manifolds of the SSW using silicon hoses.
Finally, 9828 optical fibers were plugged into the FECs.

%
%
\newcommand{\pulsertzero}{\ensuremath{t_0}\xspace}
\newcommand{\qtot}{\ensuremath{q_{\mathrm{Tot}}}\xspace}
\newcommand{\pulserqtot}{\qtot}
\newcommand{\pulserwidth}{std.~dev.\xspace}

\newcommand{\edep}{\ensuremath{E_{\mathrm{dep}}}\xspace}
\newcommand{\qe}{\ensuremath{q_{\mathrm{e}}}\xspace}
\newcommand{\qcl}{\ensuremath{q_{\mathrm{cl}}}\xspace}
\newcommand{\ipot}{\ensuremath{I_{\mathrm{pot}}}\xspace}
\newcommand{\wi}{\ensuremath{W_{\mathrm{i}}}\xspace}
\newcommand{\conversionGain}{\SI{20}{\mV\per\femto\coulomb}\xspace}
\newcommand{\dynamicRange}{\SI{2200}{\mV}\xspace}
\newcommand{\adcResolutoin}{\SI{1024}{ADC}\xspace}
\newcommand{\gain}{\ensuremath{G}\xspace}

\newcommand{\axis}[1]{\ensuremath{#1\mathrm{\mbox{-}axis}}\xspace}
\newcommand{\xaxis}{\axis{x}}
\newcommand{\yaxis}{\axis{y}}
\newcommand{\zaxis}{\axis{z}}

\newlength{\figuredefaultwidth}
\setlength{\figuredefaultwidth}{0.49\textwidth}

\subsection{Commissioning}
\label{sec:commissioning_phases}

After ROC, front-end electronics and services installation, a first commissioning phase was carried out in order to verify the proper functioning of all chambers and electronics.
Only two sectors were read out at a time due to limited availability of cooling, LV and readout infrastructure.
Various modes of data taking allowed for testing and improving the readout and reconstruction workflow.
Furthermore, the acquired data were used for calibration purposes and validation of the detector simulation.
For each sector pair (two \irocs and two \orocs), the following data sets were recorded:
\begin{itemize}
    \item \textit{Pedestal runs} are used to determine the pedestal and noise value for each readout channel.
    
    \item \textit{Pulser runs} allow for the verification of the proper functioning of each readout channel and for the study of the pulse-shaping properties of the electronics.
    
    \item \textit{\Xray runs}, recorded while operating an \xray generator with Ag target, allow for the study of the chamber stability under high load. Additionally, pad-by-pad gain calibration can be performed.

    \item \textit{Laser runs}, recording straight ionization tracks from the laser calibration system, as well as photoelectrons emitted on the central electrode, and \textit{cosmic runs}, recorded using a random trigger, are taken to demonstrate track reconstruction and develop calibration procedures.
\end{itemize}
The data sets were taken in triggered mode.
Since the front-end cards send data continuously, data selection of one TPC drift time starting with the trigger was applied in the CRU.
No data compression nor data reduction was performed at this stage.
The data were recorded as raw GBT frames as sent by the electronics.
In the following, the main results of the different tests are presented.


\subsubsection{Pedestal and noise measurements}
\label{sec:pedestal_and_noise}
\newcommand{\NoiseMean}{0.97}
\newcommand{\NoiseStdDev}{0.09}
\newcommand{\irocNoiseMean}{0.93}
\newcommand{\irocNoiseStdDev}{0.07}
\newcommand{\irocNoiseFitOffset}{0.81}
\newcommand{\irocNoiseFitSlope}{0.03}
\newcommand{\irocNoiseFitOffsetNoFEC}{0.19}
\newcommand{\irocNoiseFitOffsetNoFECPerArea}{0.59}
\newcommand{\orocOneNoiseMean}{0.97}
\newcommand{\orocOneNoiseStdDev}{0.08}
\newcommand{\orocOneNoiseFitOffset}{0.83}
\newcommand{\orocOneNoiseFitSlope}{0.03}
\newcommand{\orocOneNoiseFitOffsetNoFEC}{0.24}
\newcommand{\orocOneNoiseFitOffsetNoFECPerArea}{0.41}
\newcommand{\orocTwoNoiseMean}{0.99}
\newcommand{\orocTwoNoiseStdDev}{0.09}
\newcommand{\orocTwoNoiseFitOffset}{0.83}
\newcommand{\orocTwoNoiseFitSlope}{0.03}
\newcommand{\orocTwoNoiseFitOffsetNoFEC}{0.27}
\newcommand{\orocTwoNoiseFitOffsetNoFECPerArea}{0.38}
\newcommand{\orocThreeNoiseMean}{1.03}
\newcommand{\orocThreeNoiseStdDev}{0.10}
\newcommand{\orocThreeNoiseFitOffset}{0.87}
\newcommand{\orocThreeNoiseFitSlope}{0.03}
\newcommand{\orocThreeNoiseFitOffsetNoFEC}{0.36}
\newcommand{\orocThreeNoiseFitOffsetNoFECPerArea}{0.40}

\begin{figure}[b]
    \centering
    \includegraphics[width=\figuredefaultwidth]{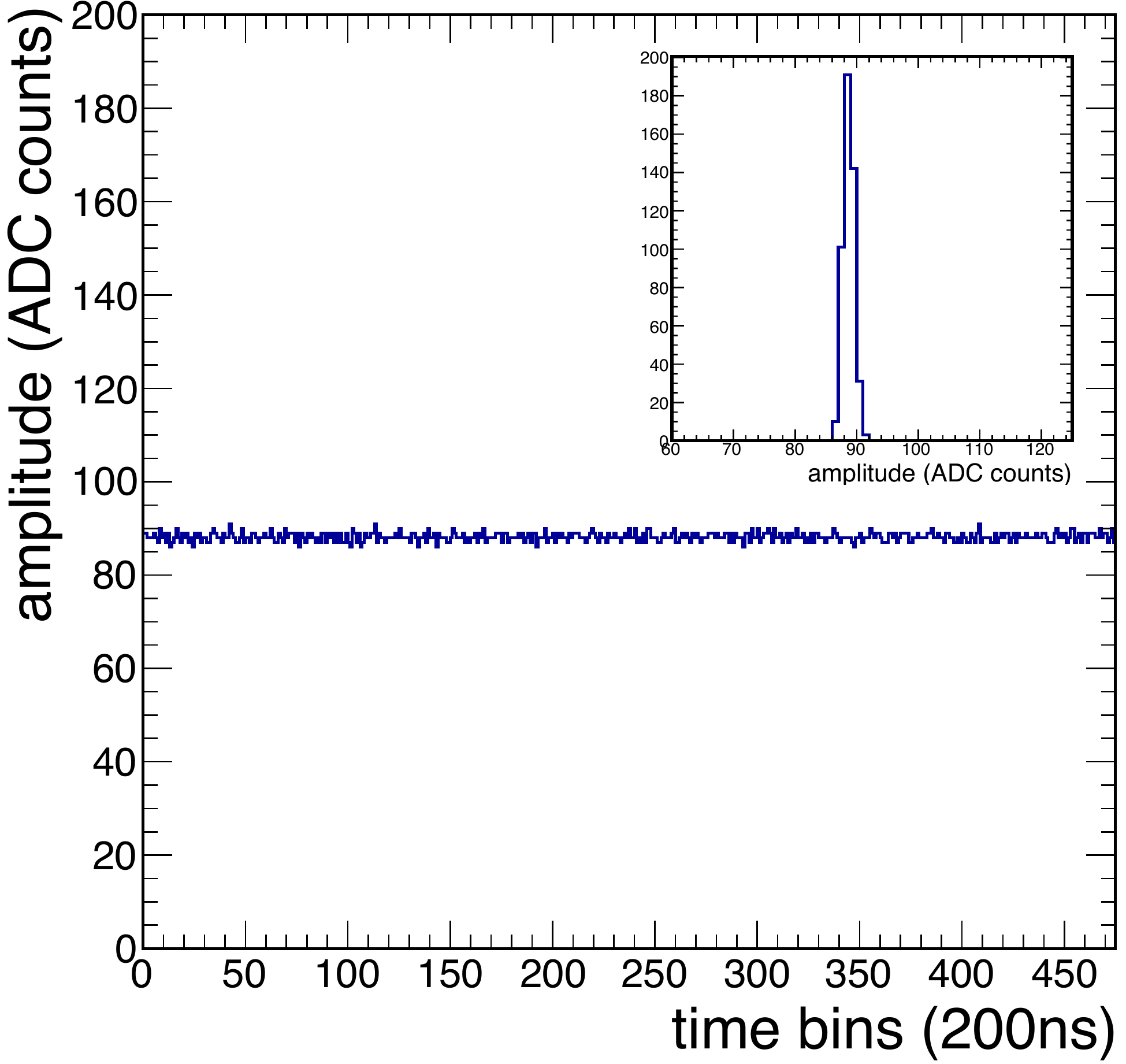}
    \caption{Typical baseline of one readout channel. The inset shows the projection onto the \axis{y}.}
    \label{fig:calcom:baseline}
\end{figure}

A typical baseline of a single readout pad (channel) is shown in \figref{fig:calcom:baseline}.
The inset shows the distribution.
Its mean value reflects the pedestal offset, while the standard deviation is used to characterize the noise.
Pedestal and noise values are measured for each pad individually.

To study the intrinsic noise of the system, care was taken to minimize the influence of external sources (e.g.\ ventilation fans, water cooling unit, electronics crates, etc.) during specific noise studies.
The results discussed in the following were obtained for one sector pair after confirming that the chosen grounding scheme (see \secref{sec:fec.fec.impl}) leads to the anticipated noise figures.
They resemble the noise level that can be obtained under close to optimal conditions.
During the standard data taking not all disturbances can be eliminated (e.g.\ clean room ventilation), leading to slightly larger noise values.

\begin{figure}[ht]
    \centering
    \includegraphics[width=\figuredefaultwidth]{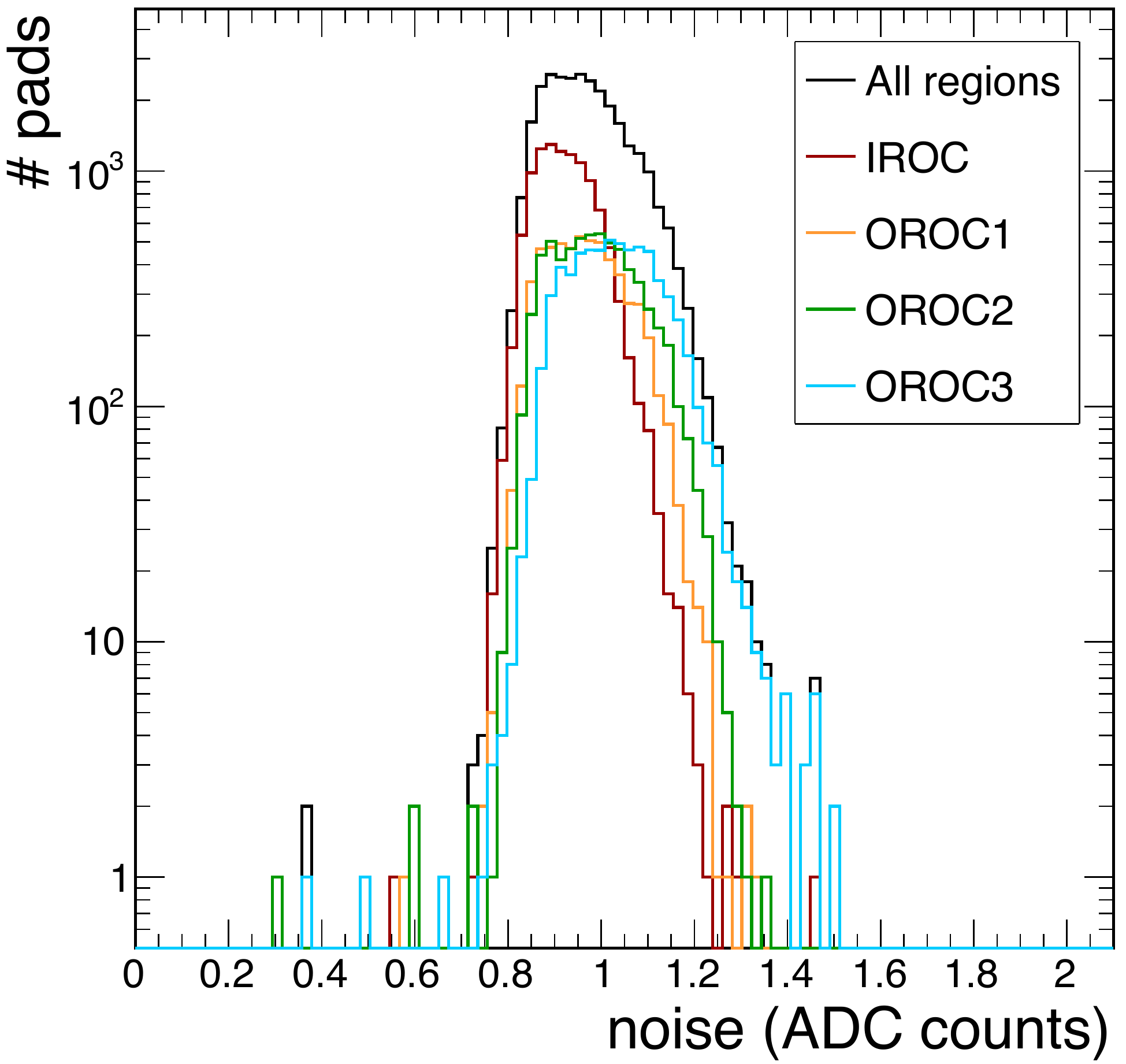}
    \caption{Noise distribution of one sector pair, separately for the \iroc (red), \orocOne (orange), \orocTwo (green), and \orocThree (blue). The total distribution is shown in black.}
    \label{fig:calcom:noise}
\end{figure}

\Figref{fig:calcom:noise} shows the noise distribution of all pads (black) as well as for the pad regions corresponding to the four GEM stacks \iroc (red), \orocOne (orange), \orocTwo (green) and \orocThree (blue).
The pads of each stack roughly have the same pad size (see \tabref{tab:roc:padplane:pads}).
The mean noise over all pads is \NoiseMean\,\adcs.
For the separate stacks, the mean noise is \irocNoiseMean, \orocOneNoiseMean, \orocTwoNoiseMean, and \orocThreeNoiseMean\,\adcs in the \iroc,  \orocOne, \orocTwo, and \orocThree, respectively.
These figures meet the design goal of a noise of 1\,\adc (see \secref{sec:fec.overview.requirements}), which corresponds to 670\,e.

\begin{figure}[ht]
    \centering
    \includegraphics[width=0.9\textwidth]{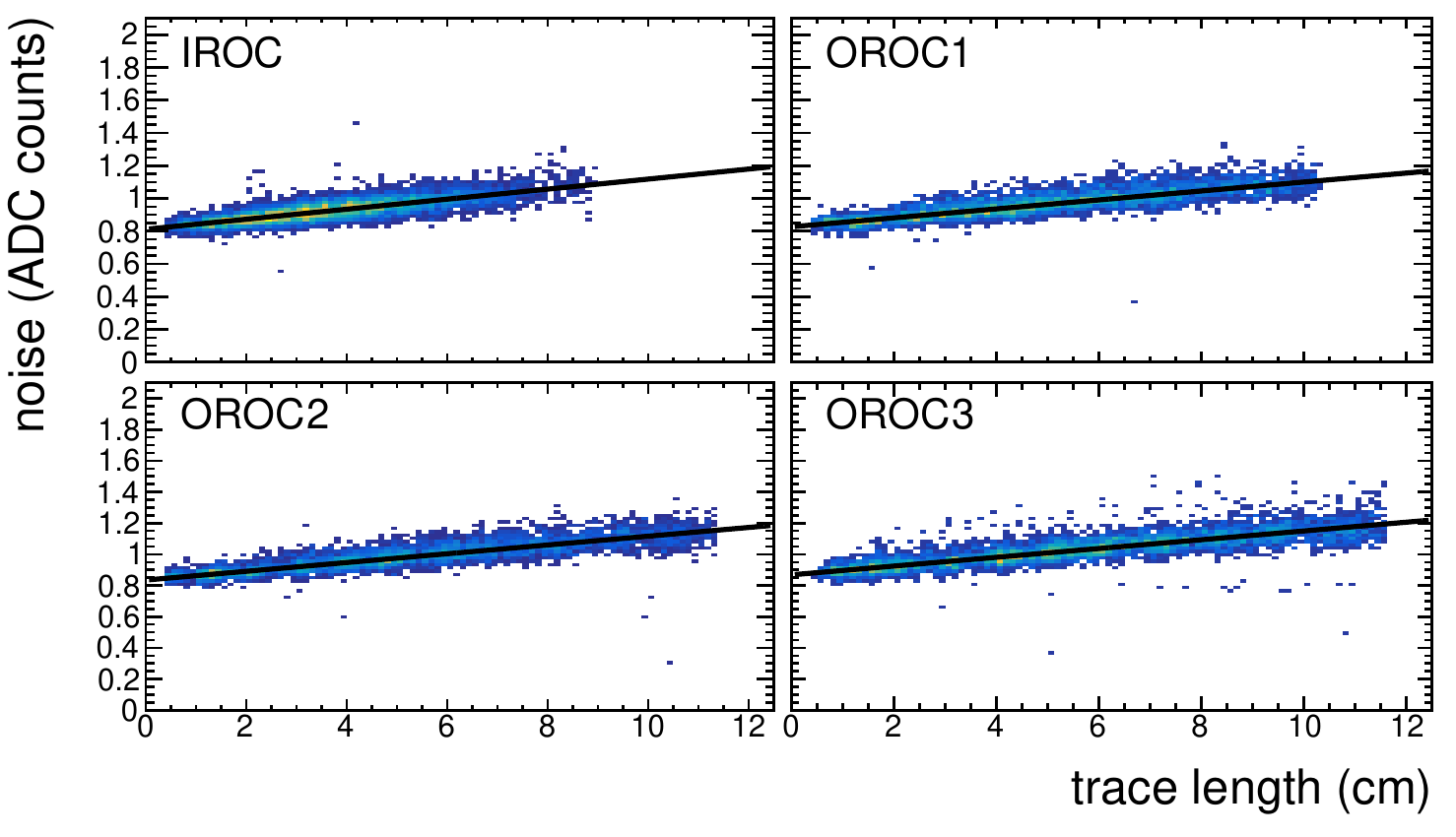}
    \caption{Correlation of the noise with the length of the trace connecting pads and connector, separately for each GEM stack. See text for details.}
    \label{fig:calcom:noise_tracelength}
\end{figure}

The distributions in all pads are slightly skewed towards higher noise values.
This is caused by the difference in length of the traces connecting each pad with the connector on the backside of the pad plane PCB (see \secref{sec:roc:padplane}).
This leads to different capacitances and, therefore, noise values for each pad.
\Figref{fig:calcom:noise_tracelength} shows the correlation of the noise with the trace length for the individual stacks.
The data of each stack is fit with a linear function.
The average noise values, as well as the fit parameters of each stack, are summarized in \tabref{tab:calcom:noise}.
The slope is identical in all stacks (\irocNoiseFitSlope\,ADC~counts\,cm$^{-1}$).
The offset value of the fit contains contributions of the intrinsic noise of the FEC itself, of the connector and of the readout pad.
The values obtained when quadratically subtracting the average noise value of the cards (\SI{0.79}{\adcs}, see \secref{sec:fec.testing}) from the fit offset are also summarized in \tabref{tab:calcom:noise} (offset no FEC).
Additionally, the values are normalized by the pad area assuming a negligible contribution from the connector ("per pad area" in \tabref{tab:calcom:noise}).
While for all \oroc pad regions the noise per area (about 0.40\,ADC~counts\,cm$^{-2}$) is similar, the value in the \iroc is slightly larger (about \irocNoiseFitOffsetNoFECPerArea\,ADC~counts\,cm$^{-2}$).

\begin{table}[ht]\footnotesize
    \caption{Average noise and standard deviation corresponding to the distributions in \figref{fig:calcom:noise}. Offset and slope of the fits from the trace-length correlation shown in \figref{fig:calcom:noise_tracelength} are given as well. 
    Additionally, the offset values are corrected for the average noise contribution of the FECs by quadratic subtraction (offset no FEC).
    The corrected offset value per pad area is also given.
    See text for details.}
    \begin{center}
    \begin{tabular}{l cc cc cc}
      \toprule
     
      \theader{1}{c}{type} & \theader{1}{c}{mean} & \theader{1}{c}{std.~dev.} & \theader{1}{c}{offset} & \theader{1}{c}{slope} & \theader{1}{c}{offset no FEC} & \theader{1}{c}{per pad area} \\
       &\theader{1}{c}{(ADC)}   & \theader{1}{c}{(ADC)} & \theader{1}{c}{(ADC)} & \theader{1}{c}{(ADC\,cm$^{-1}$)} & \theader{1}{c}{(ADC)} & \theader{1}{c}{(ADC\,cm$^{-2}$)} \\
         \midrule
         \iroc & \irocNoiseMean & \irocNoiseStdDev & \irocNoiseFitOffset & \irocNoiseFitSlope & \irocNoiseFitOffsetNoFEC & \irocNoiseFitOffsetNoFECPerArea\\
\orocOne & \orocOneNoiseMean & \orocOneNoiseStdDev & \orocOneNoiseFitOffset & \orocOneNoiseFitSlope & \orocOneNoiseFitOffsetNoFEC & \orocOneNoiseFitOffsetNoFECPerArea\\
\orocTwo & \orocTwoNoiseMean & \orocTwoNoiseStdDev & \orocTwoNoiseFitOffset & \orocTwoNoiseFitSlope & \orocTwoNoiseFitOffsetNoFEC & \orocTwoNoiseFitOffsetNoFECPerArea\\
\orocThree & \orocThreeNoiseMean & \orocThreeNoiseStdDev & \orocThreeNoiseFitOffset & \orocThreeNoiseFitSlope & \orocThreeNoiseFitOffsetNoFEC & \orocThreeNoiseFitOffsetNoFECPerArea\\

         \bottomrule
    \end{tabular}
    \end{center}
    \label{tab:calcom:noise}
\end{table}

%
%
\newcommand{\pulserwidthStdDev}{15}
\newcommand{\pulsertzeroStdDev}{5}
\newcommand{\pulserqtotRelStdDev}{12}
\newcommand{\pulsergemdist}{240}

\subsubsection{Pulse-shaping analysis}
\label{sec:pulse_shaping}
A pulse (step function) can be injected on the \gemFour bottom electrode using a calibration pulser system.
The pulse is coupled to the corresponding HV line via a \SI{30}{\pF} capacitor (see \secref{sec:hv:hv:prb}). As a consequence, a signal is induced on all pads of the readout plane by capacitive coupling.
During FEC installation, the pulser system was used to ensure proper connection of all cards, as well as the integrity of all channels of each card.
Cards with faulty channels were replaced.

\begin{figure}[ht]
    \centering
    \includegraphics[width=\figuredefaultwidth]{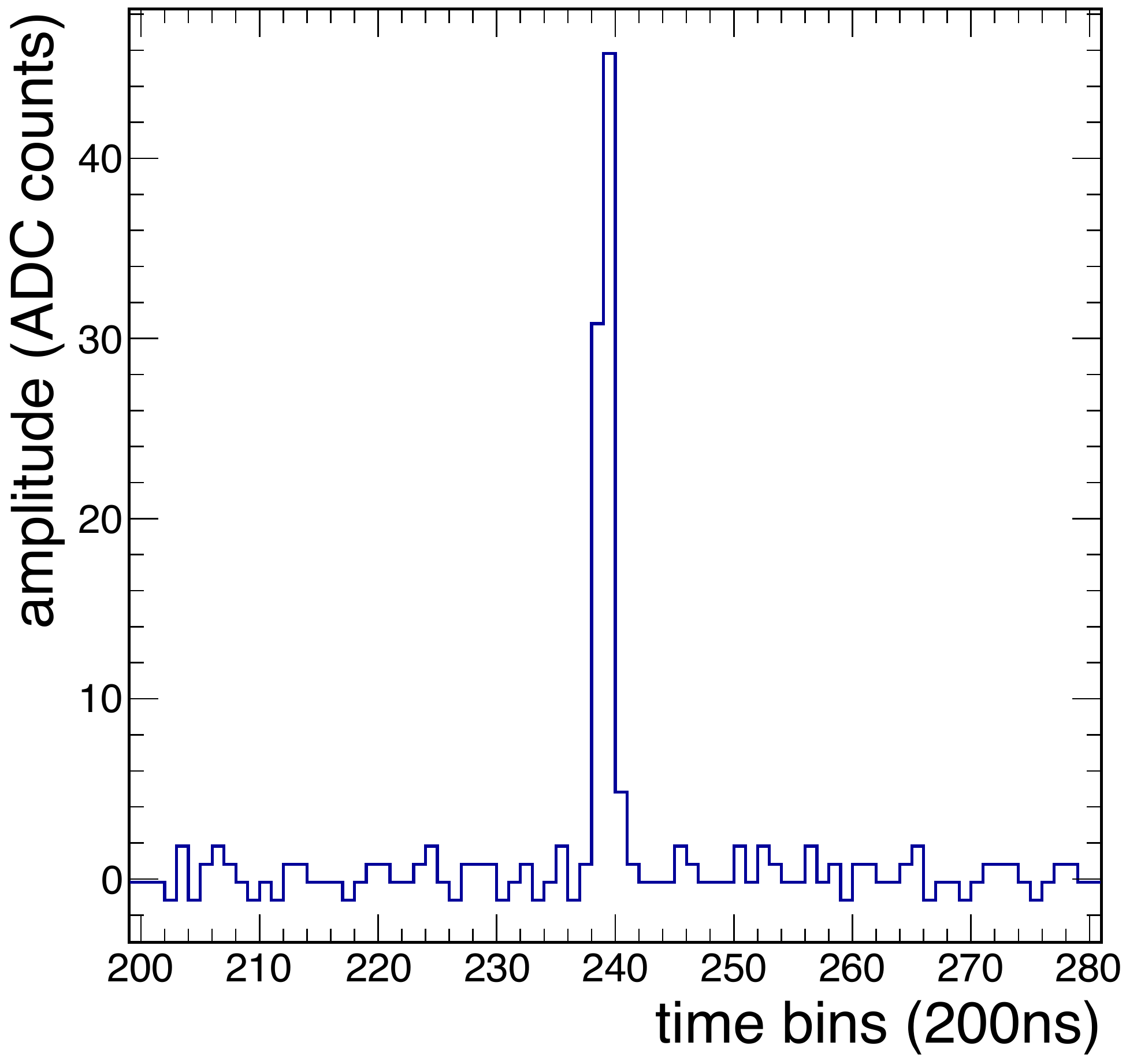}
    \caption{Typical electronics response of one pad to the pulser signal.}
    \label{fig:calcom:pulser_signal}
\end{figure}

The pulser system can be used to study the timing and shaping behavior for each electronics channel.
The typical response of a single readout channel to the pulser is displayed in \figref{fig:calcom:pulser_signal}.
Pulses are reconstructed in a $\pm 2$ time bin window around the maximum.
For the characterization of the response, the weighted mean (\pulsertzero), pulse width (\pulserwidth), as well as the pulse-charge integral (\pulserqtot) are computed for each channel.
Typically, each value is averaged over 100 events.

\begin{figure}[ht]
    \centering
    \includegraphics[width=1.3\figuredefaultwidth]{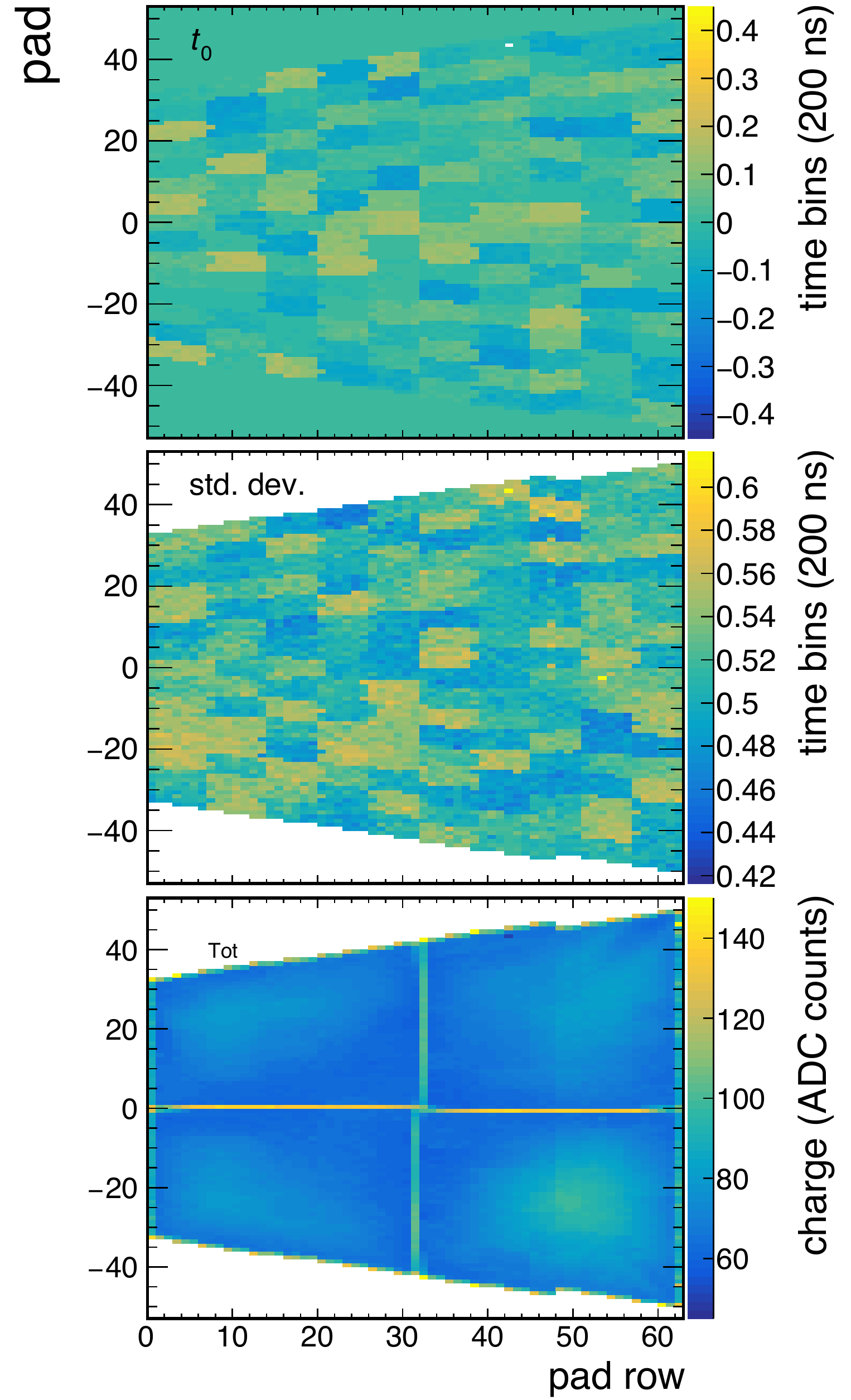}
    \caption{Typical signal response variations of the timing and shaping characteristics in an \iroc. From top to bottom the weighted mean (\pulsertzero), pulse width (\pulserwidth) and the pulse-charge integral (\pulserqtot) are shown. See text for details.}
    \label{fig:calcom:shaping}
\end{figure}

The resulting variations are exemplarily shown for one \iroc on a pad-by-pad level in \figref{fig:calcom:shaping}.
From top to bottom, the distributions of \pulsertzero, pulse width (\pulserwidth), and \pulserqtot are displayed.
A specific pattern is seen in the \pulsertzero and \pulserwidth distributions, which follows the layout of the SAMPA chip connection to the pad plane.
The variations within a chip are rather small.
Chip-by-chip variations reflect the expected production tolerances.
Typical values are about \SI{\pulsertzeroStdDev}{\ns} for the width (std.~dev.) of the \pulsertzero distribution and about \SI{\pulserwidthStdDev}{\ns} for the \pulserwidth distribution.

The \pulserqtot distribution also shows a distinct pattern.
Firstly, the influence of the spacer cross (see \secref{sec:frames}) can be observed around pad position 0 and pad-row position 32.
In addition, a geometric pattern is seen for each of the four quadrants, with higher \pulserqtot values towards the respective center.
The relative width (std.~dev.) of the \pulserqtot distribution is about \SI{\pulserqtotRelStdDev}{\percent}.
Since the pulser signal is induced by capacitive coupling, the distribution can be explained by mechanical variations of the \SI{2}{\mm} induction gap between \gemFour and the pad plane.

%
%
\subsubsection{Data taking with an \xray generator}
\label{sec:gain_calibration}

\newcommand{\GainMapStdDevI}{7.5}
\newcommand{\GainMapStdDevOone}{7.0}
\newcommand{\GainMapStdDevOtwo}{6.1}
\newcommand{\GainMapStdDevOthree}{5.9}

\begin{figure}[ht]
    \centering
    \includegraphics[width=0.7\textwidth]{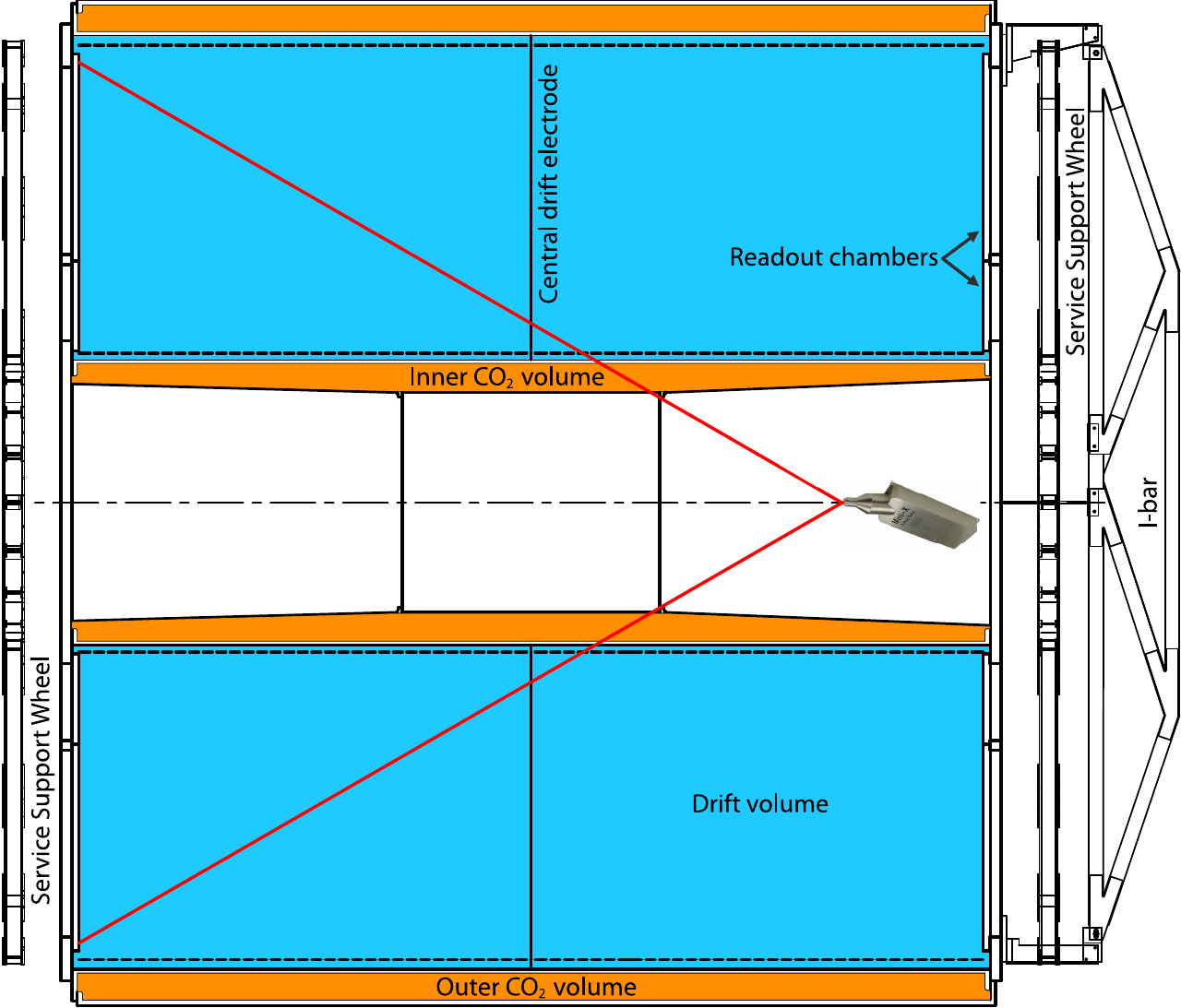}
    \caption{Positioning of the \xray generator for the irradiation of one side of the TPC.
    The red lines indicate the opening angle of \ang{\pm30}, in which the  \xray generator provides a uniform emission profile.
    The $z$-position of the  \xray generator was chosen such that the  central drum is entirely inside this region.
    The opposite side of the TPC can be irradiated as well by moving the source accordingly.
    }
    \label{fig:calcom:x-ray_pos}
\end{figure}

Data taken with an \xray generator are used to study the ROC stability under high load, to determine the average ROC gain, and to study pad-wise gain variations.
An Amptek Mini-X~\cite{amptekminix} \xray generator with Ag anode was used to
irradiate the upgraded TPC as part of the commissioning campaign.
During the measurements, the generator is typically operated at \SI{50}{kV} and a current of up to \SI{80}{\uA}.
\Figref{fig:calcom:x-ray_pos} shows the position and orientation of the \xray generator during data taking.
\Xrays from the generator mainly enter the active volume of the TPC through the central drum, where the material budget is low, whereas in the conical parts of the containment vessel, most of the incident radiation is absorbed due to thermal screens made of aluminium.
The irradiated region corresponds approximately to an opening angle of \ang{\pm30}, inside which the source has a rather uniform emission profile~\cite{amptekminix}.
In total, about \SI{2}{\percent} of the produced \xrays convert inside the active TPC volume.\\
Intense \xray irradiation, exposing simultaneously all GEM stacks of one TPC side, allows confirmation of the ROC stability under a  load that is similar to that expected for operation of the TPC in LHC Run\,3.
At the start of the campaign, a few stacks exhibited sudden current excursions, followed by trips of the power supply. Such incidents ceased after about \SI{10}{\hour} of irradiation.
After this initial phase, no further signs of rate-induced instabilities were observed during the continued campaign with an integrated irradiation time of about \SI{300}{\hour}.

To study gain variations over the active readout area, the \xray ionization clusters are reconstructed.
This is done by first finding local maxima in the recorded ADC values.
A cluster is then defined by associating further ADC values around the maximum in all three spatial directions, radial (pad row), azimuthal (pad) and drift time ($z$), in a \num{3x5x5} matrix.
The total charge of the cluster, \qtot, corresponds to the sum of all associated ADC values and is proportional to the energy deposition of the \xray photon.
This 3D approach differs from the standard 2D cluster finding algorithm which is performed in radial direction for each pad row separately, reconstructing a cluster in azimuthal (pad) and drift time ($z$) direction.
The discrete energy deposition of the \xrays may span several pad rows and, therefore, requires a dedicated algorithm.

\begin{figure}[b]
    \centering
    \includegraphics[width=0.7\textwidth]{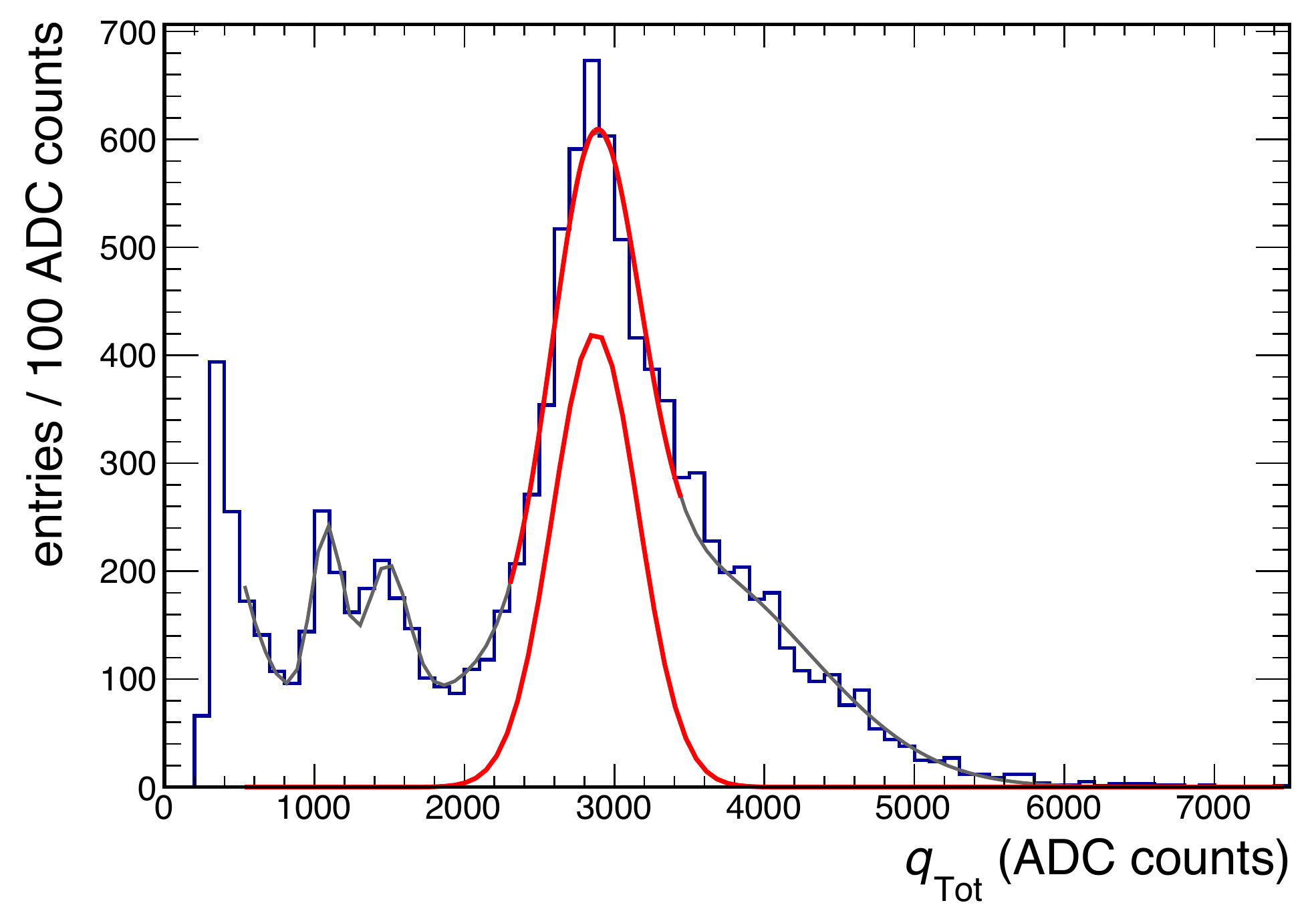}
    \caption{Reconstructed \xray spectrum in a single pad. A function describing the main and fluorescence peaks, as well as background contributions including bremsstrahlung, is fit to the spectrum. The red line shows the contribution of the main peak.}
    \label{fig:calcom:x-ray_spectrum_pad}
\end{figure}
The local gain is extracted from the \qtot distributions of the \xray clusters on a pad-by-pad level, see \figref{fig:calcom:x-ray_spectrum_pad} for an example.
The main peak at around 2800\,ADC~counts, for this specific channel, results from the \kalpha and \kbeta lines from the 
Ag anode of the \xray generator at about \SI{22}{\keV}.\newline
The two lines are not separable 
due to the limited detector resolution.
Further peaks at lower energies can be seen, which result from fluorescence in material hit by the \xrays.
The spectrum is fitted with a function describing the different contributions, including the main and fluorescence peaks, as well as the background originating mainly from the bremsstrahlung continuum.
The position of the main peak (red curve in \figref{fig:calcom:x-ray_spectrum_pad}) is used as an estimate for the gain.

\begin{figure}[ht]
    \centering
    \includegraphics[width=0.9\textwidth]{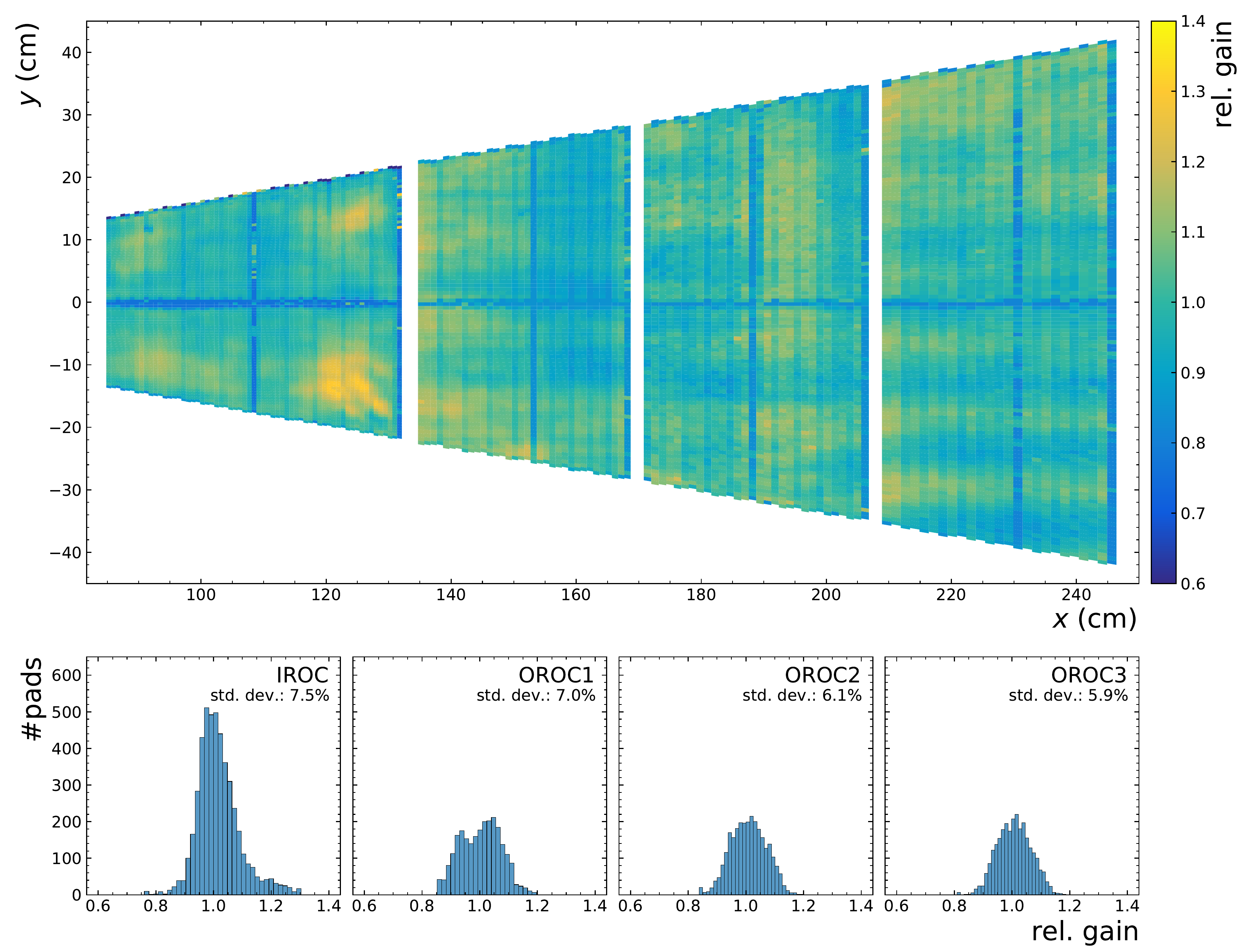}
    \caption{Two-dimensional gain map extracted from reconstructed \xray clusters in one sector (top), and one-dimensional gain distributions in the separate pad regions (bottom). For details, see text.}
    \label{fig:calcom:x-ray_gain_map}
\end{figure}

Exemplarily, the resulting relative gain variations over the pad area of one full sector are shown in \figref{fig:calcom:x-ray_gain_map} (top).
Topological gain variations over the pad area are expected due to manufacturing tolerances in the hole sizes (see \secref{sec:roc:advanced_qa}).
This can 
be seen in the \orocThree stack in \figref{fig:calcom:x-ray_gain_map} as horizontal regions with lower and higher gain.
Additionally, specific regions with lower gain can be observed.
Very prominently, the effect of the spacer cross (see \secref{sec:frames}) is visible in each pad region.
Moreover, a slightly lower gain can be observed along full pad rows ($x$-positions) resulting from the HV segment boundaries (see \secref{sec:roc:gem}).
While in the \iroc this can be observed in about every sixth pad-row, in the \orocOne, \orocTwo and \orocThree  about every fourth to second row is affected due to the increasing pad length.
The lower gain along the edges of each pad region is due to acceptance effects.
Slightly larger gain values can be observed in the corners of the four quadrants of the \iroc.
They result from field variations due to mechanical tolerances in the GEM foil spacings.
These features are already seen in the coarse gain scan during the chamber-production QA (see \secref{sec:roc:gain_uniformity}), as well as in the results of the pulser measurements (see \secref{sec:pulse_shaping}).
In the bottom plots of \figref{fig:calcom:x-ray_gain_map}, the distributions of the relative gain variations in the different pad regions are shown.
Pads in the region of the spacer cross as well as in the first and last pad row were removed.
The gain variations in this specific sector have a spread of about \SI{\GainMapStdDevI}{\percent}, \SI{\GainMapStdDevOone}{\percent}, \SI{\GainMapStdDevOtwo}{\percent} and \SI{\GainMapStdDevOthree}{\percent} (std.~dev.), in the \iroc, \orocOne, \orocTwo and \orocThree, respectively.
These variations are in accordance with the specifications.

The absolute gain of a stack can also be estimated from the spectrum of the total charge \qtot of \xray clusters in that stack, since the response of the detector, including the readout electronics, to a given energy deposition in the gas was studied in detailed microscopic simulations~\cite{TDR:tpcUpgrade}. Thus, the gain relates to the ADC value of the total charge of \SI{22}{\keV} \xrays, which is approximately the average energy of the main peak of the spectrum, by a well-determined empirical factor. In addition, the currents in all GEM electrodes were measured during the \xray irradiation campaigns. In particular, the \gemFour top currents allow a solid estimate of the average  gain for every stack.

\begin{figure}[ht]
    \centering
    \includegraphics[width=\figuredefaultwidth]{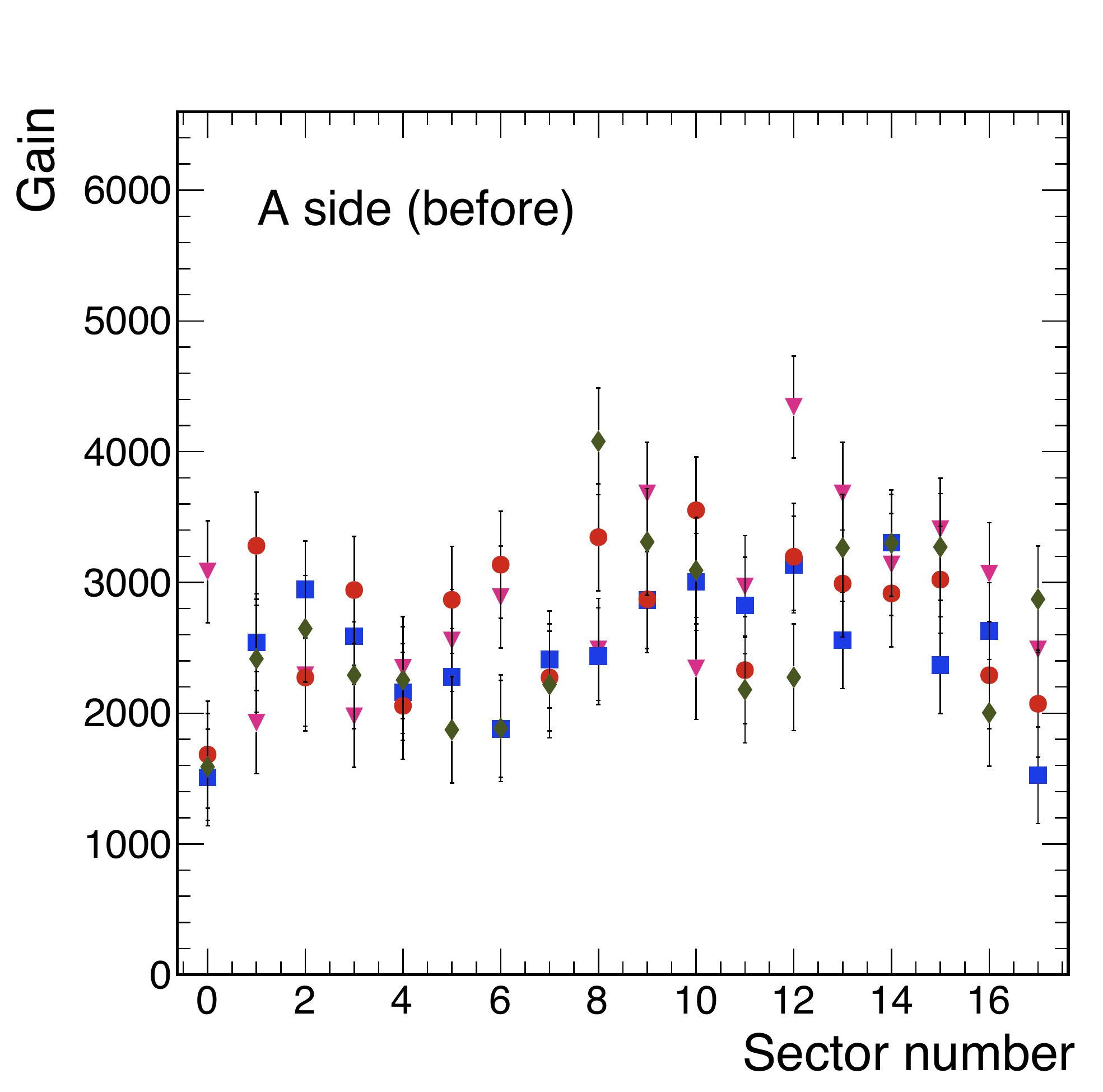}
    \includegraphics[width=\figuredefaultwidth]{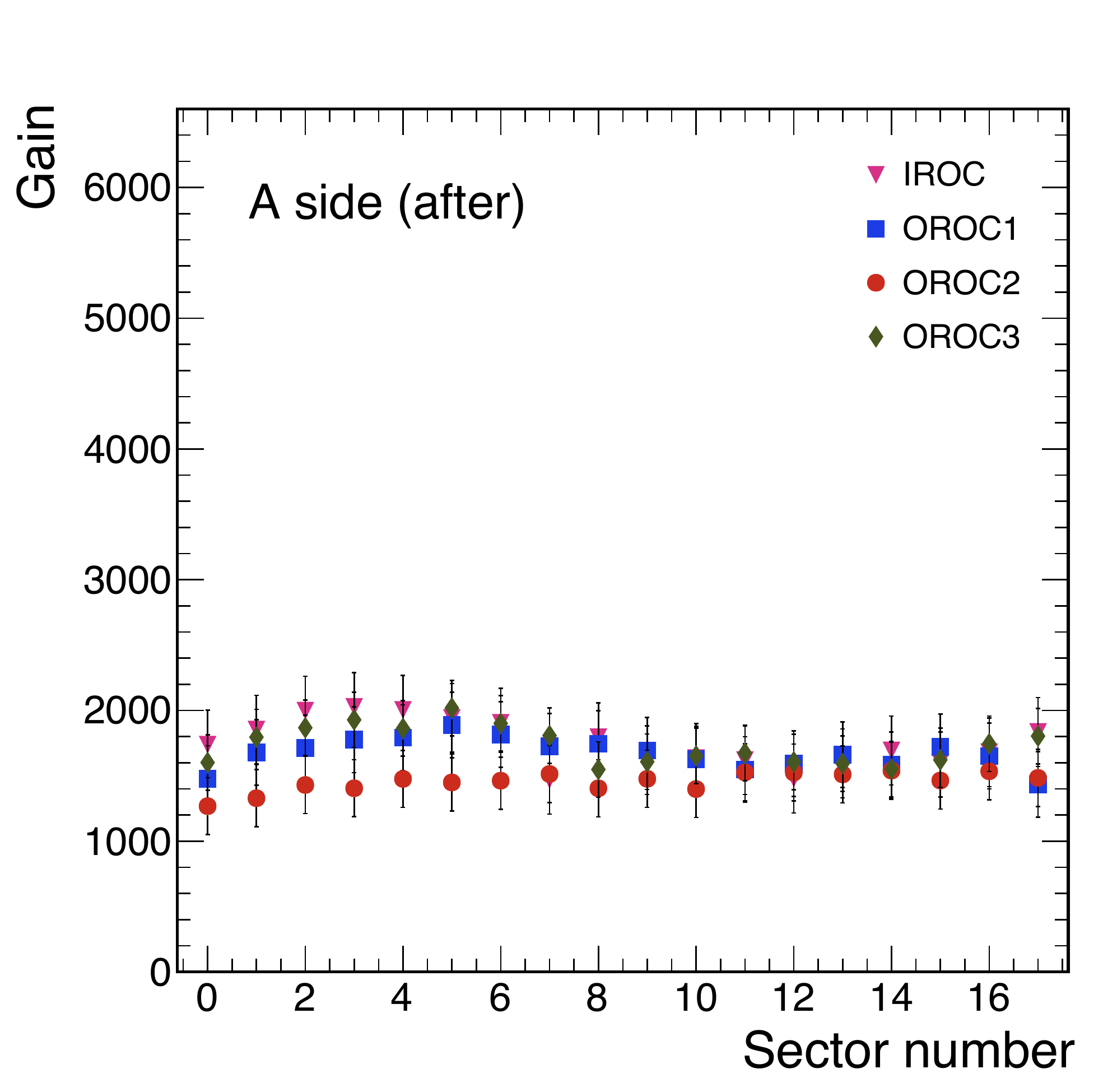}
    \caption{Gain of all stacks on the A side before (left) and after (right) gain equalization. The azimuthal modulation of the data stems from a slight misalignment of the \xray generator.}
    \label{fig:calcom:x-ray_peak_vs_current}
\end{figure}

The baseline HV settings are shown in \tabref{tab:hv:baseline}.
A first equalization of the gain was performed using the described gain calculation. The tuning of the gain on each stack was done by tuning the voltages in \gemThree and \gemFour, and the induction field, ensuring that the potential on the \gemOne top electrode remains uniform over all stacks. \Figref{fig:calcom:x-ray_peak_vs_current} shows, as an example, the estimated gain before and after the gain equalization for the A side. The gains before equalization were in general higher than nominal (2000), and show variations of up to a factor of three. The equalized gains are on average slightly lower than nominal, and superimposed with a residual sinusoidal modulation due to the imperfect alignment of the generator with the \zaxis of the TPC. A more robust gain equalization with a \kr source, which is typically performed shortly before every years data taking with beams, is expected to fully equalize the gains in all stacks to the nominal value.

\subsubsection{Data taking with the laser calibration system and with random cosmic triggers}
\label{sec:precom:laser_data_taking}
Data samples using the laser system and random triggers to select cosmic tracks were recorded during the commissioning campaign. They are useful to verify the complete data reconstruction chain from raw data to reconstructed tracks. The data are also used to develop calibration procedures, 
such as the
drift velocity calibration using reconstructed laser tracks.

\begin{figure}[ht]
    \centering
    \includegraphics[width=0.49\textwidth]{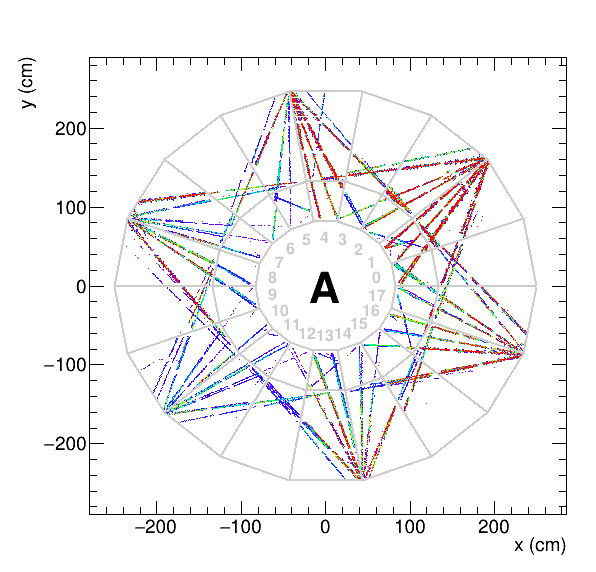}
    \includegraphics[width=0.49\textwidth]{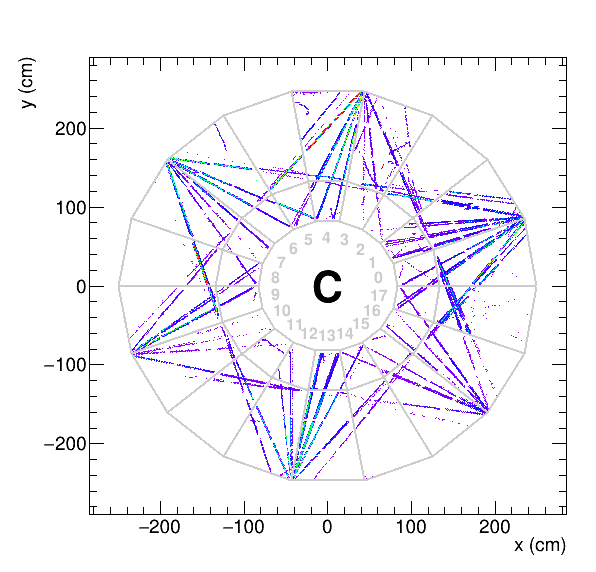}
    \caption{Reconstructed ADC values from laser tracks, left for the A~side and right for the C~side. Data taken consecutively for different sectors are merged in order to display them together.}
    \label{fig:calcom:laser_tracks}

\end{figure}

\Figref{fig:calcom:laser_tracks} shows the  measured ADC values 
of laser events
on the A~side (left) and C~side (right).
Data from different runs are merged in order to assemble a full event display.
The specific pattern created by the laser beam optics (see~\cite{TPCnim}) is clearly visible.
The same pattern is created in four planes parallel to the readout plane on each side at $z \approx \pm115, \pm820, \pm1660, \pm2440\,\mathrm{mm}$, where positive\,/\,negative positions are on the A and C~side, respectively.

\begin{figure}[ht]
    \centering
    \includegraphics[width=0.49\textwidth]{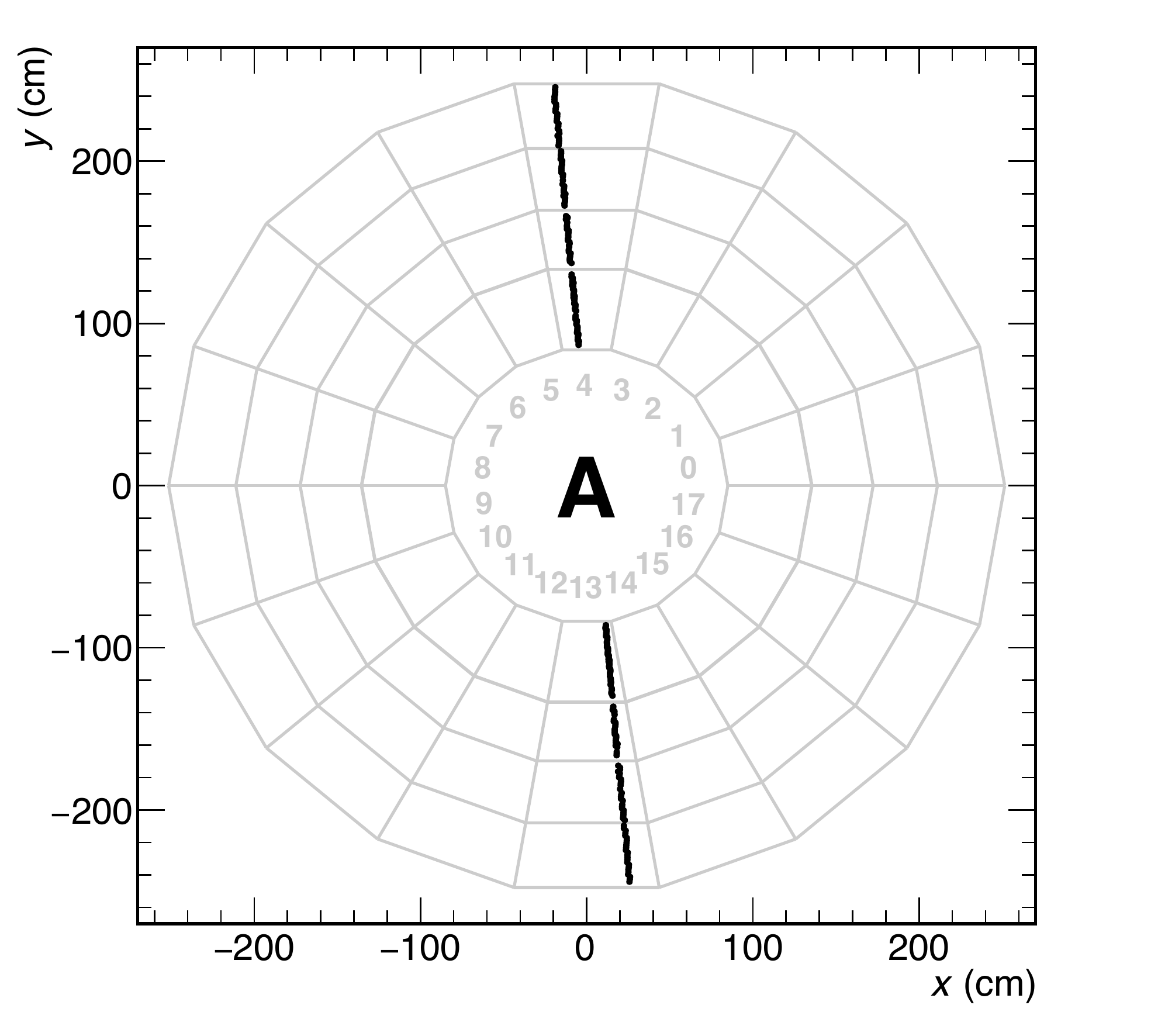}
    \includegraphics[width=0.49\textwidth]{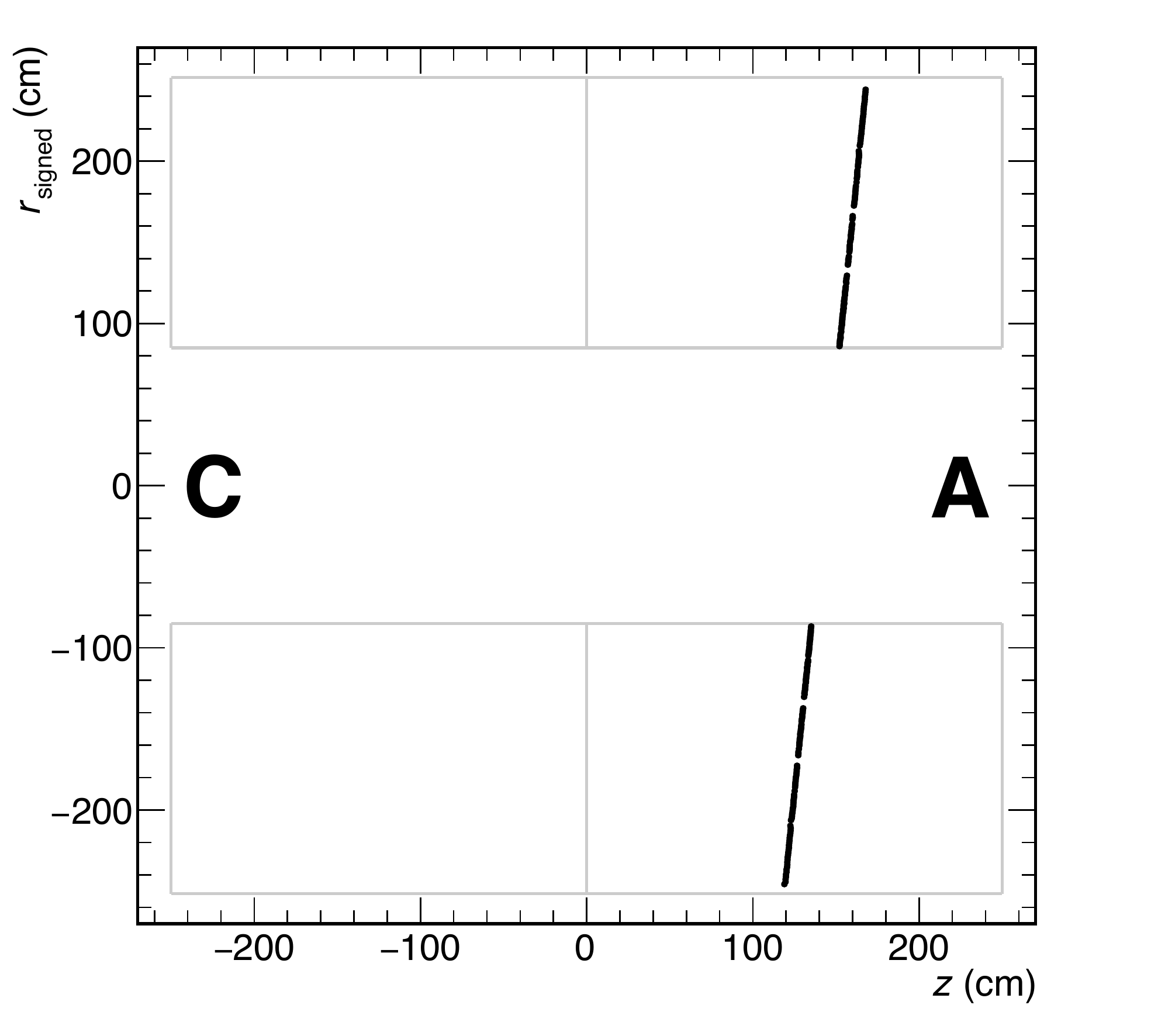}
    \caption{Clusters of a reconstructed cosmic track crossing the full TPC. The \mbox{$x$-$y$-projection} is shown on the left and  the \mbox{$r$-$z$-projection} on the right panel, where the radial position is negative for the lower half of the TPC for visualization. }
    \label{fig:calcom:cosmic}

\end{figure}

\Figref{fig:calcom:cosmic} shows clusters of a reconstructed cosmic track crossing the full TPC.
This demonstrates the 
integrity of
the 
complete
reconstruction chain.
The panels show projections in the $x$-$y$ and $r$-$z$ planes.

\section{Conclusions}
\label{sec:conclusions}
The challenges to the ALICE TPC derived from the LHC Run\,3 and Run\,4 scenario, namely continuous readout at full \PbPb luminosity, while preserving the detector performance, resulted in the choice of a new technology for the readout chambers and an entirely new readout scheme. Gas amplification structures based on GEMs, a relatively recent development providing high-rate capability, were optimized in an effort to limit the space-charge distortions in the 2.5 m long drift volumes, and at the same time preserve the particle identification of the TPC. A quadruple GEM stack with two large-pitch foils in the inner layers, and a novel field configuration is the result of this optimization work. The selected gas, \NeCOtwoNtwo, fulfills the performance requirements and provides robustness against primary discharges. A dedicated HV powering scheme, which includes hardware protection measures and advanced software interlocks, ensures a safe operation of the detectors. The newly designed pad plane provides convenient connectors to accommodate front-end cards based on the ALICE-specific SAMPA chip, and multiple optical readout through GBT technology.
The construction of the new readout chambers involved several institutes where the subsequent steps in the construction, from foil qualification to foil framing and readout chamber assembly, were preceded and followed by strict quality assurance protocols, including the irradiation of all completed readout chambers in high particle-load environments. In parallel, the readout electronics were developed and manufactured at a high yield.
During the LHC Long Shutdown 2, the TPC was brought to the surface, where the old wire chambers were removed, the field cage was adapted to new voltage settings, the new readout chambers were installed, and the new FECs connected. The upgraded TPC then underwent a thorough testing campaign in order to certify full conformity with the specifications, prior to its descending back into the experimental cavern. 
\section*{Acknowledgements}
The ALICE TPC Collaboration would like to warmly thank the ALICE Technical Coordination team for their constant support in the realisation of this upgrade; the CERN RD51 team for their close collaboration and interest in the project; the CERN PCB workshop for the efficient production of the GEM foils; the CERN accelerator teams for the operation of the Proton Synchrotron and the Super Proton Synchrotron during the test beam campaigns; the CERN GIF++ team for their help and support during the photon irradiation campaigns; the CERN gas group and the CERN cooling and ventilation group for the prompt implementation of various hardware arrangements. Special thanks are also given to the CERN transport crew for the transportation of our delicate detectors around different sites of the laboratory.

Furthermore, the ALICE TPC Collaboration acknowledges the following funding agencies for their support in the TPC Upgrade:
Funda\c{c}\~ao de Amparo \`a Pesquisa do Estado de S\~ao Paulo (FAPESP), Brasil;
Ministry of Science and Education, Croatia;
The Danish Council for Independent Research | Natural Sciences, the Carlsberg Foundation and Danish National Research Foundation (DNRF), Denmark;
Helsinki Institute of Physics (HIP) and Academy of Finland, Finland;
Bundesministerium f\"{u}r Bildung, Wissenschaft, Forschung und Technologie (BMBF), 
GSI Helmholtzzentrum f\"{u}r Schwerionenforschung GmbH, 
DFG Cluster of Excellence "Origin and Structure of the Universe", 
The Helmholtz International Center for FAIR (HIC for FAIR)
and the ExtreMe Matter Institute EMMI at the GSI Helmholtzzentrum f\"{u}r Schwerionenforschung, Germany;
National Research, Development and Innovation Office, Hungary;
Nagasaki Institute of Applied Science (IIST)
and the University of Tokyo, Japan;
Fondo de Cooperaci\'{o}n Internacional en Ciencia y Technolog\'{i}­a (FONCICYT), Mexico;
The Research Council of Norway, Norway; Ministry of Science and Higher Education and National Science Centre, Poland;
Ministry of Education and Scientific Research, Institute of Atomic Physics and Ministry of Research and Innovation, and Institute of Atomic Physics, Romania;
Ministry of Education, Science, Research and Sport of the Slovak Republic, Slovakia;
Swedish Research Council (VR), Sweden;
United States Department of Energy, Office of Nuclear Physics (DOE NP), United States of America.\\[2ex]

\bibliographystyle{JHEP_mod}
\bibliography{./bib/bibliography.bib}

\end{document}